\documentclass[a4paper,10pt]{book}

\usepackage[centertags]{amsmath}
\usepackage{amssymb}
\usepackage{enumerate}
\usepackage{epsfig}
\usepackage{array}
\usepackage{bbm}
\usepackage{color}
\usepackage{a4wide}
\usepackage{cite}
\usepackage{makeidx}
\usepackage{lscape}
\usepackage{longtable}

\newcommand{\I}{\mathrm{i}}
\newcommand{\Z}[1]{\ensuremath{\mathbbm{Z}_{#1}}} 
\newcommand{\E}[1]{\ensuremath{\mathrm{E}_{#1}}} 
\newcommand{\F}[1]{\ensuremath{\mathrm{F}_{#1}}} 
\newcommand{\G}[1]{\ensuremath{\mathrm{G}_{#1}}}
\newcommand{\SO}[1]{\ensuremath{\mathrm{SO}(#1)}}
\newcommand{\SU}[1]{\ensuremath{\mathrm{SU}(#1)}}
\newcommand{\U}[1]{\ensuremath{\mathrm{U}(#1)}}
\newcommand{\rep}[1]{\boldsymbol{#1}}
\newcommand{\crep}[1]{\overline{\boldsymbol{#1}}}
\newcommand{\tr}{\mbox{tr}}
\newcommand{\cA}{\mathcal{A}}
\newcommand{\cF}{\mathcal{F}}

\newcommand{\fA}{\mathfrak{A}}

\newcommand{\MyQSpc}{\phantom{\Large I}}

\newcommand{\LM}{\ensuremath{_\text{L}}}
\newcommand{\RM}{\ensuremath{_\text{R}}}

\newcommand{\frF}{\mathfrak{F}}
\newcommand{\frR}{\mathfrak{R}}
\newcommand{\Tra}{\mbox{Tr}}

\DeclareMathOperator{\re}{Re}

\usepackage{fancyhdr}
\newcommand{\as}{&\hspace{-4pt}}
\newcommand{\ass}{&\hspace{-10pt}}

\usepackage{titlesec}

\titleformat{\chapter}[display]
{\normalfont\huge\bfseries}{\chaptertitlename\ \thechapter}{20pt}{\Huge}
\titlespacing*{\chapter} {00pt}{0pt}{15pt}

\makeindex

\title{\Huge \bfseries Grand Unification in the Heterotic Brane World}

\author{\Large Dissertation\\\\\\\\\\\\ an der\\  Mathematisch-Naturwissenschaftlichen Fakult\"at\\der \\Rheinischen Friedrich-Wilhelms-Universit\"at\\zu Bonn\\\\\\\\\\vorgelegt von\\ Patrick Karl Simon Vaudrevange\\ aus D\"usseldorf\\\\}

\date{Bonn 2008}

\begin{document}

\maketitle

\phantom{23}
\vspace{15cm}

\begin{tabular}{ll}
\multicolumn{2}{l}{Angefertigt mit Genehmigung der Mathematisch-Naturwissenschaftlichen Fakult\"at}\\
\multicolumn{2}{l}{der Universit\"at Bonn.}\\
                   & \\
Referent:          & Prof.~Dr.~Hans Peter Nilles\\
Korreferent:       & Prof.~Dr.~Albrecht Klemm \\
Tag der Promotion: & 10. Juli 2008\\
                   & \\
\multicolumn{2}{l}{Diese Dissertation ist auf dem Hochschulschriftenserver der ULB Bonn}\\
\multicolumn{2}{l}{\texttt{http://hss.ulb.uni-bonn.de/diss\_online} elektronisch publiziert.}\\
                   & \\
Erscheinungsjahr:  & 2008
\end{tabular}
\newpage

\phantom{23}
\vspace{2cm}

\begin{center}
{\Large Abstract}
\end{center}

String theory is known to be one of the most promising candidates for a unified description of all elementary particles and their interactions. Starting from the ten-dimensional heterotic string, we study its compactification on six-dimensional orbifolds. We clarify some important technical aspects of their construction and introduce new parameters, called generalized discrete torsion. We identify intrinsic new relations between orbifolds with and without (generalized) discrete torsion. Furthermore, we perform a systematic search for MSSM-like models in the context of $\Z{6}$-II orbifolds. Using local GUTs, which naturally appear in the heterotic brane world, we construct about 200 MSSM candidates. We find that intermediate SUSY breaking through hidden sector gaugino condensation is preferred in this set of models. A specific model, the so-called benchmark model, is analyzed in detail addressing questions like the identification of a supersymmetric vacuum with a naturally small $\mu$-term and proton decay. Furthermore, as vevs of twisted fields correspond to a resolution of orbifold singularities, we analyze the resolution of $\Z{3}$ singularities in the local and in the compact case. Finally, we exemplify this procedure with the resolution of a $\Z{3}$ MSSM candidate.

\newpage

\phantom{23}
\vspace{15cm}

Vor allem m\"ochte ich mich bei Prof.~Hans Peter Nilles daf\"ur bedanken, dass ich meine Doktorarbeit bei Ihm schreiben konnte. Es war eine gro\ss artige Zeit an die ich gerne zur\"uckdenken werde. Mein Dank gilt auch der gesammten Arbeitsgruppe, den jetzigen und ehemaligen  Mitgliedern. Des Weiteren bedanke ich mich bei Takeshi Araki, Prof.~Stefan Groot Nibbelink, Denis Klevers, Prof.~Tatsuo~Kobayashi, Prof.~Jisuke~Kubo, Dr.~Oleg Lebedev, Prof.~Hans Peter Nilles, Felix Pl\"oger, Prof.~Stuart Raby, Sa\'ul Ramos-S\'anchez,  Prof.~Michael Ratz, Dr.~Michele Trapletti und Dr.~Ak\i n Wingerter f\"ur die erfolgreichen Zusammenarbeiten. F\"ur das Korrekturlesen bedanke ich mich herzlichst bei Babak Haghighat, Val\'eri L\"owen, Dr.~Andrei Micu, Dr.~Giuliano Panico und besonders bei Denis Klevers, Prof.~Michael Ratz und Dr.~Sa\'ul Ramos-S\'anchez. Mein besonderer Dank gilt meiner Familie, meinen Br\"udern Dominik und Pascal und meinen Eltern, f\"ur Ihre gro\ss artige Unterst\"utzung und Liebe.
\frontmatter

\addtolength\topmargin{-50pt}
\addtolength\textheight{105pt}

\tableofcontents

\addtolength\topmargin{50pt}
\addtolength\textheight{-105pt}

\mainmatter

\chapter{Introduction}

\section{Motivation}

\subsubsection{The General Idea of Unification}

Why do we think that string theory might be relevant for describing high energy physics? One answer is \emph{unification}\index{unification}. Unification is a concept to describe as many aspects of nature as possible within one consistent framework in order to reveal their common origin. In other words, a small, consistent set of physical laws should reproduce many observations, which at best come from various areas of physics and did not seem to be connected before. This is one of the guiding principles in physics. 

The history of physics provides many examples for unification. For example, Isaac Newton successfully described the gravitational force here on earth and the attraction of celestial objects by the same physical laws. Nowadays, it seems obvious to most people that these two forces have a common origin, but we must remember the different distance scales. It is a huge scientific step to project laws measured in small scales here on earth to the scale of the solar system. Another prominent example for unification can be found in the theory of electromagnetism by James Maxwell. Guided by experimental evidence, this theory unifies electric and magnetic forces within the framework of the so-called Maxwell equations. Finally, one can interpret Albert Einstein's theory of gravity as a kind of unification, not of different forces, but as a unified description of different aspects of nature: Einstein succeeded in unifying Newton's gravity with the observation that the speed of light is constant for \emph{any} observer. His theory changed the physicists' view of nature radically, since space and time are from these days on not just a static framework for the description of nature, but they are dynamical quantities of the theory by themselves. These examples lead us to the hope that unification of physical theories will continue in the future. 

\subsubsection{The Standard Model of Particle Physics}

Additional to the historical motivation for unification, we have several hints towards unification in some areas of particle physics today. High energy particle physics can be described successfully by the so-called \emph{Standard Model}\index{Standard Model} (SM). The Standard Model of particle physics is a quantum field theory that describes three of the four known fundamental interactions between all known elementary particles, the fundamental constituents of matter.

\begin{figure}[t]
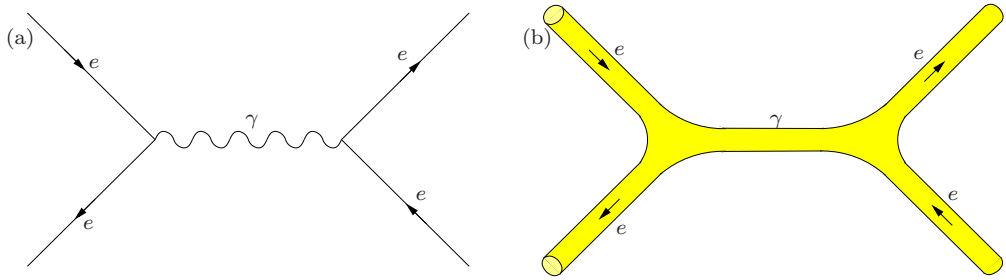

\centerline{\input feynman.pstex_t}\vspace{-0.2cm}
\caption{Electron-positron annihilation in the case of (a) QED and (b) string theory.}
\label{fig:Feynman}
\end{figure}

In this framework, all particles are point-like and the fundamental forces are mediated by the exchange of so-called \emph{gauge bosons}, bosonic particles with spin $1$, see figure~(\ref{fig:Feynman}a). The interactions are best understood in terms of the \emph{gauge group} $G_\text{SM}=\SU{3}_C\times\SU{2}_L\times\U{1}_Y$\index{Standard Model!gauge group}, describing the internal gauge symmetry of the theory. Each group factor of $G_\text{SM}$ is related to a fundamental interaction: the $\SU{3}_C$ describes quantum chromodynamics (QCD), i.e. the theory of strong interactions, by the exchange of eight (massless) gauge bosons: the gluons. The last two gauge group factors $\SU{2}_L\times\U{1}_Y$ (where $\U{1}_Y$ is named \emph{hypercharge}\index{hypercharge}) correspond to the electroweak theory, as they combine a quantum version of Maxwell's electromagnetism (known as quantum electrodynamics, or in short QED) with the weak force in a unified way. The associated interactions are mediated by the (massless) photon $\gamma$ and three (massive) gauge bosons, denoted by $W^\pm$ and $Z$. However, the fourth fundamental interaction, gravity, cannot be incorporated in this theory.

The elementary particles describing matter are fermions (with spin $1/2$). They can be characterized by their charges with respect to the three interactions (i.e. by their transformation properties under $\SU{3}_C\times\SU{2}_L\times\U{1}_Y$ gauge transformations in terms of irreducible representations). In detail, the matter fermions and their representations are
\begin{equation}
\label{eqn:onefamily}
\begin{array}{ccccccccc}
(\rep{3},\rep{2})_{1/6} & + & (\crep{3},\rep{1})_{-2/3} & + & (\crep{3},\rep{1})_{1/3} & + & (\rep{1},\rep{2})_{-1/2} & + & (\rep{1},\rep{1})_{1}\;. \\
\downarrow              &   & \downarrow                &   & \downarrow               &   & \downarrow               &   & \downarrow\\
q                       &   & \bar{u}                   &   & \bar{d}                  &   & \ell                     &   & \bar{e}
\end{array}
\end{equation}
where $q=(u,d)^T$ and $\ell = (\nu, e)^T\;$~\footnote{Note that we use the convention to write right-handed fermions in terms of left-handed ones, transforming in the complex conjugate representation. For example, the right-handed up-quark transforming as $(\rep{3},\rep{1})_{2/3}$ is expressed by its complex conjugate, denoted by $\bar{u}$. Thus, all fields in eqn.~(\ref{eqn:onefamily}) are left-handed.}. Since neutrinos are found to be extremely light but massive, a right-handed neutrino $(\rep{1},\rep{1})_{0}$ is often assumed to exist in addition, which can explain the neutrino mass scale by the so-called \emph{see-saw mechanism}. The particles of eqn.~(\ref{eqn:onefamily}) are said to form one \emph{family}\index{Standard Model!family of quarks and leptons} (or \emph{generation}) of \emph{quarks} ($q$, $\bar{u}$, $\bar{d}$) and \emph{leptons} ($\ell$, $\bar{e}$). The Standard Model contains three such families of quarks and leptons with the same charges but different masses.

\begin{figure}[t]
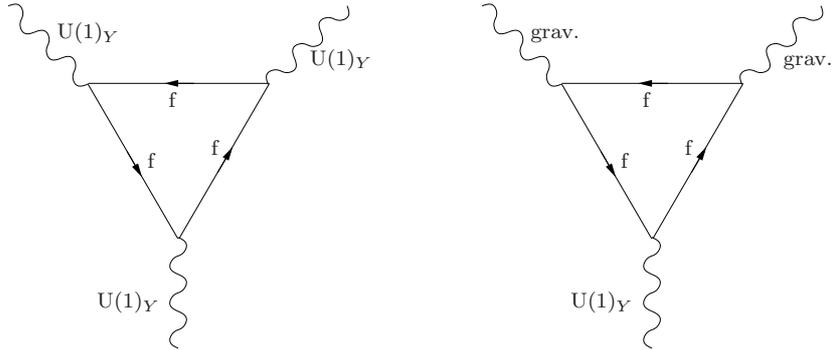

\centerline{\input anomaly.pstex_t}\vspace{-0.25cm}
\caption{Example for a possible $\U{1}_Y^3$ ($\U{1}-\text{gravity}-\text{gravity}$) anomaly. A coupling between three $\U{1}_Y$ gauge bosons (one gauge boson and two gravitons) is not consistent with the gauge symmetry. It could however be generated by quantum corrections corresponding to the first (second) triangle diagram, where all charged fermions of the theory (denoted by $\text{f}$) run in the loop. The amplitude of this diagram and therefore the anomaly is proportional to $\sum_\text{f}Y_\text{f}^3 = 0$ ($\sum_\text{f}Y_\text{f} = 0$). Consequently, using eqn.~(\ref{eqn:onefamily}), $\U{1}_Y$ is anomaly-free.}
\label{fig:Anomaly4d}
\end{figure}

The SM matter spectrum is \emph{chiral}\index{chiral spectrum}, i.e. left- and right-handed fermions transform differently under gauge transformations. A chiral spectrum has the potential to cause an inconsistency of the theory if quantum corrections violate a classical symmetry. This is called an \emph{anomaly}\index{anomaly}. Anomalies can be seen from one-loop Feynman diagrams. In the case of the Standard Model, so-called one-loop triangle diagrams can potentially violate the gauge symmetry $G_\text{SM}$ depending on the fermionic matter content of the theory, see figure~(\ref{fig:Anomaly4d}). However, luckily each generation of quarks and leptons is anomaly free by itself. Thus, the Standard Model is anomaly-free. Yet a deeper origin for this is unknown.

Beside an explanation for the interactions between the fundamental particles, the Standard Model contains a mechanism to give them masses: the so-called \emph{Higgs mechanism}\index{Standard Model!Higgs mechanism}. The mass of a fundamental particle is generated by its interaction with a scalar boson (i.e. with spin $0$), the so-called \emph{Higgs boson} $\phi=(\phi^+,\phi^0)^T$, transforming in the representation $(\rep{1},\rep{2})_{1/2}$. It has a non-trivial potential\index{$\mu$-term} $V(\phi) = \mu^2\phi^\dagger\phi + \lambda (\phi^\dagger\phi)^2$ with $\mu^2 <0$ and $\lambda > 0$. By minimizing this potential, the Higgs develops a \emph{vacuum expectation value} (vev), $\langle\phi^0\rangle=v=\sqrt{-\mu^2/(2\lambda)}$. Consequently, the $W^\pm$ and $Z$ bosons get massive and the gauge symmetry breaks $G_\text{SM} \rightarrow \SU{3}_C\times\U{1}_\text{em}$ yielding the $\U{1}_\text{em}$ responsible for QED~\footnote{We use the convention that the electric charge associated to $\U{1}_\text{em}$ is given by $Q=Y+T_{3L}$, see eqn.~(\ref{eqn:onefamily}).}. Furthermore, quarks and leptons acquire masses proportional to the Higgs' vev and to their individual interaction strengths with it (the so-called \emph{Yukawa coupling constants}).\index{Yukawa couplings}

The Standard Model has been tested extensively by experiments yielding excellent agreement between the predictions and observations. However, the Higgs boson - a fundamental ingredient - has not been observed yet.

\subsubsection{Running of the Coupling Constants and Grand Unified Theories}

But where does unification enter? The Standard Model by itself successfully unifies QED with the weak interaction by the $\SU{2}_L\times\U{1}_Y$ group structure and the Higgs mechanism.

\begin{figure}[t]
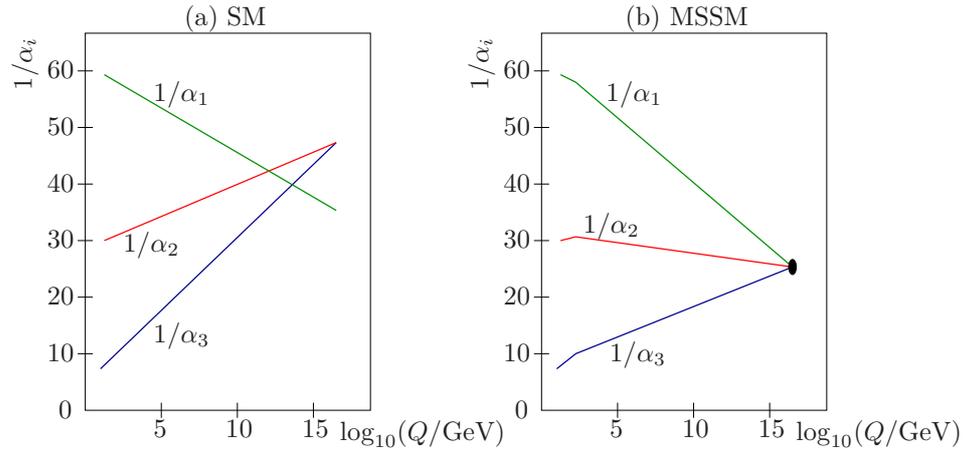

\centerline{\input CouplingsRunning.pstex_t}\vspace{-0.4cm}
\caption{Schematic plot of the gauge coupling unification in the case of (a) the Standard Model and (b) the MSSM. For the SM the couplings nearly meet at a GUT scale of about $10^{14}-10^{15}$ GeV, while for the MSSM the GUT scale is at about $3\times 10^{16}$ GeV.}
\label{fig:gaugecouplingunification}
\end{figure}

However, the electroweak force is still described by two group factors and hence by two interaction strengths, also called \emph{gauge coupling constants}. But we know from experiment and from theory that these ``constants'' in fact depend on the energy scale at which they are measured. This is a generic feature of quantum field theories, known as the \emph{running of coupling constants}. The running is specified by the \emph{renormalization group equations} (RGEs) which depend on the complete charged spectrum. It turns out, that within the Standard Model, even though the three coupling constants associated to the three group factors are very different at the electroweak scale of about $100\text{GeV}$, their values evolve in such a way that they seem to (nearly) meet at $10^{14}-10^{15}\text{GeV}$, see figure~(\ref{fig:gaugecouplingunification}a). It seems natural to assume that at the energy scale, where the gauge couplings meet, the interactions themselves are unified such that they are all described by just one gauge group factor and one coupling constant. This scenario is called \emph{grand unification}\index{Grand Unified Theories} (or GUT for grand unified theory). However, a GUT does not only unify the interactions, but automatically also the representations, i.e. quarks and leptons. The single, unified interaction is specified by the GUT gauge group and the matter by its irreducible representations. The most prominent GUT gauge groups, which we will discuss now in some detail, are $\SU{5}$ and $\SO{10}$. 

\subsubsection{$\boldsymbol{\SU{5}}$ GUT}\index{Grand Unified Theories!$\SU{5}$}

Starting with $\SU{5}$, one family of quarks and leptons is contained in a $\rep{10}$- and a $\crep{5}$-plet~\cite{Georgi:1974sy,Georgi:1974yf}. In order to see that these representations can incorporate one family, we take a look at the breaking $\SU{5} \rightarrow G_\text{SM}$ and the resulting decomposition of the $\SU{5}$ representations
\begin{equation}
\rep{10} \rightarrow (\rep{3},\rep{2})_{1/6} + (\crep{3},\rep{1})_{-2/3} + (\rep{1},\rep{1})_{1} \quad\text{and}\quad\crep{5} \rightarrow (\crep{3},\rep{1})_{1/3} + (\rep{1},\rep{2})_{-1/2}\;.
\end{equation}
Thus, the $\rep{10}$-plet contains $q$, $\bar{u}$ and $\bar{e}$, while the $\crep{5}$-plet comprises $\bar{d}$ and $\ell$. It is very important to note that $G_\text{SM}$ fits into $\SU{5}$ such that the hypercharge $\U{1}_Y$ is determined uniquely up to an overall normalization. In other words, the $\SU{5}$ GUT gives a possible explanation for the observed quantization of (hyper)charges of quarks and leptons. 

\begin{figure}[b]
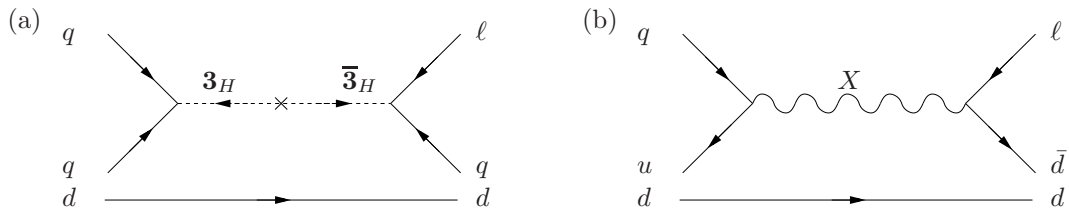

\centerline{\input protondecay.pstex_t}\vspace{-0.4cm}
\caption{Example for proton decay channel $p \rightarrow \pi^0 + \bar{e}$ by effective dimension 6 operators: (a) rapid proton decay mediated by a Higgs color-triplet. (b) proton decay by the mediation of an $X$ boson. The couplings correspond to the terms $\bar{u}^\dagger X q$ and $\ell^\dagger X \bar{d}$ originating from the covariant derivative in the kinetic terms.}
\label{fig:protondecay}
\end{figure}

Additionally to the matter representations, we have to incorporate the SM Higgs $(\rep{1},\rep{2})_{1/2}$ into the $\SU{5}$ theory. The smallest $\SU{5}$ representation containing a $(\rep{1},\rep{2})_{1/2}$ is the $\rep{5}$-plet. This, however creates a new problem, the so-called \emph{doublet-triplet splitting problem}\index{$\mu$-term!doublet-triplet splitting}. The reason is that the Higgs $\rep{5}$-plet contains an additional color-triplet $(\rep{3},\rep{1})_{-1/3}$ which can mediate fast proton decay, see figure~(\ref{fig:protondecay}a). Therefore, it must be extremely heavy in order to suppress this process and to extend the proton's lifetime\index{proton decay} above the current experimental bounds. The Higgs doublet, on the other hand, has to be light in order to provide the correct scale of the $\mu$-term in the Higgs potential. Conventional GUT theories generically suffer under this problem and do not provide a convincing solution. Furthermore, the breaking of the GUT gauge group $\SU{5}$ down to the SM can be achieved by a Higgs mechanism, where the GUT breaking scalar Higgs boson resides in an adjoint representation $\rep{24}$. As the $\rep{24}$ contains a SM singlet $(\rep{1},\rep{1})_0$, its vev can induce the desired gauge symmetry breaking. However, also here one has to take care of proton decay: when we break the $\SU{5}$ gauge group using the scalar Higgs in the $\rep{24}$ there are massive vector bosons (in the representations $(\rep{3},\rep{2})_{-5/6}$ and $(\crep{3},\rep{2})_{5/6}$, named leptoquarks $X$ and $Y$\index{leptoquarks}), beside the massless ones of the SM gauge group, which can mediate fast proton decay, see figure~(\ref{fig:protondecay}b). 

\subsubsection{$\boldsymbol{\SO{10}}$ GUT}\index{Grand Unified Theories!$\SO{10}$}

In the case of $\SO{10}$ GUTs one complete family is comprised in a 16-dimensional spinor representation of $\SO{10}$~\cite{Georgi:1974my,Fritzsch:1974nn}. Since $\SO{10}$ contains $\SU{5}$, we can write one family of quarks and leptons as $\rep{16} \rightarrow \rep{10} + \crep{5} + \rep{1}$. The additional $\SU{5}$ singlet $\rep{1}$ contained in the $\rep{16}$-plet can be interpreted as a right-handed neutrino $(\rep{1},\rep{1})_{0}$. The SM Higgs resides in a 10-dimensional vector representation of $\SO{10}$ which reads, in terms of $\SU{5}$-plets, $\rep{10} \rightarrow \crep{5} + \rep{5}$. Therefore, $\SO{10}$ GUTs naturally unify one SM family in one $\rep{16}$-plet and predict the existence of right-handed neutrinos. Furthermore, the Standard Model Higgs is contained in a $\rep{10}$-plet, thus $\SO{10}$ distinguishes between the representation for bosons ($\rep{10}$) and the one for fermions ($\rep{16}$). However, also $\SO{10}$ suffers under the notorious problems of proton decay and doublet-triplet splitting

\subsubsection{Further GUTs}

There are further GUTs, like Pati-Salam\index{Grand Unified Theories!Pati-Salam} $\SU{4}_C\times\SU{2}_L\times\SU{2}_R$, flipped $\SU{5}$\index{Grand Unified Theories!flipped $\SU{5}$} (where electromagnetism $\U{1}_\text{em}$ sits partially in an additional $\U{1}_X$) and trinification $\SU{3}_C\times\SU{3}_L\times\SU{3}_R$. However, the GUT which fits very nice into the series $\SO{10} \rightarrow\SU{5}\rightarrow G_\text{SM}$ is equipped with the exceptional group $\E{6}$. The fundamental representation of $\E{6}$\index{Grand Unified Theories!$\E{6}$} is the $\rep{27}$-plet that can be decomposed into $\SO{10}$ representations as $\rep{27} \rightarrow \rep{16} + \rep{10} + \rep{1}$. Its relevance will become clear later when we discuss supersymmetric GUTs.

\subsubsection{The Minimal Supersymmetric Standard Model (MSSM)}\index{MSSM}

Up to now we have discussed how the idea of unification can be used to partly unify the matter content of the Standard Model and their gauge interactions by means of GUTs. It is further possible to use unification to describe fermions and bosons within a unified framework, called \emph{supersymmetry}\index{supersymmetry} (SUSY). In other words, bosons and fermions are no longer distinct, but they are related by a symmetry transformation
\begin{equation}
Q|\text{boson}\rangle = |\text{fermion}\rangle \quad\text{and}\quad Q|\text{fermion}\rangle = |\text{boson}\rangle\;,
\end{equation}
where $Q$ (a two-component Weyl spinor) denotes the generator of the transformations, the so-called \emph{supercharge}\index{supercharge}. The parameter associated to an infinitesimal SUSY transformation $Q$ is denoted by $\varepsilon$, a two-component anticommuting number, which is constant for global SUSY. The number of supercharges determines the number $\mathcal{N}$ of supersymmetries, where in the case of the MSSM we have $\mathcal{N} = 1$.

Beside aesthetical reasons, the main intention to introduce supersymmetry into the Standard Model is to stabilize the Higgs mass from huge radiative corrections due to quadratic divergencies. This is the so-called \emph{hierarchy problem}\index{hierarchy problem} of the Standard Model: why is the electroweak scale so small compared to the cut-off scale (e.g. the GUT scale) entering the quadratic divergencies? If one introduces scalar partners for all chiral fermions such that each scalar has the same mass as its partner-fermion and the couplings of the scalars are chosen appropriately, then the radiative corrections to the Higgs mass vanish. The origin of the scalar partners can be explained by SUSY.

For a complete supersymmetric theory not only the fermions are accompanied by so-called \emph{superpartners}, but also the gauge bosons and the Higgs, see figure~(\ref{fig:susypartners}). One can arrange the particles and their SUSY partners into so-called \emph{supermultiplets}, such that any supersymmetry transformation maps a supermultiplet to itself. For example, in the case of global $\mathcal{N} = 1$ supersymmetry, the most important supermultiplets are the \emph{chiral multiplet} and the \emph{vector multiplet}. For $\mathcal{N} = 1$ in 4d there exists a nice representation of these supermultiplets: in the superspace formulation the four-dimensional space-time is extended by four anticommuting coordinates $\theta$ and $\bar{\theta}$ transforming as two-dimensional Weyl spinors. Then, the supermultiplets can be expressed as superfields, i.e. fields depending on the coordinates $x^\mu$, $\theta$ and $\bar{\theta}$. In this formulation, a chiral multiplet is represented by a so-called \emph{chiral superfield} $\phi_i$ containing a scalar $\varphi_i$, a Weyl spinor $\psi_i$ and an auxiliary field $F_i$~\footnote{Auxiliary fields are introduced such that the SUSY algebra closes off-shell. They have no kinetic terms. Hence, they do not propagate and they are not dynamical degrees of freedom. They can be eliminated be solving their equations of motions leading to the on-shell formulation of the SUSY algebra.}. The degrees of freedom of a vector superfield $V_a$ are a gaugino $\lambda_a$, a gauge boson $A^\mu_a$ and an auxiliary field $D_a$ (in the Wess-Zumino gauge).

Since the bosonic Higgs has now a fermionic partner (the Higgsino), the new fermionic spectrum could be anomalous. Therefore, and in order to give masses to both up- and down-type quarks, we need a second Higgs multiplet $(\rep{1},\rep{2})_{-1/2}$.

\begin{figure}[t]
\centerline{\input SusyPartners.pstex_t}
\caption{Overview of the structure of the supersymmetric particle spectrum of the MSSM. The new SUSY partners are highlighted in blue.}
\label{fig:susypartners}
\end{figure}

The so-called \emph{superpotential}\index{supersymmetry!superpotential} $\mathcal{W}(\phi_i)$ is a holomorphic function of chiral superfields $\phi_i$. In terms of the component fields $(\varphi_i,\psi_i,F_i)$\index{supersymmetry!F-term} of $\phi_i$ the superpotential yields Yukawa interactions $\psi_i\psi_j\varphi_k$ and contributes to the scalar potential $V(\varphi_i) \supset \sum_j |F_j|^2 = \sum_j |\partial \mathcal{W}(\varphi_i)/\partial \varphi_j|^2$. Additional contributions to $V(\varphi_i)$ arise from the \emph{$D$--terms}\index{supersymmetry!D-term} $D_a$ of the vector superfields. Then, the full scalar potential reads
\vspace{-0.25cm}
\begin{equation}
V(\varphi_i) = \sum_j |F_j|^2 + \sum_a |D_a|^2\;.
\end{equation}
The value of the scalar potential at its minimum gives the \emph{cosmological constant} and therefore defines the cosmological model. If $\langle V \rangle>0$, the universe is \emph{de Sitter} with accelerating expansion. If $\langle V \rangle=0$, the universe is \emph{Minkowskian}\index{Minkowski vacuum}. The third possibility $\langle V \rangle<0$ yielding an \emph{anti de Sitter} space, in which the universe immediately collapses, is obviously not possible in global SUSY. Observations indicate that the cosmological constant is tiny and positive. 

If supersymmetry is exactly realized in nature, all particles and their corresponding superpartners must have the same mass. For example, the fermionic electron and the bosonic selectron only differ by their spin, but share the same mass of $511$ keV. Since neither the selectron nor the other superpartners have been observed yet, supersymmetry must be broken. This breaking should yield a mass-splitting of the different components of the supermultiplets in order to explain the absence of superpartners. However, the breaking should be such that SUSY remains a solution to the hierarchy problem, one consequence being that the scale of SUSY breaking should be of the order $M_\text{SUSY}\approx 1$ TeV, the new cut-off scale for the SM. This can be achieved by inserting special terms to the theory that explicitly break SUSY, but do not induce quadratic divergencies in the Higgs mass. This scenario is called \emph{softly broken supersymmetry}. It can be realized for example by non-vanishing vevs of the auxiliary fields $F_i$ and $D_a$ leading to so-called \emph{spontaneous supersymmetry breaking}.\index{supersymmetry!spontaneously broken}

There is additional (theoretical) evidence for supersymmetry. In a theory with low-energy supersymmetry (with a breaking at about $1$ TeV) the gauge couplings evolve differently due to the presence of the superpartners, such that they meet at about $3\times 10^{16}$ GeV in the case of the MSSM~\cite{Dimopoulos:1981yj}. Therefore, one can say that supersymmetric theories support the idea of GUTs. Note that a supersymmetric $\E{6}$ GUT with $\rep{27}$-plets has the additional interpretation of a family-Higgs unification, as the Higgs and the SM families both reside in chiral multiplets.

\subsubsection{$\boldsymbol{R}$-Parity and Matter Parity}

Supersymmetric theories often contain so-called \emph{$R$-symmetries}\index{R-symmetry}, i.e. symmetries that do not commute with supersymmetry, such that the components within one supermultiplet transform differently under the $R$-symmetry. In the case of the MSSM, there is a discrete \emph{$R$-parity}\index{R-symmetry!R-parity} defined by $R = (-1)^{3B+L+2s}$ (with baryon number $B$, lepton number $L$ and spin $s$), see~\cite{Barbier:2004ez} for a review on $R$-parity. ``Ordinary particles'' (like quarks, leptons and Higgses) have an $R$-charge $+1$ and superpartners $-1$. $R$-parity is conserved. Therefore, superpartners need to be created in pairs such that the \emph{lightest supersymmetric particle} (the \emph{LSP}) with $R=-1$ is stable and serves as a candidate for Dark Matter (for example the neutralino, a mixture of Higgsinos and gauginos, singlet of $\SU{3}_C\times\U{1}_\text{em}$). In addition, $R$-parity forbids the ``unwanted'' terms $\bar{u}\bar{d}\bar{d}$, $\ell \ell \bar{e}$ and $q \ell \bar{d}$ in the renormalizable superpotential which induce rapid proton decay, while it allows for the ``wanted'' Yukawa couplings and the $\mu$-term. Thus, $B$ and $L$ are conserved at the renormalizable level and consequently the proton is (rather) stable.

On the other hand, \emph{matter parity}\index{matter parity} $P$ forbids and allows the same terms as $R$-parity\footnote{Hence, we will refer to $R$-parity and matter parity without distinction.}. It is defined by $P = (-1)^{3(B-L)}$, such that the matter superfields are odd ($P=-1$) and the Higgs and gauge superfields are even ($P=1$). It is a discrete $\Z{2}$ subgroup of $\U{1}_{B-L}$\index{B-L}, where the breaking $\U{1}_{B-L}\rightarrow P$ can be induced by vevs of SM singlets, denoted by $\chi$, with $B-L$ charges $3(B-L) =0\mod 2$, or equivalently $(B-L) =0\mod 2$. Note that the see-saw mechanism\index{see-saw mechanism} requires this breaking, as the right-handed neutrino (a SM singlet) is charged with respect to $B-L$ (with a $B-L$ charge of $-1$). Explicitly, in the presence of $B-L$, the Majorana mass term of the right-handed neutrino, denoted by $\bar{n}$, originates from the coupling
\begin{equation}
\bar{n}\,\bar{n}\,\chi \quad\Leftrightarrow\quad (\rep{1},\rep{1})_{(0,-1)}(\rep{1},\rep{1})_{(0,-1)}(\rep{1},\rep{1})_{(0,2)}
\end{equation}
and $\chi$ acquires a (large) vev breaking $\U{1}_{B-L}$ to matter parity ($R$-parity).

\subsubsection{Supergravity}

GUTs and SUSY are very appealing, but still do not contain the fourth force, gravity. As the name suggests, \emph{supergravity}\index{supergravity} (SUGRA)~\cite{Freedman:1976xh,Deser:1976eh} tries to address this issue (for a review, see e.g.~\cite{Nilles:1983ge}). The main assumption for SUGRA is that the parameter $\varepsilon$ of SUSY transformations becomes space-time dependent, i.e. $\varepsilon = \varepsilon(x)$, such that SUSY becomes a local symmetry. The gauge field of local SUSY turns out to be a spin $3/2$ fermion, the so-called \emph{gravitino}\index{gravitino}. Its superpartner is the \emph{graviton}, the spin $2$ messenger of the gravitational force. Both, the graviton and the gravitino, are combined in the so-called \emph{supergravity multiplet}, where the number of gravitinos is in general equal to the number $\mathcal{N}$ of supersymmetries.

$\mathcal{N}=1$ supergravity theories are described by the superpotential $\mathcal{W}(\phi_i)$, the K\"ahler potential\index{K\"ahler potential} $\mathcal{K}(\phi_i,\phi_i^\dagger)$ entering the scalars' kinetic energies and the gauge kinetic function $f(\phi_i)$ yielding the gauge coupling constant $\text{Re} f(\phi_i) = 1/g^2$. Unlike the case of global SUSY, the scalar potential can have a minimum with an anti de Sitter space-time, i.e. with negative energy,
\begin{equation}
V(\varphi^i,\varphi^*_{\bar{j}}) = e^\mathcal{K}\left((D_i \mathcal{W})(D_{\bar{j}} \mathcal{W}) G^{i\bar{j}} - 3|\mathcal{W}|^2\right),
\end{equation}\index{supergravity!scalar potential}
where $D_i$ is the covariant derivative, $G^{i\bar{j}}$ the inverse hermitian metric and $M_\text{Pl}=1$. Furthermore, the contributions from the $D$-terms have been neglected.

However, SUGRA is non-renormalizable, because the gravitational coupling is a dimensionful quantity. Therefore, one thinks that SUGRA does not serve as a good candidate of an ultraviolet complete theory of gravity and the Standard Model interactions. Nevertheless, SUGRA theories are of great importance, as they describe the low-energy effective theories obtained from string theory.

\subsubsection{Extra Dimensions and Kaluza-Klein}

\begin{figure}[b]
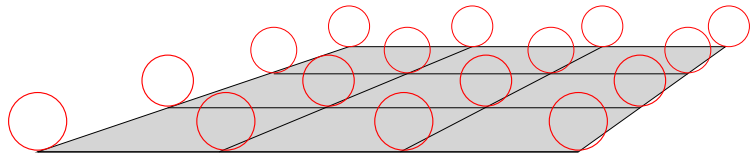

\centerline{\input M4timesCircle.pstex_t}\vspace{-0.4cm}
\caption{Five dimensional space, where the fifth dimension is compactified on a circle: the plane corresponds to our four-dimensional space-time $M_{3,1}$ and a small circle is attached to every point of $M_{3,1}$.}
\label{fig:M4timesCircle}
\end{figure}

\begin{figure}[t]
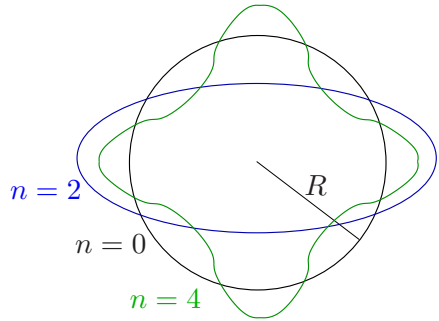

\centerline{\input KaluzaKlein.pstex_t}\vspace{-0.3cm}
\caption{Two interpretations of this illustration: (a) Compactifying a field on a circle of radius $R$ yields the zero mode with $n=0$ and massive Kaluza-Klein states, e.g. with $n=2,4$. Later, for the section on strings: (b) All elementary particles are supposed to correspond to the same string but with different excitations, e.g. $n=0,2,4$.}
\label{fig:KaluzaKlein}
\end{figure}

Initially, extra dimensions were introduced in order to unify gravity and electromagnetism by the compactification of a five dimensional theory of pure gravity on a circle, see figure~(\ref{fig:M4timesCircle}). This toy theory revealed important results common to most theories with extra dimensions:  when a $d$-dimensional theory is compactified on a compact internal space $M_{d-4}$ times a four-dimensional Minkowski space $M_{3,1}$, a field $\varphi(x^M)$ factorizes into two parts, one living solely on $M_{3,1}$ and another one on $M_{d-4}$, i.e. $\varphi(x^M) = \sum_j\alpha_j(x^\mu)\beta_j(y^i)$ where $M=1,\ldots,d$ and $i=1,\ldots,d-4$. From the 4d point of view, the massless fields (the \emph{zero modes}) arise from the harmonic fields on the compact space, i.e. $\Delta\beta_j = 0$. In addition, there is an infinite tower of massive states, the so-called \emph{Kaluza-Klein tower}, with masses proportional to the inverse compactification radius, i.e. $m_n \sim n/R$ and $n\in\mathbbm{N}$. For an illustration see figure~(\ref{fig:KaluzaKlein}a). Consequently, if the compactification radius is small, the masses of the Kaluza-Klein fields can be very high such that they cannot be detected.

\subsubsection{String Theory and Unification}

String theory is a promising candidate for a quantum theory of gravity unified with the description of all forces. In detail, it is known that the low-energy effective theory of string theory necessarily contains Einstein's theory of gravity. Furthermore, gauge theories with chiral matter spectra appear naturally from string theory. These theories are automatically free of anomalies, the reason being the internal consistency of the two-dimensional description of string theory. In addition, extra dimensions and supersymmetry arise naturally in string theory. Thus, string theory incorporates the most prominent unification mechanisms.

Unlike the point-like particles in quantum field theories, strings are extended one-dimen\-sional objects. As they propagate through d-dimensional space-time they sweep out a two-dimensional surface, the \emph{world sheet}, see figure~(\ref{fig:Feynman}b). This is analogously to a point particle running along its world line. Strings can be either closed or open, where the end-points of open strings are attached to so-called \emph{D-branes}, multi-dimensional physical objects. String theories with open strings necessarily contain closed strings, since open strings can merge yielding closed ones. The converse is not true: string theories with closed strings only can be consistent - they do not need open strings.

String theory is described by a two-dimensional quantum field theory on the world sheet with bosonic and fermionic fields. They are related by two-dimensional supersymmetry. From the world sheet point of view, the string coordinates are (bosonic) fields on the world sheet taking values in the so-called \emph{target space}. The target space is interpreted as a d-dimensional space-time which should give rise to our observable four-dimensional world. However, consistency of the world sheet field theory restricts the number of target space dimensions to $d=10$, i.e. there are ten world sheet bosons. One possibility to relate the ten-dimensional theory to our world is to assume that six string coordinates are restricted to take values in a compact space only. In other words, six spatial dimensions are compactified. If the compact space is small, these extra dimensions are not accessible from the low-energy 4d point of view. In addition, it is known that the compactification can in principle provide us with explanations for various aspects of the MSSM, like the amount of supersymmetries in 4d, the gauge group, the matter representations of one family and the number of families.

There are five consistent string theories in ten dimensions that could describe our world. They are: type I, type IIA, type IIB, heterotic $\E{8}\times\E{8}$ and heterotic $\SO{32}$. Let us briefly list some of their properties. All string theories contain closed strings, related to the fact that the graviton is a closed string. In addition, type I and IIA/B contain open ones. The strings of type I are unoriented and those of the other theories are oriented. Both type II theories have $\mathcal{N}=2$ supersymmetry in 10d, while the others have $\mathcal{N}=1$. The type I string theory has an $\SO{32}$ gauge group in 10d and the heterotic string theories have an $\E{8}\times\E{8}$ or $\SO{32}$ gauge symmetry. 

The five string theories are related by a web of \emph{dualities}. \emph{T-duality}\index{T-duality} relates for example one theory compactified on a circle with radius $R$ to another theory compactified on a circle with radius $1/R$, for example in the case of type IIA and IIB. On the other hand, \emph{S-duality}\index{S-duality} relates the weak coupling limit of one theory to the strong coupling limit of the other, for example heterotic $\SO{32}$ is related to type I and type IIB to itself. The existence of these and further dualities lead to the picture that all string theories are limits of one unique underlying theory, called \emph{M-theory}. Its low-energy effective theory can be described by 11 dimensional SUGRA, but the real nature of M-theory is not yet understood: there are proposals - for example Matrix models.

\subsubsection{The Heterotic String and Orbifold Compactifications}

The heterotic string theory~\cite{Gross:1984dd,Gross:1985fr} is special in the sense that it is the only string theory with solely closed strings. In ten-dimensional space-time it is equipped with $\mathcal{N}=1$ supersymmetry and an $\E{8}\times\E{8}$ or $\SO{32}$ gauge group. Having the aim of unification in mind, we want to relate it to the MSSM in 4d. Thus, six spatial dimensions have to be compactified.

Heterotic orbifold compactifications~\cite{Dixon:1985jw,Dixon:1986jc} provide an easy, geometrical compactification scheme. Since the orbifold space is flat everywhere except for isolated singularities at the so-called \emph{fixed points}, it is possible to perform direct string computations. However, in its simplest construction, heterotic orbifolds generically lead to four-dimensional theories with a huge gauge group (like $\E{6}$ in the standard embedding) and a large number of families (for example a net number of 27 $\rep{27}$-plets). With the development of Wilson lines~\cite{Dixon:1986jc,Ibanez:1986tp} in the context of heterotic orbifolds, one has an easy tool to break the gauge symmetry and to reduce the number of families. Within a few years the construction was so well understood that $\Z{3}$ orbifolds were constructed with the Standard Model gauge group and three generations of quarks and leptons plus vector--like exotics~\cite{Ibanez:1987sn,Ibanez:1987pj,Casas:1987us,Font:1988tp,Casas:1988se,Font:1988mm,Casas:1988hb,Font:1989aj,Casas:1989wu}. On the other hand, the compactification on more complicated $\Z{N}$ or $\Z{N}\times\Z{M}$ orbifolds was neglected for a long time, with the result that their construction was not fully clarified up to now. For example, the modular invariance conditions for $\Z{N}$ or $\Z{N}\times\Z{M}$ orbifolds and the construction of orbifold invariant states were only partially under control.

The rank of the gauge group is not reduced by a conventional orbifold compactification, such that there are many unwanted extra $\U{1}$ factors. However, two approaches have been developed in the past to address this issue. First of all, the presence of an anomalous $\U{1}$, which is canceled by terms coming from the 10d Green-Schwarz mechanism~\cite{Green:1984sg}, induces a Fayet-Iliopoulos $D$--term in heterotic orbifolds~\cite{Dine:1987xk,Atick:1987gy}. In order to retain supersymmetry (i.e. $D=0$) some fields need to develop large vevs canceling this Fayet-Iliopoulos $D$--term. These vevs induce a Higgs mechanism, such that the gauge group breaks (including a rank reduction) and some vector-like matter gets massive and therefore decouples. Secondly, the so-called \emph{rotational embedding}~\cite{Ibanez:1987xa} can lower the rank.

Very early it was realized that there exists a vast number of string compactifications~\cite{Lerche:1986cx} each serving as a \emph{vacuum} of the theory. From the string perspective, it seems that they are all equivalent, i.e. no vacuum is preferred to another. The MSSM is assumed to be just one of the possible vacua. This has recently lead to the notion of the \emph{landscape of string vacua}~\cite{Susskind:2003kw}, with the aim of making ``predictions'' from string theory by statistical analyses of the landscape. However, the actual meaning of the string landscape seems unclear. 

\begin{figure}[t]
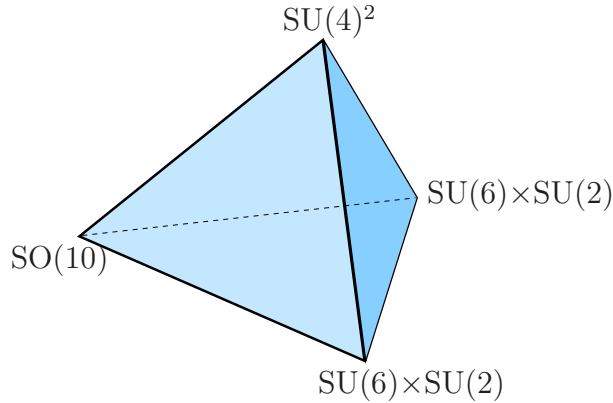

\centerline{\input tetrahedral3.pstex_t}\vspace{-0.3cm}
\caption{Visualization of two extra dimensions in the case of the heterotic brane world. The bulk gauge group is in general $\E{8}\times\E{8}$. At the singular points of the corners it is broken, such that there are local GUTs residing at the singularities. The 4d gauge group, being the common intersection of the local ones, is $\SU{3}_C\times\SU{2}_L\times\U{1}_Y$.}
\label{fig:tetra}
\end{figure}

The interest in heterotic orbifolds was renewed due to the so-called \emph{orbifold GUTs}, field theories in 5d~\cite{Kawamura:1999nj,Kawamura:2000ir} or 6d~\cite{Asaka:2001eh,Asaka:2003iy} compactified on one or two-dimensional orbifolds, respectively. They combine the benefits of GUTs (like the unification of quarks and leptons) while they have the potential to avoid their problems  (for example the doublet-triplet splitting problem). The \emph{heterotic brane world} is a stringy realization of this scenario, where \emph{local GUTs} reside on fixed points and fixed tori~\cite{Kobayashi:2004ya,Forste:2004ie,Kobayashi:2004ud,Buchmuller:2004hv,Nilles:2004ej,Buchmuller:2005jr,Buchmuller:2005sh,Buchmuller:2006ik}, see figure~(\ref{fig:tetra}). This concept of local GUTs has turned out to be one of the best guidelines for connecting the heterotic string theory to the MSSM.

\section{Outline}
The outline of this thesis is as follows:

Chapter~\ref{sec:HetOrb} is devoted to an accessible introduction to heterotic orbifold compactifications. It starts with a detailed discussion on the geometrical construction of orbifold spaces. Then, after fixing the notation for the heterotic string theory, we explain the compactification of strings on orbifolds. These sections are rather technical. The general discussion is followed by various examples ($\Z{3}$, $\Z{3}\times\Z{3}$, $\Z{6}$-II and $\Z{2}\times\Z{2}$) which will be relevant in the following chapters. Special focus lies on $\Z{6}$-II. In the next section, we briefly present the results of a classification of $\Z{N}$ orbifolds for the $\SO{32}$ heterotic string, published in~\cite{Nilles:2006np}. We conclude this chapter with the string selection rules for allowed Yukawa couplings. After explaining the selection rules, which can be understood as symmetries of the superpotential, we briefly comment on discrete anomalies, published in~\cite{Araki:2008ek}. Finally, we clarify some open questions about the so-called $\gamma$ selection rule.

In chapter~\ref{sec:DT}, we explain and generalize the concept of discrete torsion in the context of heterotic orbifolds. The results presented here have been published in~\cite{Ploger:2007iq}. The chapter starts with an unknown observation about inequivalent $\Z{N}\times\Z{M}$ orbifold models, i.e. $\Z{N}\times\Z{M}$ models whose shifts differ by lattice vectors can be inequivalent. This, previously unknown construction is named ``brother models''. Afterwards, we show how this observation can be related to discrete torsion. In the main part of this chapter, we generalize the aforementioned concepts leading to fixed point dependent (generalized) discrete torsion and generalized brother models (where shifts and Wilson lines differ by lattice vectors). It is shown that these new constructions cannot only be applied to $\Z{N}\times\Z{M}$ orbifolds, but also to the case of $\Z{N}$. We close this chapter with the observation that orbifold models with generalized discrete torsion (or likewise generalized brother models) can be equivalently described as torsionless models compactified on an orbifold with non-factorizable torus lattice.

The next chapter, chapter~\ref{sec:ML}, contains some of the most important results of this thesis, the ``Mini-Landscape'' of MSSM candidates from the heterotic $\Z{6}$-II orbifold, published in a series of papers~\cite{Lebedev:2006kn,Lebedev:2006tr,Lebedev:2007hv}. The chapter starts with an introduction to local GUTs, i.e. GUT theories which only become visible locally in the extra dimensions at an orbifold fixed point. Using the concept of local GUTs, a search strategy is developed for finding orbifold models that render some generic features of the MSSM. It turns out that a strategy based on local GUTs is very successful yielding about 200 models with the exact MSSM spectrum at low energies~\cite{Lebedev:2006kn}. Using this sample of promising models, we explore the possibility of supersymmetry breaking through hidden sector gaugino condensation and find correlations between properties of the MSSM candidates and the scale of SUSY breaking, as published in~\cite{Lebedev:2006tr}. Finally, we analyze the phenomenology of a generic model (named the ``benchmark'' model). For this specific model, we identify a $\U{1}_{B-L}$ in order to avoid rapid proton decay. It is shown how $B-L$ is broken to matter-parity by the vevs of some Standard Model singlets that carry an even $B-L$ charge. After discussing the conditions $F=\mathcal{W}=D=0$ for unbroken supersymmetry, we analyze some further phenomenological aspects of this benchmark model in detail~\cite{Lebedev:2007hv}.

Chapter~\ref{sec:blowup} is dedicated to the blow--up of $\Z{3}$ orbifold singularities, published in~\cite{Nibbelink:2008tv}. First, the blow--up procedure is discussed locally for a single fixed point and afterwards globally for the compact case. After explaining the geometry of the local singularity and its smooth resolution space, it is shown how heterotic models can be built on these spaces. Next, the transition from the singular orbifold to its smooth counterpart is explained in detail. This transition is parameterized by the vev of the so-called blow--up mode, a twisted string localized at the singularity. Starting on the singular orbifold with a zero vev for the blow--up mode, increasing the vev induces a Higgs mechanism and results in the resolution model. In the last part of the chapter this mechanism is applied to the compact $\Z{3}$ case, even when Wilson lines are present. We close this chapter with a blow--up of a well-known MSSM candidate obtained from a compact $\Z{3}$ orbifold with two Wilson lines.

In chapter~\ref{sec:conclusion} we give a brief summary and some concluding remarks. Afterwards, the appendices provide many details. Appendix~\ref{app:comput} presents some explicit computations to which various chapters of the main text refer. Some of them might help to increase the intuitive understanding of heterotic orbifolds. In appendix~\ref{app:tables} we list many tables in order to provide the details on the orbifold models which were discussed in the main text. The last appendix, appendix~\ref{app:group}, summarizes some aspects of group theory, especially the weight lattices of $\E{8}\times\E{8}$ and $\text{Spin}(32)/\Z{2}$.

\section{List of Publications}
Parts of this thesis have been published in scientific journals:

\begin{itemize}
\item S.~F\"orste, H.~P.~Nilles, P.~K.~S.~Vaudrevange and A.~Wingerter, ``Heterotic brane world,'' Phys.\ Rev.\  D {\bf 70} (2004) 106008 [arXiv:hep-th/0406208].
\item H.~P.~Nilles, S.~Ramos-S\'anchez, P.~K.~S.~Vaudrevange and A.~Wingerter, ``Exploring the SO(32) heterotic string,'' JHEP {\bf 0604} (2006) 050 [arXiv:hep-th/0603086].
\item O.~Lebedev, H.~P.~Nilles, S.~Raby, S.~Ramos-S\'anchez, M.~Ratz, P.~K.~S.~Vaudrevange and A.~Wingerter, ``A mini-landscape of exact MSSM spectra in heterotic orbifolds,'' Phys.\ Lett.\  B {\bf 645} (2007) 88 [arXiv:hep-th/0611095].
\item O.~Lebedev, H.~P.~Nilles, S.~Raby, S.~Ramos-S\'anchez, M.~Ratz, P.~K.~S.~Vaudrevange and A.~Wingerter, ``Low Energy Supersymmetry from the Heterotic Landscape,'' Phys.\ Rev.\ Lett.\  {\bf 98} (2007) 181602 [arXiv:hep-th/0611203].
\item F.~Pl\"oger, S.~Ramos-S\'anchez, M.~Ratz and P.~K.~S.~Vaudrevange, ``Mirage Torsion,'' JHEP {\bf 0704} (2007) 063 [arXiv:hep-th/0702176].
\item O.~Lebedev, H.~P.~Nilles, S.~Raby, S.~Ramos-S\'anchez, M.~Ratz, P.~K.~S.~Vaudrevange and A.~Wingerter, ``The Heterotic Road to the MSSM with R parity,'' Phys.\ Rev.\  D {\bf 77} (2008) 046013 [arXiv:0708.2691 [hep-th]].
\item S.~G.~Nibbelink, D.~Klevers, F.~Pl\"oger, M.~Trapletti and P.~K.~S.~Vaudrevange, ``Compact heterotic orbifolds in blow-up,'' JHEP04(2008)060 [arXiv:0802.2809 [hep-th]].
\item T.~Araki, T.~Kobayashi, J.~Kubo, S.~Ramos-S\'anchez, M.~Ratz, P.~K.~S.~Vaudrevange,\\``(Non-)Abelian discrete anomalies'' Nucl.Phys.B805:124-147,2008 [arXiv:0805.0207 [hep-th]].
\end{itemize}

\chapter{Heterotic Orbifolds}
\label{sec:HetOrb}

In this chapter, we describe how to compactify heterotic string theory on orbifolds (for further introductions, see e.g. \cite{Giedt:2002hw,Ramos:2008ph}). We start with a detailed review of the geometrical construction of toroidal orbifolds with Abelian point groups. Afterwards, we examine the heterotic string theory in its bosonic formulation in order to fix the notation and to prepare for the main part of this chapter, where some explicit $\mathbbm{Z}_N$ and $\mathbbm{Z}_N \times \mathbbm{Z}_M$ orbifolds are constructed. We conclude this chapter with a discussion on Yukawa couplings and string selection rules. Some parts of this chapter have been published~\cite{Nilles:2006np,Lebedev:2007hv,Araki:2008ek}.

\section{Geometry}
\label{sec:geometry}

\subsubsection{The Torus}

\begin{figure}[b]
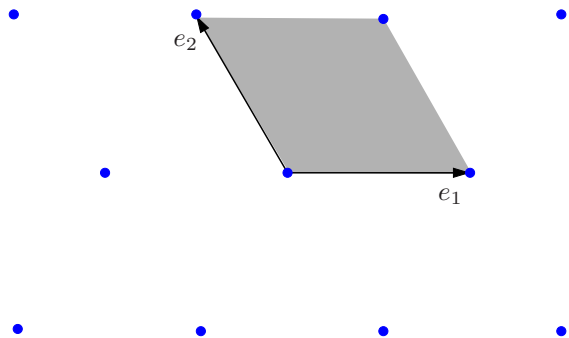

\centerline{\input SU3Torus.pstex_t}\vspace{-0.35cm}
\caption{Two-dimensional lattice $\Gamma = \text{SU}(3)$. The grey region represents the area of the torus $\mathbbm{R}^2 / \Gamma$ and is called the \emph{fundamental domain}\index{torus!fundamental domain}. Note the shorthand: $\Gamma = \text{SU}(3)$ means that $\Gamma$ can be spanned by the simple roots $e_1$ and $e_2$ of $\text{SU}(3)$.}
\label{fig:SU3Torus}
\end{figure}

A six-dimensional torus $T^6$ is chosen by specifying a six-dimensional \emph{lattice}\index{lattice}
\begin{equation}
\Gamma = \{n_\alpha e_\alpha,\; n_\alpha \in \mathbbm{Z},\; \text{ sum over $\alpha=1,\ldots,6$} \}\;,
\end{equation}
spanned by the basis vectors $e_\alpha$, $\alpha=1,\ldots,6$. Then the \emph{torus}\index{torus} is defined as the quotient space
\begin{equation}
T^6 \equiv \mathbbm{R}^6 / \Gamma\;,
\end{equation}
i.e. points of $\mathbbm{R}^6$ differing by some lattice vector of $\Gamma$ are identified. In this context, we will also call $\Gamma$ the \emph{torus lattice}. We choose it to be the root lattice of a semisimple Lie algebra. The basis vectors $e_\alpha$ of $\Gamma$ can be chosen to be the simple roots of the Lie algebra, or some other basis of the root lattice. For a 2d example $\Gamma = \text{SU}(3)$, see figure~(\ref{fig:SU3Torus}). From the basis vectors of $\Gamma$ we define the torus metric\index{torus!metric} $g$ in the absence of a nontrivial background by
\begin{equation}
g_{ij} \equiv e_i \cdot e_j\;.
\end{equation}
In the case of $e_\alpha$ being simple roots, it coincides, except for a possible normalization, with the Cartan matrix of the semisimple Lie algebra defining the lattice.

\subsubsection{The Point Group}\index{orbifold!point group}

In order to specify an orbifold, we need to choose a (finite) symmetry of the torus lattice $\Gamma$. This symmetry group is called the \emph{point group}, denoted by $P$. In the following, we restrict ourselves to the Abelian case, i.e. to cyclic groups $\mathbbm{Z}_N$ or products thereof, such that the action of the point group $P$ on the lattice can be visualized as discrete rotations mapping the lattice to itself. For example, in the case of a two-dimensional $\text{SU}(3)$ torus lattice (see figure~(\ref{fig:SU3Torus})), we can identify different symmetries that map the lattice to itself: $\mathbbm{Z}_2$, $\mathbbm{Z}_3$ and $\mathbbm{Z}_6$. In order to be a symmetry of some six-dimensional lattice $\Gamma$, the point group $P$ has to be a subgroup of the group of rotations in 6d, i.e. a subgroup of $\text{SO}(6) \simeq \text{SU}(4)$. Furthermore, $P$ sits in the Cartan subalgebra of $\text{SU}(4)$, since we want the point group to be Abelian. The rank of $\text{SU}(4)$ is three, i.e. the number of Cartan generators is three. It is convenient to choose them as
\begin{equation}
\label{eqn:J}
J_{12},\; J_{34}\;\text{and } J_{56}\;,
\end{equation}
where $J_{ij}$ generates a rotation in the plane spanned by the orthonormal basis vectors $\hat{e_i}$ and $\hat{e}_j$. In this basis an element of $P$ can be written as
\begin{equation}
\exp\left(2\pi\I \left(v^1 J_{12} + v^2 J_{34} + v^3 J_{56}\right)\right)\;,
\end{equation}
where $v^i$ specifies the rotation angle in the $i$-th plane, e.g. $v^1 = 1/3$ is a rotation about $120^\circ$ in the first plane spanned by $\hat{e}_1$ , $\hat{e}_2$. Using the three Cartan generators $J_{12}, J_{34}$ and $J_{56}$, we can have at most three independent $\mathbbm{Z}_N$ factors. Later, in section~\ref{sec:Compactification}, we will see that requiring $\mathcal{N}=1$ supersymmetry in 4d amounts to choosing point groups $P$ being in the Cartan subalgebra of $\text{SU}(3) \subset \text{SU}(4)$, which allow for at most two independent $\mathbbm{Z}_N$ factors. Therefore, in the case of an Abelian point group, $P$ is either $\mathbbm{Z}_N$ or $\mathbbm{Z}_N \times \mathbbm{Z}_M$ (for some specific values of $N$ and $M$, $N$ being a multiple of $M$, see e.g.~\cite{Katsuki:1989bf, Font:1988mk}).\\

Since $P \subset \text{SU}(3)$, it is convenient to rewrite the six-dimensional space $\mathbbm{R}^6$ in a complex basis as $\mathbbm{C}^3$, i.e. as three orthogonal complex planes. We can naturally choose the $i$-th complex plane to be spanned by $\hat{e}_{2i-1}$ and $\hat{e}_{2i}$. In this basis, elements of $P$ are complex $3\times 3$ matrices that are diagonalized simultaneously. Then, the generator $\theta$ of a $\mathbbm{Z}_N$ point group reads
\begin{equation}
\label{eqn:theta}
\theta = \text{diag}(e^{2\pi\I v^1}, e^{2\pi\I v^2}, e^{2\pi\I v^3})\;.
\end{equation}
We define the \emph{twist vector}\index{orbifold!twist vector} as
\begin{equation}
v \equiv (0, v^1, v^2, v^3)\;,
\end{equation}
where the first entry is included for later use. The order of the generator $\theta^N = \mathbbm{1}$ translates to the twist vector as $N v^i \in \mathbbm{Z}$, for $i=1,2,3$. Thus, the point group $P$ is given by
\begin{equation}
P = \big\{ \theta^k\; |\; k = 0,\ldots,N-1 \big\}\;.
\end{equation}
In order for $\theta$ to be an element of $\text{SU}(3)$ (such that $\text{det}(\theta) = 1$) the condition $v^1 + v^2 + v^3 \in \mathbbm{Z}$ has to be imposed on the twist vector. It is convenient to choose the twist vector such that this condition reads
\begin{equation}
\label{eqn:susycondition}
v^1 + v^2 + v^3 = 0\;.
\end{equation}
In the case of $\mathbbm{Z}_N \times \mathbbm{Z}_M$ the two generators $\theta$ and $\omega$ are associated to two twist vectors $v_1$ and $v_2$ satisfying
\begin{equation}
\label{eqn:susycondition2}
v_1^1 + v_1^2 + v_1^3 = 0 \quad\text{and}\quad v_2^1 + v_2^2 + v_2^3 = 0\;,
\end{equation}
which are of order $N$ and $M$, respectively.

\subsubsection{Factorized and Non-Factorized Lattices}\index{lattice!factorized}\index{lattice!non-factorized}

In the complex basis of $\mathbbm{C}^3$, where the elements of the (Abelian) point group are diagonal $3\times 3$ matrices, the underlying lattice $\Gamma$ can be aligned differently inside $\mathbbm{C}^3$. We distinguish two cases. 

In the first one, the six-dimensional lattice can be written as the product of three two-dimensional lattices and each of these two-dimensional sublattices lies inside one of the three complex planes, e.g. $\Gamma = \text{SU}(3) \times \text{SU}(3) \times \text{SU}(3)$. This lattice is called \emph{factorized}. In this case the $i$-th exponent $v^i$ of eqn.~(\ref{eqn:theta}) can easily be visualized as a rotation in the $i$-th two-dimensional sublattice.

Otherwise, the lattice is said to be \emph{non-factorized}. In this case, each basis vector of $\Gamma$ is specified by three in general non-zero complex coordinates, e.g. $\Gamma = \text{E}_6$. It is important to note that one has to specify these coordinates in order to distinguish between factorized and non-factorized lattices. For example, in the case of a $\mathbbm{Z}_2 \times \mathbbm{Z}_2$ point group of a lattice $\Gamma = \text{SU}(3) \times \text{SU}(2)^4$  one has to specify the orientation of the $\text{SU}(3)$ sublattice inside $\mathbbm{C}^3$ in order to see whether $\Gamma$ is factorized or not, see section~\ref{sec:Z2xZ2NonFactorized} for more details on this example.

\subsubsection{The Space Group}\index{orbifold!space group}

\begin{figure}[t]
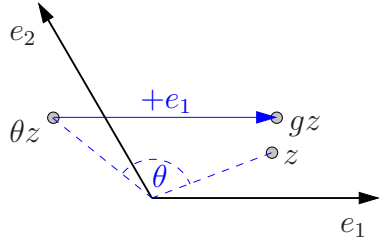

\centerline{\input Z3SpaceGroup.pstex_t}
\caption{Two-dimensional $\SU{3}$ torus lattice spanned by the simple roots $e_1$ and $e_2$. The action of $g = \left(\theta,e_1\right)$, with $\theta =  e^{2\pi\I/3}$, on some arbitrary point $z\in\mathbbm{C}$ is depicted in detail. First, $z$ is rotated to $\theta z$ and then shifted to $\theta z + e_1$.}
\label{fig:SpaceGroup}
\end{figure}

Having specified a torus lattice $\Gamma$ and a point group $P$, it is convenient to define now the \emph{space group} $S$ as the semidirect product of the point group $P$ and the translations associated to $\Gamma$. In detail, an element $g$ of $S$ can be written as
\begin{equation}
g = \left( \vartheta, n_\alpha e_\alpha \right)\;,
\end{equation}
where $\vartheta \in P$ and $n_\alpha e_\alpha \in \Gamma$, summing over $\alpha$. Then, by definition, $g$ acts on a point $z\in\mathbbm{C}^3$ as follows
\begin{equation}
gz = \left( \vartheta, n_\alpha e_\alpha \right)z = \vartheta z + n_\alpha e_\alpha\;,
\end{equation}
see figure~(\ref{fig:SpaceGroup}) for an example. Furthermore, the product of two elements of the space group $g = \left( \vartheta_1, n_\alpha e_\alpha \right)$ and $h = \left( \vartheta_2, m_\alpha e_\alpha \right)$ reads
\begin{equation}
g\; h = \left( \vartheta_1, n_\alpha e_\alpha \right)\left( \vartheta_2, m_\alpha e_\alpha \right) = \left( \vartheta_1\vartheta_2, n_\alpha e_\alpha + \vartheta_1 (m_\alpha e_\alpha) \right)\;,
\end{equation}
reflecting the properties of the semidirect product of $P$ and $\Gamma$. The inverse of $g = \left( \vartheta, n_\alpha e_\alpha \right)$ is easily found to be
\begin{equation}
g^{-1} = \left( \vartheta^{-1}, -\vartheta^{-1}(n_\alpha e_\alpha) \right)\;.
\end{equation}
It is important to notice that in general two elements of $S$ do not commute,
\begin{equation}
hg \neq gh\;,
\end{equation}
i.e. although $P$ is Abelian, the space group $S$ is not. 

\subsubsection{The Orbifold}\index{orbifold}

Now, we can define the six-dimensional (toroidal) \emph{orbifold} as the quotient space
\begin{equation}
\label{eqn:deforbifold}
\mathbbm{O} \equiv T^6 / P = \mathbbm{C}^3 / S\;,
\end{equation}
i.e. points of $\mathbbm{C}^3$ are identified in $\mathbbm{O}$ if they differ by the action of some element of the space group: $z \sim gz$ with $g \in S$~\cite{Dixon:1985jw,Dixon:1986jc}.

\begin{figure}[t]
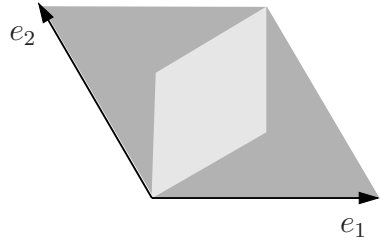

\centerline{\input SU3TorusZ3FundamentalDomain.pstex_t}\vspace{-0.25cm}
\caption{The fundamental domain\index{orbifold!fundamental domain} of the two-dimensional $\mathbbm{Z}_3$ orbifold (light grey region) is one third of the fundamental domain of the two-torus $\text{SU}(3)$ (grey region).}
\label{fig:Z3FundamentalDomain}
\end{figure}

In order to identify a fundamental domain\index{orbifold!fundamental domain} of the orbifold\footnote{Note that the fundamental domain can be represented in various ways and is therefore not unique.}, it is convenient to start from a fundamental domain of the torus $T^6$ and identify points that are mapped to each other under the action of $P$. For a two-dimensional example\index{orbifold!$\Z{3}$} see figure~(\ref{fig:Z3FundamentalDomain}).

\subsubsection{Fixed Points}\index{fixed point}

\begin{figure}[h]
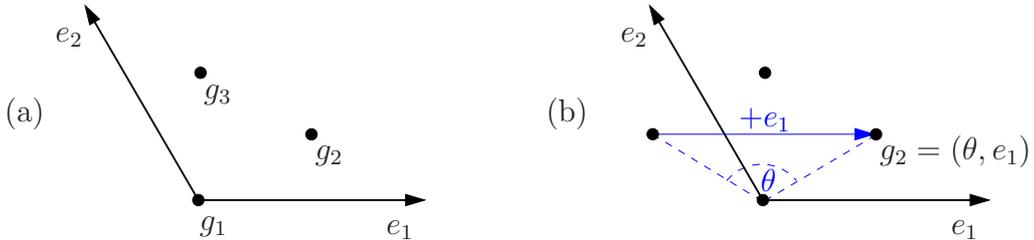

\centerline{\input SU3TorusZ3FixedPoints.pstex_t}\vspace{-0.25cm}
\caption{(a) Two-dimensional $\mathbbm{Z}_3$ orbifold with three fixed points associated to the space group elements $g_1 = \left(\theta,0\right)$,  $g_2 = \left(\theta,e_1\right)$ and  $g_3 = \left(\theta,e_1+e_2\right)$, where $\theta =  e^{2\pi\I/3}$. (b) The action of $g_2$ on the corresponding fixed point is illustrated in detail.}
\label{fig:Z3FixedPoints}
\end{figure}

In the definition of the orbifold $\mathbbm{O}$, the space group $S$ does not act freely on $\mathbbm{C}^3$. This means that there are so-called \emph{fixed points} $z_\text{f} \in \mathbbm{C}^3$, i.e. points that are invariant under the action of a nontrivial element $g = \left( \vartheta, n_\alpha e_\alpha \right) \in S$
\begin{equation}
\label{eq:fp}
gz_\text{f} = z_\text{f} \quad\Leftrightarrow\quad\text{$z_\text{f}$ is a fixed point of $g$}\;,
\end{equation}
see figure~(\ref{fig:Z3FixedPoints}) for an example\index{orbifold!$\Z{3}$}.

If the rotation $\vartheta = \text{diag}(e^{2\pi\I v^1}, e^{2\pi\I v^2}, e^{2\pi\I v^3})$ acts trivially in one of three complex directions, equation~(\ref{eq:fp}) will be solved by a whole set of fixed points, denoted as \emph{fixed torus}. For example, if $g = \left( \vartheta, 0\right)$ with $v^1 = 0$ and $v^2,v^3 \neq 0$, equation~(\ref{eq:fp}) is solved by $z_\text{f} = (z_\text{f}^1, 0, 0)$, where $z_\text{f}^1 \in \mathbbm{C}$ arbitrary, yielding a fixed torus located at the origin of the second and third complex planes.

In general, given a nontrivial element $g \in S$, it is easy to find the associated fixed point (or torus) from equation~(\ref{eq:fp}) as
\begin{equation}
\label{eqn:FixedPointCoordinate}
z_\text{f} = \left(\mathbbm{1}-\vartheta\right)^{-1} n_\alpha e_\alpha\;.
\end{equation}
Therefore, we will also denote the fixed point as $g$, i.e. by its associated space group element.

\begin{figure}[t]
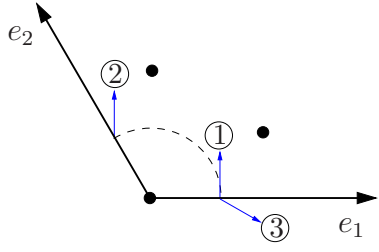

\centerline{\input SU3TorusHolonomy.pstex_t}\vspace{-0.2cm}
\caption{$\mathbbm{Z}_3$ orbifold in 2d. The vector (1) is parallel transported to (2) using the flat connection of the torus. Note that due to the orbifold this path gives a closed loop around the fixed point at the origin. Furthermore, the $\mathbbm{Z}_3$ action identifies vector (2) and (3). Thus, vector (1) is rotated by $120^\circ$ to vector (3), illustrating the non-trivial $\mathbbm{Z}_3$ holonomy group at the origin.}
\label{fig:Z3holonomy}
\end{figure}

\index{fixed point!singularity}Fixed points are \emph{curvature singularities}\index{singularity}. This can be seen from the local holonomy groups\index{holonomy group} at the various points of an orbifold. In general, the holonomy group is trivial everywhere indicating flat space (i.e. no curvature). However, at the fixed points we find a non-trivial holonomy group: $\mathbbm{Z}_N$ or a subgroup thereof. For a two-dimensional example with $\mathbbm{Z}_3$\index{orbifold!$\Z{3}$} holonomy see figure~(\ref{fig:Z3holonomy}). Since a non-trivial holonomy group is related to a non-vanishing curvature (cf. page 344ff of \cite{Nakahara:2003nw}), there are curvature singularities at the fixed points of the orbifold. In chapter~\ref{sec:blowup} it is shown how these singularities can be resolved in the context of $\mathbbm{Z}_3$ orbifolds.

\subsubsection{Inequivalent Fixed Points}\index{fixed point!equivalent}

We say that two fixed points (space group elements) $g_1$ and $g_2$ are \emph{equivalent} if they are related by conjugation, i.e.
\begin{equation}
g_1 \sim g_2 \quad\Leftrightarrow\quad g_1 = h g_2 h^{-1} \quad\text{for some }h \in S\;.
\end{equation}
In the case of $h$ being a pure translation $h = \left(\mathbbm{1}, m_\alpha e_\alpha\right)$, this can easily be interpreted as two fixed points $g_1$ and $g_2 = \left(\vartheta, n_\alpha e_\alpha\right)$ that are identified on the torus. We show this in detail:
\begin{equation}
g_1 = h g_2 h^{-1} = \left(\mathbbm{1},m_\alpha e_\alpha\right) \left(\vartheta, n_\alpha e_\alpha \right) \left(\mathbbm{1},-m_\alpha e_\alpha\right) = \left(\vartheta, (\mathbbm{1} - \vartheta)m_\alpha e_\alpha + n_\alpha e_\alpha \right)
\end{equation}
and therefore the coordinates $z_{\text{f}_2}$ and $z_{\text{f}_1}$ of the fixed points $g_2$ and $g_1$ read
\begin{eqnarray}
z_{\text{f}_2} & = & \left(\mathbbm{1}-\vartheta\right)^{-1}n_\alpha e_\alpha \\
z_{\text{f}_1} & = & \left(\mathbbm{1}-\vartheta\right)^{-1}\left[(\mathbbm{1} - \vartheta)m_\alpha e_\alpha + n_\alpha e_\alpha\right] = z_{\text{f}_2} + m_\alpha e_\alpha\;,
\end{eqnarray}
respectively. Since $z_{\text{f}_1} = z_{\text{f}_2} + m_\alpha e_\alpha$ and $m_\alpha e_\alpha$ is clearly from the torus lattice $\Gamma$, these fixed points are identified on the torus $z_{\text{f}_1} \sim z_{\text{f}_2}$. In the case of a general element $h = \left(\varphi, m_\alpha e_\alpha\right)$ this result generalizes to
\begin{equation}
z_{\text{f}_1} = \varphi z_{\text{f}_2} + m_\alpha e_\alpha = h z_{\text{f}_2} \;.
\end{equation}
This leads us to the conclusion that equivalent fixed points are identified on the orbifold $\mathbbm{O}$ and the inequivalent fixed points are given by the conjugacy classes $[g]$ of $S$. See figure~(\ref{fig:Z6IIEquivalentFP}) for a two-dimensional $\mathbbm{Z}_6$ example\index{orbifold!$\Z{6}$-II}.

\begin{figure}[h]
\centerline{\input Z6IIEquivalentFP.pstex_t}\vspace{-0.15cm}
\caption{$\mathbbm{Z}_6$ orbifold in 2d. The lattice $\Gamma = \G{2}$ admits a $\mathbbm{Z}_6$ point group with $\theta =  e^{2\pi\I/6}$. The fundamental domain\index{orbifold!fundamental domain} of the torus (grey region) is reduced on the orbifold to one sixth (light grey region). Four fixed points, corresponding to the space group elements $g_0 = \left(\theta^2,0\right)$, $g_1 = \left(\theta^2,e_1\right)$, $g_1' = \left(\theta^2,e_2-e_1\right)$ and $g_2 = \left(\theta^2,2e_1\right)$ are depicted. Note that the fixed points $g_1$ and $g_1'$ differ by a rotation with $\theta$. Furthermore, $g_1'$ and $g_2$ differ by the lattice vector $e_1$. Consequently, there are only two inequivalent fixed points (e.g. $g_0$ and $g_1$) with a $\theta^2$ point group element.}
\label{fig:Z6IIEquivalentFP}
\end{figure}

\subsubsection{The Untwisted Sector and Twisted Sectors}\index{fixed point!untwisted and twisted sector}

We group all inequivalent space group elements (and hence the associated fixed points if they exist) according to their point group elements into either the \emph{untwisted sector} or the \emph{twisted sectors}. For example, in the case of a $\mathbbm{Z}_N \times \mathbbm{Z}_M$ orbifold with generators $\theta$ and $\omega$, some element $g = \left( \theta^{k_1}\omega^{k_2}, n_\alpha e_\alpha\right) \in S$ with $0 \leq k_1 < N$ and $0 \leq k_2 < M$ is said to belong to
\begin{itemize}
\item the untwisted sector $U$ if $k_1=k_2=0$. Elements of the untwisted sector are neither associated to fixed points nor to fixed tori.
\item the twisted sector $T_{(k_1,k_2)}$ if $k_1\neq 0$ or $k_2\neq 0$.
In the case of twisted sectors, an element $g$ corresponds to a fixed point or fixed torus.
\end{itemize}
The untwisted sector $U$ and the twisted sectors $T_{(k)}$ are analogously defined in the case of a $\mathbbm{Z}_N$ orbifold.

\subsubsection{Continuous Lattice Deformations}\index{lattice!deformations}

For the definition of the orbifold, eqn.~(\ref{eqn:deforbifold}), it was crucial that the underlying torus lattice $\Gamma$ obeys the symmetry of the point group $P$. This condition strongly constrains the choice of allowed lattices. However, a given lattice $\Gamma$ always allows for some continuous deformations while keeping its point group symmetry: at least the overall size is a free parameter of the lattice which does not affect the action of the point group. 

\begin{figure}[b]
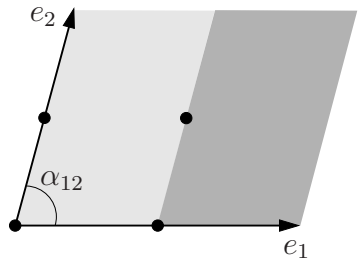

\centerline{\input Z2Moduli.pstex_t}\vspace{-0.4cm}
\caption{$\mathbbm{Z}_2$ orbifold in 2d. For any (non-degenerate) value of $|e_1|$, $|e_2|$ and $\alpha_{12} = \angle(e_1, e_2)$ the lattice admits a $\mathbbm{Z}_2$ point group with $\theta =  e^{\pi\I}$. The fundamental domain\index{orbifold!fundamental domain} of the torus (grey region) is reduced on the orbifold to one half (light grey region). Furthermore, there are four inequivalent fixed points.}
\label{fig:Z2Moduli}
\end{figure}

In order to see this in more detail we consider a specific example first. In the case of a two-dimensional $\mathbbm{Z}_2$ orbifold\index{orbifold!$\Z{2}$}, the basis vectors $e_1$ and $e_2$ of the (deformed) lattice $\Gamma = \SU{2}^2$ can be parameterized by three (real) variables $R_1$, $R_2$ and $\alpha_{12}$ as
\begin{equation}
e_1 = R_1 \quad\text{ and }\quad e_2 = R_2 e^{\I \alpha_{12}}\;,
\end{equation}
see figure~(\ref{fig:Z2Moduli}). For $R_1 \neq 0$, $R_2 \neq 0$ and $\alpha_{12} \neq n\pi$, $n \in \mathbbm{Z}$, the vectors $e_1$ and $e_2$ are linear independent and allow for the $\mathbbm{Z}_2$ point group generated by $\theta =  e^{\pi\I}$. Note that for special values the two-dimensional $\mathbbm{Z}_2$ orbifold has an enhanced symmetry~\cite{Dixon:1986qv, Kobayashi:2006wq}, e.g. for $R_1 = R_2$ and $\alpha_{12} = \frac{2\pi}{3}$ the orbifold can be visualized as a tetrahedron with its four fixed points being located at the four corners and consequently the symmetry of the orbifold is $S_4$.

In the general case, some torus lattice $\Gamma$ allows for continuous deformations as long as its torus metric $g$ is invariant under the action of the point group $P$. Let us discuss this in detail. We start with some general torus lattice $\Gamma(\alpha)$ spanned by six (real) basis vectors $e_i(\alpha)$, where $\alpha$ indicates some coordinates that describe \emph{all} deformations of the lattice. These coordinates can for example be associated to some radii or angles. We assume that for $\alpha = 0$ the lattice $\Gamma(0)$ is the root lattice of some semisimple Lie algebra which is suitable for the point group $P$. Obviously, the torus metric depends on the $\alpha$'s
\begin{equation}
g_{ij}(\alpha) = e_i(\alpha)\cdot e_j(\alpha)\;,
\end{equation}
and coincides with the Cartan matrix for $\alpha = 0$. Under the action of the point group generator $\theta$ the vectors $e_i(\alpha)$ transform as $e_i(\alpha) \mapsto \theta e_i(\alpha) = \hat\theta_{ji}(\alpha) e_j(\alpha)$. Furthermore, it is easy to see (using $\theta^T \theta = \mathbbm{1}$) that the torus metric is invariant under the action of the twist
\begin{equation}
g(\alpha) \stackrel{\theta}{\mapsto} \hat\theta^T(\alpha) g(\alpha) \hat\theta(\alpha) = g(\alpha)\;.
\end{equation}
Now, we can distinguish two cases. First of all, if $\hat\theta_{ji}(\alpha) \notin \mathbbm{Z}$, the point group is not a symmetry of the lattice. Secondly, if $\hat\theta_{ji}(\alpha) \in \mathbbm{Z}$ is constant, then it is clearly independent of the continuous parameter $\alpha$. This means that the twist $\theta$ maps the lattice to itself for any value of $\alpha$. In this case $\hat\theta$ can be identified as the so-called \emph{Coxeter element}~\cite{Markushevich:1986za}\index{Coxeter element} of the undeformed lattice $\Gamma(0)$.\footnote{The Coxeter element is an inner automorphism of the Lie lattice $\Gamma(0)$ and can be expressed by a series of Weyl reflections. It can be extended to the so-called generalized Coxeter element by including outer automorphisms (automorphisms of the Dynkin diagram). More details can be found in e.g.~\cite{Casas:1991ac} and~\cite{Reffert:2007im}.}

Consequently, starting with a torus lattice $\Gamma(0)$ and a (constant) Coxeter element $\hat\theta$, we can determine the allowed lattice deformations, corresponding to some coordinates $\alpha'$, out of all possible deformations $\alpha$ by demanding invariance of the torus metric 
\begin{equation}
\hat\theta^T g(\alpha') \hat\theta \stackrel{!}{=} g(\alpha')\;.
\end{equation}
As an example, we will discuss the deformations of the $\mathbbm{Z}_3$ orbifold later in section~\ref{sec:Z3Orbifold}.

\subsubsection{Factorizable and Non-Factorizable Lattices}\index{lattice!factorized}\index{lattice!non-factorized}

In the context of the last section, we say that an orbifold has a \emph{factorizable} lattice if it can be deformed continuously to a factorized form while keeping its point group symmetry. Otherwise, the lattice is called \emph{non-factorizable}\footnote{Note that using this nomenclature a non-factorized lattice can be factorizable.}.

\section{The Heterotic String}\index{heterotic string!bosonic construction}
\label{sec:HeteroticString}

\begin{figure}[t]
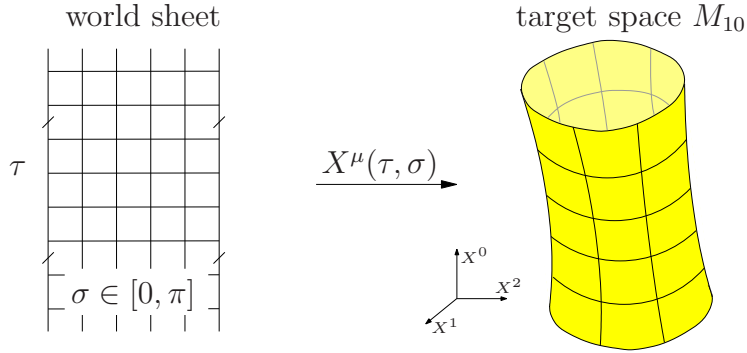

\centerline{\input stringmap.pstex_t}\vspace{-0.15cm}
\caption{The functions $X^\mu$ can be visualized as mappings from the 2d world sheet spanned by $\tau$ and $\sigma$ to the 10d target space $M_{10}$. Here, the world sheet has the topology of a cylinder.}
\label{fig:StringMap}
\end{figure}

To set the notation, we start with a brief review of the heterotic string theory~\cite{Gross:1984dd,Gross:1985fr} (for an introduction, see e.g. chapter 7 of~\cite{Becker:2007zj}). It is a theory of closed strings propagating in ten-dimensional space-time. The string is described by maps $X^\mu(\tau, \sigma)$ that embed the two-dimensional world sheet, equipped with coordinates $(\tau, \sigma)$, into the 10d target space $M_{10}$. Closed strings are subject to boundary conditions, i.e.
\begin{equation}
\label{eqn:closedstring}
X^\mu(\tau, \sigma+\pi) = X^\mu(\tau, \sigma) \quad \mu = 0,\ldots, 9\;
\end{equation}
for the bosonic degrees of freedom $X^\mu$, see figure~(\ref{fig:StringMap}). Since the heterotic string is oriented, left-movers $(\tau+\sigma)$ and right-movers $(\tau-\sigma)$ can be treated separately, e.g.
\begin{equation}
X^\mu(\tau, \sigma) = X^\mu\LM(\tau + \sigma) + X^\mu\RM(\tau - \sigma)\;.
\end{equation}

\subsubsection{The Bosonic Construction}

In the so-called bosonic construction of the heterotic string theory, the theory is described by two parts. On the one hand, the right-movers consist of the 10 dimensional superstring with bosonic and fermionic degrees of freedom $X^\mu\RM(\tau - \sigma)$ and $\Psi^\mu\RM(\tau - \sigma)$, respectively. In the context of two-dimensional conformal field theory (CFT) on the world sheet, these degrees of freedom are related by $\mathcal{N}=1$ (local) supersymmetry. On the other hand, the left-movers are described by the 26 dimensional bosonic string coordinates $X^\mu\LM(\tau + \sigma)$ and $X^I\LM(\tau + \sigma)$, where $\mu = 0,\ldots,9$ and $I = 1,\ldots, 16$. All bosonic and fermionic degrees of freedom are subject to boundary conditions analog to equation~(\ref{eqn:closedstring}), where for the fermionic degrees $\Psi^\mu\RM$ they can be either periodic (\emph{Ramond - R}) or antiperiodic (\emph{Neveu-Schwarz - NS}).

In order to match the number of dimensions of left- and right-movers to 10 dimensions, 16 bosonic left-moving degrees of freedom $X^I\LM$ are compactified on a 16-torus
\begin{equation}
\label{eqn:XIonLambda}
X^I\LM\; \sim \; X^I\LM + \pi \lambda^I\quad\text{with}\quad \lambda\in\Lambda\;,
\end{equation}
where $\Lambda$ is a 16-dimensional torus lattice. By modular invariance of the partition function, this 16-torus is required to be defined by an even and self-dual lattice $\Lambda$. In 16 dimensions there are only two choices\footnote{Cf. for example page 193ff of~\cite{Lust:1989tj} or page 286ff of~\cite{Becker:2007zj}}: the weight lattices of
\begin{equation}
\E{8}\times\E{8} \quad\text{ or }\quad \text{Spin}(32)/ \mathbbm{Z}_2\;.
\end{equation}
Some details about these lattices can be found in appendix~\ref{app:weightlattices}. As a consequence of this toroidal compactification, the 16 dimensional (internal) momentum $p$ is quantized: $p \in \Lambda$. Since the momenta $p$ are elements of the weight lattice $\Lambda$ of $\text{E}_8 \times \text{E}_8$ or $\text{Spin}(32)/ \mathbbm{Z}_2$, they have an additional interpretation as weights. In other words, strings with non-trivial internal momenta $p$ will transform non-trivially under gauge transformations - they form representations corresponding to their weights $p$. Thus, the 16 bosonic coordinates $X^I\LM$ give rise to a gauge theory and are therefore called the \emph{gauge degrees of freedom}.

\subsubsection{Quantization in the Light-Cone Gauge}

It is convenient to choose light-cone coordinates for the remaining ten dimensions. For the bosonic degrees of freedom $X^\mu$ they read
\begin{equation}
X^\pm \equiv \frac{1}{\sqrt{2}}\left( X^0 \pm X^9\right)\;,
\end{equation}
together with $X^i$ for $i = 1,\ldots, 8$ being the transversal coordinates. Then we can fix the gauge such that only the $X^i$ are physical. In this gauge only a transversal $\text{SO}(8)$ rotational group of the ten-dimensional Lorentz group $\text{SO}(9,1)$ is manifest.

The general solutions to the equations of motion for the string are given in terms of mode expansions for $X^i_{\text{L/R}}$, $X^I\LM$ and $\Psi\RM^i$ as functions of $(\tau, \sigma)$, being periodic in $\sigma$. The coefficients in these mode expansions are named \emph{oscillators}\index{oscillator}\footnote{The name ``oscillators'' for the coefficients in the mode expansion shall remind us of the harmonic oscillator, since they also fulfill the algebra of creation and annihilation operators after quantization.}. After quantization, they become creation and annihilation operators depending on their frequencies being negative or positive, respectively. For example, oscillators of the bosonic coordinates $X^i_{\text{L/R}}$ are denoted by $\tilde\alpha^i_n$ and $\alpha^i_n$, with $n$ being the frequency. 

After quantization of the heterotic string in the light-cone gauge one obtains mass equations for right- and left-movers
\begin{equation}
\label{eqn:masslessequation}
\frac{M\RM^2}{8} = \frac{q^2}{2} + N - \frac{1}{2} \quad\text{and}\quad \frac{M\LM^2}{8} = \frac{p^2}{2} + \tilde{N} - 1\;,
\end{equation}
respectively. The constants $-\tfrac{1}{2}$ and $-1$ in these mass equations are called \emph{zero-point energies}\index{zero-point energy} (or \emph{normal-ordering constants}, as they arise from the normal-ordering of the oscillators). Furthermore, oscillator excitations originating from the right-moving fermionic degrees of freedom $\Psi^i\RM$ were encoded in an $\text{SO}(8)$ weight vector $q$. This vector is also named the right-moving momentum (for details on this procedure, called ``bosonization'', see for example chapters 13,14 of~\cite{Lust:1989tj}). The GSO projection~\cite{Gliozzi:1976qd} restricts $q$ to lie either in the vector weight lattice (for $\Psi^i\RM$ from the NS sector) or spinor weight lattice (for $\Psi^i\RM$ from the R sector) of the transversal $\text{SO}(8)$. Moreover, the \emph{oscillator number}\index{oscillator!oscillator number} $\tilde{N}$ of the left-mover is defined as
\begin{equation}
\tilde{N} = \sum_{n=1}^\infty \tilde\alpha_{-n} \cdot \tilde\alpha_n\;,
\end{equation}
where the scalar product sums over the transverse modes $\tilde\alpha_n^i$ and over $\tilde\alpha_n^I$. The right-moving oscillator number $N$ is defined analogously as $N = \sum_{n=1}^\infty \alpha_{-n} \cdot \alpha_n$.

Using the commutators
\begin{equation}
[\tilde\alpha_m^i,\tilde\alpha_n^j]=[\alpha_m^i,\alpha_n^j]= m\delta^{i,j}\delta_{m+n,0}
\end{equation}
and the property of the ground state
\begin{equation}
\alpha_m^i|0\rangle\RM = \tilde\alpha_m^i|0\rangle\LM = 0 \quad\text{for } m>0\;,
\end{equation}
one sees that the oscillator numbers $\tilde{N}$ and $N$ essentially count the number of oscillators $\tilde\alpha_{-n}$ and $\alpha_{-n}$ (weighted with their frequency $n$) acting on the ground state, e.g. the eigenvalue of $\tilde{N}$ for $\tilde\alpha_{-1}^i|0\rangle\LM$ is 1.

Furthermore, in order to ensure that no point in the $\sigma$ direction of a string is preferred, we have to impose the so-called level-matching condition\index{level-matching}
\begin{equation}
\label{eqn:levelmatching}
M\RM^2 = M\LM^2\;,
\end{equation}
which removes the tachyons (i.e. states with negative $M^2$) from the spectrum.

Finally, since left- and right-movers are independent, a physical state can be written as a tensor product of a left-moving and a right-moving state
\begin{equation}
|q\rangle\RM \otimes |p\rangle\LM\;,
\end{equation}
subject to possible oscillator excitations, for example with $\tilde{\alpha}_{-1}$ for the left-mover.

\subsubsection{Representations of the Little Group}

In general, in D-dimensional space-time with Lorentz symmetry $\text{SO}(D-1,1)$ states transform on-shell in representations of the so-called \emph{little group}\index{little group}. For massive states this is $\text{SO}(D-1)$, whereas for massless ones it is $\text{SO}(D-2)$.

Therefore, in our case with $D=10$, we can identify in the massless case the transversal $\text{SO}(8)$ with the little group $\text{SO}(8)$. On the other hand, in the massive case representations of the transversal group $\SO{8}$ have to combine to representations of the little group $\SO{9}$. From here we see that everything which is related to the eight-dimensional transversal coordinates may contribute to the transformation property of a string under Lorentz transformations, i.e. the right-moving momenta $q$ and the transversal oscillators $\tilde\alpha_n^i$ define the string's representation of the little group $\SO{8}$ (or $\SO{9}$).

\subsubsection{The Massless Spectrum}\index{heterotic string!massless spectrum}

The solutions to the equation for massless right-movers (\ref{eqn:masslessequation}a) are characterized by $q^2 = 1$ (note that $N$ has integral eigenvalues, thus $N>0$ yields massive states). Since $q$ is restricted to lie either in the vector or in the spinor weight lattice of the transversal $\text{SO}(8)$, the right-movers' momenta $q$ are given by the weight vectors of either the vector or the spinor representation:
\begin{itemize}
\item The vector representation $\rep{8}_v$ of $\text{SO}(8)$ is given by $q = (\underline{\pm 1, 0, 0, 0})$ and the corresponding state describes a vector boson in 10 dimensional space-time.
\item The spinor representation $\rep{8}_s$ of $\text{SO}(8)$ is given by $q = (\pm \tfrac{1}{2}, \pm \tfrac{1}{2}, \pm \tfrac{1}{2}, \pm \tfrac{1}{2})$ (with an even number of plus signs) and the corresponding state describes a fermion in 10 dimensional space-time\footnote{For more details, see appendix~\ref{app:SO8}.}.
\end{itemize}

On the other hand, the solutions to the equation for massless left-movers (\ref{eqn:masslessequation}b) are given by either $p^2 =2$ and $\tilde{N} = 0$ or $p^2 = 0$ and $\tilde{N} = 1$. In the first case of $p^2 = 2$, the 16 dimensional internal momenta $p$ are the roots of $\E{8}\times\E{8}$ or $\SO{32}$, see appendix~\ref{app:weightlattices}. In the second case, the $\tilde{N} = 1$ oscillator states $\tilde\alpha_{-1}^K|0\rangle\LM$, where $K = i, I$, are found to be massless.

Building the tensor product of massless right- and left-movers yields the massless spectrum of the heterotic string:
\begin{eqnarray}
\label{eqn:masslesshet}
|q\rangle\RM \otimes |p\rangle\LM                    && \text{480 generators of }\E{8}\times\E{8} \text{ or }\SO{32}\\\label{eqn:16Cartan}
|q\rangle\RM \otimes \tilde\alpha_{-1}^I|0\rangle\LM && \text{16 Cartan generators } (I = 1,\ldots, 16)\\
|q\rangle\RM \otimes \tilde\alpha_{-1}^i|0\rangle\LM && \mathcal{N} = 1 \text{ SUGRA multiplet } (i = 1,\ldots,8)
\label{eqn:10dSUGRA}
\end{eqnarray}
First, we discuss the gauge quantum numbers of these states. The $480 + 16 = 496$ states in eqns.~(\ref{eqn:masslesshet}) and~(\ref{eqn:16Cartan}) transform in the adjoint representation $\rep{248} + \rep{248}$ of $\E{8}\times\E{8}$ or $\rep{496}$ of $\SO{32}$. The states of eqn.~(\ref{eqn:10dSUGRA}) are gauge singlets.

Secondly, we discuss the transformation properties under Lorentz transformation\index{$\SO{8}$}. Note that the oscillators $\tilde\alpha_{-1}^I$ with $I = 1,\ldots,16$ transform trivially under $\SO{8}$. However, the oscillators $\tilde\alpha_{-1}^i$ with $i = 1,\ldots,8$ transform in the $\rep{8}_v$ representation of $\text{SO}(8)$. Thus, we have to decompose the $\text{SO}(8)$ tensor products in eqn.~(\ref{eqn:10dSUGRA}) of weights $q$ and oscillators $\tilde\alpha_{-1}$ into irreducible representations.

We begin with the first case of $q$ being bosonic
\begin{equation}
\rep{8}_v \times \rep{8}_v =  \rep{1} + \rep{28} + \rep{35}_v
\end{equation}
giving rise to 64 bosons: the dilaton $\Phi$ ($\rep{1}$), the antisymmetric two-form $B_{ij}$ ($ \rep{28}$) and the graviton $g_{ij}$ ($\rep{35}_v$). Secondly, in the case of $q$ being fermionic, we find
\begin{equation}
\label{eqn:10dGravitino}
\rep{8}_s \times \rep{8}_v =  \rep{8}_c + \rep{56}_c
\end{equation}
corresponding to the dilatino ($\rep{8}_c$) and the gravitino ($\rep{56}_c$), in total 64 fermions.

This shows that for the massless level we have the same amount of bosonic and fermionic degrees of freedom for both, the vector multiplet of $\E{8}\times\E{8}$ or $\SO{32}$ given in eqns.~(\ref{eqn:masslesshet}) and~(\ref{eqn:16Cartan}) and the supergravity multiplet of eqn.~(\ref{eqn:10dSUGRA}), reflecting $\mathcal{N} = 1$ supersymmetry in 10 dimensions.\index{supergravity!$\mathcal{N} = 1$ in 10d}

\subsubsection{10d Anomaly Cancelation}\index{heterotic string!anomaly cancelation}\index{anomaly!10d anomaly cancelation}

The ten-dimensional heterotic string theories for both $\E{8}\times\E{8}$ and $\SO{32}$ yield anomaly free field theories. Anomalies in 10d can be understood in terms of hexagon diagrams, i.e. one--loop diagrams involving six external legs which can be gravitons and gauge bosons. The purely gravitational anomaly ``$\text{gravity}^6$'' (i.e. with 6 external gravitons) vanishes because the contribution from the dilatino ($\rep{8}_c$) and the gravitino ($\rep{56}_c$) are exactly compensated by the 496 gauginos of $\E{8}\times\E{8}$ or $\SO{32}$. The mixed ``$\text{gauge}-\text{gauge}-\text{gravity}^4$'' anomaly is more involved. It can be expressed by the 10d anomaly polynomial $I_{12}$, which does not seem to vanish at first sight. However, it factorizes for $\E{8}\times\E{8}$ and $\SO{32}$ as $I_{12} \sim X_4 \cdot X_8$, where~\cite{Green:1984sg,Green:1987mn}\footnote{$\mathfrak{R}$ denotes the 10d curvature and $\mathfrak{F}$ the 10d field strength. The trace $\mbox{tr}$ is defined in the ``fundamental'' representation. In the case of $\E{8}\times\E{8}$ it is formally defined via $\mbox{tr} =\frac{1}{30}\mbox{Tr}$, $\mbox{Tr}$ being the standard trace in the adjoint representation.} 
\begin{equation}
X_4 = \mbox{tr}\, \mathfrak{R}^2 - \mbox{tr} (\I \mathfrak{F})^2\;,
\end{equation}  
\begin{equation}
X_8 = \frac{1}{96}\left[\frac{\Tra (\I\frF)^4}{24} - \frac{(\Tra (\I\frF)^2)^2}{7200} - \frac{\Tra (\I\frF)^2 \tr \frR^2}{240} + \frac{\tr \frR^4}{8} + \frac{(\tr \frR^2)^2}{32}\right]\;.
\end{equation}
This anomaly is canceled by the counterterm $\int B \wedge X_8 $ and an anomalous variation of the antisymmetric two-form $B$ under gauge transformations.

\section{Compactification}
\label{sec:Compactification}

In order to make contact with the Standard Model of particle physics or with its minimal supersymmetric extension (the MSSM), we have to hide six spatial dimensions. In the context of the heterotic string, this is done by a compactification on a six-dimensional internal space, i.e. we choose the ten-dimensional target space $M_{10}$ as
\begin{equation}
M_{10} = M_{3,1} \times M_6\;. 
\end{equation}
Shrinking the compact dimensions of $M_6$ to unobservably small sizes leaves us with an effective theory in four-dimensional  Minkowski space-time $M_{3,1}$. Important properties of the effective theory (including the amount of supersymmetry, the gauge group and the massless matter spectrum) are directly related to the geometry and topology of the internal space. The question of low energy $\mathcal{N} = 1$ supersymmetry in 4d severely restricts possible choices for the internal space. Therefore, we will start our discussion with this topic. It will be explained why important examples for these special types of compact spaces are the Calabi-Yau manifold and the orbifold (with an additional condition on the point group).

\subsubsection{6d Compact Spaces with $\boldsymbol{\mathcal{N} = 1}$}

By compactifying on a six-dimensional space $M_6$ we clearly distinguish between our 4d Minkowski space-time and the six internal coordinates. Consequently, the transversal $\text{SO}(8)$ of the ten-dimensional Lorentz group will break. The specific form of this breaking depends on the geometry of the internal space and is, as we will see, directly related to the amount of supersymmetry in 4d. Generically, the breaking is of the form
\begin{equation}
\text{SO}(8) \rightarrow \text{SO}(2) \times \text{SO}(6)
\end{equation}
which is isomorphic to
\begin{equation}
\text{U}(1) \times \text{SU}(4)\;.
\end{equation}
The $\text{U}(1)$ is associated to the uncompactified directions of the 4d Minkowski space-time and can therefore be interpreted as the four-dimensional helicity\index{helicity}. Furthermore, from the 4d low energy point-of-view, the $\text{SU}(4)$ is rather an internal symmetry than a symmetry of space-time. As we will see in the following, 4d bosons and fermions transform differently with respect to this $\text{SU}(4)$. Hence, we identify it as an \emph{$R$-symmetry}\index{R-symmetry} (i.e. a symmetry that does not commute with supersymmetry).

Hence, we analyze the decomposition of the two eight-dimensional representations of $\text{SO}(8)$ that describe ten-dimensional bosons and fermions into representations of $\text{SU}(4)$. For details see appendix~\ref{app:SO8}. The decomposition reads
\begin{eqnarray}
\rep{8}_v & \rightarrow & \rep{6}_{0}   + \rep{1}_{1} + \rep{1}_{-1} \\
\rep{8}_s & \rightarrow & \rep{4}_{1/2} + \rep{\bar{4}}_{-1/2}\;,
\end{eqnarray}
where the subscripts state the $\text{U}(1)$ charges, i.e. the helicities. The remnants of the bosonic $\rep{8}_v$ describe six real scalars ($\rep{6}_{0}$ with spin $0$ in 4d) and one vector field (of both helicities $\rep{1}_{1} + \rep{1}_{-1}$, where the state $\rep{1}_{-1}$ is the CPT-conjugate of $\rep{1}_{1}$ with opposite helicity and thus not independent). In the case of the fermionic $\rep{8}_s$ the two four-plets $\rep{4}_{1/2} + \rep{\bar{4}}_{-1/2}$ are identified as four fermions of spin $1/2$ in 4d.

Using these decompositions for the 10d vector multiplet of $\E{8}\times\E{8}$ or $\SO{32}$ in eqn.~(\ref{eqn:masslesshet}) and~(\ref{eqn:16Cartan}), we can interpret these fields as the particle content of one $\mathcal{N} = 4$ vector multiplet in 4d space-time\footnote{The $\mathcal{N} = 4$ vector multiplet in 4d contains six real scalars, one vector field and four spin $1/2$ fermions.}. However, in order to verify $\mathcal{N} = 4$ we have to count the number of 4d gravitinos\index{gravitino}. Therefore, we decompose the 10d gravitino eqn.~(\ref{eqn:10dGravitino}) into representations of $\SU{4}$
\begin{equation}
\rep{56}_c \rightarrow \rep{4}_{3/2} + \crep{4}_{1/2} + \rep{4}_{-1/2} + \crep{4}_{-3/2} + \crep{20}_{1/2} + \rep{20}_{-1/2}\;.
\end{equation}
The two representations $\rep{4}_{3/2} + \crep{4}_{-3/2}$ correspond to the two helicity states of four spin $3/2$ fermions, the 4d gravitinos. They transform under the internal $\text{SU}(4)$ $R$-symmetry in a four-dimensional representation. Thus, from the 4d perspective we have four gravitini and consequently $\mathcal{N} = 4$ supersymmetry.\index{supergravity!$\mathcal{N} = 4$ in 4d} Since in the case of $\mathcal{N} = 4$ all fields transform in the adjoint representation, this theory is non-chiral and therefore cannot incorporate the chiral particle spectrum of the Standard Model.

\vspace{2mm}

Taking a different perspective will help us to find compact spaces allowing for $\mathcal{N} = 1$ supersymmetry in four dimensions. The internal $\SO{6}$ symmetry assumed at the beginning of this section is the symmetry of a \emph{flat} space $M_6$, in other words a symmetry of a space with trivial holonomy. We denote the parameter (field) of (\emph{local}) SUSY transformations by $\varepsilon$, being a spinor in 10d. Due to the compactification it decomposes into spinorial representations of the internal $\SO{6}$, being the $\rep{4}$ and $\crep{4}$. The $\crep{4}$-plet is the complex conjugate of the $\rep{4}$-plet with opposite chirality and therefore does not describe independent fields. We denote the field associated to the $\rep{4}$-plet by $\eta_i$ with $i=1,\ldots,4$. Since the (flat) internal space has trivial holonomy, these four fields transform as singlets of the trivial holonomy group
\begin{equation}
\rep{4} \rightarrow  \rep{1} + \rep{1} + \rep{1} + \rep{1}\;.
\end{equation}
In other words, all four spinors $\eta_i$ are covariantly constant (i.e. $\nabla_m \eta_i = 0$). The number of covariantly constant spinors on the internal space gives the number of unbroken supersymmetry charges $Q_i$. Thus, we find four unbroken supersymmetry charges $Q_i$, $i=1\ldots,4$, of $\mathcal{N} = 4$ supersymmetry in four dimensions\footnote{Beside covariantly constant spinors, this additionally requires vanishing $H$-flux, a constant dilaton and a vanishing variation of the gauginos, as we will assume.}. 

Finally, one unbroken supersymmetry $Q_1$ needs exactly one covariantly constant spinor $\eta_1$. The other three spinors $\eta_2$, $\eta_3$ and $\eta_4$ of the $\rep{4}$-plet should transform non-trivially. Then, the three associated supersymmetries are broken. This is achieved for example by compact spaces $M_6$ with $\SU{3}$ holonomy~\cite{Candelas:1985en}\index{holonomy group}, in which case the $\rep{4}$-plet decomposes in the desired way
\begin{equation}
\label{eqn:4to3p1}
\rep{4} \rightarrow  \rep{3} + \rep{1}\;,
\end{equation}
and we have a compact space that admits one covariantly constant spinor. These spaces are called \emph{Calabi-Yau} manifolds\index{Calabi-Yau}.

Another possibility to admit only one covariantly constant spinor is generic to orbifold compactifications. The generator $\theta$ of the (Abelian) point group $P$ is represented by the twist vector $v$ (see eqn.~(\ref{eqn:theta})). Its action on a spinor with weight $q$ yields in general a phase
\begin{equation}
\label{eqn:QTrafo}
\theta: |q\rangle \mapsto \exp (-2\pi\I q \cdot v) |q\rangle\;
\end{equation}
as will be discussed later in equation~(\ref{eqn:traforightmover}). We demand exactly one invariant spinor (one parameter $\eta_1$ of $\mathcal{N} = 1$ supersymmetry). We choose the corresponding weight to be
\begin{equation}
\label{eqn:ChoiceForQ}
q = \left(\frac{1}{2},\frac{1}{2},\frac{1}{2},\frac{1}{2}\right)\;.
\end{equation}
Note that if $q$ is invariant, the weight $-q$ corresponding to the CPT-conjugate of $\eta_1$ will be invariant, too. By fixing our choice for $q$ we have to impose the following condition on the twist vector
\begin{equation}
v_1 + v_2 + v_3 = 0\;,
\end{equation}
such that the transformation in eqn.~(\ref{eqn:QTrafo}) is trivial for eqn.~(\ref{eqn:ChoiceForQ}), but non-trivial for the other spinor components. Since additionally $v^0 = 0$ we see that the point group $P$ is a subgroup of $\SU{3} \subset \SU{4}$.

\subsubsection{Remark: Right-Moving Momenta in Index Notation}

Sometimes, it is convenient to use an index notation for the right-moving momenta $q$. This notation is also used in the literature quite frequently (e.g.~\cite{Green:1987sp,Becker:2007zj}). Thus, we will review it briefly, but specialize to the cases we will need later. Consider the bosonic states $|q\rangle\RM$ with weights $q = (0,\underline{\pm 1,0,0})$, corresponding to the six compactified dimensions of $M_6$.  To each $q$ we can associate either a holomorphic index $i$ or an anti-holomorphic index $\bar{i}$. In detail, this reads
\begin{equation}
|q_{(i)}\rangle\RM \Leftrightarrow |i\rangle\RM \quad\text{and}\quad |q_{(\bar{i})}\rangle\RM \Leftrightarrow |\bar{i}\rangle\RM
\end{equation}
with $i = 1,2,3$ and $\bar{i} = \bar{1},\bar{2},\bar{3}$, and 
\begin{equation}
\label{eqn:QIndex}
\begin{array}{ll}
q_{(1)} = (0,-1,0,0)\;, & q_{(\bar{1})} = (0,1,0,0)\;, \\
q_{(2)} = (0,0,-1,0)\;, & q_{(\bar{2})} = (0,0,1,0)\;, \\
q_{(3)} = (0,0,0,-1)\;, & q_{(\bar{3})} = (0,0,0,1)\;. \\
\end{array}
\end{equation}
Thus, an anti-holomorphic index $\bar{i}$ transforms in the complex conjugate representation of the holomorphic index $i$, compare to eqn.~(\ref{eqn:QTrafo}).

\section{Strings on Orbifolds}\index{strings on orbifolds}
\label{sec:HetOnOrb}

Now we are prepared to compactify heterotic strings on orbifolds~\cite{Dixon:1985jw,Dixon:1986jc}. In other words, we choose the internal part of the ten-dimensional target space $M_{10}$ as the quotient space of an orbifold
\begin{equation}
M_6 =  \mathbbm{C}^3 / S\;.
\end{equation}
We restrict ourselves to toroidal, Abelian and symmetric orbifolds. Examples for asymmetric orbifolds, where right- and left-moving degrees of freedom are compactified on different six-dimensional spaces, are considered in~\cite{Narain:1986qm,Ibanez:1987pj}. Furthermore, examples of non-Abelian orbifolds having non-Abelian point groups like $A_4$ can be found in e.g.~\cite{Bailin:1987fw}. We start our discussion with the investigation of boundary conditions for closed strings on orbifolds. In this section, the main equations which are relevant for the computation of the massless spectrum are highlighted by boxes.

\subsubsection{Boundary Conditions}

\begin{figure}[t]
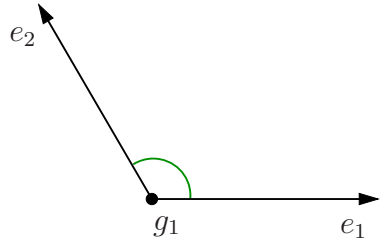

\centerline{\input SU3Z3TwistedString.pstex_t}\vspace{-0.2cm}
\caption{Two-dimensional $\Z{3}$ orbifold with point group generator $\theta = e^{2\pi \I/3}$. A twisted string with constructing element $g_1=\left(\theta,0\right)$ localized at the origin is depicted.}
\label{fig:Z3TwistedString}
\end{figure}

On the orbifold there are more boundary conditions that lead to closed strings than on flat ten-dimensional Minkowski space, i.e. there are new closed strings that are closed only up to the action of some space group element. These new boundary conditions read in the case of the bosonic degrees of freedom $X^i(\tau, \sigma+\pi) = (g X)^i(\tau, \sigma)$ with $i = 1, \ldots, 6$, or equivalently in the complex basis
\begin{equation}
\label{eqn:twistedbc}
Z^i(\tau, \sigma+\pi) = (g\, Z)^i(\tau, \sigma) \quad i = 1,2,3\;.
\end{equation}\index{strings on orbifolds!constructing element}
$g \in S$ is called the \emph{constructing element} of the string. It is the element of the space group that maps the one end of the string to the other one, yielding a closed string on the quotient space of the orbifold, see figure~(\ref{fig:Z3TwistedString}). In the case of $g = \left(\mathbbm{1}, 0 \right)$ the boundary condition eqn.~(\ref{eqn:twistedbc}) reduces to the boundary condition of the uncompactified heterotic string eqn.~(\ref{eqn:closedstring}). According to the grouping into untwisted and twisted sectors this string belongs to the untwisted sector. On the other hand, strings with non-trivial constructing elements belong to some twisted sectors and are thus called \emph{twisted strings}. 

\subsubsection{Twisted Mode Expansion}

Since the orbifold is a flat space, except for the singularities at the fixed points, we can write down the general solutions to the equations of motion in terms of simple mode expansions, like in the case of the uncompactified heterotic string, see e.g.~\cite{Nilles:1987uy,Ibanez:1987dw}.

First, we choose a constructing element $g = \left(\theta^{k_1} \omega^{k_2}, n_\alpha e_\alpha\right) \in S$. The corresponding twisted string is subject to the so-called \emph{local twist} $v_g \equiv k_1 v_1 + k_2 v_2$~\footnote{Analogously in the case of $\Z{N}$ orbifolds: if $g = \left(\theta^k, n_\alpha e_\alpha\right) \in S$ then the local twist reads $v_g \equiv k v$.}. From the bosonic mode expansion on can infer that the center of mass of a twisted string is attached to the fixed point (or fixed torus) associated to $g$. The twisted string cannot move apart; it is \emph{localized} at the fixed point. This implies that the components of the internal momentum corresponding to the directions of the fixed point vanish. In the case of a fixed torus only those components of the internal momentum can be non-zero that point into the directions of the torus.

Furthermore, due to the non-trivial twist $v_g$, the mode expansions of a twisted string contain \emph{twisted oscillators}\index{oscillator!twisted}, i.e. oscillators with fractional frequencies: for example $\tilde\alpha_{-1/3}$ for the bosonic coordinates on a $\Z{3}$ orbifold. As these oscillators originate from the mode expansions of the bosonic coordinates $Z^i$ or their complex conjugates $Z^{\bar{i}} \equiv (Z^i)^*$, they are equipped with holomorphic indices $i=1,2,3$ or antiholomorphic ones $\bar{i} = \bar{1},\bar{2},\bar{3}$, respectively. Figuratively speaking, an oscillator with index $i$ or $\bar{i}$ acts in the $i$-th complex plane. In general, the twisted (bosonic) oscillators read $\tilde\alpha^i_{n-\omega_i}$ or $\tilde\alpha^{\bar{i}}_{n+\omega_i}$, where $n \in \Z{}$ and
\begin{equation}
\label{eqn:omega}
\omega_i = (v_g)_i \mod 1 \quad\text{such that}\quad 0 \leq \omega_i < 1\;.
\end{equation}
Note that $\omega_i$ will appear quite frequently in the following. As usual, an oscillator with negative frequency corresponds to a creation operator and an oscillator with positive frequency to an annihilation operator. As the zero-point energies\index{zero-point energy} of the left- or right-moving sectors depend on the respective oscillator contents, the presence of twisted oscillators modifies them. However, it turns out that the shift $\delta c$ in the zero-point energies is the same in both sectors and reads
\begin{equation}
\label{eqn:deltac}
\fbox{$ \displaystyle \delta c = \frac{1}{2}\sum_{i=1}^3 w_i (1 - w_i)\;. $}
\end{equation}

Furthermore, due to the presence of twisted oscillators in the mode expansion of the fermionic degrees of freedom $\Psi\RM^i$, the right-moving momenta $q$ are shifted by the local twist $v_g$. Consequently, the equation for massless right-movers reads
\begin{equation}
\label{eqn:masslesstwistedrightmover}
\fbox{$ \displaystyle \frac{(q+v_g)^2}{2} - \frac{1}{2} + \delta c = 0\;. $}
\end{equation}
As before, $q$ is restricted to be from the vector or from the spinor weight lattice of $\text{SO}(8)$. Furthermore, we define the \emph{shifted right-moving momentum} $q_\text{sh} = q + v_g$. 

However, we cannot write down the mass equation for the left-movers now. The reason being that the left-moving sector of the compactified heterotic string has to be changed more than just by the six-dimensional orbifold in order to lead to a consistent theory in 4d. We will discuss this in detail later, after we considered the transformation properties of right-movers and twisted oscillators.

\subsubsection{Transformation of Right-Movers and Oscillators under the Space Group}

On the orbifold, right-moving states transform under the action of the space group. In order to see this, we express a given state $|q_\text{sh}\rangle\RM$ by its corresponding vertex operator (see e.g. appendix C of~\cite{Kobayashi:2004ya})
\begin{equation}
e^{-2 q_\text{sh}\cdot H}\;,
\end{equation}
where $H^i$, $i=1,\ldots,4$ are the four bosonized coordinates. Under the action of some space group element $h\in S$, they are shifted according to $H \mapsto H + \pi v_h$. Consequently the right-moving state $|q_\text{sh}\rangle\RM$ acquires a phase
\begin{eqnarray}
\label{eqn:traforightmover}
|q_\text{sh}\rangle\RM & \stackrel{h}{\mapsto} & e^{-2\pi\I q_\text{sh} \cdot v_h} |q_\text{sh}\rangle\RM\;.
\end{eqnarray}

Also the bosonic oscillators transform under the action of the space group. Since we will need this only for the left-moving oscillators $\tilde\alpha^i_{n-\omega_i}$ and $\tilde\alpha^{\bar{i}}_{n+\omega_i}$, we will restrict to this case in the following. As a given oscillator carries either a holomorphic index $i$ or an antiholomorphic one $\bar{i}$, it transforms according to the rotation $v_h^i$ in the $i$-th complex plane, i.e.
\begin{eqnarray}
\label{eqn:OsciTrafo}
\tilde\alpha^i_{n-\omega_i}         & \stackrel{h}{\mapsto} & e^{+2\pi\I v_h^i} \; \tilde\alpha^i_{n-\omega_i} \\
\tilde\alpha^{\bar{i}}_{n+\omega_i} & \stackrel{h}{\mapsto} & e^{-2\pi\I v_h^i} \; \tilde\alpha^{\bar{i}}_{n+\omega_i} \nonumber\;.
\end{eqnarray}
Now, we can turn back to the question of the orbifolded left-movers.

\subsubsection{Gauge Embedding}

Modular invariance of the theory (see section~\ref{sec:SummaryConditions}) forbids the orbifold action to be restricted only to the six-dimensional compact space $M_6$. It is necessary to extend the action of the orbifold to the gauge degrees of freedom $X^I\LM$ of the left-moving sector. Thus, the space group $S$ is not enough to define a consistent heterotic orbifold model. We need to define a group, the so-called \emph{gauge twisting group}\index{strings on orbifolds!gauge twisting group} denoted by $G$, that acts on the 16 gauge degrees of freedom $X^I\LM$, i.e. $X\LM \stackrel{h}{\mapsto} h\,X\LM$ with $h \in G$ and points that are mapped to each other are identified $X\LM \sim hX\LM$. Therefore, one can understand $G$ as a group that defines a 16-dimensional orbifold on the internal coordinates $X^I\LM$. In general, $G$ corresponds to an automorphism of the Lie algebra of $\E{8}\times\E{8}$ or $\text{Spin}(32)/ \mathbbm{Z}_2$. However, it is known that any (inner) automorphism can be realized as a \emph{shift}\index{strings on orbifolds!shift}~\cite{Dixon:1986jc}
\begin{equation}
X\LM \stackrel{h}{\mapsto} X\LM + \pi V\;,
\end{equation}
which acts freely. Consequently, the 16-dimensional orbifold in the gauge degrees of freedom has neither fixed points nor fixed tori. 

The space group $S$ (defining the orbifold of the six-dimensional space) and the gauge twisting group $G$ (defining the one of the gauge degrees of freedom) are forced to act simultaneously due to modular invariance. In other words, $G$ is an embedding of the space group $S$ acting in the gauge degrees of freedom
\begin{equation}
S \hookrightarrow G\;.
\end{equation}
Explicitly, the simultaneous action of $S$ and $G$ on the 3 complex and 16 real coordinates reads
\begin{equation}
Z    \;\; \stackrel{h}{\mapsto} \;\; \theta Z \quad\text{ and }\quad X\LM \;\; \stackrel{h}{\mapsto} \;\; X\LM + \pi V\;,
\end{equation}
for a transformation under the twist $h = \left(\theta,0;\; V, 0\right)$. Since the shift $V$ is the embedding of the twist $\theta$, it needs to be of the same order $N$: since $\theta^N = \mathbbm{1}$ is the identity on $M_6$, $NV$ must also act trivially on the gauge degrees of freedom. We will discuss this in detail later. We name the full, consistent group that acts on both, the three complex-dimensional space $M_6$ and the gauge degrees of freedom $X^I\LM$, as the \emph{orbifold group}\index{strings on orbifolds!orbifold group} $O$. An element of $O$, in the case of a $\mathbbm{Z}_N \times \mathbbm{Z}_M$ point group, is of the form
\begin{equation}
\label{eqn:OrbifoldGroupElement}
h = \left(\theta^{t_1} \omega^{t_2}, m_\alpha e_\alpha;\; t_1 V_1 + t_2 V_2, m_\alpha A_\alpha\right) \in O\;,
\end{equation}
and acts on the coordinates according to
\begin{eqnarray}
Z    & \stackrel{h}{\mapsto} & \theta^{t_1} \omega^{t_2} Z + m_\alpha e_\alpha \\
X\LM & \stackrel{h}{\mapsto} & X\LM + \pi \left(t_1 V_1 + t_2 V_2 + m_\alpha A_\alpha\right)\;.
\end{eqnarray}
That is, a twist $\theta^{t_1} \omega^{t_2}$ is accompanied by a shift $t_1 V_1 + t_2 V_2$ and a torus lattice vector $m_\alpha e_\alpha$ by $m_\alpha A_\alpha$. In detail, whenever we go along a torus direction $e_\alpha$ the gauge degrees of freedom are shifted by $A_\alpha$. This induces a phase $e^{2\pi\I p\cdot A_\alpha}$ which depends on the momentum $p$, i.e.\footnote{A state $|p\rangle\LM$ with momentum $p$ corresponds to a vertex operator $e^{2\I p \cdot X\LM}$, see e.g. appendix C of~\cite{Kobayashi:2004ya}.}
\begin{equation}
\label{eqn:WLPhase}
e^{2\I p \cdot X\LM}  \; \stackrel{e_\alpha}{\mapsto}\;  e^{2\I p\cdot (X\LM + \pi A_\alpha)} = e^{2\pi\I p\cdot A_\alpha} e^{2\I p \cdot X\LM} \quad\Rightarrow\quad |p\rangle\LM \; \stackrel{e_\alpha}{\mapsto}\; e^{2\pi\I p\cdot A_\alpha}|p\rangle\LM  \;.
\end{equation}
Hence, the shifts $A_\alpha$ are called \emph{Wilson lines}\index{Wilson line}~\cite{Dixon:1986jc,Ibanez:1986tp}.

Analogously to eqn.~(\ref{eqn:WLPhase}), we see that under the action of some general orbifold group element $h$ of eqn.~(\ref{eqn:OrbifoldGroupElement}) a left-mover with momentum $p$ acquires a phase
\begin{equation}
e^{2\pi\I p \cdot (t_1 V_1 + t_2 V_2 + m_\alpha A_\alpha)}\;.
\end{equation}

\subsubsection{Twisted Mode Expansion II}

Now, we have defined two equivalence relations on the gauge degrees of freedom. First of all, the coordinates $X\LM$ are compactified on the 16-torus $\Lambda$: $X\LM \sim X\LM + \pi p$ for $p \in \Lambda$. Secondly, the gauge twisting group $G$ defines an orbifold of the 16 gauge degrees of freedom, namely
\begin{equation}
X\LM \sim X\LM + \pi \left(k_1 V_1 + k_2 V_2 + n_\alpha A_\alpha\right)\;.
\end{equation}
The combination of both yields new boundary conditions for strings being closed in the gauge degrees of freedom
\begin{eqnarray}
\label{eqn:TwistedBoundaryConditionXL}
X\LM(\tau + \sigma + \pi) & = & g\; X\LM(\tau + \sigma) + \pi p\\
                          & = & X\LM(\tau + \sigma) + \pi\left(p + k_1 V_1 + k_2 V_2 + n_\alpha A_\alpha\right)\;,
\end{eqnarray}
i.e. they are only closed up to the action of the element $g = \left(k_1 V_1 + k_2 V_2, n_\alpha A_\alpha\right) \in G$ and a lattice shift with $p \in \Lambda$. We define
\begin{equation}
V_g \equiv k_1 V_1 + k_2 V_2 + n_\alpha A_\alpha \quad\text{and}\quad p_\text{sh} \equiv p + V_g\;
\end{equation}
as the \emph{local shift} $V_g$ associated to $g$ and the \emph{shifted momentum} $p_\text{sh}$, respectively. The mode expansion for a twisted string, $X\LM(\tau + \sigma) = x + p_\text{sh}(\tau + \sigma) + \text{oscillators}$, gives the general solution of the equation of motion, which is compatible with the boundary condition eqn.~(\ref{eqn:TwistedBoundaryConditionXL}). There, we see that $p_\text{sh}$ defines the internal momentum of this twisted state. Being from the gauge degrees of freedom, $p_\text{sh}$ has the additional interpretation as the \emph{weight} defining the representation under gauge transformations.

Having defined a twisted state $|p_\text{sh}\rangle\LM$ using the boundary conditions $g = \left(k_1 V_1 + k_2 V_2, \right.$ $\left. n_\alpha A_\alpha\right) \in G$, we can transform it along the direction $h = \left(t_1 V_1 + t_2 V_2, m_\alpha A_\alpha\right) \in G$. From the vertex operator $e^{2\I p_\text{sh}\cdot X\LM}$, we see that it acquires a phase
\begin{equation}
|p_\text{sh}\rangle\LM \; \stackrel{h}{\mapsto}\; e^{2\pi\I p_\text{sh} \cdot V_h}\; |p_\text{sh}\rangle\LM\;.
\end{equation}

Furthermore, the equation for massless left-movers from a twisted sector with constructing element $g$ reads
\begin{equation}
\label{eqn:masslesstwistedleftmover}
\fbox{$ \displaystyle \frac{(p+V_g)^2}{2} + \tilde{N} - 1 + \delta c = 0\;, $}
\end{equation}
where $V_g$ is the local shift and $\delta c$ as defined in eqn.~(\ref{eqn:deltac}). Note that the oscillator number $\tilde{N}$ now sums over the six twisted oscillators of the six compactified dimensions. Therefore, it can have fractional eigenvalues, e.g. the eigenvalue of $\tilde{N}$ for $\tilde\alpha_{-1/3}^i$ is $1/3$.

For computational reasons, we want to know which frequencies $n-\omega_i$ or $-n+\omega_i$ with $n\in\mathbbm{N}$ of twisted oscillators can in principle appear in the massless spectrum. Since $\tilde{N} \geq 0$ and $\delta c > 0$ for some non-trivial constructing element $g \in S$, we know that the solutions to eqn.~(\ref{eqn:masslesstwistedleftmover}) are constrained by
\begin{equation}
(p + V_g)^2 < 2 \quad\text{and}\quad \tilde{N} \leq 1 - \delta c\;.
\end{equation}
Consequently, only combinations of the twisted oscillators (having negative frequencies)
\begin{equation}
\fbox{$ \displaystyle \tilde\alpha^i_{-\omega_i} \quad\text{and}\quad\tilde\alpha^{\bar{i}}_{-1+\omega_i} $}
\end{equation}
for $i=1,2,3$ and $\bar{i}=\bar{1},\bar{2},\bar{3}$ can potentially yield massless excited left-movers.

\subsubsection{Transformation Phase}\index{strings on orbifolds!transformation phase}

From the previous discussions we know the transformation properties of each part of a twisted state under the action of the Orbifold group. Here, we want to summarize these results and complete them by introducing the so-called \emph{vacuum phase}.\index{strings on orbifolds!vacuum phase}

We start with a state $|q_\text{sh}\rangle\RM \otimes \tilde\alpha |p_\text{sh}\rangle\LM$ that describes a closed string with constructing element $g\in S$ and is possibly excited by some oscillators $\tilde\alpha$. Under the action of some element $h\in S$, the state transforms with a phase
\begin{equation}
|q_\text{sh}\rangle\RM \otimes \tilde\alpha|p_\text{sh}\rangle\LM \; \stackrel{h}{\mapsto}\; \Phi |q_\text{sh}\rangle\RM \otimes \tilde\alpha|p_\text{sh}\rangle\LM\;.
\end{equation}
The transformation phase $\Phi$ reads in detail
\begin{equation}
\label{eq:transformationphase}
\fbox{$ \displaystyle \Phi ~\equiv~ e^{2\pi\I\,[p_\text{sh}\cdot V_h - R \cdot v_h ]}\, \Phi_\text{vac}\;. $}
\end{equation}
The last term of this equation, the vacuum phase $\Phi_\text{vac}$, is given by\footnote{More details about this extra phase can be found for example at the end of section 3 in reference~\cite{Ibanez:1987pj} or in section 5 of reference~\cite{Kobayashi:1991rp}.}
\begin{equation}
\label{eqn:vacuumphase}
\fbox{$ \displaystyle \Phi_\text{vac} ~=~ e^{2\pi\I\,[-\frac{1}{2}(V_g\cdot V_h - v_g\cdot v_h)]}\;, $}
\end{equation}
compare to appendix~\ref{app:invariant} and appendix A of~\cite{Ploger:2007iq}. Furthermore, in order to summarize the transformation properties of $q_\text{sh}$ and of the oscillators we have introduced the so-called \emph{R--charge}\index{R--charge}. It is defined as
\begin{equation}
\fbox{$ \displaystyle R^i~\equiv~q_\text{sh}^i - \tilde{N}^i + \tilde{N}^{*i} \;. $}
\end{equation}
$\tilde{N}^i$ and $\tilde{N}^{*i}$, $i=0,\ldots,3$, are integer oscillator numbers, counting the number of oscillators $\tilde{\alpha}^i$ and $\tilde{\alpha}^{\bar i}$ acting on the ground state $|p\rangle\LM$, respectively. In detail, they are given by splitting the eigenvalues of the number operator $\tilde{N}$ according to
\begin{equation}
\label{eqn:oscillatornumbersplit}
\tilde{N} = \omega_i \tilde{N}^i + \bar{\omega}_i \tilde{N}^{*i}\;,
\end{equation}
where $\omega_i = (v_g)_i \mod 1$ and  $\bar{\omega}_i = -(v_g)_i \mod 1$ such that $0 \leq \omega_i, \bar{\omega}_i < 1$.

\subsection{Physical States}
\label{sec:physstates}

Since the massless string is completely specified by its constructing element $g$, its left- and right--moving shifted momenta $p_\text{sh}$ and $q_\text{sh}$ and possible oscillator excitations $\tilde{\alpha}$, we write down a first ansatz for a physical state from the Hilbert space ${\mathcal H}_g$ on an orbifold
\begin{equation} \label{eqn:physicalstate1}
|\text{phys}\rangle~\sim~ |q_\text{sh}\rangle\RM \otimes \tilde{\alpha} |p_\mathrm{sh}\rangle\LM \otimes |g\rangle\;.
\end{equation}

Up to now it is not guaranteed that a physical state is actually compatible with the orbifold. To ensure this compatibility, invariance of $|\text{phys}\rangle$ under the action of all elements of the orbifold group $O \subset S \otimes G$ must be imposed. To do so, the boundary condition for twisted strings eqn.~(\ref{eqn:twistedbc}) is multiplied by an arbitrary element $h \in S$:
\begin{eqnarray}
                 h\,Z(\tau, \sigma + \pi) & = & h\,g\,Z(\tau, \sigma) \\
\Leftrightarrow  h\,Z(\tau, \sigma + \pi) & = & h\,g\,h^{-1}\,h\,Z(\tau, \sigma) \label{eqn:boundarytimesh}
\end{eqnarray}

For keeping the expressions simple, we choose shifts and Wilson lines such that the vacuum phase\index{strings on orbifolds!vacuum phase} $\Phi_\text{vac} = 1$ vanishes. Now, we can distinguish two cases:

\subsubsection{Commuting Elements: $\boldsymbol{[h,g] = 0}$}

First, let us consider the transformation property of $|\text{phys}\rangle$ with respect to a commuting element $h$. Note that geometrically a commuting element can be interpreted either as being associated to the same fixed point as $g$ or as acting in directions orthogonal to the ones in which $g$ acts, see appendix~\ref{app:commuting}. In the case of commuting elements, the boundary condition eqn.~(\ref{eqn:boundarytimesh}) reads
\begin{equation}
h\,Z(\tau, \sigma + \pi)~=~g\,h\,Z(\tau, \sigma)\;,
\end{equation}
i.e., the constructing element $g$ is invariant under the action of $h$,
\begin{equation} \label{eqn:constr_element_case1} 
|g\rangle  \ \stackrel{h}{\rightarrow} \ |h\,g\,h^{-1}\rangle~=~|g\rangle \;.
\end{equation}
$hZ$ closes under the same constructing element $g$ as $Z$. Thus, both give rise to the same Hilbert space ${\mathcal H}_g \ \stackrel{h}{\rightarrow} \ {\mathcal H}_{hgh^{-1}} = {\mathcal H}_g$. Furthermore, on the orbifold space ${\mathbb C}^3/S$ the string coordinates $hZ$ and $Z$ are identified. Thus, $hZ$ and $Z$ describe the same physical state.

In summary, provided a constructing element $g$, we have shown that for commuting elements $h$, $h\,Z$ and $Z$ give rise to the same physical states from the same Hilbert space. Since $h$ has to act as the identity on $|\text{phys}\rangle$, the following condition follows using eqns.~(\ref{eqn:physicalstate1}), (\ref{eq:transformationphase}) and~(\ref{eqn:constr_element_case1}):
\begin{equation}
\label{eqn:projectout}
\fbox{$ \displaystyle  p_\text{sh}\cdot V_h - R\cdot v_h  ~\stackrel{!}{=}~0 \mod 1\phantom{I_q}. $}
\end{equation}
Note that in the general case the contribution from the vacuum phase eqn.~(\ref{eqn:vacuumphase}) has to be included here. If the state $|\text{phys}\rangle$ does not fulfill the invariance condition eqn.~(\ref{eqn:projectout}) it is not ``compatible'' with the orbifold space and hence needs to be removed from the spectrum: non-invariant states are projected out.

In other words, the total vertex operator of the state with boundary condition $g$ has to be single-valued when transported along $h$ if $h$ is an allowed loop $[h,g] = 0$. 

\subsubsection{Non--Commuting Elements: $\boldsymbol{[h,g] \neq 0}$}

Next, considering a non--commuting element $h$ eqn.~(\ref{eqn:boundarytimesh}) yields
\begin{equation}
h\,Z(\tau, \sigma + \pi) ~=~ \left(h\,g\,h^{-1}\right)\,h\, Z(\tau, \sigma)\;,
\end{equation}
i.e., the constructing element $g$ is not invariant under the action of $h$,
\begin{equation} \label{eqn:mapgtohgh}
|g\rangle  ~ \xrightarrow{h}~ |h\,g\,h^{-1}\rangle \neq |g\rangle\;.
\end{equation}
In the upstairs picture, i.e. in the covering space ${\mathbbm C}^3$ of the orbifold ${\mathbbm C}^3/S$, one has different Hilbert spaces for the states with boundary conditions $g$ and $h\,g\,h^{-1}$. In this picture, eqn.~(\ref{eqn:mapgtohgh}) says that $h$ maps states from a given Hilbert space $\mathcal{H}_g$ onto a different Hilbert space $\mathcal{H}_{h\,g\,h^{-1}}$. Subsequent application of $h$ then leads to the sequence \footnote{Note that in all ${\mathcal H}_{h^ngh^{-n}}$ the left--moving momenta $p_\text{sh}$ of equivalent states are identical. The same holds for $q_\text{sh}$ and $R$.}
\begin{equation}
\mathcal{H}_g ~ \xrightarrow{h}~ \mathcal{H}_{h\,g\,h^{-1}}~ \xrightarrow{h}~ \mathcal{H}_{h^2\,g\,h^{-2}}~ \xrightarrow{h}~ \mathcal{H}_{h^3\,g\,h^{-3}} ~ \xrightarrow{h}~ \ldots\;.
\end{equation}
The crucial point is now that on the orbifold $h\,Z$ and $Z$ are identified. This means that, on the orbifold, the different Hilbert spaces $\mathcal{H}_{h^n\,g\,h^{-n}}$ of the upstairs picture are to be combined into a single orbifold Hilbert space  $\mathcal{H}_{[g]}$. Invariant states are then linear combinations of states from all $\mathcal{H}_{h^n\,g\,h^{-n}}$. Such linear combinations do, in general, involve relative phase factors (often called \emph{gamma--phase}\index{strings on orbifolds!gamma-phase} $\gamma$). So, the new ansatz for a physical state reads:
\begin{eqnarray}
|\text{phys}\rangle &\sim& \sum_{n} \left( e^{-2\pi \I\, n\, \gamma}\, |q_\text{sh}\rangle\RM \otimes \tilde{\alpha}|p_\text{sh}\rangle\LM \otimes |h^n\,g\,h^{-n}\rangle\right) \nonumber\\
                    &  = & |q_\text{sh}\rangle\RM \otimes \tilde{\alpha} |p_\mathrm{sh}\rangle\LM \otimes \left(\sum_{n} e^{-2\pi \I\, n\, \gamma}\ |h^n\,g\,h^{-n}\rangle\right)\;,
\label{eqn:physicalstate2}
\end{eqnarray}
where $\gamma = \text{integer}/N$, $N$ being the order of the orbifold. The geometrical part of the linear combination transforms non--trivially under $h$
\begin{equation}
\label{eqn:constr_element_case2}
\sum_{n} e^{-2\pi \I n \gamma}\ |h^n\,g\,h^{-n}\rangle\, \stackrel{h}{\rightarrow} \ e^{2\pi \I\,\gamma} \sum_{n} e^{-2\pi \I\, n\, \gamma}\ |h^n\,g\,h^{-n}\rangle\;.
\end{equation}
Since $h$ has to act as the identity on $|\text{phys}\rangle$, the following condition follows using eqns.~(\ref{eq:transformationphase}), (\ref{eqn:physicalstate2}) and~(\ref{eqn:constr_element_case2}) for non--commuting elements:
\begin{equation}
\label{eqn:projectionwithgamma} 
p_\text{sh}\cdot V_h - R\cdot v_h + \gamma~\stackrel{!}{=}~0 \mod 1\;.
\end{equation}
Notice that $\gamma$ depends on $h$. Thus, we can always choose $\gamma (h)$ such that this condition is satisfied\footnote{In this sense, building linear combinations and computing the $\gamma$ phase is not a projection condition. Note that $\gamma (h)$ is well--defined: if $h_1gh_1^{-1} = h_2gh_2^{-1}$ then $\gamma (h_1) = \gamma (h_2)$.}. In principle, these steps have to be repeated for all non--commuting elements in order to ensure invariance of the physical state under the action of the whole orbifold group $O \subset S \otimes G$. The result for $|\text{phys}\rangle$ reads
\begin{equation}
|\text{phys}\rangle ~=~ |q_\text{sh}\rangle\RM \otimes \tilde{\alpha}|p_\text{sh}\rangle\LM \otimes \left( \sum_{h = \mathbbm{1} \text{ or } [h,g]\neq 0} e^{-2\pi \I \gamma(h)}\ |h\,g\,h^{-1}\rangle \right)\;,
\end{equation}
where the summation over $h$ is such that each term $|h\,g\,h^{-1}\rangle$ appears only once. Note that the summation over $h$ can be understood as a summation over all representatives of the conjugacy class of $g$.

\subsubsection{Example}

\begin{figure}[t!]
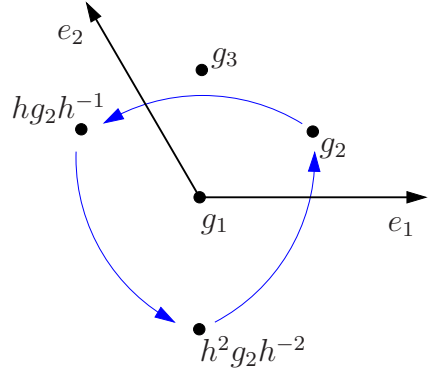

\centerline{\input SU3TorusLinearCombination.pstex_t}\vspace{-0.3cm}
\caption{Illustration of the $\gamma$--factor. The fixed point associated with the space group element $g_2=(\theta,\,e_1)$ is invariant under $(\theta,\,e_1)$, but transforms into equivalent fixed points outside the fundamental domain under $h=(\theta,0)$. To form an eigenstate of $(\theta,0)$, one needs to build linear combinations of  the equivalent fixed points. The corresponding eigenvalues can be $1, e^{\pm 2\pi \I/3}$.}
\label{fig:gammarule}
\end{figure}

To illustrate the construction of physical states, let us consider an example in the twisted sector of a two-dimensional $\Z{3}$ orbifold\index{orbifold!$\Z{3}$}. In the SU(3) lattice spanned by $e_1$ and $e_2$, there are three inequivalent fixed points associated to the constructing elements $g_1=(\theta,\, 0)$, $g_2=(\theta,\, e_1)$ and $g_3=(\theta,\, e_1+e_2)$, or analogously $g_i = (\theta,\, a_i\, e_1 + b_i\, e_2)$ for $i=1,2,3$ with $a_i=(0,\,1,\,1)$ and $b_i=(0,\,0,\,1)$ (compare to figure~(\ref{fig:Z3FixedPoints})). Then, using $h=(\mathbbm{1},ne_1+me_2)$, the geometrical part of a physical state can be written as
\begin{equation} \label{eqn:SU3_linear_combination}
\sum_{n,m} e^{-2\pi \I (n+m) \gamma}\ \big|(\theta,(n+m+a_i)\,e_1+(2m-n+b_i)\,e_2)\big\rangle \;.
\end{equation}
Since the action of $\theta$ has order 3, the only possible $\theta$--eigenvalues of eqn.~(\ref{eqn:SU3_linear_combination}) have $\gamma=0,\,\pm\frac{1}{3}$. In the case of $\gamma=0$, eqn.~(\ref{eqn:SU3_linear_combination}) is invariant under all rotations and translations for all three $g_i$. However, if $\gamma=\pm\frac{1}{3}$,  the eigenvalue of eqn.~(\ref{eqn:SU3_linear_combination}) depends on $g_i$: for the fixed point at the origin associated to $g_1$, eqn.~(\ref{eqn:SU3_linear_combination}) is invariant under $\theta$, but has an eigenvalue
\begin{equation}
e^{2 \pi \I\,\gamma\,(k+l)} \quad\text{under}\quad \left(\mathbbm{1}, ke_1+le_2\right)\;.
\end{equation}
Similarly, for the fixed points away from the origin, corresponding to $g_i$ ($i \neq 1$), eqn.~(\ref{eqn:SU3_linear_combination}) picks up a phase
\begin{equation}
e^{- 2 \pi \I\,\gamma\,(a_i + b_i)} \quad\text{under}\quad \theta\;, 
\end{equation}
see figure~\ref{fig:gammarule}. It can be shown that for physical states $\gamma \neq 0$ is only possible in the presence of a Wilson line in the $e_1$ and $e_2$ directions.

\subsubsection{In Practice only Commuting Elements: $\boldsymbol{[h,g] = 0}$}

As we have seen in the last sections, it is necessary to enforce invariance of physical states under the action of the whole orbifold group $O$. For strings from the untwisted sector this condition cannot be weakened: all elements $h\in S$ commute with the constructing element $\left(\mathbbm{1},0\right)$ of the untwisted sector and hence we must project on the $h$-invariant subspace of the untwisted sector.

However, in the case of twisted strings we can weaken this condition. The first step, to project out states not invariant under commuting elements, remains unchanged. The second step, to build linear combinations involving the gamma-phase $\gamma$ can be omitted. The gamma-phases are not needed for the computation of the massless spectrum. It is always possible to \emph{choose} the gamma-phase $\gamma(h)$ for a given state such that this state remains invariant under $h$. From this perspective, there is no projection involved in building linear combinations. Furthermore, we will show later in section~\ref{sec:Yukawa} that also for the computation of the allowed Yukawa couplings the gamma-phases are redundant.

In summary, the procedure to compute the massless spectrum reads: for each inequivalent constructing element $g$ we first have to solve the equations for massless strings (eqns.~(\ref{eqn:masslesstwistedrightmover}) and~(\ref{eqn:masslesstwistedleftmover})). Next, we have to identify the commuting elements $h$ (for example by the method presented in appendix~\ref{app:commuting}) and finally project out the non-invariant states (according to eqn.~(\ref{eqn:projectout})).

\subsection{Conditions on the Gauge Embeddings}
\label{sec:ConditionsGaugeEmbedding}

The following considerations will lead us to some consistency conditions on the gauge embeddings, i.e. the shift $V$ and the Wilson lines $A_\alpha$. We will derive them explicitly for the case of $\mathbbm{Z}_N$ orbifolds, but just state the results for the $\mathbbm{Z}_N \times \mathbbm{Z}_M$ case, which can be derived analogously.

\subsubsection{The Order of the Shift}

Consider a $\mathbbm{Z}_N$ orbifold with $\theta$ being the order $N$ generator of the point group. Now, act $N$ times with an orbifold group element $g =\left(\theta,0;V,\; 0\right)$ on the 3 complex + 16 real bosonic coordinates $Z$ and $X\LM$.
\begin{equation}
\label{eqn:ShiftOrder}
\begin{array}{lll}
Z    & \stackrel{g^N}{\mapsto} & \theta^N Z\;,\\
X\LM & \stackrel{g^N}{\mapsto} & X\LM + \pi NV\;.
\end{array}
\end{equation}
We know that $\theta^N = \mathbbm{1}$ is the identity on the three orbifolded coordinates $Z$. We furthermore see that $\theta^N$ is embedded as $NV$ into the gauge degrees of freedom $X\LM$. Consequently, we have to demand that the identity operation $\theta^N$ is embedded as a ``trivial'' element acting on the $X\LM$. Since, the gauge degrees of freedom are already compactified on the lattice $\Lambda$, see eqn.~(\ref{eqn:XIonLambda}), we find that this trivial element can be a lattice vector, that is
\begin{equation}
\label{eqn:ShiftLatticeConditions}
NV \in\Lambda\;.
\end{equation}
In summary, the twist $\theta$ of order $N$ is transmitted to a shift $V$ of the same order. In the case of a $\mathbbm{Z}_N \times \mathbbm{Z}_M$ point group, the shifts $V_1$ and $V_2$ have to be of order $N$ and $M$, respectively.

\subsubsection{The Order of the Wilson Line}\index{Wilson line!order}

This time we act with $g =\left(\theta,e_\alpha;\; V, A_\alpha\right)$ on the 3+16 bosonic degrees of freedom $Z$ and $X\LM$. However, in general we do not do this $N$ times, but only $N_\alpha$ times, where $N_\alpha$ is determined by the equation
\begin{equation}
\label{eqn:Nalpha}
\underbrace{\theta^{N_\alpha-1} e_\alpha + \theta^{N_\alpha-2} e_\alpha + \;\ldots\; + e_\alpha}_{N_\alpha\text{ terms}} = 0\;,
\end{equation}
and $N$ is a multiple of $N_\alpha$. Knowing the value of $N_\alpha$, we can transform the bosonic coordinates with $g^{N_\alpha}$ as follows
\begin{equation}
\label{eqn:WLOrder}
\begin{array}{lll}
Z    & \stackrel{g^{N_\alpha}}{\mapsto} & \theta^{N_\alpha} Z + \underbrace{\theta^{N_\alpha-1} e_\alpha + \;\ldots\; + e_\alpha}_{=0\text{ from eqn.~(\ref{eqn:Nalpha})}}\;\\
X\LM & \stackrel{g^{N_\alpha}}{\mapsto} & X\LM + \pi N_\alpha V + \pi N_\alpha A_\alpha\;.
\end{array}
\end{equation}
We denote the space group part of $g$ as $g_s \equiv \left(\theta,e_\alpha\right)$. From eqn.~(\ref{eqn:WLOrder}) we see that $g_s^{N_\alpha} = \left(\theta^{N_\alpha},0\right)$ does not contain any lattice vector $e_\alpha$ of the torus. Therefore, $g_s^{N_\alpha}$ should be embedded into the gauge degrees of freedom as a pure shift without Wilson lines, i.e. as $N_\alpha V$. This is achieved by demanding that the Wilson line $A_\alpha$ is of order $N_\alpha$, i.e.
\begin{equation}
N_\alpha A_\alpha \in\Lambda\;,
\end{equation}
no summation over the index $\alpha$. Hence, the name \emph{discrete Wilson line}.

In general, the Wilson lines $A_\alpha$ are not independent on the orbifold. In detail, a Wilson line $A_\alpha$ is associated to a torus lattice vector $e_\alpha$. On the orbifold however, some torus vectors are no longer inequivalent, i.e. they are identified. Thus, the associated Wilson lines of equivalent torus vectors have to be identified, too. This leads us to the condition
\begin{equation}
\label{eqn:WLIdentified1}
\theta e_\alpha = n_\alpha e_\alpha \quad\Rightarrow\quad A_\alpha = n_\alpha A_\alpha + \lambda\;,
\end{equation}
where $\lambda \in \Lambda$ and the sum runs over $\alpha = 1,\ldots,6$, such that the Wilson lines associated to $e_\alpha$ and $\theta e_\alpha$ are the same. We can derive this condition more explicitly by transforming the bosonic coordinates with $g =\left(\theta,e_\alpha;\; V, A_\alpha\right)$ twice
\begin{equation}
\label{eqn:WLIdentified2}
\begin{array}{lll}
Z    & \stackrel{g^2}{\mapsto} & \theta^2 Z + \theta e_\alpha + e_\alpha\;\\
X\LM & \stackrel{g^2}{\mapsto} & X\LM + \pi 2V + \pi 2A_\alpha\;.
\end{array}
\end{equation}
From this, we get the same conclusion: the Wilson lines associated to the torus lattice vectors $e_\alpha$ and $\theta e_\alpha$ must be the same.

We give two short {\bf examples}. First, consider the case of a two-dimensional $\Z{3}$ orbifold\index{orbifold!$\Z{3}$} on an $\SU{3}$ lattice. In this case, following eqn.~(\ref{eqn:WLIdentified1}), the torus lattice vectors $e_1$ and $e_2$ are related by the twist and the associated Wilson lines $A_1$ and $A_2$ have to be equal~\cite{Ibanez:1986tp}, i.e.
\begin{equation}
\theta e_1 = e_2 \quad\Rightarrow\quad A_1 = A_2\;,
\end{equation}
up to a lattice vector which conventionally is set to zero. Furthermore, from $\theta^2 e_1 + \theta e_1 + e_1 = 0$ we find that the Wilson line must be of order 3
\begin{equation}
3 A_1 = 3 A_2 \in \Lambda\;,
\end{equation}
as explained in eqns.~(\ref{eqn:Nalpha}) and~(\ref{eqn:WLOrder}).

The second example concerns the six-dimensional $\Z{6}$-II orbifold\index{orbifold!$\Z{6}$-II} on an $\SU{6}\times\SU{2}$ lattice. In reference~\cite{Ramos:2008ph} it is shown that this lattice allows for two independent Wilson lines: one of order 6 and one of order 2. 
\begin{eqnarray}
A_1 = A_2 = A_3 = A_4 = A_5 & \text{with}& 6A_1 \in \Lambda\quad\text{and}\\
A_6                         & \text{with}& 2A_6 \in \Lambda\;.
\end{eqnarray}

The conditions on the shift and Wilson lines derived here together with further conditions arising from modular invariance will be summarized later in section~\ref{sec:SummaryConditions}, for both types of orbifolds $\mathbbm{Z}_N$ and $\mathbbm{Z}_N \times \mathbbm{Z}_M$.

\subsubsection{Remark: Rotation Embedding and Continuous Wilson Lines} \index{strings on orbifolds!rotation embedding}

As a remark, we briefly discuss the possibility to embed the twist $\theta$ as a rotation $\Theta$ into the gauge degrees of freedom $X\LM$~\cite{Ibanez:1987xa}, i.e.
\begin{equation}
X\LM \stackrel{\theta}{\mapsto} \Theta X\LM\;.
\end{equation}
Following the steps of eqn.~(\ref{eqn:ShiftOrder}) in this case leads to the obvious requirement that also $\Theta$ is of order $N$, that is
\begin{equation}
\Theta^N = \mathbbm{1}\;.
\end{equation}
Furthermore, there are important consequences in the case of nontrivial Wilson lines. Thus, we analyze the action of an orbifold group element containing a torus shift and its associated Wilson line, for example take $g=\left(\theta,e_\alpha;\; \Theta, A_\alpha\right)$ and take the $N'$-th power of it, i.e. $g^{N'}$. As we will see, it is important to distinguish between two cases for $N'$:
\begin{enumerate}
\item In the first case, the Wilson line is invariant under $\Theta$, i.e. $\Theta A_\alpha = A_\alpha$. It will turn out that setting $N' = N_\alpha$ using the definition of eqn.~(\ref{eqn:Nalpha}) is useful in this case.
\item In the second case, the Wilson line transforms non-trivially under $\Theta$, i.e. $\Theta A_\alpha \neq A_\alpha$ and it is convenient to set $N' = N$.
\end{enumerate}
Now, we transform the 3+16 bosonic degrees of freedom with $g^{N'}$. Restricting to the gauge degrees of freedom $X\LM$, this yields
\begin{equation}
\begin{array}{lll}
X\LM & \stackrel{g^{N'}}{\mapsto} & \Theta^{N'} X\LM + \underbrace{\Theta^{N'-1} A_\alpha + \Theta^{N'-2} A_\alpha + \;\ldots\; + A_\alpha}_{N'\text{ terms}}\;.
\end{array}
\end{equation}
Again, we discuss the two cases separately:
\begin{enumerate}
\item Since $\Theta A_\alpha = A_\alpha$, we see that $\Theta^{N_\alpha-1} A_\alpha + \;\ldots\; + A_\alpha = N_\alpha A_\alpha$ and the Wilson line must be of order $N_\alpha$. Compare to eqn.~(\ref{eqn:WLOrder}).
\item In this case $\Theta^{N-1} A_\alpha + \Theta^{N-2} A_\alpha\;\ldots\; + A_\alpha = 0$ vanishes automatically, since $\Theta$ is a rotation of order $N$. Thus, there is no restriction on the length of the Wilson line. Hence, the name \emph{continuous Wilson line}\index{Wilson line!continuous}.
\end{enumerate}
Continuous Wilson lines are known to break the rank of the gauge group. This rank reduction by continuous Wilson lines can be interpreted alternatively as a Higgs mechanism induced by an untwisted field obtaining a vev~\cite{Ibanez:1987xa,Font:1988tp,Forste:2005gc,Forste:2005rs,Wingerter:2005ph,LoaizaBrito:2005fa}.

\subsection{Modular Invariance}\index{modular invariance!of shifts and Wilson lines}
\label{sec:SummaryConditions}

Modular invariance of one--loop amplitudes imposes strong conditions on the shifts and Wilson lines. In $\Z{N}$ orbifolds, the order $N$ shift $V$ and the twist $v$ must fulfill~\cite{Dixon:1986jc,Vafa:1986wx}:
\begin{equation}\label{eq:znmodularinv}
N \,\left(V^2 - v^2\right)~=~0 \mod 2 \,.
\end{equation}
For $\Z{N}\times\Z{M}$ orbifolds with Wilson lines, modular invariance, including some consistency requirements as discussed in appendix~\ref{app:ModularInvariance}, yields the conditions~\cite{Ploger:2007iq}
\begin{subequations}\label{eq:newmodularinv}
\begin{eqnarray}
  N\,\left(V_{1}^{2} - v_{1}^{2} \right)  & = & 0 \mod 2\;, \\
  \label{eq:fsmi1}
  M\,\left(V_{2}^{2} - v_{2}^{2} \right)  & = & 0 \mod 2\;, \\
  \label{eq:fsmi2}
  M\,\left(V_{1}\cdot V_{2} - v_{1}\cdot v_{2} \right)  & = & 0 \mod 2\;, \\
  \label{eq:fsmi3}
  N_\alpha\,\left(A_{\alpha}\cdot V_{i}\right)  & = & 0 \mod 2\;, \\
  \label{eq:fsmi4}
  N_\alpha\,\left(A_{\alpha}^2\right)  & = & 0 \mod 2\;, \\
  \label{eq:fsmi5}
  Q_{\alpha\beta}\,\left(A_{\alpha}\cdot A_{\beta}\right)  & = & 0 \mod 2 \quad (\alpha \neq \beta)\;,
  \label{eq:fsmi6}
\end{eqnarray}
\end{subequations}
where $N_\alpha$ is the order of $A_\alpha$ and $Q_{\alpha\beta}\equiv\text{gcd}(N_{\alpha},N_{\beta})$ denotes the greatest common divisor of $N_{\alpha}$ and $N_{\beta}$ and, as before, $N$ is a multiple of $M$.\footnote{In the case of two different \Z2 Wilson lines we find that \eqref{eq:fsmi5} can be relaxed, i.e.\ $\text{gcd}(N_{\alpha},N_{\beta})$ can be replaced by $N_{\alpha}\,N_{\beta}=4$, provided there exists no  $g \in P$ with the property $g\,e_\alpha~\neq~e_\alpha$ but $g\,e_\beta~=~e_\beta$. Imposing the weaker condition leads, as we find, to anomaly-free spectra.}

\subsection{The Untwisted Sector}

After the general discussion on strings on orbifolds, we can now turn to more explicit calculations. First, we will analyze the untwisted sector in this section. Strings in the untwisted sector fulfill the trivial boundary conditions of eqn.~(\ref{eqn:closedstring}), i.e. their constructing elements are $g = \left(\mathbbm{1}, 0\right)$. Consequently, we know the solutions to the equations for massless untwisted strings. They are given by the 10d spectrum of the heterotic string: the 10d SUGRA multiplet and the 10d vector multiplets corresponding to the 16+480 generators of the 10d gauge group, as discussed in section~\ref{sec:HeteroticString}. However, these states will in general not be invariant under the action of the orbifold. As explained in section~\ref{sec:physstates}, we have to project out non-invariant states in order to find the physical states of the orbifold model.

\subsubsection{The SUGRA multiplet and Moduli}\index{supergravity!$\mathcal{N} = 1$ in 4d}
\label{sec:SUGRA10dcompactification}

The 10d supergravity multiplet specified in eqn.~(\ref{eqn:10dSUGRA}) is a gauge singlet (with $p = 0$). Hence, its compactification only depends on the twist, but is independent of the choices for shifts and Wilson lines\footnote{Note that in the untwisted sector the vacuum phase is trivial, as $v_g=0$ and $V_g = 0$.}. 

If the twist $\theta$ (or $\theta$ and $\omega$ in the case of a $\Z{N}\times\Z{M}$ point group) fulfills the $\mathcal{N} = 1$ condition, eqn.~(\ref{eqn:susycondition}), it is easy to see that the following components of the 10d supergravity multiplet, eqn.~(\ref{eqn:10dSUGRA}), are invariant
\begin{equation}
|q\rangle\RM \otimes \tilde{\alpha}_{-1}^\mu |0\rangle\LM\quad\text{with}\quad\mu=0,1\quad\text{and}\quad q = \pm \left(\tfrac{1}{2},\tfrac{1}{2},\tfrac{1}{2},\tfrac{1}{2}\right) \quad\text{or}\quad q = \left(\pm 1,0,0,0\right)\;.
\end{equation}
As usual, we can read off the transformation properties under 4d Lorentz transformation from the right-mover. The combination of the fermionic right-moving momentum $q$ and the oscillator $\tilde{\alpha}_{-1}^\mu$ transforming as a 4d space-time vector boson gives rise to one spin $3/2$ fermion plus its CPT conjugate, the gravitino\index{gravitino}. On the other hand, combining the bosonic $q$ with this oscillator yields a spin $2$ boson plus its CPT conjugate, the graviton. Thus, in summary, these states correspond to the 4d SUGRA multiplet for $\mathcal{N} = 1$.

Beside the 4d supergravity multiplet, there are further invariant components of its 10d version. Here, we restrict to the bosonic states and use the index notation of section~\ref{sec:HeteroticString} for the right-moving momenta $q$. First, we have to discuss briefly how right-movers $|i\rangle\RM$ or $|\bar{i}\rangle\RM$ with holomorphic or anti-holomorphic indices transform under the action of the orbifold group. From eqn.~(\ref{eqn:QIndex}) we can see that an index $i$, $\bar{i}$ transforms with a phase 
\begin{equation}
\label{eqn:IndexTrafo}
\begin{array}{llcll}
e^{-2\pi\I q_{(i)}\cdot v} = e^{2\pi\I v^i}        & \text{for } i\;\;\text{ and} & & e^{-2\pi\I q_{(\bar{i})}\cdot v} = e^{-2\pi\I v^i} & \text{for } \bar{i}\;,
\end{array}
\end{equation}
under the action of the twist $\theta$, respectively. Now, we can summarize all ``internal'' components of the 10d supergravity multiplet
\begin{equation}
\label{eqn:AllModuli}
N_{i\bar{j}} \equiv |i\rangle\RM \otimes \tilde\alpha_{-1}^{\bar{j}}|0\rangle\LM  \quad\text{and}\quad N_{ij} \equiv |i\rangle\RM \otimes \tilde\alpha_{-1}^j|0\rangle\LM\;,
\end{equation}
with $i,j = 1,2,3$ and $\bar{j} = \bar{1},\bar{2},\bar{3}$. The states $N_{\bar{i}j}$ and $N_{\bar{i}\bar{j}}$ are the CPT partners of the above-mentioned ones and are therefore not listed. Independently of the choice of the point group $P$, we see from the transformation properties given in eqns.~(\ref{eqn:OsciTrafo}) and~(\ref{eqn:IndexTrafo}) that $N_{1\bar{1}}$, $N_{2\bar{2}}$ and $N_{3\bar{3}}$ are invariant under the action of $P$. Furthermore, a state $N_{ij}$ can only be invariant if the twist $\theta$ contains at least one rotation by $180^\circ$ in one of the three complex dimensions, i.e. $v_i = \tfrac{1}{2}$ for some $i$. Depending on the specific choice of $\theta$ further components of~(\ref{eqn:AllModuli}) can be invariant for some combinations of $i,j = 1,2,3$ and $\bar{j} = \bar{1},\bar{2},\bar{3}$, see e.g.~\cite{Molera:1989zr,Ibanez:1992hc}. 

These states are \emph{moduli}\index{modulus} and describe metric variations of the type $\delta g_{i\bar{j}}$ and $\delta g_{ij}$, respectively. The states $N_{i\bar{j}}$ are real $(1,1)$ moduli describing variations of the K\"ahler\index{modulus!K\"ahler modulus} structure and the $N_{ij}$ are complex $(1,2)$ moduli corresponding to variations of the complex structure\index{modulus!complex structure modulus}. It is important to note that the number of untwisted moduli is a topological quantity and therefore does not depend on the torus lattice $\Gamma$, but only on the point group $P$. On the other hand the interpretation of a given modulus as some radius or angle depends on $\Gamma$.\footnote{For a $\mathbbm{Z}_4$ example consult reference~\cite{Casas:1991ac} and compare eqn.~(53) and (64) therein, where the deformation degrees of freedom are listed for the $\SO{4}^3$ and the $\SU{4}^2$ torus lattice, respectively.}

\subsubsection{The 16 Cartan Generators}

The states of eqn.~(\ref{eqn:16Cartan}) correspond to the 16 Cartan generators. In detail, they read
\begin{equation}
|q\rangle\RM \otimes \tilde\alpha_{-1}^I|0\rangle\LM\;.
\end{equation}
Their left-moving momenta are trivial $p=0$. However they are not gauge singlets, as they are excited by oscillators $\tilde\alpha_{-1}^I$ in the 16 gauge degrees of freedom. It is important to note that these left-movers are unaffected by shifts in the gauge degrees of freedom, i.e. by shifts with the shift vector $V$ or with Wilson lines $A_\alpha$. Therefore, the left-movers are invariant under the action of the orbifold group. On the other hand, the right-movers acquire phases such that
\begin{eqnarray}
|q\rangle\RM \otimes \tilde\alpha_{-1}^I|0\rangle\LM & \stackrel{\theta}{\rightarrow} & e^{-2\pi\I q \cdot v}\; |q\rangle\RM \otimes \tilde\alpha_{-1}^I|0\rangle\LM\;.
\end{eqnarray}
Thus, only those 10d states are invariant on the orbifold that have $q=(-1,0,0,0)$, $q=(-\tfrac{1}{2},-\tfrac{1}{2},$ $-\tfrac{1}{2},-\tfrac{1}{2})$ or the CPT conjugates thereof, i.e. $-q$. This yields 4d vector bosons and 4d Weyl spinors and consequently vector multiplets in 4d. The other states have to be projected out.

In summary, we find 16 4d vector multiplets associated to the 16 Cartan generators. The rank of the 4d gauge group cannot be reduced by shift embeddings; it is always 16.

\subsubsection{480 generators of $\boldsymbol{\E{8}\times\E{8}}$ or $\boldsymbol{\SO{32}}$}

The 480 states associated to the charged generators of $\E{8}\times\E{8}$ or $\SO{32}$ transform under a general orbifold transformation $h\in O$ with a phase
\begin{eqnarray}
\label{eqn:trafo480}
|q\rangle\RM \otimes |p\rangle\LM & \rightarrow & e^{2\pi\I(p\cdot V_h - q\cdot v_h)}\; |q\rangle\RM \otimes |p\rangle\LM\;.
\end{eqnarray}
The destiny of these states depend on their right-movers. For a given state the right-mover can be either invariant under the action of the orbifold group, or not.

In the first case, $q\cdot v_h = 0$, the invariant right-mover carries a momentum $q=(-1,0,0,0), q=(-\tfrac{1}{2},-\tfrac{1}{2},-\tfrac{1}{2},-\tfrac{1}{2})$ or a CPT conjugate thereof, as in the case of the Cartans. But what does this imply for the left-mover? In order for the whole state to be invariant, the left-moving momentum has to satisfy the condition
\begin{equation}
\label{eqn:4dGaugeGroupPhase}
p\cdot V_h = 0 \quad\Leftrightarrow\quad p \cdot V = 0 \quad\text{ and }\quad p \cdot A_\alpha = 0\quad\text{for }\alpha=1,\ldots,6\;.
\end{equation}
The set of invariant states in this category gives rise to the 4d gauge group. Their left-moving momenta $p$ are the roots of the corresponding adjoint representation.

\begin{figure}[t]
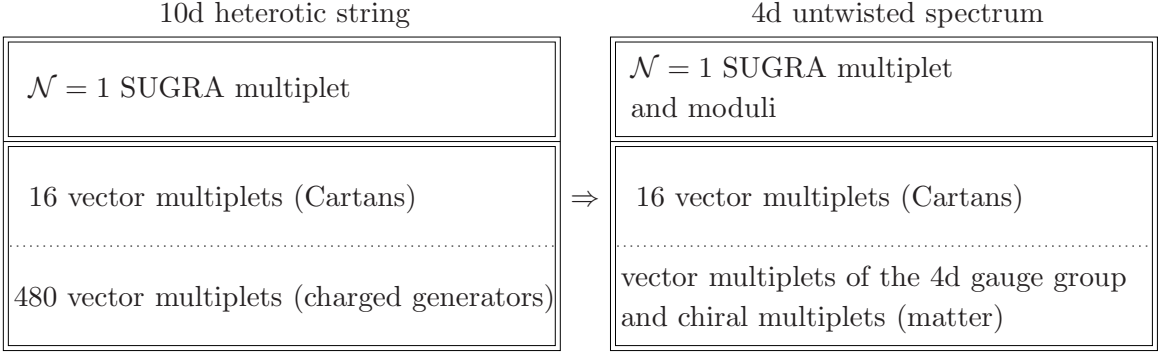

\centerline{\input UntwistedSpectrum.pstex_t}\vspace{-0.3cm}
\caption{General overview of the spectrum of the untwisted sector.}
\label{fig:UntwistedSpectrum}
\end{figure}

In the second case, $q\cdot v_h \neq 0$, the right-mover carries a momentum $q=(0,\underline{-1,0,0})$ or $q=(-\tfrac{1}{2},\underline{-\frac{1}{2},\frac{1}{2},\frac{1}{2}})$ (or a CPT conjugate thereof). From the 4d perspective, such a right-mover corresponds to a scalar or to a Weyl spinor giving rise to a chiral multiplet in 4d. If a given state with $q\cdot v_h \neq 0$ does not fulfill the projection condition eqn.~(\ref{eqn:trafo480})
\begin{equation}
p\cdot V_h - q\cdot v_h \stackrel{!}{=} 0 
\end{equation}
it is not invariant and hence has to be removed from the 4d spectrum. The set of invariant states gives rise to matter fields, transforming in representations specified by their weights $p$.

In summary, the untwisted sector of the 4d orbifold model contains the SUGRA multiplet, the moduli associated to the geometry of the internal orbifold space, the 4d gauge group and matter representations, see fig.~(\ref{fig:UntwistedSpectrum}).

\subsection{The Twisted Sectors}

As discussed before, the twisted sectors are constructed by the following procedure:
\begin{itemize}
\item choose the inequivalent constructing elements $g$,
\item solve the equations~(\ref{eqn:masslesstwistedrightmover}) and~(\ref{eqn:masslesstwistedleftmover}) for massless right- and left-movers, and finally
\item use the orbifold projection eqn.~(\ref{eqn:projectout}) to get the invariant states.
\end{itemize}
Here, we want to discuss briefly the interpretation of these twisted states as matter representation. The condition for massless right-movers eqn.~(\ref{eqn:masslesstwistedrightmover}) reads
\begin{equation}
\frac{(q+v_g)^2}{2} - \frac{1}{2} + \delta c = 0\;.
\end{equation}
For a non-trivial constructing element $g$ the zero-point energy is shifted by $\delta c > 0$. Consequently, $(q+v_g)^2 < 1$ and we see that twisted right-movers cannot give rise to vector bosons, i.e. $q\neq (\pm 1,0,0,0)$. The twisted sectors cannot provide additional gauge group factors, but only matter representations in the form of chiral multiplets. Such a chiral multiplet contains in general a state from $g$ and its CPT conjugate from $g^{-1}$:  if $q_\text{sh}$ is a solution for $v_g$, $-q_\text{sh}$ is a solution for $v_{g^{-1}} = -v_g$ and if $p_\text{sh}$ is a solution for $V_g$, $-p_\text{sh}$ is one for $V_{g^{-1}} = -V_g + \lambda$ with $\lambda \in \Lambda$. If furthermore $g=g^{-1}$, both (the state and its CPT partner) appear in the same sector. Having this in mind, we will from now on concentrate on the left-chiral states (with $q^0 = -\tfrac{1}{2}$) and their bosonic SUSY partners. We will not list the right-chiral CPT conjugates (with $q^0 = \tfrac{1}{2}$) and their SUSY partners any more.

\subsection[Anomalous $\U{1}$]{Anomalous $\boldsymbol{\U{1}}$}
\label{sec:anomalousU1}

The 4d gauge group generically contains many $\U{1}$ factors (in rare cases even 16 $\U{1}$'s and no non-Abelian gauge group factor). It is well known that at most one $\U{1}$ factor can appear to be anomalous for heterotic orbifold models~\cite{Green:1984sg,Witten:1984dg,Dine:1987xk,Casas:1987us,Kobayashi:1996pb}. This factor is denoted by $\U{1}_\text{anom}$. All other gauge group factors (Abelian and non-Abelian) are anomaly-free. However, if we start with an arbitrary basis of $\U{1}$ generators and compute their anomalies it turns out that more than one $\U{1}$ appear to be ``anomalous'' in general. Nevertheless, we can always perform a basis change such that the anomaly is rotated to a single direction. In the following we will discuss in detail how to construct the anomalous $\U{1}$ gauge group and what are the conditions it has to obey.

\subsubsection{Identifying $\boldsymbol{\U{1}_\text{anom}}$}

The Cartan generators of $\E{8}\times\E{8}$ or $\SO{32}$ are conventionally denoted by $H_I$, $I=1,\ldots,16$. By definition they act on left-moving states as
\begin{equation}
H^I|p_\text{sh}\rangle\LM = p_\text{sh}^I |p_\text{sh}\rangle\LM\;.
\end{equation}
Suppose we have $n$ Abelian gauge group factors $\U{1}^{(i)}$, $i=1,\ldots,n$, in the 4d gauge group. They are generated by linear combinations of the Cartans, i.e.
\begin{equation}
Q_i = t_i \cdot H = t_i^I H^I\;,
\end{equation}
where we sum over $I=1,\ldots,16$. The coefficients $t_i^I$ could be from $\mathbbm{R}$, but it is more convenient to choose them as $t_i^I \in \mathbbm{Q}$. They have to be orthogonal to all simple roots $\alpha_j$ of the non-Abelian gauge group factors, i.e. $t_i \cdot \alpha_j = 0$. Furthermore, we choose them to be orthogonal among each other, $t_i \cdot t_j = 0$. Consequently, the $i$-th $\U{1}$ charge of the state $|p_\text{sh}\rangle\LM$ reads
\begin{equation}
Q_i |p_\text{sh}\rangle\LM = (t_i \cdot H) |p_\text{sh}\rangle\LM = (t_i \cdot p_\text{sh}) |p_\text{sh}\rangle\LM\;,
\end{equation}
in short form, $Q_i = t_i \cdot p_\text{sh}$. Hence, we will also denote $t_i$ as the generator of the $i$-th $\U{1}$ factor. In principle, this $\U{1}^{(i)}$ could be anomalous. To see this, we have to evaluate for example the $\U{1}^{(i)}-\text{grav}-\text{grav}$ anomaly. It is proportional to the ``trace'' of $Q_i$ over all fermionic states with momenta $p_\text{sh}^{(f)}$
\begin{equation}
A_i ~\equiv~ \text{Tr}\, Q_i = \sum_f Q_i^{(f)} = \sum_f t_i \cdot p_\text{sh}^{(f)}\;.
\end{equation}
If this does not vanish the associated $i$-th $\U{1}$ factor is anomalous. Generically, starting with some arbitrary $\U{1}$ generators $t_i$, many $\U{1}$ factors seem to be anomalous. However, we can construct the unique (up to a rescaling) generator $t_\text{anom}$ of the anomalous $\U{1}_\text{anom}$ by using the anomaly coefficients $A_i$. In detail this reads\footnote{We are thankful to Prof. Michael Ratz for pointing this out.}
\begin{equation}
\fbox{$ \displaystyle t_\text{anom} \equiv \sum_{i=1}^n \frac{A_i}{t_i \cdot t_i}\,t_i\;. $}
\end{equation}\index{anomaly!generator of $\U{1}_\text{anom}$}
Since the generators $t_i$ are chosen to be orthogonal, we can easily read off the anomalous generator in terms of the shifted momenta $p_\text{sh}$~\cite{Araki:2008ek}
\begin{equation}
t_\text{anom} = \sum_{i,f} \frac{t_i \cdot p_\text{sh}^{(f)}}{t_i \cdot t_i}\,t_i \quad \Rightarrow\quad t_\text{anom} = \frac{1}{12}\sum_f p_\text{sh}^{(f)}\;,
\end{equation}
where the rescaling with the factor $\tfrac{1}{12}$ is the convention used in section~\ref{sec:discreteanomalies}. From here we can check that this is the desired result: the anomaly corresponding to $t_\text{anom}$ is non-vanishing
\begin{equation}
A_\text{anom} = \sum_f t_\text{anom} \cdot p_\text{sh}^{(f)} = 12 |t_\text{anom}|^2 \neq 0
\end{equation}
and the ones corresponding to some orthogonal directions $\tilde{t}_i \cdot t_\text{anom} = 0$ for $i=2,\ldots,n$ vanish,
\begin{equation}
\tilde{A}_i = \sum_f \tilde{t}_i \cdot p_\text{sh}^{(f)} = \tilde{t}_i \cdot \big(\sum_f p_\text{sh}^{(f)} \big) = 12\, \tilde{t}_i \cdot t_\text{anom} = 0\;.
\end{equation}

\subsubsection{Conditions on $\boldsymbol{\U{1}_\text{anom}}$}\index{anomaly!cancelation of $\U{1}_\text{anom}$}

For a $\U{1}$ gauge factor there are several possible anomalies:
\begin{equation}
\begin{array}{lll}
\U{1}^{(i)}-\text{grav}-\text{grav}, & \qquad    & \U{1}^{(i)}-\U{1}^{(i)}-\U{1}^{(i)},\\
\U{1}^{(i)}-G-G,                     &\text{and} &  \U{1}^{(i)}-\U{1}^{(j)}-\U{1}^{(j)},\phantom{\big(\big)^{I^I}}
\end{array}
\end{equation}
where $G$ denotes a non-Abelian gauge group factor (like $\SU{2}$) and $i,j=1,\ldots,n$.

In the preferred basis where only the first $\U{1}$ is anomalous, the $\U{1}$'s satisfy the following conditions~\cite{Casas:1987us,Kobayashi:2004ya}
\begin{equation}
\label{eq:Anomaly_conditions}
\frac{1}{24}\text{Tr}\,Q_i = \frac{1}{6 |t_i|^2}\text{Tr}\,Q_i^3 = \text{Tr}\,\ell Q_i = \frac{1}{2|t_j|^2}\text{Tr}\,Q_j^2 Q_i = \left\{ \begin{array}{ll} \frac{1}{2}|t_\text{anom}|^2 \neq 0 \quad\text{\MyQSpc} & \text{if } i = 1\text{, i.e. anom}, \\ 0\text{\MyQSpc} & \text{otherwise}\end{array} \right.
\end{equation}
where $i\neq j$ and $l$ denotes the Dynkin index\footnote{The Dynkin index $\ell(\boldsymbol{r}^{(f)})$ of some representation $\rep{r}^f$ is defined by $\ell(\rep{r}^{(f)})\,\delta_{ab}=\tr(\mathsf{t}_a(\rep{r}^{(f)})\,\mathsf{t}_b(\rep{r}^{(f)}))$, using the generator $\mathsf{t}_a$ of $G$ in the representation $\rep{r}^f$. The conventions are such that $\ell(\rep{M})=1/2$ for $\SU{M}$ and $\ell(\rep{M})=1$ for $\SO{M}$} with respect to the non-Abelian gauge group factor $G$.

In the case when eqn.~(\ref{eq:Anomaly_conditions}) does not vanish, these conditions guarantee that the anomalous $\U{1}$ is canceled by the generalized Green-Schwarz mechanism~\cite{Green:1984sg} (i.e. by an anomalous variation of the $B$ field, compare to section~\ref{sec:HeteroticString}). It is important to note that an anomalous $\U{1}$ induces a so-called \emph{Fayet--Iliopoulos $D$--term}\index{Fayet--Iliopoulos D-term}~\cite{Dine:1987xk,Atick:1987gy}, a constant in $D_\text{anom}$ which is proportional to the anomaly $\text{Tr}\, Q_\text{anom}$, i.e.
\begin{equation}
\label{eqn:AnomU1FI}
D_\text{anom} \simeq \frac{g}{192\pi^2}\text{Tr}\, Q_\text{anom}\;.
\end{equation}

\section{Some Orbifolds}

In this section we will analyze the following orbifolds in some detail: $\Z{3}$, $\Z{3}\times\Z{3}$, $\Z{6}$-II and $\Z{2}\times\Z{2}$. As they will recur in the following chapters, we will discuss them here and refer to this section later on. The focus lies on geometrical aspects such as the torus lattice and its deformations, the fixed point structure and, in the case of the $\Z{2}\times\Z{2}$ orbifold, on the difference between factorizable and non-factorizable torus lattices.

\subsection[The $\Z{3}$ Orbifold]{The $\boldsymbol{\Z{3}}$ Orbifold}\index{orbifold!$\Z{3}$}
\label{sec:Z3Orbifold}

\begin{figure}[t]
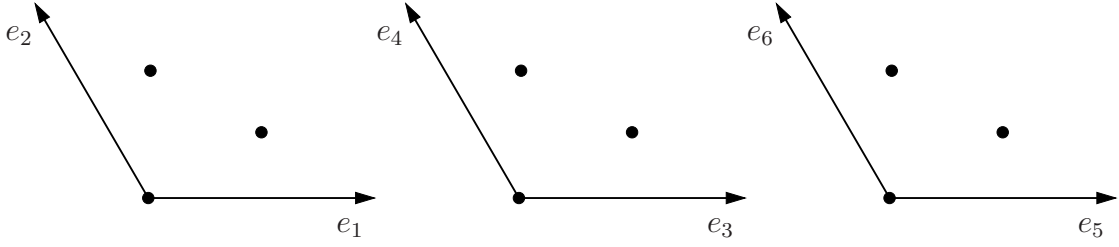

\centerline{\input Z3Orbifold.pstex_t}\vspace{-0.2cm}
\caption{The six-dimensional $\Z{3}$ orbifold on a factorized  $\Gamma = \SU{3}\times\SU{3}\times\SU{3}$ lattice with $3\times 3\times 3 = 27$ fixed points.}
\label{fig:Z3Orbifold}
\end{figure}

As the $\Z{3}$ orbifold can be seen as the simplest one, it has attracted a lot of attention in the literature~\cite{Ibanez:1987sn,Ibanez:1987pj,Casas:1987us,Font:1988tp,Casas:1988se,Font:1988mm,Casas:1988hb,Font:1989aj,Giedt:2000bi,Giedt:2001zw,Giedt:2003an,Giedt:2004wd,Giedt:2005vx}. We will mainly use it in chapter~\ref{sec:blowup} where we discuss the blow-up of $\Z{3}$ singularities. However, as it is instructive, we will discuss $\Z{3}$ in detail here. 

The six-dimensional torus lattice is chosen to be $\Gamma = \SU{3}\times\SU{3}\times\SU{3}$, where each $\SU{3}$ sublattice lies inside one of the three complex planes, see figure~(\ref{fig:Z3Orbifold}). Furthermore, the generator $\theta$ of the $\Z{3}$ point group is specified by the twist vector
\begin{equation}
v = \left(0, \tfrac{1}{3}, \tfrac{1}{3},-\tfrac{2}{3} \right)\;,
\end{equation}
fulfilling the condition eqn.~(\ref{eqn:susycondition}) for $\mathcal{N}=1$ supersymmetry in 4d: $v^1+v^2+v^3=0$.

\subsubsection{The 4d SUGRA Multiplet and Moduli}\index{modulus}

As explained in the general case in section~\ref{sec:SUGRA10dcompactification}, the invariant components of the ten-dimensional $\mathcal{N} = 1$ SUGRA multiplet eqn.~(\ref{eqn:10dSUGRA}) are first of all
\begin{equation}
|q\rangle\RM \otimes \tilde{\alpha}_{-1}^\mu |0\rangle\LM\quad\text{with}\quad\mu=0,1\quad\text{and}\quad q=\left(-\tfrac{1}{2},-\tfrac{1}{2},-\tfrac{1}{2},-\tfrac{1}{2}\right)\;,
\end{equation}
corresponding to the 4d $\mathcal{N} = 1$ SUGRA multiplet, represented by its single gravitino, and additionally nine $(1,1)$ moduli $N_{i\bar{j}}$ with $i=1,2,3$ and $\bar{j} = \bar{1}, \bar{2}, \bar{3}$
\begin{equation}
|q\rangle\RM \otimes \tilde{\alpha}_{-1}^{\bar{j}} |0\rangle\LM \quad\text{with }\;q = \left(-\tfrac{1}{2},\underline{-\tfrac{1}{2},\tfrac{1}{2},\tfrac{1}{2}}\right)\;\text{ and }\;\bar{j} = \bar{1},\bar{2},\bar{3}\;,
\end{equation}
plus their scalar partners. These moduli describe continuous deformations of the torus lattice $\Gamma$, where each real $(1,1)$ modulus corresponds to one geometric parameter. Obviously, we need to identify nine geometric parameters that deform the torus lattice in such a way that the twist $\theta$ remains a symmetry. These parameters are~\cite{Kobayashi:1991rp,Casas:1991ac}
\begin{equation}
\label{eqn:Z3parameters}
R_1, R_3, R_5 \quad\text{and}\quad \alpha_{13},\alpha_{15},\alpha_{35},\alpha_{14},\alpha_{16},\alpha_{36}\;.
\end{equation}
Their geometric meaning as the three radii of the three $\SU{3}$ tori and six angles is presented in the appendix~\ref{sec:Z3Deform}. Furthermore, there it is shown that the twist $\theta$ remains a symmetry of the deformed torus lattices.

\subsubsection{Breaking of the 10d Gauge Group}

As noted before, the 4d components of the 16 ten-dimensional Cartan generators are invariant. These states are $|q\rangle\RM \otimes \tilde\alpha_{-1}^I|0\rangle\LM$ with $q = (-\tfrac{1}{2},-\tfrac{1}{2},-\tfrac{1}{2},-\tfrac{1}{2})$ for the gauginos and $q=(-1,0,0,0)$ for the gauge bosons. Thus, the rank of the 4d gauge group remains 16.

\begin{table}[t]
\begin{center}
\begin{tabular}{|c|c|c|}
\hline
shift $V$                              & gauge group from $\E{8}\times\E{8}$ & label\\
\hline
\hline
$\phantom{I^{I^I}}\left(0^8\right)\left(0^8\right)\phantom{I^{I^I}}$& $\E{8}\times\E{8}$                  & T \\
$\tfrac{1}{3}\left(-2,1^2,0^5\right)\left(0^8\right)$               & $\E{6}\times\SU{3}\times\E{8}$      & A \\
$\tfrac{1}{3}\left(-2,1^2,0^5\right)\left(-2,1^2,0^5\right)$        & $\big[\E{6}\times\SU{3}\big]^2$     & B \\
$\tfrac{1}{3}\left(1^2,0^6\right)\left(-2,0^7\right)$               & $\E{7}\times\SO{14}\times\U{1}^2$   & C \\
$\tfrac{1}{3}\left(-2,1^4,0^3\right)\left(-2,0^7\right)$            & $\SU{9}\times\SO{14}\times\U{1}$    & D \\
\hline
\hline
shift $V$                                                    & gauge group from $\SO{32}$       & label\\
\hline\hline
$\phantom{I^{I^I}}\left(0^{16}\right)\phantom{I^{I^I}}$                  & $\SO{32}$                        & T32 \\
$\tfrac{1}{3}\left(-2,1^2,0^{13}\right)$                                 & $\SO{26}\times\SU{3}\times\U{1}$ & A32 \\
$\left(\frac{1}{2}^2,-\frac{1}{6}^6,\frac{1}{2}^{8}\right)$              & $\SO{20}\times\SU{6}\times\U{1}$ & B32 \\
$\left(\frac{1}{2}^2,\frac{1}{6},-\frac{1}{6}^8,\frac{1}{2}^{5}\right)$  & $\SO{14}\times\SU{9}\times\U{1}$ & C32 \\
$\left(\frac{1}{2}^2,-\frac{1}{6}^{12},-\frac{1}{2}^2\right)$            & $\SO{8}\times\SU{12}\times\U{1}$ & D32 \\
$\left(\frac{1}{3},\frac{2}{3},0,-\frac{1}{3}^{13}\right)$               & $\SU{15}\times\U{1}^2$           & E32 \\
\hline
\end{tabular}
\caption{\label{orbshift}
All inequivalent shifts for the $\mathbbm{Z}_3$ orbifold and the corresponding gauge groups in the case of $\E{8}\times\E{8}$ and $\SO{32}$ heterotic string.}
\end{center}
\end{table}

The compactification of the charged gauge bosons and gauginos of the 10d gauge group $\E{8}\times\E{8}$ or $\SO{32}$ depends on the choices of the shift and the Wilson lines. As we have seen in section~\ref{sec:HetOnOrb} they are subject to modular invariance conditions. In the case without Wilson lines (i.e. $A_\alpha = 0$) and $\E{8}\times\E{8}$ (or $\SO{32}$) gauge group in 10d, there are only five (or six) inequivalent shifts leading to five (six) inequivalent $\Z{3}$ orbifolds, respectively. They are listed in table~(\ref{orbshift}) together with their resulting 4d gauge groups. Model A of $\E{8}\times\E{8}$ and model A32 of $\SO{32}$ are the so-called \emph{standard embedding}\index{strings on orbifolds!standard embedding} models. For them the twist $v= (0,v^1,v^2,v^3)$ is embedded into the shift as $V=(v^1,v^2,v^3,0^{13})$ such that the modular invariance condition $N(V^2-v^2) = 0 \mod 2$ is trivially fulfilled\footnote{In the case of a Calabi-Yau manifold the standard embedding is given by taking the gauge connection to be equal to the spin connection~\cite{Candelas:1985en}. In terms of the curvature this means $\I \mathcal{F}=\mathcal{R}$, see section~\ref{sec:blowupgeometry}}.

As an example, we will explain model C in some detail. The roots of the unbroken gauge group are determined by the condition $p\cdot V = 0$. Explicitly, they are given by
\begin{equation}
\begin{array}{|l|rl|}
\hline
\left(0,0,\underline{\pm 1,\pm 1, 0^4}\right)\left(0^8\right)\phantom{I^{I^I}}                  &60         & \\[1mm]
\left(\pm(1,-1),0^6\right)\left(0^8\right)                                                      & 2         & \\[1mm]
\left(\pm\left(\frac{1}{2},-\frac{1}{2}\right),\left(\pm\frac{1}{2}\right)^6\right)\Big(0^8\Big)& 2\times 32& \Rightarrow \text{126 roots} + \text{7 Cartans} \Rightarrow \rep{133} \text{ of } $\E{7}$\\[1mm]
\hline
\left(0^8\right)\left(0,\underline{\pm 1,\pm 1, 0^5}\right)\phantom{I^{I^I}}                     & 84        & \Rightarrow \phantom{0}\text{84 roots} + \text{7 Cartans} \Rightarrow \phantom{0}\rep{91} \text{ of } $\SO{14}$\\[1mm]
\hline
\phantom{I^{I^I}}                                                                                &           & \qquad\qquad\qquad\;\;\text{2 Cartans}\; \Rightarrow \U{1}^2\;.\\
\hline
\end{array}\nonumber
\end{equation}
As indicated, they form the adjoint representations of $\E{7}\times\SO{14}$. Furthermore, the two remaining Cartans give rise to $\U{1}^2$. Their generators can be chosen as
\begin{equation}
t_\text{anom} = \big(-\tfrac{3}{2},-\tfrac{3}{2},0^6\big)\big(0^8\big) \quad\text{and}\quad t_2 = \big(0^8\big)\big(\tfrac{3}{2},0^7\big)\;,
\end{equation}
where the first one turns out to be anomalous.

In the case of the $\Z{3}$ orbifold, the three right-moving momenta $q=(0,\underline{-1,0,0})$ have the same transformation property under the action of the orbifold group, i.e. $q\cdot v =-\tfrac{1}{3}\mod 1$. Hence, states in the untwisted sector always appear with a multiplicity of three. We list the left-moving momenta $p$ which combine with these right-movers (or their CPT conjugate) and form invariant matter states of the untwisted sector:

\begin{equation}
\begin{array}{|l|rl|}
\hline
\big(\pm\big(\frac{1}{2},\frac{1}{2}\big),\big(\pm\frac{1}{2}\big)^6\big)\big(0^8\big) & 2\times 2^5 & \phantom{\big(\big)^{I^I}}\;\\[1mm]
\left(\underline{\pm 1,0},\underline{\pm 1,0^5}\right)\left(0^8\right)                 & 48          & \Rightarrow \text{112 weights} \Rightarrow (\rep{56},\rep{1})_{(\tfrac{3}{2},0)} + (\crep{56},\rep{1})_{(-\tfrac{3}{2},0)} \\[2mm]
\hline
\left(\pm\left(1,1\right),0^6\right)\left(0^8\right)\phantom{\big(\big)^{I^I}}         & 2           & \Rightarrow \phantom{00}\text{2 weights} \Rightarrow (\rep{1},\rep{1})_{(-3,0)} + (\rep{1},\rep{1})_{(3,0)}\\[1mm]
\hline
\left(0^8\right)\left(\pm 1, \underline{\pm 1,0^7}\right)\phantom{\big(\big)^{I^I}}    & 2\times 14  & \Rightarrow \phantom{0}\text{28 weights} \Rightarrow (\rep{1},\rep{14})_{(0,-\tfrac{3}{2})} + (\rep{1},\rep{14})_{(0,\tfrac{3}{2})}\\[2mm]
\hline
\big(0^8\big)\big(\big(\pm\frac{1}{2}\big)^8\big)\phantom{\big(\big)^{I^I}}            & 2^7          & \Rightarrow \text{128 weights} \Rightarrow (\rep{1},\crep{64})_{(0,\tfrac{3}{4})} + (\rep{1},\rep{64})_{(0,-\tfrac{3}{4})}\\[2mm]
\hline
\end{array}\nonumber
\end{equation}
In summary, the left-chiral charged matter spectrum of the untwisted sector reads
\begin{equation}
3(\rep{56},\rep{1})_{(\tfrac{3}{2},0)} + 3(\rep{1},\rep{1})_{(-3,0)} + 3(\rep{1},\rep{14})_{(0,-\tfrac{3}{2})} + 3(\rep{1},\crep{64})_{(0,\tfrac{3}{4})}\;.
\end{equation}

\subsubsection{The Twisted Sector}

For the twisted sectors, we will give only a few details and mainly state the results. Since the first twisted sector $T_{(1)}$ contains only the right-chiral CPT conjugates of the states from $T_{(2)}$, we can focus on the later one. As there are no Wilson lines all 27 fixed points are degenerate, i.e. have an identical copy of the massless spectrum. Thus, we can start with some constructing element, say $\left(\theta^2,0\right)$. There, we compute the massless spectrum and multiply the result by 27 yielding the complete twisted matter spectrum. An efficient procedure to solve the equations for massless right- and left-movers is described in appendix B.3 of~\cite{Vaudrevange:2005dp}. The result reads
\begin{equation}
27 (\rep{1},\rep{14})_{(1,-\tfrac{1}{2})} + 27 (\rep{1},\rep{1})_{(-2,1)} + 81 (\rep{1},\rep{1})_{(1,1)}\;.
\end{equation}

\subsection[The $\Z{3}\times\Z{3}$ Orbifold]{The $\boldsymbol{\Z{3}\times\Z{3}}$ Orbifold}\index{orbifold!$\Z{3}\times\Z{3}$}
\label{sec:Z3xZ3Orbifold}

Like in the case of the $\Z{3}$ orbifold, we choose the six-dimensional lattice to be factorized as $\Gamma = \SU{3}\times\SU{3}\times\SU{3}$. This torus allows for a $\Z{3}\times\Z{3}$ point group, represented by the twist vectors
\begin{equation}
v_1 = \left(0,\tfrac{1}{3},0,-\tfrac{1}{3}\right) \quad\text{and}\quad v_2 = \left(0, 0, \tfrac{1}{3},-\tfrac{1}{3} \right)\;.
\end{equation}
From eqn.~(\ref{eqn:susycondition2}) we know that this choice will retain $\mathcal{N}=1$ supersymmetry in 4d. Furthermore, only three moduli are invariant: 
\begin{equation}
N_{1\bar{1}}, N_{2\bar{2}} \quad\text{and}\quad N_{3\bar{3}}\;.
\end{equation}
They correspond to deformations of three radii
\begin{equation}
R_1=R_2, R_3=R_4 \quad\text{and}\quad R_5=R_6\;.
\end{equation}

The number of inequivalent shift vectors has been classified in~\cite{Ploger:2007iq} taking into account that for $\Z{N}\times\Z{M}$ orbifolds shift vectors differing by lattice vectors can lead to inequivalent models. We will explain this in detail in section~\ref{sec:DiscreteTorsion}. All shifts and the resulting models for $\E{8}\times\E{8}$ and $\SO{32}$ can be found in~\cite{WebTables:2007z3}. For more details on $\Z{3}\times\Z{3}$, we refer to e.g.~\cite{Font:1989aj,Ploger:2006dp}.

\subsection[The $\Z{6}$-II Orbifold]{The $\boldsymbol{\Z{6}}$-II Orbifold}\index{orbifold!$\Z{6}$-II}
\label{sec:Z6IIOrbifold}

The $\Z{6}$-II orbifold has recently become very popular~\cite{Kobayashi:2004ya,Kobayashi:2004ud,Buchmuller:2004hv,Buchmuller:2005jr,Buchmuller:2005sh,Buchmuller:2006ik,Lebedev:2006kn,Lebedev:2006tr,Buchmuller:2007zd,Buchmuller:2007qf,Lebedev:2007hv}. Its point group is defined by the twist vector
\vspace{-0.35cm}
\begin{equation}
\label{eqn:Z6IITwist}
v = \left(0, \tfrac{1}{6},\tfrac{1}{3},-\tfrac{1}{2} \right)\;.
\end{equation}
This is in contrast to $\Z{6}$-I, where $v= \left(0, \tfrac{1}{6},\tfrac{1}{6},-\tfrac{1}{3} \right)$. The invariant moduli for $\Z{6}$-II are three real $(1,1)$ moduli $N_{1\bar{1}}$, $N_{2\bar{2}}$, $N_{3\bar{3}}$ and one complex $(1,2)$ modulus $N_{33}$ allowing for continuous deformations of in total five geometric parameters. From the form of the twist vector $v$ we see that the first and fifth twisted sectors $T_{(1)}$ and $T_{(5)}$ contain fixed points (since $v_i \neq 0$ for $i=1,2,3$), whereas $T_{(2)}$ and $T_{(4)}$ have fixed tori (due to $2v_3 = 4v_3 = 0\mod 1$) and $T_{(3)}$  has fixed tori as well ($3v_2 = 0\mod 1$). Their number and localization depend on the lattice.

\begin{figure}[ht]
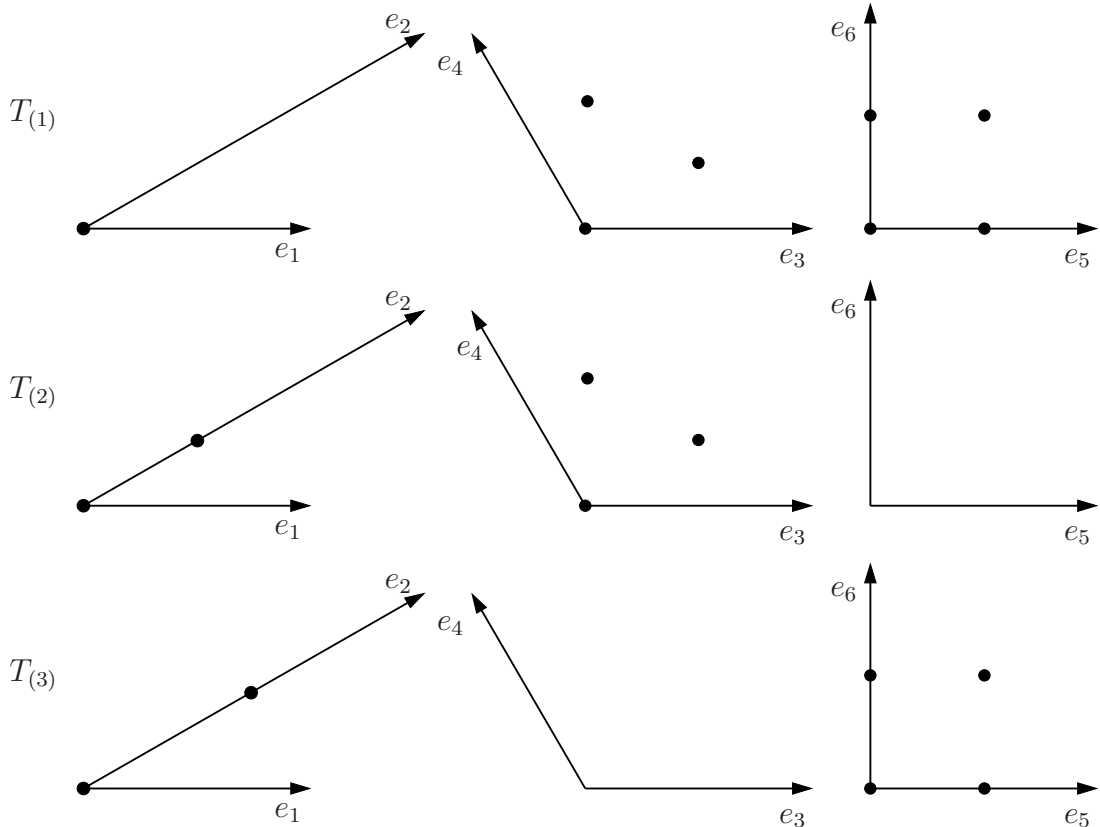

\centerline{\input Z6IIOrbifold.pstex_t}
\vspace{-0.45cm}
\caption{$\Z{6}$-II orbifold on the factorized lattice $\G{2}\times\SU{3}\times\SU{2}^2$. The inequivalent fixed points and fixed tori of the $T_{(1)}$, $T_{(2)}$ and $T_{(3)}$ twisted sectors are depicted. Strings in the second (third) twisted sector are free to move in the third (second) complex plane.}
\label{fig:Z6IIFix}
\end{figure}

We choose the factorized torus lattice: it reads
\vspace{-0.1cm}
\begin{equation}
\Gamma = \G{2}\times\SU{3}\times\SU{2}^2\;,
\end{equation}
that is depicted in fig.~(\ref{fig:Z6IIFix}), including the fixed point structure for the twisted sectors $T_{(1)}$, $T_{(2)}$ and $T_{(3)}$. The sectors $T_{(4)}$, $T_{(5)}$ are equivalent to $T_{(2)}$, $T_{(1)}$, respectively. For this lattice the five moduli correspond to deformations of four independent radii $R_1 = R_2$, $R_3 = R_4$, $R_5$ and $R_6$ and one angle $\alpha_{56}$. All inequivalent shift embeddings for $\E{8}\times\E{8}$ have been classified leading to 61 different choices~\cite{Katsuki:1989cs}. 

\subsection[The $\Z{2}\times\Z{2}$ Orbifold]{The $\boldsymbol{\Z{2}\times\Z{2}}$ Orbifold}\index{orbifold!$\Z{2}\times\Z{2}$}
\label{sec:Z2xZ2Orbifold}

The $\Z{2}\times\Z{2}$ point group is generated by the twists $\theta$ and $\omega$ whose action on $\mathbbm{C}^3$ is defined by the corresponding twist vectors
\begin{equation}
\label{eqn:Z2xZ2Twists}
v_1 = \left(0, \tfrac{1}{2},-\tfrac{1}{2},0 \right) \quad\text{and}\quad v_2 = \left(0, 0, \tfrac{1}{2},-\tfrac{1}{2} \right)\;.
\end{equation}
From eqn.~(\ref{eqn:susycondition2}) we know that this choice will retain $\mathcal{N}=1$ supersymmetry in 4d. Before choosing a lattice, we can see that three real $(1,1)$ moduli $N_{1\bar{1}}$, $N_{2\bar{2}}$, $N_{3\bar{3}}$ are invariant, as usual. In addition, there are three complex $(1,2)$ moduli $N_{11}$, $N_{22}$, $N_{33}$ giving rise to in total nine geometric deformation parameters. Their interpretation can only be specified after choosing a torus lattice. The four elements of the point group $\{\mathbbm{1}, \theta, \omega, \theta\omega\}$ generate four sectors: the untwisted sector $U$ and three twisted sectors, denoted by $T_{(1,0)}$, $T_{(0,1)}$ and $T_{(1,1)}$. Since the associated twist vectors $v_1$, $v_2$ and $v_3 = v_1+v_2$ have a zero-entry, there will be only fixed tori and no fixed points. The number of fixed tori per twisted sector, however, depends on the torus lattice. In the following we will discuss two examples for the torus lattice: a factorized one and a non-factorizable one.

\subsubsection{Factorized Lattice}

The prototype for a factorized lattice in the context of $\Z{2}\times\Z{2}$ orbifolds is $\Gamma = \text{SU}(2)^6$. In this case the moduli correspond to six radii $R_i=|e_i|$ for $i=1,\ldots,6$ and three angles $\alpha_{12}$, $\alpha_{23}$ and $\alpha_{56}$. Each twisted sector contains 16 fixed tori, in total $3\times 16 = 48$. Further details on $\Z{2}\times\Z{2}$ orbifolds on this lattice can be found in~\cite{Forste:2004ie,Vaudrevange:2005dp}.

\subsubsection{Non-Factorizable Lattice}
\label{sec:Z2xZ2NonFactorized}

\begin{figure}[t]
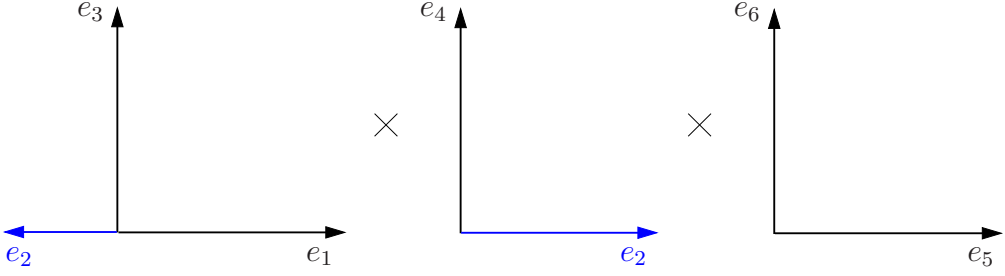

\centerline{\input Z2xZ2_A2Lattice.pstex_t}
\caption{The alignment of the non-factorizable lattice $\Gamma = \SU{3}\times\SU{2}^4$ inside the three orthogonal complex planes $\mathbbm{C}\times\mathbbm{C}\times\mathbbm{C}$. Note that the $\SU{3}$ sublattice spanned by $e_1$ and $e_2$ lies skew inside $\mathbbm{C}^3$. This lattice allows for a point group $\Z{2}\times\Z{2}\subset \SU{3}$.}
\label{fig:Z2xZ2_A2Lattice}
\end{figure}

All inequivalent non-factorizable lattices $\Gamma$ for the $\Z{2}\times\Z{2}$ orbifold have been classified~\cite{Forste:2006wq} leading to a total number of eight distinct lattices, denoted by A.1 to A.8. As an example, we choose the lattice $\Gamma = \SU{3}\times\SU{2}^4$ named A.2. In order to see that this lattice is non-factorizable we have to specify the vectors spanning the lattice $\Gamma$ in $\mathbbm{C}^3$
\begin{equation}
\begin{array}{lll}
e_1 = (\sqrt{2},0,0),   & e_2 = (-\tfrac{1}{\sqrt{2}},\sqrt{\tfrac{2}{3}},0), & e_3 = (\sqrt{2}\I,0,0),  \\
e_4 = (0,\sqrt{2}\I,0), & e_5 = (0,0,\sqrt{2}),                               & e_6 = (0,0,\sqrt{2}\I),
\end{array}
\end{equation}
where we have fixed all radii and angles, as depicted in fig.~(\ref{fig:Z2xZ2_A2Lattice}). This lattice allows for the $\Z{2}\times\Z{2}$ point group generated by the twist vectors eqn.~(\ref{eqn:Z2xZ2Twists}). 
Using the nine deformation parameters\footnote{Using $e_i = (e_i^1,e_i^2,e_i^3)\in\mathbbm{C}^3$.} $R_1 = |e_1| = 2|e_2^1|$, $\tilde{R}_2 \equiv |e_2^2|$, $R_i = |e_i|$ for $i=3,\ldots,6$, $\alpha_{13}$, $\alpha_{24}$ and $\alpha_{56}$ it cannot be deformed continuously to a factorized form. Thus, it is non-factorizable. Due to the special alignment of $\Gamma$ inside $\mathbbm{C}^3$, the number of inequivalent fixed tori changes compared to the factorized lattice. For example, in $T_{(1,0)}$ the constructing twist $\theta$ leaves in total $16$ fixed tori invariant as in the factorized case. However, now $\theta\omega$ identifies some of them leading to $12$ inequivalent fixed tori, see fig.~(\ref{fig:Z2xZ2_A2SU3SubLattice}) for an illustration. Also in the other two twisted sectors $T_{(0,1)}$ and $T_{(1,1)}$ the number of inequivalent fixed tori is reduced, from $16$ to $8$ in both sectors. All $12+8+8$ fixed tori are depicted in fig.~(\ref{fig:Z2xZ2_A2FixedTori}). For a model on this lattice see section~\ref{sec:Z2xZ2}.

\begin{figure}[ht!]
\centerline{\input Z2xZ2_A2SU3Sublattice.pstex_t}
\caption{Fixed points in the $\SU{3}$ sublattice of $T_{(1,0)}$. $\theta = \text{diag}(-1,-1)$ (in the real basis) leaves four points invariant. However, under the action of $\theta\omega = \text{diag}(-1,1)$ the points corresponding to $\left(\theta, e_2\right)$ and $\left(\theta, e_1 + e_2\right)$ are identified resulting in three inequivalent ones.}
\label{fig:Z2xZ2_A2SU3SubLattice}

\vspace{0.35cm}

\centerline{\input Z2xZ2_A2FixedTori.pstex_t}
\caption{Localization of the fixed tori of the $\Z{2}\times\Z{2}$ orbifold on the non-factorizable lattice $\Gamma = \SU{3}\times\SU{2}^4$ separated by twisted sectors $T_{(k,l)}$. The coordinates of the fixed tori in $\mathbbm{C}^3$ are marked by circles and triangles. There are $(2 \times 2) + (4 \times 2) = 12$ inequivalent fixed tori in the sector $T_{(1,0)}$, $2 \times 4 = 8$ in $T_{(0,1)}$ and finally $2 \times 4 = 8$ in $T_{(1,1)}$.}
\label{fig:Z2xZ2_A2FixedTori}
\end{figure}

\section[Classification of $\Z{N}$ Orbifolds for $\SO{32}$]{Classification of $\boldsymbol{\Z{N}}$ Orbifolds for $\boldsymbol{\SO{32}}$}
\label{sec:SO32}

In this section we summarize the results from a classification of $\Z{N}$ orbifold compactifications of the $\SO{32}$ heterotic string without Wilson lines~\cite{Nilles:2006np}. For each orbifold we choose a specific torus lattice. In detail, they read $\SU{3}^3$, $\SO{5}^2\times\SO{4}$, $\G{2}^2\times\SU{3}$, $\G{2}\times\SU{3}\times\SO{4}$, $\SU{7}$, $\SO{9}\times\SO{5}$, $\SO{9}\times\SO{4}$, $\F{4}\times\SU{3}$ and $\F{4}\times\SO{4}$, listed according to the order of the $\Z{N}$ point groups in table~(\ref{tab:compareSO32andE8xE8}). 

For the classification of the shift vectors $V$ we use Dynkin diagram techniques, as described for example in~\cite{Kac:1986gs,Choi:2003pqa,Wingerter:2005ph,Nilles:2006np,Ramos:2008ph}. The results are summarized in table~(\ref{tab:compareSO32andE8xE8}a). From there, we see that generically there are more models in the $\SO{32}$ case than in $\E{8}\times\E{8}$ and the difference in the amount of respective models increases with increasing order $N$. More details on the $\SO{32}$ models, including the shift vectors $V$, the twist vectors $v$, the gauge groups and the resulting massless matter spectra can be found in~\cite{WebTables:2006SO}.

Finally, we analyze the appearance of massless spinors of $\SO{10}$ and $\SO{12}$ for our $\SO{32}$ models\footnote{It was shown in~\cite{Nilles:2006np} that massless spinors of bigger groups do not appear in orbifold models of the $\SO{32}$ heterotic string. For related work on $\SO{32}$ see~\cite{Choi:2004wn}.}. It is a well-known fact that by breaking the adjoint of $\SO{32}$ we cannot obtain spinor representations of $\SO{2M}$. This can be seen from the length squares of the respective weights: the roots of $\SO{32}$ correspond to weights $p$ of $\text{Spin}(32)/\Z{2}$ with $p^2 = 2$; on the other hand spinorial weights $p$ of $\text{Spin}(32)/\Z{2}$ have a length squared $p^2 = 4$. Thus, the latter one cannot be massless in the untwisted sector, see appendix~\ref{app:weightlattices}. However, they might be massless in some twisted sectors. In fact, in section 3 of~\cite{Nilles:2006np} it is shown that for all orbifolds $\Z{N}$ with $N > 3$ there exists a shift vector yielding a spinor $\rep{16}$ of $\SO{10}$ in the first twisted sector. These models might be excellent starting points for a ``Mini-Landscape'' of $\SO{32}$ heterotic orbifolds, compare to chapter~\ref{sec:ML}. In addition, there are models where the spinor appears in higher twisted sectors. To summarize that, table~(\ref{tab:compareSO32andE8xE8}b) lists the number of models equipped with spinor representations of $\SO{10}$ or $\SO{12}$ from any twisted sector.

\begin{table}[ht]
\begin{center}
\begin{tabular}{|l|r|r|}
\multicolumn{3}{l}{(a)$\phantom{I_{I_I}}$}\\
\multicolumn{3}{l}{$\phantom{I_{I_I}}$}\\\hline
\hline
           & \multicolumn{2}{c|}{\# ineq. models in}\\
$\Z{N}$    & $\SO{32}$ & $\E{8}\times\E{8}$\\
\hline
\hline
$\Z{3}$    &    6      &    5      \\ 
$\Z{4}$    &   16      &   12      \\
$\Z{6}$-I  &   80      &   58      \\
$\Z{6}$-II &   75      &   61      \\
$\Z{7}$    &   56      &   40      \\
$\Z{8}$-I  &  196      &  145      \\
$\Z{8}$-II &  194      &  146      \\
$\Z{12}$-I & 2295      & 1669      \\
$\Z{12}$-II& 2223      & 1663      \\
\hline
\end{tabular}
\hspace{1cm}
\begin{tabular}{|r|r|r|}
\multicolumn{3}{l}{(b)$\phantom{I_{I_I}}$}\\
\multicolumn{3}{l}{$\phantom{I_{I_I}}$}\\\hline
models with        & \multicolumn{2}{c|}{\# models with}\\
anomalous $\U{1}$  & $\rep{16}$ of $\SO{10}$ & $\rep{32}$ of $\SO{12}$\\
\hline
\hline
    5              &  0                      &  0 \\ 
   12              &  2                      &  0 \\
   76              &  4                      &  4 \\
   65              & 10                      &  3 \\
   55              &  2                      &  0 \\
  193              & 12                      &  0 \\
  166              & 11                      &  7 \\
 2269              & 80                      & 36 \\
 2097              &116                      & 10 \\
\hline
\end{tabular}
\end{center}
\caption{(a) Comparison between the number of inequivalent $\Z{N}$ models for $\SO{32}$~\cite{Nilles:2006np} and $\E{8}\times\E{8}$ heterotic string~\cite{Ramos:2008ph}. Note that the $\E{8}\times\E{8}$ results obtained in~\cite{Ramos:2008ph} coincide with the older ones in~\cite{Katsuki:1989bf} for point groups $\Z{N}$ with $N < 8$, but differ in the other cases. (b) Numbers of inequivalent $\Z{N}$ models containing at least one spinor of $\SO{10}$ or $\SO{12}$. We also present the number of $\SO{32}$ models having an anomalous $\U{1}$.}
\label{tab:compareSO32andE8xE8}
\end{table}

\section{Yukawa Couplings}\index{Yukawa couplings}
\label{sec:Yukawa}

Consider the $n$--point correlation function of two fermions and $n-2$ bosons~\cite{Dixon:1986qv, Hamidi:1986vh}
\begin{equation}
\label{eqn:correlation}
\langle\text{F\,F\,B}\ldots\text{B}\rangle\;.
\end{equation}
The corresponding physical states shall be denoted by $\Psi_r$, $r = 1,\ldots,n$. Then, in the field theory limit, a non--vanishing correlation function implies the following term in the superpotential
\begin{equation}
\mathcal{W} ~\supset~ \Psi_1\, \Psi_2\, \Psi_3 \ldots \Psi_n\;,
\end{equation}
where the $\Psi_r$'s now denote the corresponding chiral superfields. A complete evaluation of eqn.~(\ref{eqn:correlation}) has only been performed for 3--point couplings and yields a moduli dependent coupling strength~\cite{Dixon:1986qv,Hamidi:1986vh,Casas:1991ac,Erler:1992gt}.

On the other hand, symmetries of eqn.~\eqref{eqn:correlation} give rise to the so--called string selection rules. These rules determine whether a given coupling vanishes or not. We use the following notation: the constructing elements of $\Psi_r$ are denoted by $g_r \in S$, their left-moving shifted momenta by $p_{\text{sh},r}$ and finally their R--charges by $R_r$, respectively. Then, the string selection rules read:

\subsection{Gauge Invariance}\index{Yukawa couplings!gauge invariance}
Invariance of eqn.~(\ref{eqn:correlation}) under variations in the gauge degrees of freedom result in a condition on the associated momenta $p_\text{sh}$: the sum over all left--moving shifted momenta $p_{\mathrm{sh},r}$ must vanish:
\begin{equation}
\label{eqn:gaugeinvariance} 
\sum_r p_{\mathrm{sh},r} = 0
\end{equation}
This translates to the field theoretic requirement of gauge invariance for allowed terms in the superpotential.

\subsubsection{Anti-Symmetry under Particle Exchange}

If some field appears more than once in a given coupling, it is important to take care of the symmetry / antisymmetry properties under the exchange of identical particles.

Let us see this by an easy example. Consider a theory with gauge group $\SU{2}$ and a particle content including some doublets and singlets. Furthermore, assume that the coupling $\rep{2} \times \rep{2} \times \rep{1}$ is allowed by all selection rules. Indeed, from gauge invariance it is clearly allowed, since the tensor product of two doublets contains a singlet, i.e.
\begin{equation}
\rep{2} \times \rep{2} = \rep{3}_s + \rep{1}_a\;,
\end{equation}
where $s$ and $a$ denote whether the representation is symmetric or antisymmetric under the exchange of the $\rep{2}$'s. Thus, if both doublets correspond to the same particle the singlet vanishes. We can also see this in the index notation, where $A^i$ and $B^j$ denote the components of the doublets. Then
\begin{equation}
A^i B^j = \underbrace{\frac{1}{2}\left(A^i B^j + A^j B^i\right)}_{\rep{3}_s} + \underbrace{\frac{1}{2}\left(A^i B^j - A^j B^i\right)}_{\rep{1}_a}
\end{equation}
and the singlet $\rep{1}_a$ vanishes if both doublets are the same $A=B$. In appendix~\ref{app:gaugeinvariancesymmetry}, we give a long but not complete list of the most frequent such cases.

\subsection{Space Group Invariance}\index{Yukawa couplings!space group invariance}
\label{sec:spacegroupselection}

The product of constructing elements $g_r$ must be the identity:
\begin{equation}
\label{eqn:spacegroupselectionrule}
\prod_r g_r = (\mathbbm{1}, 0) \ .
\end{equation}

\begin{figure}[b]
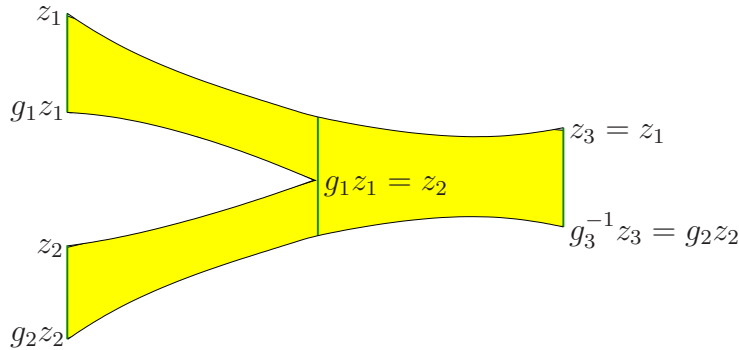

\centerline{\input spacegroup.pstex_t}
\vspace{-0.3cm}
\caption{Illustration of the space group selection rule in the covering space $\mathbbm{C}^3$ of the orbifold $\mathbbm{C}^3/S$. The closed strings with boundary conditions  $g_1$, $g_2$ and $g_3^{-1}$ look like open ones in $\mathbbm{C}^3$. In order to merge, the constructing elements have to multiply to the identity. Note that the boundary conditions have to be evaluated at the time of the strings' merging.}
\label{fig:spacegroup}
\end{figure}

This selection rule can be visualized as the geometrical ability of twisted strings to join. Consider the example of a 3--point coupling of strings with constructing elements $g_1$, $g_2$ and $g_3^{-1}$ (choosing $g_3^{-1}$ will turn out to be more convenient than $g_3$). In the covering space $\mathbbm{C}^3$ of the orbifold $\mathbbm{C}^3/S$ the three strings look like open strings. Two ``open'' strings merge in order to give the third one. However, in order to be able to merge the boundary conditions have to multiply to the identity, see figure~(\ref{fig:spacegroup}) for an illustration. We can evaluate the three boundary conditions at the time of merging,
\begin{eqnarray}
                     z_3 & = & g_3g_2 z_2 = g_3 g_2 g_1 z_1 = g_3 g_2 g_1 z_3 \\
\Rightarrow  g_1 g_2 g_3 & = & \mathbbm{1}\;.
\end{eqnarray}

For computational purposes this condition is not very practical. For a given twisted string with constructing element $g_r$ this condition depends on the specific choice of $g_r \in [g_r]$. A given coupling might seem forbidden by the space group selection rule using $g_r \in [g_r]$, but with a different choice $g_r' \in [g_r]$ it could turn out to be allowed. Thus, it is more convenient to reformulate this condition in terms of conjugate elements $h_r g_r h_r^{-1}$
\begin{equation}
\prod_r h_r g_r h_r^{-1} = (\mathbbm{1}, v)
\end{equation}
with $v \in \sum_r(\mathbbm{1} - \theta^{k_r})\Lambda$, see for example~\cite{Kobayashi:1990mc}. Then, one can take \emph{any} representative $g_r \in [g_r]$ and each constructing element $g_r$ is conjugated independently by $h_r$. Due to the definition of $v$ it is enough to use pure rotations for $h_r$, e.g. $h_1 = (\theta,0)$, $h_2 = (\theta^3,0)$, $h_3 = (\mathbbm{1},0)$.

\subsection[$R$--Charge conservation]{$\boldsymbol{R}$--Charge conservation}\index{Yukawa couplings!$R$--charge conservation}

The conditions for $R$--charge conservation read~\cite{Kobayashi:2004ya}
\begin{equation}
\label{eqn:rchargeconservation} 
\sum_r R_r^i~=~0 \mod  N^i \quad\text{ for }i=1,2,3\;,
\end{equation}
 where $N^i$ denotes the order of the $i$--th component of the twist vector, i.e.\ $N^i v^i \in \mathbb{Z}$ (no summation). Here, two of the $R_r$ are fermionic and the rest are bosonic. For computational purposes, it is more convenient to use the purely bosonic notation, where eqn.~(\ref{eqn:rchargeconservation}) becomes $\sum_r R_r^i = -1 \text{ mod } N^i$ and all $R_r$ are bosonic.

$R$--charge conservation can be understood as a remnant of ten-dimensional Lorentz invariance. In 10d couplings have to be invariant under the full Lorentz group $\SO{9,1}$. This includes invariance under the action of
\begin{equation}
\label{eqn:continuousLorentz}
e^{2\pi\I \alpha_1 J_{12}}\;, e^{2\pi\I \alpha_2 J_{34}} \;\text{and}\; e^{2\pi\I \alpha_3 J_{56}}\;,
\end{equation}
where $\alpha_i \in\mathbbm{R}$ and $J_{i,i+1}$ denote the Cartan generators as defined in eqn.~(\ref{eqn:J}). By the compactification, the $\SO{9,1}$ breaks to the 4d Lorentz group $\SO{3,1}$ times some ``internal part''. In section~\ref{sec:Compactification} we discussed that for a generic compactification the internal part is $\SU{4}$, the $R$ symmetry group of $\mathcal{N} = 4$ supersymmetry. For Abelian orbifolds the $R$ symmetry $\SU{4}$ is in general broken to a discrete subgroup of $\U{1}^3\subset\SU{4}$ and the $\U{1}^3$ corresponds to the three elements listed in eqn.~(\ref{eqn:continuousLorentz}). More specifically, there are unbroken discrete symmetries of the six-dimensional orbifold space. As they remain a symmetry of the space, they should also remain a symmetry of the theory, i.e. of the couplings. In the literature, these symmetries are often called \emph{sublattice rotations}~\cite{Hamidi:1986vh,Font:1988tp,Araki:2007ss}.

For factorizable torus lattices, the discrete remnants of the $\SU{4}$ $R$--symmetry are generated by three elements
\begin{equation}
\label{eqn:discreteLorentz}
e^{2\pi\I J_{12}/N^1}\;, e^{2\pi\I J_{34}/N^2} \;\text{and}\; e^{2\pi\I J_{56}/N^3}\;,
\end{equation}
where $N^i$ denotes the order of $v^i$, as before. These elements generate an Abelian discrete $R$--symmetry group
\begin{equation}
\Z{N^1}\times\Z{N^2}\times\Z{N^3}\;.
\end{equation}
Next, we have to clarify how this group acts on states compactified on the orbifold. Consider some generic state with possible oscillator excitations $\tilde{\alpha}$, $|q_\text{sh}\rangle\RM \otimes \tilde{\alpha} |p_\mathrm{sh}\rangle\LM$. Under a sublattice rotation in the $i$-th plane, it picks up a phase $e^{2\pi\I R^i/N^i}$, compare to eqns.~(\ref{eqn:traforightmover}) and~(\ref{eqn:OsciTrafo}). Consequently, demanding invariance of the $n$-point coupling under these discrete transformations results in the conditions stated in eqn.~(\ref{eqn:rchargeconservation}).

In the non-factorizable case, the situation seems unclear. For example, for the $\Z{6}$-II orbifold with a $\Gamma = \SU{6}\times\SU{2}$ torus lattice the only discrete subgroup of $\U{1}^3$ that maps the torus lattice $\Gamma$ to itself is $\theta = e^{2\pi\I (v_1 J_{12} + v_2 J_{34} + v_3 J_{56})}$, compare to~\cite{Ramos:2008ph}. Note that in this case $\theta$ is not an $R$ symmetry in the strict sense as it commutes with the generator of supersymmetry and therefore all components of a SUSY multiplet carry the same $R$--charges. On the other hand, the non-factorizable torus lattice discussed in section~\ref{sec:Z2xZ2NonFactorized} in the case of a $\Z{2}\times\Z{2}$ orbifold allows for normal sublattice rotations. One can check that using the visualization of the lattice in fig.~(\ref{fig:Z2xZ2_A2Lattice}). In summary, it remains an open question how $R$--charge conservation is defined in the non-factorizable case.

\subsection{Discrete Anomalies}\index{discrete anomaly}
\label{sec:discreteanomalies}

The string selection rules listed in the previous sections clearly restrict the form of the superpotential. Therefore, these rules can be understood in terms of discrete symmetries (obviously except for gauge invariance). Very much like continuous symmetries, discrete symmetries can be broken by quantum effects, i.e.\ have an anomaly~\cite{Krauss:1988zc}. If this is the case, one expects that the corresponding conservation laws be violated through non-perturbative effects. The criteria for discrete symmetries to be non-anomalous, and thus to be exact, have been studied in the Abelian ($\Z{N}$) case~\cite{Ibanez:1991hv,Banks:1991xj}.

In~\cite{Araki:2008ek} it is shown how to rederive the anomaly constraints using the path integral approach~\cite{Fujikawa:1979ay,Fujikawa:1980eg} for Abelian and non-Abelian discrete symmetries. However, here we restrict to the case of a discrete $\Z{N}$ symmetry and the anomalies $\Z{N}-G-G$ and $\Z{N}$-gravity-gravity are determined by
\begin{eqnarray}
A_{\Z{N}-G-G}                        &=&\frac{1}{N}\sum_{\rep{r}^{(f)}} q^{(f)} \, \big(2\,\ell(\rep{r}^{(f)})\big) ~\in~\Z{} \;,\label{eq:A_Z_N-G-G}\\
A_{\Z{N}-\mathrm{grav}-\mathrm{grav}}&=&\frac{2}{N}\sum_{m} q^{(m)} \, \dim\rep{R}^{(m)} ~\in~\Z{}\;,\label{eq:A_Z_N-grav-grav}
\end{eqnarray}
compare to figure~(\ref{fig:Anomaly4d}) in the introduction. If $A_{\Z{N}-G-G}$ and $A_{\Z{N}-\mathrm{grav}-\mathrm{grav}}$ are both integer the discrete symmetry is non-anomalous\footnote{In the first equation, eqn.~(\ref{eq:A_Z_N-G-G}), the sum extends over all representations $\rep{r}^f$ of the non-Abelian gauge group factor $G$ which carry integer $\Z{N}$ charges $q^{(f)}$. The Dynkin index $\ell(\boldsymbol{r}^{(f)})$ of some representation $\rep{r}^f$ is defined as in section.~\ref{sec:anomalousU1}. In the second equation, eqn.~(\ref{eq:A_Z_N-grav-grav}), the sum extends over all representations $\rep{R}^{(m)}$, where $\rep{R}^{(m)}$ denotes the representation under \emph{all} non-Abelian gauge group factors, e.g. $\rep{R}^1 = (\rep{3},\rep{2})$ of $\SU{3}\times\SU{2}$ which yields $\dim\rep{R}^1 = 6$.}.

\subsubsection{Discrete Anomalies for $\boldsymbol{\Z{6}}$-II Orbifolds}

As an example, we check whether the discrete symmetries of $\Z{6}$-II orbifold models with torus lattice $\G{2}\times\SU{3}\times\SU{2}^2$ are anomalous. For details on $\Z{6}$-II see section~\ref{sec:Z6IIOrbifold}. In this case, the space group selection rule~(\ref{eqn:spacegroupselectionrule}) can be interpreted as a $\Z{6}\times\Z{2}\times\Z{2}'\times\Z{3}$ symmetry, denoted as $\Z{6}^k\times\Z{2}^\text{flavor}\times\Z{2}^{\text{flavor}\,\prime}\times\Z{3}^\text{flavor}$. In detail, the point group selection rule reads
\begin{equation}\label{eq:k-rule}
 \sum_{r=1}^n k^{(r)}~=~0 \mod 6 \;,
\end{equation}
where the sum runs over the $n$ states involved in the coupling. Furthermore, the translational part can be rewritten as 
\begin{subequations}\label{eq:SpaceGroupRules}
\begin{eqnarray}
\SU{2}^2\text{ plane} & : &  \sum\limits_{r=1}^n k^{(r)}\, n_2^{(r)}         ~=~ 0\mod2\;,\\*
                      &   &  \sum\limits_{r=1}^n k^{(r)}\, n_2^{(r)\,\prime} ~=~ 0\mod2\;,\\* \label{eq:n2primeRule}
\SU{3}\text{ plane}   & : &  \sum\limits_{r=1}^n k^{(r)}\, n_3^{(r)}         ~=~ 0\mod3\;.
\end{eqnarray}
\end{subequations}
The quantum numbers $n_3^{(r)}$, $n_2^{(r)}$ and $n_2^{(r)\,\prime}$ specify the localization of the states on the orbifold; here, we follow the conventions of \cite{Buchmuller:2006ik}. Under this symmetry, each state comes with one $\Z{6}$ charge $k$, two $\Z{2}$ charges $q_2$,$q_2'$ and one $\Z{3}$ charge $q_3$, all being integer and defined modulo the order of the respective $\Z{N}$, i.e.
\begin{eqnarray}
\Z2^\text{flavor}           & : & q_2 ~=~k\,n_2  \mod 2\;,\\
\Z2^{\text{flavor}\,\prime} & : & q_2'~=~k\,n_2' \mod 2\;,\\
\Z3^\text{flavor}           & : & q_3 ~=~k\,n_3  \mod 3\;.
\end{eqnarray}
Now, for a given model, the $\Z{N}-G-G$ anomalies are computed according to
\begin{eqnarray}
 A_{\Z6^k-G-G}               & = & \frac{1}{6}\,\sum_{\rep{r}^{(f)}} k^{(f)}\,2\,\ell(\rep{r}^{(f)})\;, \label{eq:A_Z6-G-G} \\
 A_{\Z{n}^\text{flavor}-G-G} & = & \frac{1}{n}\,\sum_{\rep{r}^{(f)}} q_n^{(f)}\,2\,\ell(\rep{r}^{(f)})\;.\label{eq:FlavorAnomaly}
\end{eqnarray}
These anomalies turn out to be universal for the different gauge group factors $G$ of a given model. However, generically they do not vanish. The discrete symmetries arising from the space group selection rule are consequently anomalous. We define the integer quantities $k_\mathrm{anom}$, $n_3^\mathrm{anom}$ and $n_2^\mathrm{anom}$ by
\begin{eqnarray}
 A_{\Z6^k-G-G}             & = & \frac{1}{6}k_\text{anom}   \mod 1\;,\\
 A_{\Z3^\text{flavor}-G-G} & = & \frac{1}{3}n_3^\text{anom} \mod 1\;,\\
 A_{\Z2^\text{flavor}-G-G} & = & \frac{1}{2}n_2^\text{anom} \mod 1\;,
\end{eqnarray}
such that these discrete anomalies can be encoded in the so-called \emph{anomalous space group element}\index{anomalous space group element} $g^\mathrm{anom}=(\theta^{k^\mathrm{anom}},n_\alpha^\mathrm{anom}\,e_\alpha)$.

\subsubsection{Anomalous Space Group Element vs. Generator of the Anomalous $\boldsymbol{\U{1}_\text{anom}}$}

Interestingly, the anomalous space group element is closely related to the anomalous $\U{1}_\text{anom}$ present in most orbifold models. Using the normalization
\begin{equation}
t_\text{anom}~=~\frac{1}{12} \sum_i p_\text{sh}^{(i)}\;.
\end{equation}
for the generator of $\U{1}_\text{anom}$, compare to section~\ref{sec:anomalousU1}, we make the following observation
\begin{equation}
t_\text{anom}~=~ k^\text{anom}\,V + \sum_\alpha n_\alpha^\text{anom}\,A_\alpha +\lambda\;,
\end{equation}
where $\lambda \in \Lambda$. In other words, the anomalous $\U{1}$ generator can be expressed in terms of shift and Wilson lines, where the coefficients are given by the discrete anomalies of the space group selection rule. For more details and further relations among discrete anomalies see~\cite{Araki:2008ek}.

\subsection[A Note on the $\gamma$ Rule]{A Note on the $\boldsymbol{\gamma}$ Rule}\index{Yukawa couplings!no need for a $\gamma$ rule}
\label{sec:GammaShort}\index{strings on orbifolds!gamma-phase}

In the literature, there exists an additional selection rule, here referred to as the $\gamma$ rule. In our notation, it reads~\cite{Casas:1991ac,Kobayashi:2004ya}
\begin{equation}
\sum_i \gamma_i = 0 \text{ mod }1\;,
\end{equation}
where $\gamma_i$ denotes the gamma--phase of $\Psi_i$. In this section, it is argued that, in contrast to previous statements, a fully consistent approach yields to automatic fulfillment of the $\gamma$ rule~\cite{Lebedev:2007hv}. The discussion here is restricted to the case where $\Phi_\text{vac} = 1$\index{strings on orbifolds!vacuum phase}, for the general case with non-vanishing vacuum phase see appendix~\ref{app:gamma}.

The correlation function corresponding to the coupling
\begin{equation}
\label{eqn:couplingpsi}
 \Psi_1\, \Psi_2 \dots \Psi_n
\end{equation}
should be invariant under the action of the full space group. Let us assume first that the states $\Psi_i$ corresponded to linear combinations of equivalent fixed points within the fundamental domain of the torus (see e.g.~\cite{Casas:1991ac,Kobayashi:1991rp,Kobayashi:2004ya}). For example, in the case of the $\mathbb{Z}_6$--II orbifold only fixed points in the $G_2$ lattice could form linear combinations. Under this assumption, different states $\Psi_i$ would be eigenstates with respect to \emph{different} space group elements. So one could not transform the coupling eqn.~(\ref{eqn:couplingpsi}) with a given $h = (\theta^l,m_\alpha e_\alpha)$.

Thus, the fully consistent approach for building invariant linear combinations, as presented in section~\ref{sec:physstates}, is necessary. In this case, we can compute the gamma--phase for all states $\Psi_i$ from eqn.~(\ref{eqn:projectionwithgamma}), i.e. $\gamma_i = \gamma_i (h)$ for arbitrary $h=(\theta^l,m_\alpha e_\alpha)$. But since allowed couplings already fulfill the selection rules eqns.~(\ref{eqn:gaugeinvariance}) and~(\ref{eqn:rchargeconservation}), the $\gamma$ rule is satisfied trivially
\begin{eqnarray}
\label{eqn:GammaShort}
\gamma_i(h)                    & = & R_i \cdot v_h - p_{\text{sh},i} \cdot V_h\;, \\
\Rightarrow \sum_i \gamma_i(h) & = & \underbrace{\left(\sum_i R_i\right)}_{\sim 0 \text{ see eqn.~(\ref{eqn:rchargeconservation})}}\cdot v_h 
                                   - \underbrace{\left(\sum_i p_{\text{sh},i}\right)}_{=0 \text{ see eqn.~(\ref{eqn:gaugeinvariance})}}\cdot V_h\;, \\
                              & = & 0 \mod  1\;.\nonumber
\end{eqnarray}
Thus, the $\gamma$ rule in the fully consistent approach is \emph{not} a selection rule. It is a consequence of other selection rules and invariance of the states. We therefore conclude that the coupling must only satisfy gauge invariance, $R$--charge conservation and the space group selection rule.

This has important consequences. For example, in the model A1 of~\cite{Kobayashi:2004ya}, there is no mass term for the exotics $\bar{q}_2 q_2$ up to order 9 in singlets. However, it was found in~\cite{Lebedev:2007hv} that the coupling $\bar{q}_2 q_2 S_9 S_{15} S_{22} S_{33}$ is allowed by the selection rules of section~\ref{sec:Yukawa}. Further, using the prescription of section~\ref{sec:physstates}, the gamma--phases of the corresponding physical states are $\gamma_i = (\tfrac{1}{2},\,0,\,0,\,\tfrac{5}{6},\,\tfrac{2}{3},\,0)$ for $h = (\theta,0)$ , which sum up to $2$. This is in contrast to~\cite{Kobayashi:2004ya}, where $\gamma_i = (0,\,0,\,0,\,\tfrac{1}{2},\,\tfrac{2}{3},\,0)$ and linear combinations were built differently.

\chapter{Discrete Torsion}
\label{sec:DT}

Discrete torsion is an elegant way to extend the orbifold construction~\cite{Vafa:1986wx,Font:1988mk,Vafa:1994rv,Sharpe:2000ki,Gaberdiel:2004vx}. It yields new, consistent models, which were thought not to be accessible from the torsionless construction. Early work mainly concentrated on discrete torsion in the case of $\mathbbm{Z}_N \times \mathbbm{Z}_M$ orbifolds~\cite{Font:1988mk}. However, its importance for model--building was underestimated compared to other parameters of the theory, like Wilson lines.

In reference~\cite{Ploger:2007iq}, it was shown that discrete torsion is much deeper connected to the standard construction of orbifolds than expected. New relations to the shift-embedding and to the torus lattice were discovered. That is, the effect of discrete torsion can likewise be mimicked by a torsionless model which has either a different shift-embedding or a different, non-factorizable torus lattice. Furthermore, it turns out that discrete torsion can likewise be applied to the case of $\mathbbm{Z}_N$ orbifolds.

The results of reference~\cite{Ploger:2007iq} are presented in this chapter. More details, especially for the examples, and some additional discussions are included in order to make discrete torsion accessible to a broader audience.

\section{Brother Models and Discrete Torsion}
\label{sec:DiscreteTorsion}

In this section we start by examining a new possibility to find inequivalent models. We discuss under what circumstances models with shifts differing by lattice vectors have different spectra and are thus inequivalent. Then we review the concept of discrete torsion, and clarify its relation to models in which shifts differ by lattice vectors.

\subsection[Brother Models]{Brother Models}
\label{sec:brothermodel}

Let us start by clarifying under which conditions two $\mathbbm{Z}_N \times \mathbbm{Z}_M$ orbifold models M and M$'$ are equivalent. First, we restrict to the case without Wilson lines, where the models M and M$'$ are described by the set of shifts $(V_1,V_2)$ and $(V_1',V_2')$, respectively. Clearly, if the shifts are related by Weyl reflections\index{Weyl reflection}, i.e.
\begin{equation}\label{eq:Weyl}
 (V_1',V_2')~=~(W\,V_1,W\,V_2)\;,
\end{equation}
where $W$ represents a series of Weyl reflections, one does obtain equivalent models. Let us now turn to comparing the spectra of two models M and M$'$, where
\begin{equation}\label{eq:BrotherDef}
 (V_1',V_2')~=~(V_1+\Delta V_1,V_2+\Delta V_2)\;,
\end{equation}
with $\Delta V_1,\Delta V_2 \in \Lambda$ being some lattice vectors. For future reference, we call models related by eqn.~(\ref{eq:BrotherDef}) \emph{brother models}.\index{brother model}\index{brother model!\"off \"off}

The brother model described by the set of shifts $(V_1',V_2')$ is also subject to modular invariance constraints. For the sake of keeping the expressions simple, we restrict here to models fulfilling the following (stronger) conditions
\begin{subequations}\label{eq:strongmodularinv2}
\begin{eqnarray}
  V_{i}^{2} - v_{i}^{2} & = & 0 \mod 2  \quad (i = 1, 2)\;,\label{eq:strong1} \\
  V_{1} \cdot V_{2} - v_{1} \cdot v_{2} & = & 0 \mod 2\;.   \label{eq:strong2}
\end{eqnarray}
\end{subequations}
Later, in section~\ref{sec:GeneralizationDTAndBrothers} we will relax these conditions and use the conventional ones of section~\ref{sec:SummaryConditions}. Equations~(\ref{eq:strongmodularinv2}) imply that the vacuum phase $\Phi_\mathrm{vac}~=~1 $ is trivial in the transformation phase eqn.~(\ref{eq:transformationphase}). The requirement that the shifts $(V'_1,\,V'_2)$ fulfill the modular invariance conditions of eqn.~(\ref{eq:strongmodularinv2}) leads to constraints on the lattice vectors $(\Delta V_1,\Delta V_2)$. Thus, they are not arbitrary but have to fulfill the following conditions
\begin{subequations}\label{eq:modinvconditions}
\begin{eqnarray}
  V_{i} \cdot \Delta V_{i} & = & 0 \mod 1 \quad \quad i = 1, 2 \;,\\
  V_{1} \cdot \Delta V_{2} + \Delta V_{1} \cdot V_{2} + \Delta V_1\cdot\Delta V_2& = & 0 \mod 2\;.
\end{eqnarray}
\end{subequations}

\subsubsection{Massless Spectra for the Models M and M$'$}

By considering the massless spectra of twisted strings corresponding to the constructing element
\begin{equation}
 g~=~(\theta^{k_1} \omega^{k_2}, n_\alpha e_\alpha)~\in~S
\end{equation}
of the models M and M$'$, we will see why they can be different. For simplicity, we restrict our attention to non-oscillator states. Physical states arise from tensoring together left- and right-moving solutions of the masslessness conditions, eqns.~(\ref{eqn:masslesstwistedrightmover}) and (\ref{eqn:masslesstwistedleftmover}),
\begin{eqnarray}
|q + k_1 v_1 + k_2 v_2\rangle\RM \, \otimes\, |p + k_1 V_1 + k_2 V_2\rangle\LM     &&\text{for M}\;,\\
|q + k_1 v_1 + k_2 v_2\rangle\RM \, \otimes\, |p' + k_1 V'_1 + k_2 V'_2\rangle\LM  &&\text{for M}' \;,
\end{eqnarray}
where $p' = p - k_1\Delta V_1 - k_2 \Delta V_2$ and the shifted momenta $p_\text{sh}$ and $p_\text{sh}'$ of the left-movers are identical for M and M$'$. According to the transformation phase eqn.~(\ref{eq:transformationphase}) with $\Phi_\mathrm{vac}=1$, these massless states transform under the action of a commuting element
\begin{equation}
 h~=~(\theta^{t_1} \omega^{t_2}, m_\alpha e_\alpha) \in S \quad\text{with}\quad
 [h,g]~=~0
\end{equation}
with the phases
\begin{eqnarray}
\Phi  & = & e^{2\pi \I\, \left[(p + k_1 V_1 + k_2 V_2)\cdot (t_1 V_1 + t_2 V_2) - (q + k_1 v_1 + k_2 v_2)\cdot (t_1  v_1 + t_2 v_2)\right]} \qquad\text{for M}\;,\nonumber\\
\Phi' & = & e^{2\pi \I\, \left[(p' + k_1 V'_1 + k_2 V'_2)\cdot (t_1 V'_1 + t_2 V'_2) - (q + k_1 v_1 + k_2 v_2 )\cdot (t_1 v_1 + t_2 v_2)\right]} \ \quad\text{for M}'\;.\nonumber
\end{eqnarray}
By using the constraints eqn.~(\ref{eq:modinvconditions}) and the properties of an integral lattice, $p\cdot \Delta V_i \in {\mathbb Z}$ for $p, \Delta V_i \in \Lambda$, the mismatch between the phases can be simplified to
\begin{equation}
\Phi'~=~\Phi \, e^{-2\pi \I\, (k_1 t_2 - k_2 t_1)V_2\cdot\Delta V_1}\;.
\end{equation}

\subsubsection{The Brother Phase}

That is, the transformation phase of states in model M$'$ differs from the transformation phase of states in model M by a relative phase
\begin{equation}\label{eq:brotherphase}
 \widetilde{\varepsilon}~\equiv~e^{-2\pi \I (k_1 t_2 - k_2 t_1)V_2\cdot\Delta V_1}\;.
\end{equation}
According to the nomenclature `brother models', the relative phase $\widetilde{\varepsilon}$ will be referred to as \emph{brother phase}. It is straightforward to see that the same relative phase occurs for oscillator states, and the derivation can be repeated for shifts satisfying the ``normal'' modular invariance conditions eqn.~(\ref{eq:newmodularinv}) rather than eqn.~(\ref{eq:strongmodularinv2}), yielding the same qualitative result.

The (brother) phase $\widetilde{\varepsilon}$ has certain properties and the fact that it can be non-trivial has important consequences. First of all, $\widetilde{\varepsilon}$ depends on the definition of the model M$'$, i.e. on the lattice vectors $(\Delta V_1,\Delta V_2)$. Furthermore, it clearly depends on the constructing element $g$ and on the commuting element $h$, 
\begin{equation}
\widetilde{\varepsilon}~=~ \widetilde{\varepsilon}(g,h)\;.
\end{equation}
From here we see that the brother phase vanishes for $g = (\mathbbm{1}, 0)$, i.e. for the untwisted sector. Thus, the gauge group and the untwisted matter coincide for brother models. On the other hand, since the brother phase does not vanish in general, the brother models M and M$'$ may have different twisted sectors, and therefore be inequivalent. This result extends also to the case where we subject the shifts only to the weaker constraints eqn.~(\ref{eq:newmodularinv}). For further details, see also~\cite{Ploger:2006dp}.

\subsubsection{A $\boldsymbol{\Z{3}\times\Z{3}}$ Example}\index{orbifold!$\Z{3}\times\Z{3}$}

Let us now study an example to illustrate the results obtained so far. Consider a $\Z{3}\times\Z{3}$ orbifold of $\E{8}\times\E{8}$ with standard embedding, compare to section~\ref{sec:Z3xZ3Orbifold}. Model M is defined by
\begin{equation}
V_1~=~\frac{1}{3}\left(1,0,-1,0^{5} \right) \left( 0^{8} \right) \quad\text{and}\quad 
V_2~=~\frac{1}{3}\left(0,1,-1,0^{5} \right) \left( 0^{8} \right)\;.
\end{equation}
The resulting model has an $\E{6}\times\U{1}^2\times\E{8}$ gauge group, 84 $(\crep{27},\rep{1})$ and 243 non-Abelian singlets, charged under the $\U{1}$'s. Now define the brother model M$'$ by
\begin{equation}
\Delta V_1~=~\left(0,-1,0,1,0^{4}\right) \left( 0^{8} \right) \quad\text{and}\quad 
\Delta V_2~=~\left(1,0,0,0,1,0^{3}\right) \left( 0^{8} \right)\;,
\end{equation}
which fulfill the conditions eqn.~(\ref{eq:modinvconditions}). From eqn.~(\ref{eq:brotherphase}) we find the following non-trivial brother phase
\begin{equation}\label{eq:brotherphasez3xz3}
\widetilde{\varepsilon}(g,h) ~=~ \widetilde{\varepsilon}(\theta^{k_1}\omega^{k_2},\theta^{t_1}\omega^{t_2}) ~=~e^{\frac{2\pi\I}{3}\, (k_1\, t_2 - k_2\, t_1)}.
\end{equation}
As expected, the gauge group and the untwisted matter of model M$'$ remain the same as in model M. However, the twisted sectors get modified. The total number of generations is reduced to 3 $(\crep{27},\rep{1})$ and 27 $(\rep{27},\rep{1})$. The number of singlets remains the same as before. For the detailed spectra of the models M and M$'$ see table~(\ref{tab:Z3xZ3Example}) in the appendix.

Model M$'$ is not an unknown construction, but has been studied in the literature in the context of $\Z{3}\times\Z{3}$ orbifolds with discrete torsion \cite{Font:1988mk}. As we shall see, the brother phase, eqn.~(\ref{eq:brotherphasez3xz3}), is nothing but the discrete torsion phase (eqn.~(4) in reference~\cite{Font:1988mk}). To make this statement more precise, we review discrete torsion in detail in section~\ref{sec:discretetorsionphase}, and analyze its relation to the brother phase in section~\ref{sec:comparison}.

\subsection{One--loop Partition Function}\index{partition function}
\label{sec:partitionfunction}

In order to review discrete torsion in orbifolds, we start with some basics about the partition function $\mathcal{Z}$ following Vafa~\cite{Vafa:1986wx}. The partition function can be written in terms of a genus expansion of the string vacuum to vacuum amplitude. We will mainly focus on the one--loop contribution to the partition function which is given by the one--loop vacuum to vacuum amplitude. In this case, the world sheet has the topology of a torus. This torus is embedded into the target space by the string coordinates. As we will see there are various different embeddings and the partition function sums over all amplitudes corresponding to these different embeddings. In addition, some aspects of the two--loop contributions to the partition function will be needed in order to derive sufficient constraints for discrete torsion.

However, before we can return to the partition function we need to summarize some facts about the world sheet torus and its symmetries, the group of modular transformations (see e.g. page 92ff of~\cite{Becker:2007zj}).

\subsubsection{World Sheet Torus}
\begin{figure}[t]
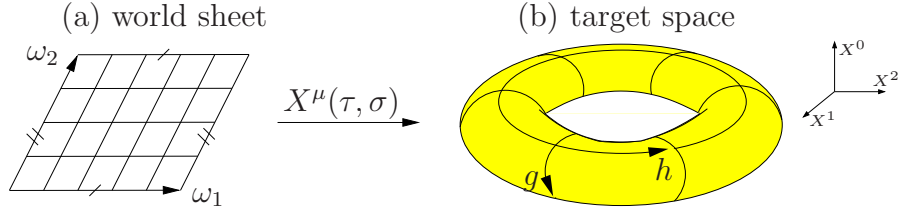

\centerline{\input stringmaptorus.pstex_t}\vspace{-0.3cm}
\caption{(a) In the case of the one--loop vacuum to vacuum amplitude the world sheet has the topology of a torus which is spanned by $\omega_1$ and $\omega_2$. (b) Here, its embedding into the ten-dimensional target space is illustrated for the case of trivial boundary conditions $g=h= \left(\mathbbm{1} ,0\right)$.}
\label{fig:WorldsheetTorus}
\end{figure}

It is convenient to use complex coordinates on the world sheet. Then, the world sheet torus is described by a two-dimensional lattice $\Gamma_2$ which is spanned by two complex vectors. We denote them by
\begin{equation}
\omega_1,\; \omega_2 \quad\text{with}\quad \frac{\omega_1}{\omega_2} \notin \mathbbm{R}\;,
\end{equation}
such that they are linearly independent, see figure~(\ref{fig:WorldsheetTorus}a). Since we are working with a conformal field theory on the world sheet, we can rescale this lattice by a conformal transformation to an equivalent one which is spanned by
\begin{equation}
\tau \equiv \frac{\omega_1}{\omega_2} \quad\text{and}\quad 1\;,
\end{equation}
where we have introduced $\tau$, the \emph{modular parameter} of the torus. In order to avoid confusion, we will denote the (real) world sheet coordinates by $(\sigma_1,\sigma_2)$ in this section. Then a point on the world sheet torus is described by the complex number
\begin{equation}
\label{eqn:worldsheetparamtorus}
\sigma_1 + \tau \sigma_2 \quad\text{with } \sigma_1,\sigma_2 \in [0,\pi]\;.
\end{equation}

\subsubsection{Modular Group $\text{SL}(2,\mathbbm{Z})$}\index{modular group $\text{SL}(2,\mathbbm{Z})$}

The world sheet torus is not uniquely described by the lattice $\Gamma_2 = \{\tau, 1\}$. For example, the basis vectors $\{\tau+1,1\}$ and $\{1, -\tau\}$ span the same lattice and therefore define the same torus. In general, two sets of basis vectors define the same two-dimensional torus and are thus considered to be equivalent if they are related by an element of the \emph{modular group} $\text{SL}(2,\mathbbm{Z})$, i.e.
\begin{equation}
\left(\begin{array}{c} \tau \\ 1\end{array}\right) \sim 
\left(\begin{array}{cc}a & b \\c & d\end{array}\right)
\left(\begin{array}{c} \tau \\ 1\end{array}\right)\quad\text{with}\quad ad-bc = 1\;,
\end{equation}
with $a,b,c,d \in\Z{}$, see e.g. page 309ff of \cite{Nakahara:2003nw}. The group $\text{SL}(2,\mathbbm{Z})$ is generated by two elements, denoted by $S$ and $T$. They are conventionally chosen to be of the form
\begin{eqnarray}
\label{eqn:S}
S = \left(\begin{array}{cc} 0 & 1 \\-1 & 0\end{array}\right)\;, & \qquad & S \left(\begin{array}{c} \tau \\ 1\end{array}\right) = \left(\begin{array}{c} 1 \\ -\tau\end{array}\right) \quad\text{and} \\
T = \left(\begin{array}{cc} 1 & 1 \\ 0 & 1\end{array}\right)\;, & \qquad & T \left(\begin{array}{c} \tau \\ 1\end{array}\right) = \left(\begin{array}{c} \tau + 1 \\ 1 \end{array}\right)\;.
\label{eqn:T}
\end{eqnarray}

\begin{figure}[ht]
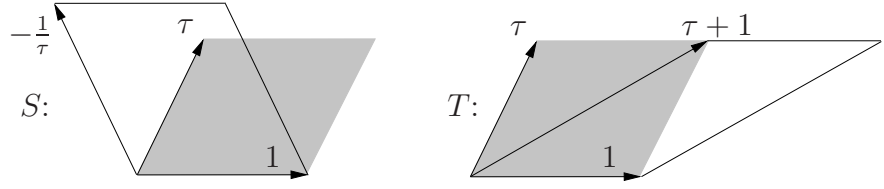

\centerline{\input ModularGroupST.pstex_t}\vspace{-0.25cm}
\caption{The actions of $S$ and $T$ on the modular parameter $\tau$ are depicted. The resulting tori are (conformally) equivalent to the original one (grey region).}
\label{fig:ST}
\end{figure}

As a remark, we note that the group that relates equivalent world sheet tori can also be represented by the following transformation of the modular parameter $\tau$
\begin{equation}
\label{eqn:PSL}
\tau \;\mapsto\; \frac{a\tau + b}{c\tau + d} \quad\text{where}\quad \left(\begin{array}{cc}a & b \\c & d\end{array}\right) \in \text{PSL}(2, \mathbbm{Z})\;.
\end{equation}
Here, we are actually defining a $\text{PSL}(2,\mathbbm{Z}) = \text{SL}(2,\mathbbm{Z}) / \mathbbm{Z}_2$ transformation, since a $\text{SL}(2,\mathbbm{Z})$ matrix and its negative are the same for the mapping on $\tau$, i.e. if $A\in\text{SL}(2,\mathbbm{Z})$ then $A$ and $-A$ are identified in $\text{PSL}(2,\mathbbm{Z})$. Now, the generators $S$ and $T$ are represented by
\begin{equation}
\tau \;\stackrel{S}{\mapsto}\; -\frac{1}{\tau} \quad\text{and}\quad \tau \;\stackrel{T}{\mapsto}\; \tau + 1\;.
\end{equation}
Their action on the modular parameter $\tau$ is illustrated in figure~(\ref{fig:ST}). The $T$ transformation given here is obviously the same as the one of eqn.~(\ref{eqn:T}). The $S$ transformation is conformally equivalent to the one of eqn.~(\ref{eqn:S}), i.e.
\begin{equation}
S \left(\begin{array}{c} \tau \\ 1\end{array}\right) = -\tau \left(\begin{array}{c} -\frac{1}{\tau} \\ 1\end{array}\right) \quad\Rightarrow\quad \tau \;\stackrel{S}{\mapsto}\; -\frac{1}{\tau}
\end{equation}
up to a scaling factor $-\tau$. The generators $S$ and $T$ of $\text{PSL}(2,\mathbbm{Z})$ obey the relations
\begin{equation}
S^2 = 1 \quad\text{and}\quad(ST)^3 = 1\;.
\end{equation}

\subsubsection{The Structure of the Partition Function}\index{partition function!one--loop}

By using different boundary conditions for closed strings we can embed the world sheet torus into the target space in different ways. Each embedding will give some contribution to the one-loop vacuum to vacuum amplitude and for this reason to the one--loop partition function. Considering the bosonic coordinates $Z(\sigma_1, \sigma_2)\in \mathbbm{C}^3$ in the complex basis we find that for each closed string with constructing element $g$
\begin{equation}
\label{eqn:bc_sigma2}
Z(\sigma_1, \sigma_2+\pi) = g\; Z(\sigma_1, \sigma_2)\;,
\end{equation}
there are several possibilities to close also in the $\sigma_1$ direction
\begin{equation}
\label{eqn:bc_sigma1}
Z(\sigma_1 + \pi, \sigma_2) = h\; Z(\sigma_1, \sigma_2)\;.
\end{equation}
Note that the string needs to be closed in both directions in order to describe the process of vacuum to vacuum transition. The easiest possibility is $g = h = \left(\mathbbm{1} ,0\right)$, see figure~(\ref{fig:WorldsheetTorus}b). However, there are many more. From the following consideration
\begin{eqnarray}
Z(\sigma_1+\pi, \sigma_2+\pi) & = & g\; Z(\sigma_1+\pi, \sigma_2) \\
                              & = & g h\; Z(\sigma_1, \sigma_2) \quad\text{and}\\
Z(\sigma_1+\pi, \sigma_2+\pi) & = & h\; Z(\sigma_1, \sigma_2 + \pi) \\
                              & = & h g\; Z(\sigma_1, \sigma_2)
\end{eqnarray}
we see that we can choose any two \emph{commuting} space group elements $g$ and $h$\;\footnote{For a geometrical interpretation of commuting space group elements see appendix~\ref{app:commuting}.}. Thus, an embedding of the world sheet torus into the target space is characterized by a pair of (commuting) boundary conditions $(g,h)$. Each pair of boundary conditions $(g,h)$ contributes to the partition function with a term denoted by $\mathcal{Z}(g,h)$. Hence, the one--loop partition function $\mathcal{Z}$ has the overall structure
\begin{equation}
\label{eqn:partitionfunction}
 \mathcal{Z}~=~\sum_{\substack{[g,h]=0}} \mathcal{Z}(g,h)\;,
\end{equation}
where the sum runs over pairs of commuting space group elements $g,h\in\mbox{S}$ and the integration over the modular parameter is included in $\mathcal{Z}(g,h)$. Here, we are not interested in the specific form of $\mathcal{Z}(g,h)$, but refer to the literature, e.g.~\cite{Ibanez:1987pj,Bailin:1987dm,Senda:1987pf,Minahan:1987ha,Nilles:1987uy,Gaberdiel:2004vx}.

\subsubsection{Modular Transformations of the Partition Function}\index{modular invariance!of the partition function}

Next, we analyze the action of modular transformations generated by $S$ and $T$ on the boundary conditions eqns.~(\ref{eqn:bc_sigma2}) and (\ref{eqn:bc_sigma1}). First, we need to know how $S$ and $T$ act on the world sheet coordinates $(\sigma_1, \sigma_2)$. Using eqns.~(\ref{eqn:S}) and (\ref{eqn:T}) we find
\begin{eqnarray}
1\; \sigma_1 + \tau \sigma_2 & \stackrel{S}{\mapsto} & -\tau \sigma_1 +        1\; \sigma_2 \;\;\;\qquad = 1\;\sigma_2 + (- \tau) \sigma_1 \\
1\; \sigma_1 + \tau \sigma_2 & \stackrel{T}{\mapsto} &   1\; \sigma_1 + (\tau + 1) \sigma_2 \quad  = 1\;(\sigma_1 + \sigma_2) + \tau\sigma_2\;
\end{eqnarray}
Comparing to eqn.~(\ref{eqn:worldsheetparamtorus}) we see that the world sheet coordinates transform as
\begin{eqnarray}
\label{eqn:SonSigma}
(\sigma_1, \sigma_2) & \stackrel{S}{\mapsto} & (\sigma_2,-\sigma_1) \\
(\sigma_1, \sigma_2) & \stackrel{T}{\mapsto} & (\sigma_1 + \sigma_2, \sigma_2)\;.
\label{eqn:TonSigma}
\end{eqnarray}
Now we can act with the generators of $\text{SL}(2,\mathbbm{Z})$ on the boundary conditions. We start with generator $S$ acting on eqns.~(\ref{eqn:bc_sigma2}) and (\ref{eqn:bc_sigma1}). This yields
\begin{equation}
\label{eqn:SonBC}
\begin{array}{rclcrcl}
Z(\sigma_1, \sigma_2 + \pi) & = & g\; Z(\sigma_1, \sigma_2) \quad&\stackrel{S}{\mapsto}&\quad Z(\sigma_2 + \pi, -\sigma_1) & = & \tilde{g}\; Z(\sigma_2,-\sigma_1) \\
Z(\sigma_1 + \pi, \sigma_2) & = & h\; Z(\sigma_1, \sigma_2) \quad&\stackrel{S}{\mapsto}&\quad \phantom{I^{I^I}}Z(\sigma_2, -\sigma_1 - \pi) & = & \tilde{h}\; Z(\sigma_2, -\sigma_1)\;,
\end{array}
\end{equation}
where we have used the transformation properties of the world sheet parameters eqn.~(\ref{eqn:SonSigma}). Furthermore, we have introduced transformed boundary conditions $(\tilde{g}, \tilde{h})$. They can easily be expressed in terms of the original elements $(g,h)$ by evaluating the right-hand side of eqn.~(\ref{eqn:SonBC}) using the eqns.~(\ref{eqn:bc_sigma2}) and~(\ref{eqn:bc_sigma1})
\begin{eqnarray}
\tilde{g}\; Z(\sigma_2,-\sigma_1) & = & h\; Z(\sigma_2,-\sigma_1) \\
\tilde{h}\; Z(\sigma_2,-\sigma_1) & = & g^{-1}\; Z(\sigma_2,-\sigma_1)\;.
\end{eqnarray}
This is the desired result: it states that the boundary conditions of the target space torus are transformed under the action of the world sheet transformation $S$ according to
\begin{equation}
(g,h) \;\stackrel{S}{\mapsto}\; (\tilde{g}, \tilde{h}) = (h,g^{-1})\;.
\end{equation}
These steps can be repeated for the generator $T$ and result in
\begin{equation}
(g,h) \;\stackrel{T}{\mapsto}\; (\tilde{g}, \tilde{h}) = (gh,h)
\end{equation}
In summary, an embedding of the world sheet torus into the target space using the boundary conditions $(g,h)$ is transformed by a modular transformation according to
\begin{equation}
(g,h) \;\stackrel{SL(2, \mathbbm{Z})}{\mapsto}\; (g^a h^b, g^c h^d)\;.
\end{equation}

Since modular transformations map the world sheet torus to an equivalent one, the partition function eqn.~(\ref{eqn:partitionfunction}) must be \emph{modular invariant}, i.e. invariant under modular transformations. Therefore, we know that two summands of the partition function have to be equal if they are related by a $\text{SL}(2,\mathbbm{Z})$ transformation,
\begin{equation}
\label{eqn:ZforModTrafo}
\mathcal{Z}(g,h) =  \mathcal{Z}(g^a h^b, g^c h^d) \qquad\text{for}\qquad 
\left(\begin{array}{cc} a & b\\ c& d\\\end{array}\right) \in \text{SL}(2, \mathbbm{Z})\;.
\end{equation}

\subsubsection{Remark: Higher Genus Contributions}\index{partition function!two--loop}

\begin{figure}[ht]
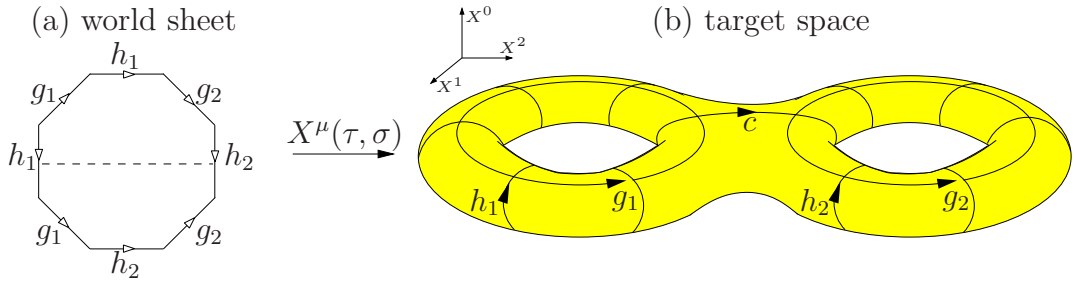

\centerline{\input stringmaptorusg2.pstex_t}
\caption{(a) In the case of the two--loop amplitude the world sheet has the topology of a genus 2 torus (after identifying the sides according to the labels and arrows; first, fold at the dashed line and then identify both $g_1$'s and both $g_2$'s). (b) Here, its embedding into the ten-dimensional target space is illustrated for the case of trivial boundary conditions $g_i=h_i=(\mathbbm{1},0)$. $c$ denotes the cycle connecting the two handles with associated boundary condition $h_2h_1^{-1}$.}
\label{fig:genus2}
\end{figure}

The two-loop contributions to the partition function $\mathcal{Z}$ are given by the different embeddings of the genus 2 world sheet into the target space. This embedding is characterized by four boundary conditions $(g_1,h_1;g_2,h_2)$ defined along the four inequivalent cycles on the genus 2 surface. There is a fifth cycle, denoted by $c$, which connects the two handles. The situation is depicted in figure~(\ref{fig:genus2}). The genus 2 surface is supposed to be factorizable into a pair of one--loop diagrams touching at a point. In order for the resulting one--loop diagrams to be well-defined we have to demand $[g_1,h_1] = [g_2,h_2] =0$, as before. Furthermore, a Dehn twist\index{Dehn twist} along $c$ changes the four boundary conditions according to\footnote{A Dehn twist is a homeomorphism of the surface to itself, generated by cutting the surface along $c$ twisting the cut by $2\pi$ and gluing it back together.}
\begin{equation}
\label{eqn:Dehn2Loop}
(g_1,h_1;g_2,h_2) ~\mapsto~ (g_1h_2h_1^{-1},h_{1};g_2h_1h_2^{-1},h_2)\;.
\end{equation}

\subsection[Discrete Torsion Phase for $\Z{N}\times\Z{M}$ Orbifolds]{Discrete Torsion Phase for $\boldsymbol{\Z{N}\times\Z{M}}$ Orbifolds}
\label{sec:discretetorsionphase}

Following the idea of Vafa~\cite{Vafa:1986wx}, we can introduce relative phases $\varepsilon (g,h)$\index{discrete torsion} between the different terms in the partition function
\begin{equation}
 \mathcal{Z}~=~\sum_{\substack{[g,h]=0}} \varepsilon(g,h)\; \mathcal{Z}(g,h)\;,
\end{equation}
where different assignments of phases lead, in general, to different orbifold models. Since these phases can only take discrete values, as we will see later, $\varepsilon(g,h)$ is called \emph{discrete torsion phase}. We use the convention that the phase of the term $(g,g)$ is trivial, i.e.
\begin{equation}
\varepsilon(g,g) ~=~ 1 \; . 
\end{equation} 
Modular transformations interchange the terms $\varepsilon(g,h)\; \mathcal{Z}(g,h)$ in the partition function, according to eqn.~(\ref{eqn:ZforModTrafo}). Therefore, the corresponding phases have to be identical
\begin{equation}
\varepsilon(g,h) = \varepsilon(g^a h^b, g^c h^d)\;.
\end{equation}
On the other hand, modular transformations do not mix \emph{all} terms of the partition function in general, thus some non-trivial phases will remain to be allowed. 

At two--loop, the partition function allows to switch on analogous phases, $\varepsilon(g_{1},h_{1};g_{2},h_{2})$. From the requirement of factorizability of the two--loop vacuum to vacuum amplitude into two one--loop diagrams touching at a point, one infers~\cite{Vafa:1986wx} 
\begin{equation}
 \varepsilon(g_1,h_1;g_2,h_2) ~=~\varepsilon(g_1,h_1)\,\varepsilon(g_2,h_2)\;.
\end{equation}

Furthermore, the Dehn twist along $c$ interchanges the boundary conditions according to eqn.~(\ref{eqn:Dehn2Loop}). Thus, the corresponding torsion phases have to match
\begin{equation}
\varepsilon(g_1,h_1;g_2,h_2) ~=~ \varepsilon(g_1h_2h_1^{-1},h_{1};g_2h_1h_2^{-1},h_2)\;.
\end{equation}
These conditions can be rephrased into~\cite{Vafa:1986wx} \index{modular invariance!of discrete torsion}
\begin{subequations}\label{eq:phaseconstraints}
\begin{eqnarray}
 \varepsilon(g_{1} g_{2},g_{3}) &=& \varepsilon(g_{1},g_{3})\, \varepsilon(g_{2},g_{3})\;,  \label{eq:phaseconstraints1} \\
 \varepsilon(g_{1},g_{2})       &=& \varepsilon(g_{2},g_{1})^{-1}\;,                        \label{eq:phaseconstraints2} \\
 \varepsilon(g_{1},g_{1})       &=& 1\;.                                                    \label{eq:phaseconstraints3} 
\end{eqnarray} 
\end{subequations}

\subsubsection{Old Interpretation}

Following the discussion of reference~\cite{Font:1988mk}, for orbifolds without Wilson lines $g,h$ are chosen to be elements of the point group $P$. In $\mathbbm{Z}_N$ orbifolds, due to this choice and eqns.~(\ref{eq:phaseconstraints}) the phases have to be trivial, $\varepsilon(g,h)=1$ for all $g,h \in P$. Therefore, in the case of $\mathbbm{Z}_{N}$ orbifolds without Wilson lines, non-trivial discrete torsion cannot be introduced.

In $\mathbbm{Z}_{N}\times \mathbbm{Z}_{M}$ orbifolds, still without Wilson lines, the situation is different because there are independent pairs of elements (such that the first element is not a power of the second) which commute with each other.  If we take two point group elements $g=\theta^{k_1}\omega^{k_2}$ and $h=\theta^{t_1}\omega^{t_2}$, the eqns.~\eqref{eq:phaseconstraints} determine the shape of the corresponding phase, 
\begin{equation}\label{eq:torsionphase2}
\varepsilon(g,h) ~=~ \varepsilon(\theta^{k_1}\omega^{k_2},\theta^{t_1}\omega^{t_2}) ~=~ e^{\frac{2\pi\I\, m}{M}(k_1 t_2-k_2 t_1)}\,\,,
\end{equation}
where $m\in \mathbbm{Z}$ \cite{Font:1988mk}. In particular, there are only $M$ inequivalent assignments of $\varepsilon$. Later, in section~\ref{sec:GeneralizedDiscreteTorsion} we will give a \emph{new interpretation} of discrete torsion allowing for more possibilities.

The most important consequence of non-trivial $\varepsilon$-phases for our discussion is that they modify the transformation phase of twisted states and thus change the twisted spectrum, i.e. the transformation phase of eqn.~(\ref{eq:transformationphase}) is modified according to 
\begin{equation}\label{eq:modifiedtransformationphase}
 \Phi \longmapsto \varepsilon(g,h) \,\Phi\;.
\end{equation}

\subsection{Brother Models versus Discrete Torsion}
\label{sec:comparison}

\begin{figure}[b]
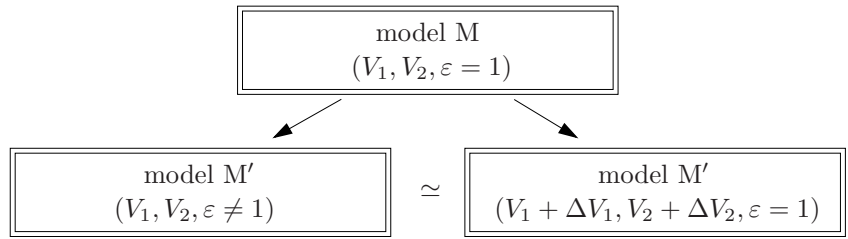

\centerline{\input DTMatching.pstex_t}\vspace{-0.3cm}
\caption{Models with non-trivial discrete torsion have an equivalent description as models with trivial discrete torsion but a different gauge embedding.}
\label{fig:MirageTorsion}
\end{figure}

Let us now come back to the task of establishing the relation between the discrete torsion phase and the brother phase as introduced in section~\ref{sec:brothermodel}. From eqns.~(\ref{eq:brotherphase}) and~(\ref{eq:torsionphase2}) it is clear that both phases can be made to coincide. More precisely, since $V_2$ can be written as $V_{2}=\frac{\lambda_{2}}{M}$ with $\lambda_{2} \in \Lambda$ (cf.\ eqn.~(\ref{eqn:ShiftLatticeConditions})), one can achieve
\begin{equation}
\label{eq:mdef}
 -V_{2} \cdot \Delta V_{1} ~=~ \frac{m}{M} 
\end{equation}
for an appropriate choice of $\Delta V_1\in\Lambda$. Since the solutions to the mass equations and the projection conditions are the same in a model with discrete torsion and a brother model, whose associated phases fulfill eqn.~(\ref{eq:mdef}), the spectra of both models coincide. We will therefore regard both models as \emph{equivalent}. This means that introducing a discrete torsion phase, eqn.~(\ref{eq:torsionphase2}), is equivalent to changing the gauge embedding according to 
\begin{equation}
 (V_1,V_2) ~ \to ~ (V_1+\Delta V_1,V_2+\Delta V_2)
\end{equation}
with $\Delta V_i\in\Lambda$ and $-V_{2} \cdot \Delta V_{1}=m/M$. In particular, the assignment of discrete torsion to a given $\mathbbm{Z}_N\times\mathbbm{Z}_M$ model is a `gauge-dependent' statement in the sense that torsion can be traded for changing the gauge embedding (see figure~(\ref{fig:MirageTorsion})).

\subsubsection{Example: Standard Embeddings for all $\mathbbm{Z}_N \times \mathbbm{Z}_M$}

To illustrate the results, we construct the standard embedding models for $\mathbbm{Z}_N \times \mathbbm{Z}_M$ orbifolds with an $\E{8}\times\E{8}$ lattice of reference~\cite{Font:1988mk} first with discrete torsion and secondly in terms of non-standard embedding shifts without discrete torsion (brother models). We use the following recipe to construct brother models, i.e.\ mimic models with discrete torsion:\\ 
For a given set of shifts $V_1$ and $V_2$ fulfilling the modular invariance conditions, find a new set of shifts $V_1'=V_1+\Delta V_1$ and $V_2'=V_2+\Delta V_2$ with the following properties:
\renewcommand{\labelenumi}{(\roman{enumi})}
\begin{enumerate}
 \item the new shifts differ from the original set only by lattice vectors, i.e.\ $\Delta V_1,\Delta V_2\in\Lambda$
 \item the new shifts also fulfill the modular invariance conditions, and
 \item the `interference term' $V_2\cdot \Delta V_1$ is not an integer.
\end{enumerate}
In practice (and for any $N,M$), the above properties can be expressed in terms of linear Diophantine equations for which we always find solutions.

Possible choices for the shifts $(V_1+\Delta V_1,V_2+\Delta V_2)$ are shown in table~(\ref{tab:DiscreteTorsionBrothers}) in the appendix, where we list the shifts of torsionless models equivalent to the discrete torsion models of reference~\cite{Font:1988mk}. 

The results obtained so far in this section have important consequences for the classification of $\Z{N}\times\Z{M}$ orbifolds. Introducing a discrete torsion phase in the sense of reference~\cite{Font:1988mk} does not lead to new models. That is, all models with this discrete torsion can be equivalently obtained by scanning over torsionless models only. 

It is also instructive to interpret the equivalence between discrete torsion and changing the gauge embedding in terms of geometry. Discrete torsion can be regarded as a property of the 6D compact space while changing the gauge embedding affects the (left-moving) coordinates of the gauge lattice only. Hence one might argue that discrete torsion and choosing a different gauge embedding are two different features of orthogonal dimensions. However, by embedding the `spatial' twist in the gauge degrees of freedom, these features get combined in such a way that it is no longer possible to make a clear separation. Using a more technical language one might rephrase this statement by saying that, since physical states arise from tensoring left- and right-movers together, the phases $\varepsilon$ and $\widetilde{\varepsilon}$ cannot be distinguished. Consequently, properties of the zero-modes can be ascribed neither to the gauge embedding alone nor to the presence of discrete torsion, but only to both.

\section{Generalized Discrete Torsion}
\label{sec:GeneralizationDTAndBrothers}

The results of the previous section can be generalized. To see this, we first generalize the brother phase of section~\ref{sec:brothermodel} for orbifolds with Wilson lines. It will turn out that generalized brother models do also exist for $\Z{N}$ orbifolds with Wilson lines. Also discrete torsion can be generalized~\cite{Gaberdiel:2004vx}. Thus, in a second step, we write down the most general ansatz for a generalized discrete torsion phase consistent with the modular invariance conditions. Finally, we compare the generalized brother phases to the generalized discrete torsion phases. As before, we can relate both phases.

\subsection{Generalized Brother Models}
\label{sec:GeneralizedBrothers}

Let us turn to the discussion of orbifolds with Wilson lines~\cite{Ibanez:1986tp}. A (torsionless) model M is defined by $(V_1,V_2,A_\alpha)$. A brother model M$'$ appears by adding lattice vectors to the shifts and Wilson lines, i.e.\ M$'$ is defined by 
\begin{equation}
 (V'_1, V'_2, A'_\alpha)~=~(V_1+\Delta V_1, V_2+\Delta V_2, A_\alpha+\Delta A_\alpha)\;,
\end{equation}
with $\Delta V_i,\Delta A_\alpha \in \Lambda$. From the modular invariance conditions~(\ref{eq:newmodularinv}), the choice of lattice vectors $(\Delta V_i,\Delta A_\alpha)$ is constrained by
\begin{subequations}\label{eq:newModularInvforDelta}
\begin{eqnarray}
M \left(V_1\cdot\Delta V_2 + V_2\cdot\Delta V_1 + \Delta V_1 \cdot\Delta V_2 \right) & = & 0\text{ mod }
2~\equiv~ 2\,x \;,\label{eq:newModularInvforDeltaVV}\\
N_\alpha \left(V_i\cdot\Delta A_\alpha + A_\alpha\cdot\Delta V_i + \Delta V_i \cdot\Delta A_\alpha
\right) & = &  0\text{ mod }
2~\equiv~ 2\,y_{i\alpha} \;,\label{eq:newModularInvforDeltaVA}\\
Q_{\alpha\beta} \left(A_\alpha\cdot\Delta A_\beta + A_\beta\cdot\Delta A_\alpha + \Delta A_\alpha\cdot\Delta A_\beta
\right) & = &  0\text{ mod }
2~\equiv~ 2\,z_{\alpha\beta}\;,\label{eq:newModularInvforDeltaAA}
\end{eqnarray}
\end{subequations}
where $x,\,y_{i\alpha},\,z_{\alpha\beta}\,\in{\mathbb Z}$. Repeating the steps of section~\ref{sec:brothermodel} one arrives at a \emph{generalized brother phase}\index{brother model!generalized case}
\begin{eqnarray}
 \widetilde{\varepsilon}&=&\exp\left\{-2\pi \I\,\left[ (k_1\, t_2-k_2\, t_1)\left(V_{2} \cdot \Delta V_{1} - \frac{x}{M}\right) + (k_1\, m_{\alpha} - t_1\, n_{\alpha})\left(A_{\alpha} \cdot \Delta V_{1}-\frac{y_{1\alpha}}{N_\alpha}\right)\right.\right. \nonumber\\*
 & &  \hphantom{\exp\left\{\right.}{}\left.\left. + (k_2\, m_{\alpha}-t_2\, n_{\alpha})\left(A_{\alpha} \cdot \Delta V_{2}-\frac{y_{2\alpha}}{N_\alpha}\right)  + n_{\alpha}\, m_{\beta}\left( A_{\beta} \cdot \Delta  A_{\alpha}-\frac{z_{\alpha\beta}}{Q_{\alpha\beta}}\right)\right] \right\}\;,
 \label{eq:generalbrotherphase} 
\end{eqnarray}
corresponding to the constructing element $g=(\theta^{k_1}\omega^{k_2},n_{\alpha}e_{\alpha})$ and the commuting element $h=(\theta^{t_1}\omega^{t_2},m_{\alpha}e_{\alpha})$. One can see that $D_{\alpha\beta}\equiv A_\beta\cdot\Delta A_\alpha - z_{\alpha\beta}/Q_{\alpha\beta}$ is (almost) antisymmetric in $\alpha,\,\beta$,
\begin{equation}
D_{\alpha\beta} = -D_{\beta\alpha}\text{ mod }1\,.
\end{equation}

Notice that also in the case of orbifolds with \emph{lattice-valued Wilson lines}\index{Wilson line!lattice-valued}, $A_\alpha\in\Lambda$, the last three terms of eqn.~(\ref{eq:generalbrotherphase}) can be non-trivial, giving rise to new brother models.

\subsubsection{Brother Models in $\boldsymbol{\Z{N}}$ Orbifolds}

From eqn.~(\ref{eq:generalbrotherphase}), it is clear that the generalized brother phase is also important for $\Z{N}$ orbifolds. More precisely, in $\Z{N}$ orbifolds with Wilson lines, the second term ($A_{\alpha} \cdot \Delta V_{1}$) and the fourth term ($A_{\beta} \cdot \Delta  A_{\alpha}$) of eqn.~(\ref{eq:generalbrotherphase}) are not always trivial and thus also lead to brother models. 

Let us illustrate this with an example in $\Z{4}$ with twist\index{orbifold!$\Z{4}$} $v=\frac{1}{4}(0,\,-2,\,1,\,1)$ acting on the compactification lattice $\Gamma= \SO{4}^3$, and standard embedding~\cite{Katsuki:1989bf,Katsuki:1989qz}. The gauge group is $\E{6}\times\SU{2}\times\U{1}\times\E{8}$. By turning on the lattice-valued Wilson lines 
\begin{equation}
\label{eqn:Z4BrotherModelWL}
 A_1=\left(0^8 \right)\left(1^2,0^6\right), \quad\quad\quad A_5=A_6=\left(0^8 \right)\left(0,1^2,0^5\right),
\end{equation}
a non-trivial generalized brother phase with $D_{15}=D_{16}=-\frac{1}{2}$ is introduced. The untwisted and first twisted sectors remain unchanged, but the number of (anti-) families in the second twisted sector is reduced from 10 $(\crep{27},\rep{1},\rep{1})$ + 6 $(\rep{27},\rep{1},\rep{1})$ to 6 $(\crep{27},\rep{1},\rep{1})$ + 2 $(\rep{27},\rep{1},\rep{1})$. The detailed spectra of both models are given in table~(\ref{tab:Z4LatticeWL}) in the appendix.

\subsection{Generalized Discrete Torsion}
\label{sec:GeneralizedDiscreteTorsion}

In section~\ref{sec:discretetorsionphase} we have discussed the discrete torsion phase as introduced in reference~\cite{Font:1988mk}. More recently, this concept has been extended by introducing a generalized discrete torsion phase in the context of type IIA/B string theory~\cite{Gaberdiel:2004vx}. This generalized torsion phase depends on the fixed points of the orbifold. It weights differently terms in the partition function corresponding to the same twisted sector but different fixed points, and is constrained by modular invariance. 

Following the steps of section~\ref{sec:discretetorsionphase} and considering $g,h\in S$, we write down the general solution of eqns.~(\ref{eq:phaseconstraints}) for the discrete torsion phase\index{discrete torsion!generalized case} as\footnote{Note that we employ the stronger constraints~(\ref{eq:phaseconstraints}) rather than the conditions presented in~\cite{Gaberdiel:2004vx}. It might be possible to relax condition~(\ref{eq:phaseconstraints2}), in which case additional possibilities could arise. We ignore this possibility in the present study.}
\begin{equation}\label{eq:generalizedtorsionphase}
\varepsilon(g,h)~=~e^{2 \pi \I\,[a\, (k_1\, t_2 - k_2\, t_1) + b_{\alpha}\, (k_1\, m_{\alpha} - t_1\, n_{\alpha}) + c_{\alpha}\, (k_2\, m_{\alpha} - t_2\, n_{\alpha}) + d_{\alpha \beta}\, n_{\alpha}\, m_{\beta}]}\;.
\end{equation}
Modular invariance constrains the values of $a,b_{\alpha},c_{\alpha},d_{\alpha\beta}$, $\alpha,\beta=1,\ldots,6$. Therefore, $a=\widetilde{a}/M,\,b_\alpha = \widetilde{b}_\alpha/N_\alpha$, $c_\alpha = \widetilde{c}_\alpha/N_\alpha$, $d_{\alpha\beta}=\widetilde{d}_{\alpha\beta}/N_{\alpha\beta}$ with $\widetilde{a},\,\widetilde{b}_\alpha$, $\widetilde{c}_\alpha$, $\widetilde{d}_{\alpha\beta} \in \mathbbm{Z}$, $N_{\alpha\beta}$ being the greatest common divisor of $N_\alpha$ and $N_\beta$. In addition, $d_{\alpha\beta}$ must be antisymmetric in $\alpha,\,\beta$.

The parameters $b_\alpha$, $c_\alpha$, $d_{\alpha\beta}$ are additionally constrained by the geometry of the orbifold. It is not hard to see that if $e_\alpha \simeq e_\beta$ on the orbifold, then $b_\alpha=b_\beta,\, c_\alpha=c_\beta$ and $d_{\alpha\beta}=0$ must hold (cf.\ the examples below).

The generalized discrete torsion is not restricted only to $\Z{N}\times\Z{M}$ orbifolds, as the usual discrete torsion was, but will likewise appear in the $\Z{N}$ case. Clearly, since in $\Z{N}$ orbifolds there is only one twist, the parameters $a$ and $c_\alpha$ vanish.

\subsubsection{Examples}

Let us consider the $\Z{3}\times\Z{3}$\index{orbifold!$\Z{3}\times\Z{3}$} orbifold compactified on an $\SU3^3$ lattice, see section~\ref{sec:Z3xZ3Orbifold}. In this case we have $e_1 \simeq e_2,\, e_3 \simeq e_4$ and $e_5 \simeq e_6$ on the orbifold. This implies that there are only three independent $b_\alpha$, namely $b_1,\,b_3,\,b_5$, while $b_2 = b_1,\,b_4 = b_3,\,b_6 = b_5$. Analogously, only $c_1,\,c_3,\,c_5$ are independent. Further, the antisymmetric matrix $d_{\alpha\beta}$ takes the form
\begin{equation}
 d_{\alpha\beta}~=~ \left(
 \begin{array}{cccccc}
 0&0&d_1&d_1&d_2&d_2\\
 0&0&d_1&d_1&d_2&d_2\\
 -d_1&-d_1&0&0&d_3&d_3\\
 -d_1&-d_1&0&0&d_3&d_3\\
 -d_2&-d_2&-d_3&-d_3&0&0\\
 -d_2&-d_2&-d_3&-d_3&0&0
 \end{array}
 \right)\;.
 \label{eq:dmatrixZ3xZ3}
\end{equation}
Including the parameter $a$, there are 10 independent discrete torsion parameters, which can take values $0$, $\tfrac{1}{3}$ or $\tfrac{2}{3}$.

For the $\Z{2}\times\Z{2}$\index{orbifold!$\Z{2}\times\Z{2}$} orbifold on an $\SU2^6$ lattice (as discussed in section~\ref{sec:Z2xZ2Orbifold}) an analogous consideration shows that there are no restrictions for the discrete torsion parameters. Therefore, there are $1+6+6+15=28$ independent parameters $a,\,b_\alpha,\,c_\alpha,\,d_{\alpha\beta}$, with values either 0 or $\tfrac{1}{2}$. However, since the coefficients $n_\alpha m_\beta$ of $d_{\alpha\beta}$ for $(\alpha,\,\beta)\in\{(1,2),\,(3,4),\,(5,6)\}$ vanish, the corresponding $d_{\alpha\beta}$ are not physical, leading to 25 effective parameters.
 
In the case of the $\Z{6}$-II orbifold\index{orbifold!$\Z{6}$-II} on a $\G{2}\times\SU{3}\times\SU{2}^2$ torus lattice (cf. section~\ref{sec:Z6IIOrbifold}) the following discrete torsion parameters can in principle be non-vanishing: $b_3 = b_4 = 0,\tfrac{1}{3},\tfrac{2}3{}$, $b_5, b_6 = 0,\tfrac{1}{2}$ and $d_{56} = 0,\tfrac{1}{2}$. However, the corresponding coefficients in the torsion phase eqn.~(\ref{eq:generalizedtorsionphase}) vanish. Thus, generalized discrete torsion has no effect in $\Z{6}$-II orbifolds.

\subsubsection{Generalized Discrete Torsion and Local Spectra}

In order to understand the action of the generalized discrete torsion, let us consider the following example. We start with the $\Z{3}\times\Z{3}$\index{orbifold!$\Z{3}\times\Z{3}$} standard embedding without Wilson lines, $A_\alpha=0$, and switch on the discrete torsion phase, eqn.~(\ref{eq:generalizedtorsionphase}), with $b_3=b_4=\frac{1}{3}$. The total number of families is reduced from 84 $(\crep{27},\rep{1})$ to 24 $(\crep{27},\rep{1})$ and 12 $(\rep{27},\rep{1})$. For details on the spectrum see table~(\ref{tab:Z3xZ3ExampleBNotZero}) in the appendix.

Due to its form, the discrete torsion phase $\varepsilon=e^{2\pi\I\,b_{\alpha}\,(k_1\, m_{\alpha}-t_1\,n_\alpha)}$ distinguishes between different fixed points of a particular twisted sector. That is, generalized discrete torsion can be thought of as a local feature. In general, the additional phase at a given fixed point coincides with a brother phase of the torsionless model (cf.\ first term of eqn.~(\ref{eq:generalbrotherphase})), i.e.\ locally one can find $\Delta V_i$ such that
\begin{equation}\label{eq:epsilonloc}
\varepsilon~=~e^{2\pi\I\,b_{\alpha}\,(k_1\, m_{\alpha}-t_1\,n_\alpha)} ~=~e^{-2\pi\I\,(k_1\, t_2-k_2\, t_1) \left(V_2\cdot\Delta V_1 - \frac{x}{3}\right)}
\end{equation}
with appropriate $x$. Then, each local spectrum coincides with the local spectrum of some brother model. The interpretation of generalized discrete torsion in terms of `localized discrete torsion' parallels the concept of local shifts (cf.\ \cite{Gmeiner:2002es,Forste:2004ie,Buchmuller:2004hv,Buchmuller:2006ik}) in orbifolds with Wilson lines.

Note that $\Delta V_i$ as in eqn.~(\ref{eq:epsilonloc}) cannot be found for twisted sectors where $b_\alpha$ corresponds to a direction $e_\alpha$ of a fixed torus, where $b_\alpha$ projects out all states of the sector.

\begin{figure}[!t!]
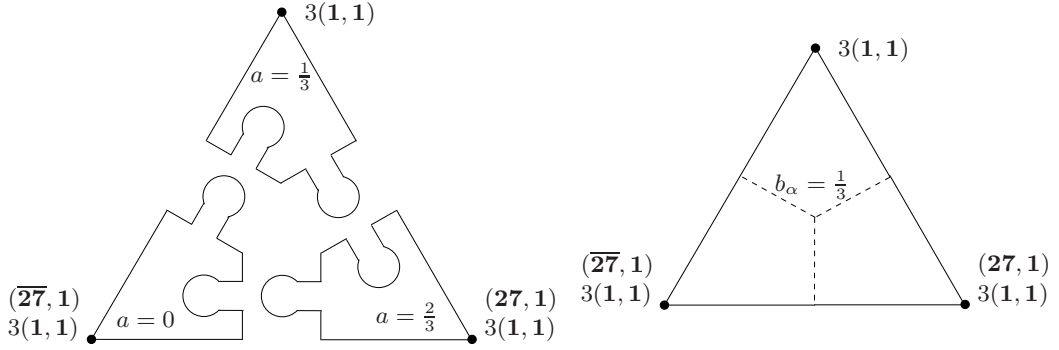

\centerline{\input LocalTorsion.pstex_t}\vspace{-0.2cm}
\caption{Sketch of a (2D) $\SU{3}$ plane of a $\Z{3}\times\Z{3}$ orbifold (the second plane in the example). On the left: parts (`corners') from different brother models can be `sewed together' to a model in which the torsion phase differs for different fixed points. This is equivalent to switching on the generalized discrete torsion phase $b_\alpha$, as depicted on the right.}
\label{fig:puzzle}
\end{figure}

For concreteness, we first focus on the three fixed points in the second torus of the $T_{(0,1)}$ twisted sector. As depicted in figure~(\ref{fig:puzzle}), the local spectra of the three brother models,
\begin{equation}
a\equiv -\left(V_2\cdot\Delta V_1 - \frac{x}{3}\right)=0,\,\frac{1}{3},\,\frac{2}{3}\;,
\end{equation}
can be combined consistently into one model with $b_3=b_4=\frac{1}{3}$. On the other hand, in the $T_{(1,0)}$ twisted sector there is a fixed torus in the directions $e_3,\,e_4$; thus the sector is empty.

This procedure can also be applied to the terms $c_\alpha$ and $d_{\alpha\beta}$ of the generalized discrete torsion phase, eqn.~(\ref{eq:generalizedtorsionphase}).

\subsubsection{Generalized Brother Models versus Generalized Discrete Torsion}

As in our previous discussion in section~\ref{sec:DiscreteTorsion}, also the generalized versions of the discrete torsion phase and the brother phase have a very similar form. Indeed, whenever there are non-trivial solutions to the eqns.~(\ref{eq:newModularInvforDelta}), one can equivalently describe models with generalized discrete torsion phase in terms of generalized brother models. This is the generic case.

However, there are exceptions. Namely, as we will explain below, models with $d_{\alpha\beta}\neq 0$ in $\Z{3}\times\Z{3}$ orbifolds\index{orbifold!$\Z{2}\times\Z{2}$} without Wilson lines cannot be interpreted in terms of brother models.

Consider the fourth part of the generalized discrete torsion phase of eqn.~(\ref{eq:generalizedtorsionphase}),
\begin{equation}\label{eq:dtermdiscretetorsion}
\varepsilon~=~e^{2\pi\I\,d_{\alpha\beta}\,n_{\alpha} m_{\beta}}\;,
\end{equation}
with $d_{\alpha\beta}\in \left\lbrace 0,\tfrac{1}{3},\tfrac{2}{3}\right\rbrace$. An analogous term appears in the generalized brother phase as
\begin{equation}\label{eq:dtermbrotherphase}
\widetilde{\varepsilon}~=~\exp\left[-2 \pi \I\, n_{\alpha}m_{\beta}\left(A_\beta \cdot \Delta A_{\alpha}-\frac{z_{\alpha\beta}}{Q_{\alpha\beta}} \right)\right]\;,
\end{equation}
where $Q_{\alpha\beta}~=~3$, since the Wilson lines have order 3. In general, both phases can be made coincide by choosing $\Delta A_{\alpha}\in \Lambda$ such that 
\begin{equation}
-\left(A_\beta \cdot \Delta A_{\alpha}-\frac{z_{\alpha\beta}}{3} \right)=d_{\alpha\beta}\;.
\end{equation}
On the other hand, in the case when $A_\alpha=0$ and $\Delta A_\alpha\neq 0$, eqn.~(\ref{eq:dtermbrotherphase}) simplifies to 
\begin{equation}\label{eq:dtermbrotherphaseZ3xZ3}
 \widetilde{\varepsilon}
 ~=~e^{2\pi \I\, n_{\alpha}m_{\beta}\left(\frac{z_{\alpha\beta}}{3}\right)}
 ~=~e^{2\pi \I\, n_{\alpha}m_{\beta}\left(\frac{\Delta A_\alpha\cdot\Delta A_\beta}{2} \right)}\;,
\end{equation}
where the second equality follows from the definition of $z_{\alpha\beta}$, eqn.~(\ref{eq:newModularInvforDeltaAA}). As $\Delta A_\alpha$ are lattice vectors, this equality can only hold if $z_{\alpha\beta}=0\text{ mod }3$, which implies that the brother phase eqn.~(\ref{eq:dtermbrotherphaseZ3xZ3}) is trivial. Thus, in this case, the generalized discrete torsion phase leads to models which cannot arise by adding lattice vectors to shifts and Wilson lines.

In summary, the generalized discrete torsion phases admit more possible assignments than the generalized brother phases. Nevertheless, a large class of the models with generalized discrete torsion has an equivalent description in terms of models with a modified gauge embedding.

These results have important implications. By introducing generalized discrete torsion, or lattice-valued Wilson lines, one can control the local spectra. One can therefore expect that introducing generalized discrete torsion, or alternatively shifting the Wilson lines by lattice vectors, will gain a similar importance as discrete Wilson lines~\cite{Ibanez:1986tp} for orbifold model building.

As stated above, switching on generalized discrete torsion can lead to the disappearance of complete local spectra. This raises the question of how to interpret this fact in terms of geometry. Some of the localized zero-modes can be viewed as blow-up modes which allow to resolve the orbifold singularity associated to a given fixed point~\cite{Hamidi:1986vh,Dixon:1986qv,Walton:1987bu,Aspinwall:1994ev,Vafa:1994rv} (see chapter~\ref{sec:blowup} and \cite{Lust:2006zh,Honecker:2006qz,Nibbelink:2007rd,Nibbelink:2008tv} for recent developments). If at a given fixed point there are no zero modes, one might argue that, therefore, the associated singularity cannot be `blown up'. In what follows, we shall advertise an alternative interpretation.

\section{Connection to Non-factorizable Orbifolds}
\label{sec:Z2xZ2}

In this section it is shown that in many cases orbifold models M with certain geometry, i.e.\ torus lattice $\Gamma$, and generalized discrete torsion switched on are equivalent to torsionless models M$'$ based on a different lattice $\Gamma'$. Model M$'$ has less fixed points than M, and the mismatch turns out to constitute precisely the `empty' fixed points of model M.

The simplest examples are based on $\Z{2}\times\Z{2}$\index{orbifold!$\Z{2}\times\Z{2}$} orbifolds with standard embedding and without Wilson lines. As compactification lattice $\Gamma$, we choose an $\SU2^6$ lattice, as discussed in section~\ref{sec:Z2xZ2Orbifold}. As we have seen in section~\ref{sec:GeneralizedDiscreteTorsion}, in this case there are 25 physical parameters for generalized discrete torsion, with values either 0 or $\frac{1}{2}$. For concreteness, we restrict to the 12 $d_{\alpha\beta}$ parameters and scan over all $2^{12}$ models.

\begin{table}[t!]
{\small{
\begin{center}
\begin{tabular}{|c||c|c|c|c|r||c||l|}
\hline
& $T_{(1,0)}$  & $T_{(0,1)}$  & $T_{(1,1)}$  & total            & $\#_S$ & $d_{\alpha\beta} = \tfrac{1}{2}$ & $A_\alpha \neq 0$\\
\hline
\hline
A.1   & $(16,0)$     & $(16,0)$     & $(16,0)$     & $(51,3)$         & 246  & $-$                & $-$ \\
\hline
A.2   & $(12,4)$     & $(8,0)$      & $(8,0)$      & $(31,7)$         & 166  & $d_{24}$           & $A_2=(S)(0^8)$, $A_4 =(V)(0^8)$ \\
\hline
A.3   & $(10,6)$     & $(4,0)$      & $(4,0)$      & $(21,9)$         & 126  & $d_{14}, d_{23}$   & $A_1=(S)(0^8)$, $A_2 =(0^8)(S)$,\\
      &              &              &              &                  &      &                    & $A_3=(0^8)(V)$, $A_4 =(V)(0^8)$ \\
\hline
A.4   & $(8,0)$      & $(8,0)$      & $(8,0)$      & $(27,3)$         & 126  & $d_{26}, d_{46}$   & $A_2 =(V')(0^8)$, $A_4 =(V)(0^8)$,\\
      &              &              &              &                  &      &                    & $A_6 =(S)(0^8)$ \\
\hline
A.5   & $(6,2)$      & $(6,2)$      & $(4,0)$      & $(19,7)$         & 106  & $d_{24}, d_{36}$   & $A_2 =(S)(0^8)$, $A_3 =(0^8)(S)$,\\
      &              &              &              &                  &      &                    & $A_4 =(V)(0^8)$, $A_6 =(0^8)(V)$\\
\hline
A.6   & $(6,2)$      & $(4,0)$      & $(4,0)$      & $(17,5)$         &  86  & $d_{16}, d_{24},$  & $A_1 =(V)(0^8)$, $A_2 =(0^8)(V)$,\\
      &              &              &              &                  &      & $d_{36}$           & $A_3 =(V')(0^8)$, $A_4 =(0^8)(S)$,\\
      &              &              &              &                  &      &                    & $A_6 =(S)(0^8)$ \\
\hline
A.7   & $(4,0)$      & $(4,0)$      & $(4,0)$      & $(15,3)$         &  66  & $d_{16}, d_{25},$  & $A_1=(V)(0^8)$, $A_2 =(0^8)(V)$,\\
      &              &              &              &                  &      & $d_{36}, d_{45}$   & $A_3 =(V')(0^8)$, $A_4 =(0^8)(V')$,\\
      &              &              &              &                  &      &                    & $A_5 =(0^8)(S)$, $A_6 =(S)(0^8)$\\
\hline
A.8   & $(3,1)$      & $(3,1)$      & $(3,1)$      & $(12,6)$         &  66  & $d_{16}, d_{24},$  & $A_1 =(W_1)(0^8)$, $A_2 =(0^8)(W_1)$,\\
      &              &              &              &                  &      & $d_{35}$           & $A_3 =(0^8)(W_1')$, $A_4 =(0^8)(W_2)$,\\
      &              &              &              &                  &      &                    & $A_5 =(0^8)(W_2')$, $A_6 =(W_2)(0^8)$\\
\hline
\end{tabular}
\end{center}\vspace{-0.4cm}
\caption{Examples of $\Z2\times\Z2$ orbifolds with generalized discrete torsion. The $2^\mathrm{nd}$--$4^\mathrm{th}$ columns list the number of anti-families and families, respectively, for the various sectors $T_{(k_1,k_2)}$. In all models, the untwisted sector gives a contribution of $(3,3)$ (anti-)families. $\#_S$ denotes the total number of singlets. These spectra can either be obtained by turning on generalized discrete torsion $d_{\alpha\beta}$ as specified in the next-to-last column, or by using lattice-valued Wilson lines $A_\alpha$ as listed in the last column, compare to eqn.~(\ref{eq:buildingblocks}).}
\label{tab:Z2xZ2}
}}
\end{table}

Beside other models with a net number of zero families\footnote{More details on models with a net number of zero families can be found in the table~(\ref{tab:Z2xZ2ExampleZERO}) in the appendix.}, we find eight models (and their mirrors, i.e.\ models where families and anti-families are exchanged). They are listed in table~(\ref{tab:Z2xZ2}), where we present the number of (anti-)families for each twisted sector and the total number of singlets. As discussed in section~\ref{sec:GeneralizedDiscreteTorsion}, models with non-trivial $d_{\alpha\beta}$ are equivalent to torsionless models with lattice-valued Wilson lines. Possible representatives of these Wilson lines can be composed out of the building blocks
\begin{align}
 W_1&  =~(0^6,1,1)\;,
 & W_2 &=~(0^5,1,1,0)\;,
 &  W_1'& =~(1,1,0^6) \;, 
 & W_2' & =~(0,1,1,0^5)\;, \nonumber \\
 S&=~(\tfrac{1}{2}^8)\;, 
 & V & =~(0^7,2)\;, 
 & V' & =~(0^6,2,0)\;,\label{eq:buildingblocks}
 & & \phantom{I^I}
\end{align}
and are listed in the last column of table~(\ref{tab:Z2xZ2}).

\subsubsection{Relations between different $\boldsymbol{\Z{2}\times\Z{2}}$ Constructions}

Models leading to spectra coinciding with the ones in table~(\ref{tab:Z2xZ2}) have already been discussed in the literature. They appeared first in reference~\cite{Donagi:2004ht} in the context of free fermionic string models related to the $\Z{2}\times\Z{2}$ orbifold with an additional freely acting shift (see also~\cite{Donagi:2008xy,Kiritsis:2008mu}). More recently, new $\Z{2}\times\Z{2}$ orbifold constructions have been found in studying orbifolds of non-factorizable six-tori~\cite{Faraggi:2006bs,Forste:2006wq}. We find that for each model M of table~(\ref{tab:Z2xZ2}) there is a corresponding `non-factorizable' model M$'$ with the following properties:
\begin{enumerate}
\item Each `non-empty' fixed point, i.e.\ each fixed point with local zero-modes, in the model M can be mapped to a fixed point with the same spectrum in model M$'$. 
\item The number of `non-empty' fixed points in M coincides with the total number of fixed points in M$'$.
\end{enumerate}

\subsubsection{Generalization to $\boldsymbol{\Z{N}\times\Z{M}}$ and Interpretation}

These relations are not limited to $\Z{2}\times\Z{2}$ orbifolds, rather we find an analogous connection also in other $\Z{N}\times\Z{M}$ cases, see table~(\ref{tab:Z2xZ4ExampleGDT}) in the appendix ($\Z{N}\times\Z{M}$ orbifolds based on non-factorizable compactification lattices have recently been  discussed in~\cite{Takahashi:2007qc}). This result hints towards an intriguing impact of generalized discrete torsion on the interpretation of orbifold geometry. What the (zero-mode) spectra concerns, introducing generalized discrete torsion (or considering generalized brother models) is equivalent to changing the geometry of the underlying compact space, $\Gamma\to\Gamma'$. To establish complete equivalence between these models would require to prove that the couplings of the corresponding states are the same, which is beyond the scope of the present study. It is, however, tempting to speculate that non-resolvable singularities, as discussed above, do not `really' exist as one can always choose (for a given spectrum) the compactification lattice $\Gamma$ in such a way that there are no `empty' fixed points.

\section[How to classify $\Z{N}\times\Z{M}$ Orbifolds]{How to classify $\boldsymbol{\Z{N}\times\Z{M}}$ Orbifolds}
\label{sec:ClassifyZNxZM}

In\cite{Ploger:2007iq} it is shown how to use the knowledge obtained from discrete torsion to classify $\Z{N}\times\Z{M}$ orbifolds using the concept of brother models. The strategy is exemplified for the case of $\Z{3}\times\Z{3}$ orbifolds resulting in 120 and 131 inequivalent shift-embeddings in the case of $\E{8}\times\E{8}$ and $\SO{32}$, respectively. More details can be found in~\cite{Ploger:2007iq,WebTables:2007z3,Ramos:2008ph}.


\chapter{The Mini-Landscape}
\label{sec:ML}

In the previous chapters we have obtained a detailed understanding of the construction of heterotic orbifolds. Now, we can turn to more phenomenological questions: the aim of this chapter is to establish connections between the heterotic string on the one side and particle physics on the other, where the link is assumed to be the orbifold compactification.

In detail, since the MSSM is one of the most promising candidates for describing particle physics at the LHC, we will try to obtain MSSM candidates as 4d low-energy effective theories from the compactification of the heterotic string on orbifolds. For doing so, we have the main tools at hand. Starting from the input parameters of the heterotic orbifold (the torus lattice $\Gamma$, the point group $P$, the shift $V$ and the Wilson lines $A_\alpha$), we can compute the 4d gauge group and the massless matter spectrum (see section~\ref{sec:HetOnOrb}). Furthermore, with the help of the string selection rules we can derive the form of the superpotential (see section~\ref{sec:Yukawa}). Additional information about mass hierarchies for example can be deduced from geometrical aspects of the orbifold, i.e. from the localization of the MSSM fields in the extra dimensions.

However, the main problem we are facing now is that the input parameters of the compactification seem to be arbitrary, yielding a vast number of different 4d models~\cite{Lerche:1986cx}. We do not know any selection mechanism which prefers one compactification scheme to another. In other words, all orbifold compactifications seem to be equal from the string theory point of view. In contrast, from the low energy point of view we have a model in favor, the MSSM. So, how can we resolve this clash? In this chapter, we will use the landscape~\cite{Susskind:2003kw} as a tool, not to build all models, but to find as many good ones as possible. Then we can hope to find similarities among the good models that might help us to understand better the actual meaning of the landscape. The guiding principle which will lead us to the allocated regions of the landscape of heterotic orbifolds will be the concept of \emph{local GUTs}~\cite{Kobayashi:2004ya,Forste:2004ie,Kobayashi:2004ud,Buchmuller:2004hv,Nilles:2004ej,Buchmuller:2005jr,Buchmuller:2005sh,Buchmuller:2006ik}.

The work presented in this chapter has been published in a series of papers~\cite{Lebedev:2006kn,Lebedev:2006tr,Lebedev:2007hv}.

\section{Local GUTs}
\label{sec:localGUT}

In order not to get lost in the landscape of string vacua by building a huge number of phenomenologically invalid models, we need a guideline that points us to promising regions of the landscape. In general, when we choose a search strategy we can only know its success after we have actually performed the search. However, there are some hints towards a promising strategy coming from models already constructed in the heterotic landscape, e.g.~\cite{Kobayashi:2004ya,Buchmuller:2005jr,Kim:2006hv}. These constructions have in some sense the idea of GUTs in common. Starting from a big gauge group $\E{8}\times\E{8}$ or $\SO{32}$ in 10d it seems natural to encounter a GUT on the way down to 4d. It is possible to encounter the GUT as an intermediate step of the compactification (e.g. in 5d or 6d\cite{Buchmuller:2007qf}), or as an ordinary 4d GUT. In addition, there is the possibility that the GUT only becomes manifest in some special region of the compact space. This is the concept of local GUTs.

\subsubsection{A Local GUT for every Fixed Point}\index{local GUT}

In the context of heterotic orbifolds, the 10d gauge group is broken at the fixed points to local GUT gauge groups. In the following, we will discuss this in detail starting with fixed points from the first twisted sector of $\Z{N}$ orbifolds.

\begin{figure}[t]
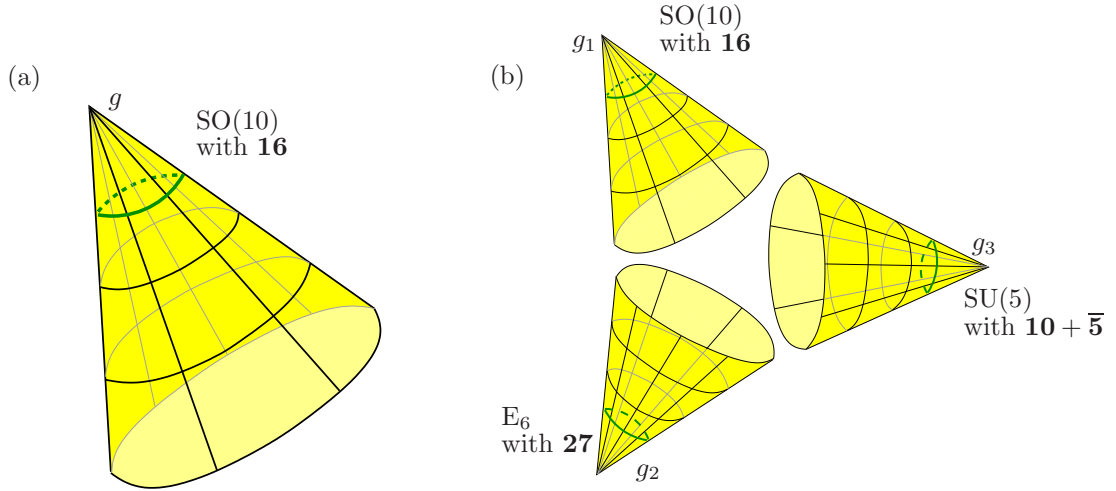

\centerline{\input orbifoldcone1.pstex_t \hspace{1.4cm}\input orbifoldcone2.pstex_t}\vspace{-0.3cm}
\caption{Visualization of the local GUT picture: (a) In the zoom-in picture at the fixed point $g$ a local $\SO{10}$ GUT becomes visible. (b) At every fixed point $g_i$ there is a local GUT induced by the local shifts $V_{g_i}$: $\SO{10}$, $\E{6}$ and $\SU{5}$, respectively. At the respective fixed points matter forms representations under its local GUT gauge group.}
\label{fig:LocalGUTs}
\end{figure}

We begin with some $\Z{N}$ orbifold and consider one of its fixed points from the first twisted sector $T_{(1)}$ in detail. By $g$ we denote the space group element associated to this fixed point. Now, we zoom-in to the fixed point $g$. Since $g$ is from the $T_{(1)}$ sector, the local orbifold space looks like $\mathbbm{C}^3/g$ and is not modded out further by some other element of the full space group\footnote{At the fixed point $g$, the local orbifold space is determined by all commuting elements $[g,h]=0$. Since $g$ is assumed to be from the first twisted sector, the commuting elements are trivial, see appendix~\ref{app:commuting}.}. Only $g$ describes the local space. If we now ``compactify'' the heterotic string on this local orbifold, we have to project on $g$ invariant states only. Consequently, the 10d gauge symmetry breaks according to the local shift $V_g$, i.e. only those roots $p$ of $\E{8}\times\E{8}$ or $\SO{32}$ remain unbroken that fulfill the condition
\begin{equation}
p\cdot V_g = 0 \text{ mod } 1\;,
\end{equation}
and hence give rise to the local GUT gauge group, see eqn.~(\ref{eqn:4dGaugeGroupPhase}). Moreover, $g$ defines twisted boundary conditions that give rise to twisted strings. Since there are no further space group elements in the zoom-in picture that define the local orbifold $\mathbbm{C}^3/g$, we do not have to ensure invariance of the twisted strings in addition\footnote{Twisted states from the first twisted sector are automatically invariant under the orbifold action, see appendix~\ref{app:T1IsInvariant}.}. Thus, we have found the local matter spectrum. Obviously, the matter spectrum at $g$ has to form representations with respect to its local GUT gauge group, see figure~(\ref{fig:LocalGUTs}a). However, if we zoom-in to different fixed points we have different local shifts with different local GUTs and hence matter representations under various local gauge groups. For a visualization of this situation see figure~(\ref{fig:LocalGUTs}b).

\subsubsection{The 4d Perspective}
\begin{figure}[t]
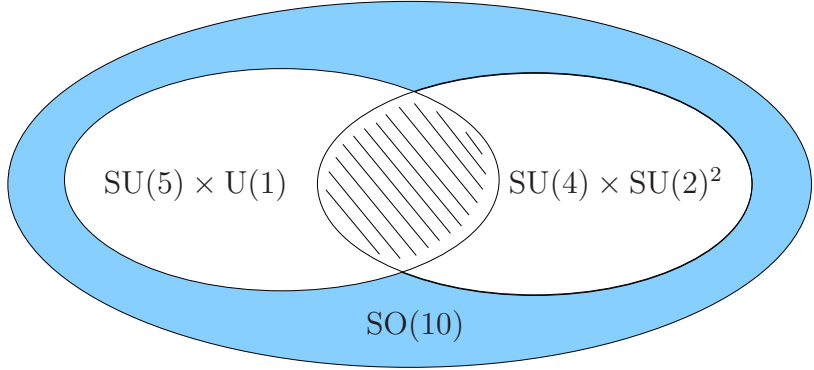

\centerline{\input GaugeGroupIntersection.pstex_t}\vspace{-0.2cm}
\caption{The 4d gauge group of the Standard Model $\SU{3}\times\SU{2}\times\U{1}_Y\times\U{1}_X$ (dashed region) containing $\U{1}_{B-L}$ can be obtained from $\SO{10}$ as the intersection of a local $\SU{5}$ GUT and a local Pati-Salam GUT, as explained in~\cite{Asaka:2001eh}.}
\label{fig:4dGGIntersection}
\end{figure}

Now, we zoom-out from the local picture to the 4d one. First, consider the local GUT gauge group. The associated untwisted sector states are free to move on the full orbifold space now. Thus, from the four-dimensional point of view, only those roots $p$ of $\E{8}\times\E{8}$ or $\SO{32}$ remain unbroken that fulfill all local conditions simultaneously, i.e.
\begin{equation}
p\cdot V_{g_i} = 0 \text{ mod } 1\; \quad\text{for all elements } g_i,
\end{equation}
giving rise to the 4d gauge group (compare to eqn.~(\ref{eqn:4dGaugeGroupPhase})). In other words, the 4d gauge group arises as the common intersection of all the local ones, see figure~(\ref{fig:4dGGIntersection}) for an example. As mentioned before, there are no further projection conditions for the twisted matter states from $T_{(1)}$ when we zoom-out to the 4d picture. Thus, the $T_{(1)}$ twisted matter representations of the local GUT gauge group just branch group theoretically into representations of the 4d gauge group. Let us consider an example.

\subsubsection{Local $\boldsymbol{\SO{10}}$ GUT}\index{Grand Unified Theories!$\SO{10}$}

If we have a local $\SO{10}$ gauge group with a local $\rep{16}$-plet at some fixed point of the first twisted sector and if additionally the 4d gauge group is $\SU{3}\times\SU{2}\times\U{1}_Y$, the spinor of $\SO{10}$ decomposes according to\cite{Georgi:1974my,Fritzsch:1974nn},
\begin{equation}
\rep{16}~=~(\rep{3},\rep{2})_{1/6}+(\crep{3},\rep{1})_{-2/3}+(\crep{3},\rep{1})_{1/3}+(\rep{1},\rep{2})_{-1/2}+(\rep{1},\rep{1})_{1}+(\rep{1},\rep{1})_{0}\;,
\end{equation}
giving rise to one generation of the SM matter plus a right--handed neutrino
\begin{equation}
\rep{16}~\rightarrow~ q + \bar{u} + \bar{d} + \ell + \bar{e} + \bar{n}\;.
\end{equation}
Thus, using the concept of local GUTs for the first twisted sector provides an easy mechanism to obtain complete generations of quarks and leptons in terms of single (local) GUT representations, like $\rep{16}$-plets. The important difference to conventional GUTs is that the local GUT gauge group breaks to the Standard Model one by the orbifold compactification to 4d and not via a Higgs mechanism.

\subsubsection{Split Multiplets for the Higgses and Local GUTs}

As discussed in the previous section, quarks and leptons can be combined to form a complete representation of the GUT gauge group. On the other hand, the Higgs fields $\phi$ and $\bar{\phi}$ of the MSSM do not form complete GUT multiplets. This leads to the famous doublet-triplet splitting problem\index{$\mu$-term!doublet-triplet splitting} of conventional GUTs~\cite{Dimopoulos:1981yj,Sakai:1981gr,Sakai:1981pk}. For example, in $\SO{10}$ the smallest representation that can incorporate the Higgses is the ten-dimensional vector representation $\rep{10}$. In terms of the Standard Model gauge group it decomposes according to
\begin{equation}
\rep{10}~=~ (\crep{3},\rep{1})_{1/3} + (\rep{1},\rep{2})_{-1/2} + (\rep{3},\rep{1})_{-1/3} + (\rep{1},\rep{2})_{1/2}\;.
\end{equation}
The MSSM Higgses $\phi$ and $\bar{\phi}$ in the representations $(\rep{1},\rep{2})_{-1/2}$ and $(\rep{1},\rep{2})_{1/2}$, respectively, shall be light if the $\mu$-term 
\begin{equation}
\mu\bar{\phi}\phi
\end{equation}
is small. On the other hand, the Higgs-triplets $(\crep{3},\rep{1})_{1/3}$ and $(\rep{3},\rep{1})_{-1/3}$ are not observed and more importantly can mediate fast proton decay due to dimension 5 operators. Furthermore, they can alter gauge coupling unification. Therefore, they must be heavy with a mass presumably at the GUT scale or higher. So, what mechanism distinguishes between the light doublets and the heavy triplets?

Local GUTs can also provide an intuitive answer here. Up to now, we have restricted to local GUTs from the first twisted sector of $\Z{N}$ orbifolds, since then no further orbifold projections could remove local matter representations or parts thereof. But now, this is actually the desired feature: by zooming-out from the local picture to the 4d one, the Higgs representation decomposes the into Standard Model ones containing doublets and triplets and the orbifold can potentially project out the unwanted triplet states.

\subsubsection{Remark: Local GUTs of Higher Twisted Sectors}\index{local GUT}

Local GUTs from the first twisted sector provide \emph{complete} GUT multiplets decomposed into representations of the 4d gauge group. The origin of this feature is the absence of orbifold projection conditions for matter from the first twisted sector. Note, however, that this situation is not unique to $T_{(1)}$. In principle, it applies additionally to some fixed points of higher twisted sectors. Geometrically, these are fixed points that \emph{only} arise in some higher twisted sectors $T_{(k)}$ and its ``anti-sector'' $T_{(N-k)}$ such that the orbifold space can be written locally as $\mathbbm{C}^3/g$ (with $g$ from the twisted sector $T_{(k)}$).

For example, in the case of the $\Z{6}$-II orbifold with torus lattice $\G{2}\times\SU{3}\times\SU{2}^2$, the fixed points of the second (and fourth) twisted sector, away from the origin in the $\G{2}$ plane, are not present in the other twisted sectors. In figure~(\ref{fig:Z6IICommuting2}) in the appendix they are denoted by $g_2$. Only Wilson lines that go along the directions of the fixed torus induce orbifold projections for matter states originating from these fixed points. This breaks the local GUT and projects out some matter states resulting in a ``smaller'' GUT. However, the matter representations with respect to this smaller local GUT are not affected further when we zoom-out to the 4d picture, since further orbifold projections have to be carried out using the constructing element itself (and powers thereof) and are thus trivial. However, due to the presence of fixed tori in higher twisted sectors, twisted matter forms $\mathcal{N}=2$ hypermultiplets. Hence, this matter is non-chiral.

In summary, also higher twisted sectors allow in principle for local GUTs such that \emph{complete} GUT multiplets just decompose into representations of the 4d gauge group.

\section[The $\mathbbm{Z}_6$-II Landscape]{The $\boldsymbol{\mathbbm{Z}_6}$-II Landscape}\index{local GUT}

Having the guideline of local GUTs in mind, we perform a search for MSSM candidates in the framework of $\Z{6}$-II orbifold\index{orbifold!$\Z{6}$-II} compactifications of the heterotic $\E8\times\E8$ string. The underlying torus lattice is chosen to be $\G{2}\times\SU{3}\times\SU{2}^2$, as discussed in section~\ref{sec:Z6IIOrbifold}.

Next, we choose the local GUT by choosing the shift. Looking through the list of all 61 inequivalent shifts and their massless matter spectra for $\Z{6}$-II~\cite{Katsuki:1989cs}, we select local $\SO{10}$ and local $\E{6}$ GUTs as the most promising. In detail, there are two gauge shifts leading to a local $\SO{10}$ GUT with $\rep{16}$-plets in the first twisted sector,
\begin{eqnarray}\label{eq:so10shifts}
V^{ \SO{10},1} = & \left(\tfrac{1}{3},\,\tfrac{1}{2},\,\tfrac{1}{2},\,0,\,0,\,0,\,0,\,0\right)&\left(\tfrac{1}{3},\,0,\,0,\,0,\,0,\,0,\,0,\,0\right) \;, \\
V^{ \SO{10},2 }= & \left(\tfrac{1}{3},\,\tfrac{1}{3},\,\tfrac{1}{3},\,0,\,0,\,0,\,0,\,0\right)&\left(\tfrac{1}{6},\,\tfrac{1}{6},\,0,\,0,\,0,\,0,\,0,\,0\right) \;,\nonumber
\end{eqnarray}
and two shifts leading to a local $\E{6}$ GUT with $\rep{27}$-plets in the first twisted sector,
\begin{eqnarray}\label{eq:e6shifts}
 V^{\E6 , 1}= & \left(\tfrac{1}{2},\,\tfrac{1}{3},\,\tfrac{1}{6},\,0,\,0,\,0,\,0,\,0\right)&\left(0,\,0,\,0,\,0,\,0,\,0,\,0,\,0\right)\;,\\
 V^{ \E6 ,2}= & \left(\tfrac{2}{3},\,\tfrac{1}{3},\,\tfrac{1}{3},\,0,\,0,\,0,\,0,\,0\right)&\left(\tfrac{1}{6},\,\tfrac{1}{6},\,0,\,0,\,0,\,0,\,0,\,0\right)\;, \nonumber
\end{eqnarray}
where $V^{\E6 , 1}$ corresponds to the standard embedding. Since these shifts provide $\rep{16}$- or $\rep{27}$-plets in the first twisted sectors and therefore complete generations of quarks and leptons, we can hope to find MSSM candidates with three generations using the strategy described in the following. In other words, these four shifts are the starting points for our journey through the Mini-Landscape. By turning on Wilson lines we will explore the corresponding regions in detail now.

\subsection{The Search Strategy}

Starting from the four local GUT shifts, a search strategy for finding phenomenologically interesting models is developed. The strategy reads:

\begin{itemize}
 \item[1.] Generate all two Wilson line models.
 \item[2.] Identify ``inequivalent'' models.
 \item[3.] Select models with $G_\text{SM} \subset \SU5 \subset \SO{10}$ or $G_\text{SM} \subset \SU5 \subset \E{6}$ 
 \item[4.] Select models with three net $(\rep{3},\rep{2})$.
 \item[5.] Select models with non--anomalous $\U1_{Y} \subset \SU5$.
 \item[6.] Select models with net 3 SM generations + Higgses + vector--like.
\end{itemize}

In detail, the steps are as follows: At {\bf step 1}, one of the four GUT--shifts, eqns.~(\ref{eq:so10shifts}) or ~(\ref{eq:e6shifts}), is chosen and all two Wilson line models are constructed using the methods described in~\cite{Giedt:2000bi,Wingerter:2005ph} and in the appendix of~\cite{Nilles:2006np}. These models can be separated into two cases: either these two Wilson lines are of order two ($A_5$ and $A_6$) or one Wilson line is of order three ($A_3 = A_4$) and one of order two ($A_5$). Since the \Z6-II orbifold neither allows for discrete torsion nor for brother models (see section~\ref{sec:GeneralizationDTAndBrothers}), the ansatz of~\cite{Giedt:2000bi,Wingerter:2005ph,Nilles:2006np} is sufficient to construct all different models (for an extension to three Wilson lines, see~\cite{Lebedev:2008un}).

However, this ansatz for constructing models has redundancies and thus many models are equivalent. To determine the inequivalent models at {\bf step 2}, a simple and fast comparison-method is used: two models are considered to be equivalent if they have the same gauge group, the same non-Abelian matter spectrum and the same amount of non-Abelian singlets. Thus, models differing only in $\U{1}$ charges are treated as equivalent. In addition, some models differ only by the localization of states on the various fixed points. We know that these ambiguities occur and it is likely that in some cases Yukawa couplings are affected. Hence this criterion may underestimate the number of truly inequivalent models.

At {\bf step 3}, those models are retained that have a Standard Model gauge group $\SU{3}\times\SU{2}\times\U{1}_Y$ originating from an $\SU{5}$. This $\SU{5}$ in turn shall lie inside the local GUT under consideration, i.e. $\SO{10}$ or $\E{6}$. Technically, this is done by demanding that the simple roots of $\SU{3}\times\SU{2}$ can be written as a linear combination of the ones of $\SU{5}$. This choice of the SM gauge group ensures that we can use the standard $\SU{5}$ GUT hypercharge\index{hypercharge} generator. Generically it is of the form\index{hypercharge}
\begin{equation}
t_Y = \left(0,\,0,\,0,\,\tfrac{1}{2},\,\tfrac{1}{2},\,-\tfrac{1}{3},\,-\tfrac{1}{3},\,-\tfrac{1}{3}\right)\left(0,\,0,\,0,\,0,\,0,\,0,\,0,\,0\right)\;,
\end{equation}
which is appropriate for gauge coupling unification and yields a weak mixing angle $\sin\theta_W = 3/8$ at the GUT scale. 

The next search criterion at {\bf step 4} selects models having a net number of three left-handed quark doublets $(\rep{3},\rep{2})$, where net number means that we also allow for situations like $4(\rep{3},\rep{2})$ plus $(\crep{3},\rep{2})$, for example. 

{\bf Step 5} ensures that the hypercharge chosen previously in step 3 is non--anomalous. Technically, this is achieved by demanding that the generators of $\U{1}_Y$ and $\U{1}_\text{anom}$ are orthogonal, i.e. $t_Y \cdot t_\text{anom} = 0$. A non--anomalous hypercharge is necessary, because an anomalous one would be broken at the high scale due to the presence of the Fayet--Iliopoulos D-term\index{Fayet--Iliopoulos D-term}, resulting in a catastrophe for electroweak symmetry breaking.

In the last step models are selected if they have the chiral matter content of the MSSM, i.e. three generations of quarks and leptons and at least one pair of Higgses. In addition, models at {\bf step 6} are allowed to have vector-like exotics. In order for some exotics to be vector-like with respect to the SM gauge group, they either have to form real representations of $\SU{3}\times\SU{2}\times\U{1}_Y$ or they have to come in pairs of some representations plus their complex conjugates. Then, it is in principle possible to write down a mass term for these exotics with a very high mass such that the exotics decouple from the low energy effective theory. Note, however, that the couplings in the superpotential relevant for the mass terms can not be put in by hand, but they have to be derived from string theory, as explained later.

\subsection{The Results - Part I}

\begin{table}[t]
\centerline{
\begin{tabular}{l||l|l||l|l}
 criterion & $V^{\SO{10},1}$ & $V^{\SO{10},2}$ & $V^{\E6,1}$ & $V^{\E6,2}$\\
\hline
\hline
 2. inequivalent models with 2 Wilson lines                         &$22,000$ & $7,800$ & $680$ &$1,700$ \\[0.1cm]
 3. SM gauge group $\subset \SU{5} \subset \SO{10}$  ({\rm or}~\E6) & $3,563$ & $1,163$ &  $27$ &   $63$ \\[0.1cm]
 4. 3 net $(\rep{3},\rep{2})$                                       & $1,170$ &   $492$ &   $3$ &   $32$ \\[0.1cm]
 5. non--anomalous $\U1_{Y}\subset \SU5 $                           &   $528$ &   $234$ &   $3$ &   $22$ \\[0.1cm]
 6. spectrum $=$ 3 generations $+$ vector-like                      &   $128$ &    $90$ &   $3$ &    $2$ \\
\hline
\end{tabular}
}\vspace{-0.3cm}
\caption{Statistics of \Z6-II orbifolds based on the shifts $V^{\SO{10},1},V^{\SO{10},2},V^{\E6,1},V^{\E6,2}$ with two Wilson lines.}
\label{tab:SummaryZ6IIML}
\end{table}

\begin{figure}[b]
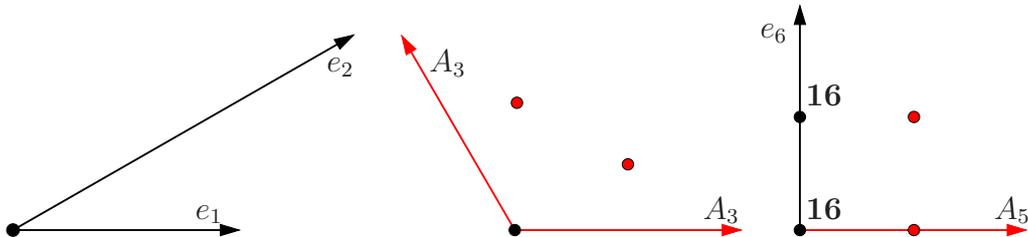
 
\centerline{\input Z6IIOrbifold2FP.pstex_t}\vspace{-0.4cm}
\caption{First twisted sector $T_{(1)}$ of the $\Z{6}$-II orbifold on the lattice $\G{2}\times\SU{3}\times\SU{2}^2$ with one Wilson line $A_3$ along the $e_3$ and $e_4$ direction and another one $A_5$ along the $e_5$ direction. The Wilson lines lift the fixed point degeneracy such that only at the $1\times 1\times 2 = 2$ black fixed points the local $\SO{10}$ GUTs with local $\rep{16}$-plets remain.}
\label{fig:Z6IIOrbifold2FP}
\end{figure}

The results of this search strategy are summarized in table~(\ref{tab:SummaryZ6IIML}). It leads to $128+90+3+2=223$ models having the chiral matter content of the MSSM~\cite{Lebedev:2006kn}\index{MSSM!candidates from orbifolds}. Surprisingly, all of these $223$ MSSM candidate models have one order three Wilson line and one of order two~\cite{Buchmuller:2006ik}. This leads to the situation that two SM generations of quarks and leptons originate from the local GUT structure of $\SO{10}$ or $\E{6}$, as depicted in figure~(\ref{fig:Z6IIOrbifold2FP}), and the components of the third generation are localized at various twisted or untwisted sectors. None of the good models is based on two order two Wilson lines. Many details of these models, including the Wilson lines, the (hidden sector) gauge group, the massless matter and their localization, are listed in a web page~\cite{WebTables:2006ml}.

It is instructive to compare this model-scan to others. In certain types of intersecting D--brane models, it was found that the probability of obtaining the SM gauge group and three generations of quarks and leptons (in some cases with chiral exotics) is at best $10^{-9}$ \cite{Gmeiner:2005vz,Douglas:2006xy,Gmeiner:2007we}. The criterion which comes closest to the requirements imposed in \cite{Gmeiner:2005vz,Douglas:2006xy} is step 4. We find that within the sample presented here the corresponding probability is 5\,\%. In \cite{Dijkstra:2004cc,Anastasopoulos:2006da}, orientifolds of Gepner models were scanned for chiral MSSM matter spectra, and it was found that the fraction of such models is $10^{-14}$. In table~(\ref{tab:SummaryZ6IIML}) the corresponding probability, i.e.\ the fraction of models passing criterion 6, is of order 1\,\%. Note also that, in all $223$ models, hypercharge is normalized as in standard GUTs and thus consistent with gauge coupling unification. This comparison shows that this sample of heterotic orbifolds is unusually ``fertile'' compared to other constructions.  The probability of finding something close to the MSSM is much higher than that in other patches of the landscape analyzed so far.

If one relaxes the constraints of step 3 (i.e. SM $\subset \SU{5} \subset \SO{10}$ {\rm or}~\E6) and step 5 (i.e. $\U1_{Y}\subset \SU5$) the number of models increases by a factor of about 10~\cite{Raby:2007yc}. However the additional constraint that $\sin^2\theta_W = 3/8$ reduces this number by 90\% so that there are only a handful of additional models. It suggests that in order to find the MSSM, one may need to require local GUTs.

In the following, we will investigate the 223 MSSM candidates further. Each model allows for a vast number of different vacuum configurations, i.e. different choices for the fields that develop vevs. Most of these choices will break the SM gauge group and are therefore clearly not relevant. Other choices that keep $\SU{3}\times\SU{2}\times\U{1}_Y$ might only differ in the structure of the Yukawa couplings for quarks and leptons. However, it is neither possible nor desirable to analyze all vacuum configurations for a given model. An analysis of some general properties seems more promising.

\subsection{Properties of the MSSM Candidates}
\label{sec:PropMSSM}

\subsubsection{Heavy Top and Decoupling of the Exotics}

The next task is to find out whether the exotics of the 223 MSSM candidates can be made heavy according to the string selection rules for the superpotential. As the computation of the superpotential terms relevant for the decoupling of the exotics is a very time consuming issue, an intermediate step (``heavy top'') was included in the model search in~\cite{Lebedev:2006kn}\footnote{However, here we will follow both methods, with and without heavy top. The first method will be denoted by an (a), the second one by a (b).}. In {\bf step (7a)}, we require a renormalizable $\mathcal{O}(1)$ Yukawa coupling $(\rep{3}, \rep{2})_{1/6}\,  (\crep{3}, \rep{1})_{-2/3} \, (\rep{1}, \rep{2})_{1/2}$, i.e. one of the following types
\begin{equation}
 U\,U\,U\;,\quad U\,T\,T\;,\quad T\,T\,T \;,
\end{equation}
where $U$ and $T$ denote generic untwisted and twisted fields, respectively. The $U\,U\,U$ coupling is given by the gauge coupling,  $U\,T\,T$ is a local coupling and thus is unsuppressed, while the $T\,T\,T$ coupling is significant only when the twisted fields are localized at the same fixed point. Models in which the above couplings are absent or suppressed are analyzed in {\bf step (7b)}.

Using the string selection rules of section~\ref{sec:Yukawa}, all terms entering the superpotential which are relevant for the masses of the exotics are computed up to order 8 in the fields. In detail, such a term looks like
\begin{equation}
\bar{X} X s \ldots s\;,
\end{equation}
with at most six (in general different) singlets $s \in \{s_i\}$ and $\bar{X}$ transforms in the complex conjugate representation of $X$. Note that the singlets $s$ are only demanded to be singlets with respect to the Standard Model gauge group, i.e. they transform as $(\rep{1},\rep{1})_0$ but are charged under the extra $\U{1}$'s and possibly also under some hidden non-Abelian gauge factor. Consequently, the vevs of these fields induce a gauge symmetry breaking. For many models, all extra $\U{1}$ factors are broken, but some hidden sector non-Abelian gauge group factors remain unbroken. This results in most cases in a separation of hidden and observable sectors, i.e. particles from the hidden and the observable sector are not simultaneously charged with respect to any gauge group factor. This situation yields the possibility of SUSY breaking by hidden sector gaugino condensation as discussed in section~\ref{sec:lowsusy}. Furthermore, $\bar{X}$ and $X$ are complex conjugate only with respect to $\SU{3}\times\SU{2}\times\U{1}_Y$.

Having the relevant terms of the superpotential, we assume that all singlets $s_i$ develop vevs. Only those models are selected at {\bf step 8} where all mass matrices of all exotics have maximal rank such that all exotics are heavy and decouple from the respective low-energy effective theory.

\subsubsection{D--flatness}

Next, at {\bf step 9} of the search strategy it is checked whether for a given model the exotics decouple along D--flat directions in order to be in agreement with $\mathcal{N} = 1$ supersymmetry. In fact, this requires additionally F--flatness. However, F--flatness is more involved, since the whole superpotential including the coupling strengths is needed and not only the terms relevant for the masses of the exotics. Therefore, F-flatness will only be explored for certain examples, see section~\ref{sec:RParity}, and we restrict to D--flatness here.

D--flatness\index{D--flatness} is ensured by specifying a gauge invariant monomial\footnote{In general, one has to specify an analytic gauge invariant polynomial $I(z)$ such that $\partial I(z)/\partial z^a |_{z = \xi} = C \bar{\xi}_a$ defines a $D_a=0$ solution, where $C$ is a complex constant $C\neq 0$ and $\xi_a$ denotes the vev of the field $z_a$~\cite{Buccella:1982nx}.} of the fields that shall develop vevs~\cite{Buccella:1981ib,Buccella:1982nx}. For example, consider some fields $s_i$ and assume that the following monomial $I(s_i)$ is gauge invariant
\begin{equation}
\label{eqn:DExample}
I(s_i) = (s_1)^3\, (s_2)^1\, (s_3)^2\, (s_4)^2\;.
\end{equation}
A field involved in this monomial attains a vev which is related to the power to which it appears in $I$. Explicitly, for eqn.~(\ref{eqn:DExample}),
\begin{equation}
\frac{|\langle s_1\rangle|}{\sqrt{3}} = \frac{|\langle s_2\rangle|}{\sqrt{1}} = \frac{|\langle s_3\rangle|}{\sqrt{2}} = \frac{|\langle s_4\rangle|}{\sqrt{2}}\;,
\end{equation}
and the vevs of all other fields $s_5, s_6, \ldots$ vanish. 

The situation changes slightly in the presence of anomalous $\U{1}$'s, which are in fact contained in all $223$ MSSM candidates of the Mini-Landscape. The $D$--term of such an anomalous $\U{1}$ gauge symmetry includes the Fayet--Iliopoulos term\index{Fayet--Iliopoulos D-term}~\cite{Dine:1987xk,Atick:1987gy}, a constant which is proportional to the anomaly $\text{Tr}\, Q_\text{anom}$, see section~\ref{sec:anomalousU1}. Then, the monomial $I(s_i)$ has to be gauge invariant with respect to all gauge factors except for the anomalous $\U{1}$. Furthermore, the total charge $\sum_i Q_{\text{anom},i}$ of the monomial has to cancel the Fayet--Iliopoulos term, i.e.
\begin{equation}
\label{eqn:AnomU1}
D_\text{anom} \simeq \frac{g}{192\pi^2}\text{Tr}\, Q_\text{anom} + g \sum_i Q_{\text{anom},i} |\langle s_i\rangle|^2 \stackrel{!}{=}0\;,
\end{equation}
From this we infer that the vevs of at least some fields $s_i$ contributing to the cancellation of the Fayet--Iliopoulos term are required to be at a high scale. Further details on $D=0$ can be found in appendix (B) of~\cite{Lebedev:2007hv}.

\subsection{The Results - Part II}

Following the steps 7 to 9 for both cases, with and without heavy top, we obtain results as summarized in table~(\ref{tab:Summary2Z6IIML})~\cite{Lebedev:2006kn,Lebedev:2007hv}. For most of the models the exotics decouple at order 8 (or less). It is likely that if we go to higher orders in the superpotential, the exotics of more MSSM candidates (maybe even of all) will decouple. In addition, it turns out that $D=0$ is not a severe constraint. That is, for nearly all models, we find gauge invariant monomials that contain almost all Standard Model singlets such that they can develop vevs. 
\begin{table}[!ht]
\centerline{
\begin{tabular}{l||l|l||l|l}
 criterion & $V^{\SO{10},1}$ & $V^{\SO{10},2}$ & $V^{\E6,1}$ & $V^{\E6,2}$\\
\hline
\hline
 7a. heavy top                    & $72$ & $37$ & $3$ & $2$ \\
 8a. exotics decouple at order 8  & $56$ & $32$ & $3$ & $2$ \\
 9a. D--flat direction            & $55$ & $32$ &     &     \\
\hline
\hline
 7b. no heavy top                 & $56$ & $53$ &     &     \\
 8b. exotics decouple at order 8  & $50$ & $53$ &     &     \\
 9b. D--flat direction            & $50$ & $53$ &     &     \\
\hline
\end{tabular}
}
\caption{Further analysis of the $223$ MSSM candidates, either along method (a) or (b), as discussed in the text. For these steps we concentrated on the $\SO{10}$ shifts.}
\label{tab:Summary2Z6IIML}
\end{table}

\section{Low-Energy SUSY Breaking}
\label{sec:lowsusy}

As briefly mentioned in the last section, the MSSM candidates have the necessary ingredients for supersymmetry breaking via gaugino condensation\index{gaugino condensation} in the hidden sector~\cite{Nilles:1982ik,Ferrara:1982qs,Derendinger:1985kk,Dine:1985rz}, published in~\cite{Lebedev:2006tr}. In particular, the models contain non-Abelian gauge group factors beside the Standard Model with little or no matter. The corresponding gauge interactions become strong at some intermediate scale such that the corresponding gauginos condensate $\langle \lambda \lambda \rangle$. This can lead to spontaneous supersymmetry breakdown in the hidden sector, communicated to the observable sector by gravity~\cite{Nilles:1982ik}. The specifics of the SUSY breaking depend on the moduli stabilization mechanism, but the main features such as the scale of supersymmetry breaking hold more generally. In particular, the gravitino mass\index{gravitino} is related to the gaugino condensation scale $\Lambda \equiv \langle \lambda \lambda \rangle^{1/3} $ by
\begin{equation}
m_{3/2} \sim {\Lambda^3 \over M_\mathrm{Pl}^2} \;,
\end{equation}
while the proportionality constant is model--dependent. The gaugino condensation scale in turn is given by the renormalization group (RG) invariant scale of the condensing gauge group,
\begin{equation}
\label{eqn:Lambda}
\Lambda ~\sim~  M_\mathrm{GUT}\,\exp \left( -\frac{1}{2\beta}\,\frac{1}{g^2(M_\mathrm{GUT})} \right) \;,
\end{equation}
where $\beta$ is the beta--function and $g$ is the gauge coupling constant related to the dilaton $S$ by $1/g^2 =\re S$. This translates into a superpotential for the dilaton, $W\sim \exp(-3 S /2\beta )$, which suffers from the notorious ``run--away'' problem, i.e.\ the vacuum of this system is at $S\rightarrow \infty$. For a discussion on the stabilization of the dilaton (at the realistic value $\re S \simeq 2$) and of the $T$--modulus see~\cite{Binetruy:1996xj,Casas:1996zi,Font:1990nt,Nilles:1990jv} and~\cite{Lebedev:2006tr}. Finally, for a gaugino condensation scale of about $\Lambda\sim 10^{13}\text{GeV}$ the gravitino mass is in the TeV region which is favored by phenomenology.

Obviously, in order to retain a hidden sector non-Abelian gauge group we have to modify step 8 of the search strategy: only fields neutral under both, the Standard Model \emph{and} the hidden sector non-Abelian gauge group, are allowed to acquire vevs. This yields in general less fields with vev than before. Consequently, the number of models with decoupled vector--like exotics is reduced, see table~(\ref{tab:SummarySUSY}).

\begin{table}[b]
\centerline{
\begin{tabular}{l||l|l||l|l}
 criterion & $V^{\SO{10},1}$ & $V^{\SO{10},2}$ & $V^{\E6,1}$ & $V^{\E6,2}$\\
\hline
\hline
 2. inequivalent models with 2 Wilson lines                         &$22,000$ & $7,800$ & $680$ &$1,700$ \\[0.1cm]
 3. SM gauge group $\subset$ SU(5) $\subset$ SO(10)  ({\rm or}~\E6) & $3,563$ & $1,163$ &  $27$ &   $63$ \\[0.1cm]
 4. 3 net $(\rep{3},\rep{2})$                                       & $1,170$ &   $492$ &   $3$ &   $32$ \\[0.1cm]
 5. non--anomalous $\U1_{Y}\subset \SU5 $                           &   $528$ &   $234$ &   $3$ &   $22$ \\[0.1cm]
 6. spectrum $=$ 3 generations $+$ vector-like                      &   $128$ &    $90$ &   $3$ &    $2$ \\[0.1cm]
 7. heavy top                                                       &    $72$ &    $37$ &   $3$ &    $2$ \\[0.1cm]
 8. exotics decouple $+$ gaugino condensation                       &    $47$ &    $25$ &   $3$ &    $2$ \\
\hline
\end{tabular}
}\vspace{-0.2cm}
\caption{Statistics of \Z6-II orbifolds based on the shifts $V^{\SO{10},1},V^{\SO{10},2},V^{\E6,1},V^{\E6,2}$ with two Wilson lines for the study of gaugino condensation.}
\label{tab:SummarySUSY}
\end{table}

\begin{figure}[t]
\centerline{\includegraphics{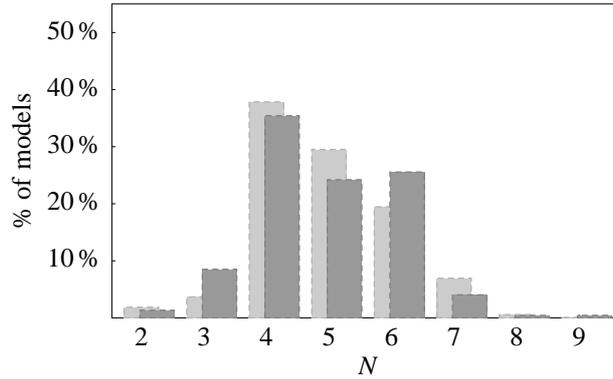}}
\vspace{-0.4cm}
\caption{Number of models vs.\ the size of the largest gauge group in the hidden sector. $N$ labels  $\SU{N}$, $\SO{2N}$, $\E{N}$ groups. The background corresponds to step 2, while the foreground corresponds to step 6.}
\label{fig:histogram1}
\end{figure}

\begin{figure}[b]
\centerline{\includegraphics{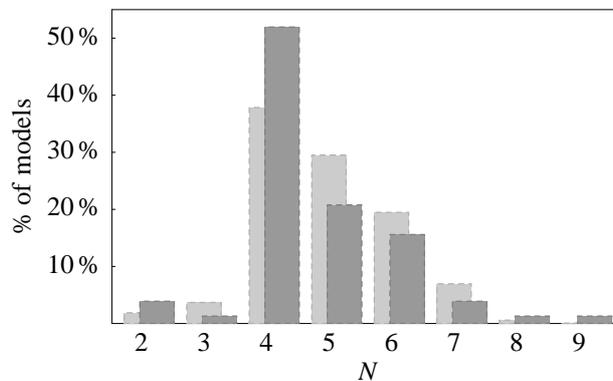}}
\vspace{-0.4cm}
\caption{As in figure~(\ref{fig:histogram1}) but with models of step 8 in the foreground.}
\label{fig:histogram2}
\end{figure}

As the beta--function depends on the hidden sector gauge group (and its light matter representations), we start with a discussion on the various hidden sectors that appear in the MSSM candidates. Figure~(\ref{fig:histogram1}) displays the frequency of occurrence of various gauge groups in the hidden sector (see~\cite{Dienes:2006ut} for a related study). The preferred size ($N$) of the hidden sector gauge groups depends on the conditions imposed on the spectrum. When all inequivalent models with 2 Wilson lines are considered, $N=4,5,6$ appear with similar likelihood and $N=4$ is somewhat preferred. If we require the massless spectrum to be the MSSM + vector--like matter, the fractions of models with $N=4,5,6$ become even closer. However, if we further require a heavy top quark and the decoupling of exotics at order 8, $N=4$ is clearly preferred (see figure~(\ref{fig:histogram2})). In this case, $\SU{4}$ and $\SO{8}$ groups provide the dominant contribution. Since all or almost all matter charged under these groups is decoupled, this leads to hidden sector gaugino condensation at an intermediate scale. (We note that before step 8, gaugino condensation does not occur in many cases due to the presence of hidden sector matter.) 

\begin{figure}[!ht!]
\centerline{\includegraphics{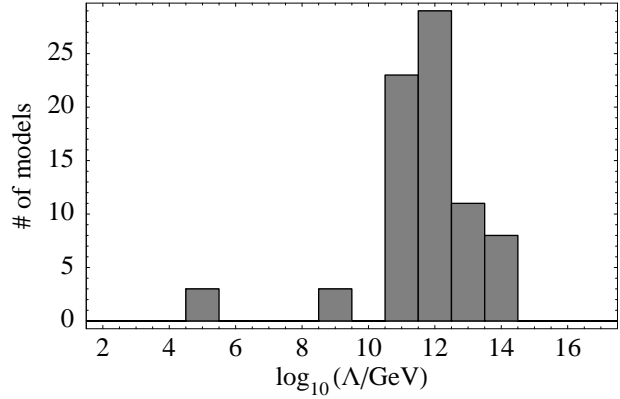}}
\vspace{-0.4cm}
\caption{Number of models vs.\ scale of gaugino condensation~\cite{Lebedev:2006tr}.}
\label{fig:gcresult}
\end{figure}

Possible scales of gaugino condensation are shown in figure~(\ref{fig:gcresult}). These are obtained from eqn.~(\ref{eqn:Lambda}) by computing the beta--functions for each case and using  $g^2(M_\mathrm{GUT}) \simeq 1/2$. The correlation between the observable and hidden sectors is a result of the fact that modular invariance constrains the gauge shifts and Wilson lines in the two sectors. 

We see that among the promising models, intermediate scale supersymmetry breaking is preferred. The underlying reason is that realistic spectra require complicated Wilson lines, which break the hidden sector gauge group. The surviving gauge factors are not too big (unlike in Calabi--Yau compactifications with the standard embedding), nor too small.

There are significant uncertainties in the estimation of the supersymmetry breaking scale. First, the identification of $\langle \lambda \lambda \rangle^{1/3} $ with the RG invariant scale  is not precise. A factor of a few uncertainty in this relation leads to 2 orders of magnitude uncertainty in $m_{3/2}$. Also, there could be significant string threshold corrections which can affect the estimate. Thus, the resulting ``prediction'' for the superpartner masses should be understood within 2-3 orders of magnitude.

\section{R-parity}\index{MSSM!benchmark model}
\label{sec:RParity}

As a detailed analysis of all 223 MSSM candidates is very time-consuming, we will only choose one of them in this section and analyze its phenomenological implications in detail. We will call it the \emph{benchmark} model, as some of its properties are likely to be generic for other models of the Mini-Landscape, too. We construct a supersymmetric configuration $F=D=\mathcal{W}=0$ for the benchmark model which additionally yields a possible solution to the $\mu$ problem for free. Furthermore, in order to avoid rapid proton decay we identify a $\U{1}_{B-L}$ symmetry. By breaking $\U{1}_{B-L}$ by the vevs of fields with even $B-L$ charges, we obtain a discrete symmetry, $R$--parity (or matter-parity), that forbids dangerous dimension-4 baryon/ lepton number violating operators. In addition, the see-saw mechanism for light neutrino masses and the structure of the Yukawa couplings for quarks and leptons are analyzed. The results presented here have been published in reference~\cite{Lebedev:2007hv}\footnote{The benchmark model discussed here is named ``benchmark model 1B'' in~\cite{Lebedev:2007hv}}.

\subsubsection{Technical Details of the Benchmark Model}

The benchmark model is defined by the shifts and Wilson lines
\begin{subequations}
\begin{eqnarray}
V   &=& \left( \tfrac{1}{3},-\tfrac{1}{2},-\tfrac{1}{2},0^5\right)\,\left(\tfrac{1}{2},-\tfrac{1}{6},-\tfrac{1}{2},-\tfrac{1}{2},-\tfrac{1}{2},-\tfrac{1}{2},-\tfrac{1}{2},\tfrac{1}{2}\right)\;, \\
A_5 &=& \left( 0,-\tfrac{1}{2},-\tfrac{1}{2},-\tfrac{1}{2},\tfrac{1}{2},0,0,0\right)\,\left(4,-3,-\tfrac{7}{2},-4,-3,-\tfrac{7}{2},-\tfrac{9}{2},\tfrac{7}{2}\right)\;, \\
A_3 &=& \left(-\tfrac{1}{2},-\tfrac{1}{2},\tfrac{1}{6},\tfrac{1}{6},\tfrac{1}{6},\tfrac{1}{6},\tfrac{1}{6},\tfrac{1}{6}\right)\,\left( \tfrac{1}{3},0,0,\tfrac{2}{3},0,\tfrac{5}{3},-2,0\right)\;.
\end{eqnarray}
\end{subequations}
A possible second order 2 Wilson line is set to zero ($A_6 = 0$). The gauge group after compactification is $\SU{3}\times\SU{2}\times[\SU{4}\times\SU{2}]\times\U{1}^9$, where the nine $\U{1}$ generators can be chosen as
\begin{equation}
t_1 ~=~ t_Y ~=~ \left(0,0,0,-\tfrac{1}{2},-\tfrac{1}{2},\tfrac{1}{3},\tfrac{1}{3},\tfrac{1}{3}, 0^8\right)
\end{equation}
and

\begin{equation}
\begin{array}{lll}
t_2 = (-1,0^{15}),      & t_3 = (0,-1,0^{14}), & t_4 = (0,0,-1,0^{13}), \\
t_5 = (0^3,(-1)^5,0^8), & t_6 = (0^9,-1,0^6),  & t_7 = (0^8,1,0,0,-1,0^4),\\
t_8 = (0^{12},-1,0^3),  & t_9 = (0^{13},-1,0,0) \;.&
\end{array}
\end{equation}
Therefore, the benchmark model contains the gauge group $\SU{3}\times\SU{2}\times\U{1}_Y$ of the Standard Model. The generator of the anomalous $\U{1}$ reads
\begin{equation}
t_\text{anom}~=~\sum\limits_{i=1}^9\alpha_i\,\mathsf{t}_i\;,\quad\text{where}\:\{\alpha_i\}~=~\left\{0,\tfrac{2}{3},0,-\tfrac{5}{3},\tfrac{1}{3},-\tfrac{1}{3},\tfrac{1}{3},2,\tfrac{1}{3}\right\}\;.
\end{equation}
Since $\alpha_1 = 0$, hypercharge is non-anomalous. The details of the spectrum are given in table~\ref{tab:benchmark1} in the appendix.

\subsubsection{Supersymmetric Minkowski Vacuum}\index{Minkowski vacuum}

Now consider the vacuum configuration where the fields 

\begin{eqnarray}
\label{eqn:vacuumconf1b}
\{\widetilde{s}_i\} = & \{\chi_{1}, \chi_{2}, \chi_{3}, \chi_{4},
h_{1}, h_{2}, h_{3}, h_{4}, h_{5}, h_{6}, h_{7}, h_{8}, h_{9},
h_{10}, s^0_{1},s^0_{2},s^0_{3},
s^0_{4}, &\nonumber\\
& \quad s^0_{5}, s^0_{6},s^0_{7},s^0_{8}, s^0_{9},s^0_{10},
s^0_{11},s^0_{12}, s^0_{13},s^0_{14},
s^0_{15},s^0_{16}, s^0_{17}, s^0_{18}, s^0_{20}, & \\
 & \quad s^0_{21}, s^0_{22}, s^0_{23},s^0_{24}, s^0_{25}, s^0_{26}, s^0_{27},s^0_{28},s^0_{29},
s^0_{30}, s^0_{31},s^0_{32} \} & \nonumber
\end{eqnarray}
develop a vev while the expectation values of all other fields vanish. By an appropriate choice of the vevs this configuration yields $D=0$. The corresponding gauge invariant monomial is given in~\cite{WebTables:2006ml}.

In order to get a supersymmetric Minkowski vacuum we have to demand $F=\mathcal{W}=0$ in addition. An interesting feature of this model is that the superpotential of the SM singlets $\chi$, $h$ and $s^0$ factorizes into polynomials of $D_4$ doublets and $D_4$ singlets\footnote{The $D_4$ family symmetry is a consequence of the space group selection rule in the $\SU{2}^2$ torus lattice and the trivial Wilson line $A_6 = 0$~\cite{Kobayashi:2004ya,Kobayashi:2006wq}. States sitting at the two vertical fixed points on the $\SU{2}^2$ torus transform as doublets under $D_4$, compare to figure~\ref{fig:Z6IIOrbifold2FP}.}, i.e. 
\begin{equation}
\mathcal{W}(\chi,h,s^0) = \sum_i  P_i(\widetilde{D}) \ \tilde P_i(S) \;.
\end{equation}
$P_i(\widetilde{D})$ denotes polynomials in SM singlet fields which transform as $D_4$ doublets $\widetilde{D}$ and $\tilde P_i(S)$ denotes polynomials in SM $\times D_4$ singlets $S$. In particular, the $D_4$ doublets which enter $\mathcal{W}(\widetilde{s})$ up to order six in fields are
\begin{equation}
\begin{array}{lll}
\widetilde{D}_1 = (s^0_3, s^0_9)    & \widetilde{D}_2 = (s^0_4, s^0_{10}) & \widetilde{D}_3 = (s^0_5, s^0_{11}) \\
\widetilde{D}_4 = (s^0_6, s^0_{12}) & \widetilde{D}_5 = (s^0_7, s^0_{13}) & \widetilde{D}_6 = (s^0_8, s^0_{14})\;.
\end{array}
\end{equation}
The polynomial in $D_4$ doublets is, to order six, quadratic in doublets and is given by the $D_4$ invariant scalar product, for example,
\begin{equation}
\widetilde{D}_1 \cdot \widetilde{D}_2 = (s^0_3 \ s^0_4 + s^0_9 \ s^0_{10}) \ .
\end{equation}
We then find (up to calculable dimensionful coefficients in units of the string scale)
\begin{eqnarray}
\mathcal{W}(\chi,h,s^0) &=& \left(\widetilde{D}_1 \cdot \widetilde{D}_2\right)\ \left(s^0_{26} + s^0_{29} + (s^0_{26} s^0_{26} + s^0_{26} s^0_{29} + s^0_{29} s^0_{29}) (s^0_{15} + s^0_{16})\right)\\
&& {} + \left(\widetilde{D}_1 \cdot \widetilde{D}_6 + \widetilde{D}_2\cdot \widetilde{D}_5\right)\ s^0_{30}\ \biggl[ s^0_{30} (s^0_{15} + s^0_{16}) + s^0_{17} (s^0_{25} + s^0_{28}) \nonumber\\
&& {} + s^0_{18} (s^0_{24} + s^0_{27})+ s^0_{31} (s^0_{20} + s^0_{21}) + s^0_{32} (s^0_{22} + s^0_{23})  \nonumber \\
&& {} + (s^0_{19} + s^0_1 s^0_{18}+ s^0_2 s^0_{17}) (s^0_{26} + s^0_{29})+ h_1 (h_8 + h_{10}) + h_2 (h_7 + h_9)\biggr] \nonumber \\ 
&& {} + \left(\widetilde{D}_3 \cdot \widetilde{D}_4\right)\ s^0_{19} s^0_{30}\ \left( s^0_{17} s^0_{18} + h_1 h_2 + (s^0_{20} + s^0_{21}) (s^0_{22} + s^0_{23})\right)\;. \nonumber
\label{eq:W}
\end{eqnarray}
Thus, to order 6 in SM and hidden $\SU4$ singlets, the polynomials $P_i(\widetilde{D})$ are completely determined by the $D_4$ symmetry, while the polynomials $\tilde P_i(S)$ are non-trivial for all $i$. One particular $F= D =0$ solution is given by the roots of
\begin{equation}
\langle P_i(\widetilde{D}) \rangle = \langle \tilde P_i(S) \rangle = 0
\end{equation}
for all polynomials $i$ (compare to e.g.~\cite{Buchmuller:2006ik,Lebedev:2007hv}).  Hence, the superpotential up to order six in SM singlets vanishes
\begin{equation}
 \left\langle  \mathcal{W}(\chi,h,s^0) \right\rangle~=~0 \;.
\end{equation}
Therefore, the total superpotential is given solely by its non--perturbative part. This is expected to be very small and thus a small gravitino mass and a small cosmological constant can in principle be achieved\footnote{Further analysis of this model has revealed the existence of an approximate $R$-symmetry responsible for a hierarchically small superpotential at the minimum~\cite{Kappl:2008ie}.}.

\subsubsection{Identifying a $\boldsymbol{\U{1}_{B-L}}$ Symmetry}

In the MSSM, lepton doublets $\ell$ and Higgs doublets $\phi$ carry the same charges with respect to the SM gauge group. Consequently the superpotential contains the operators 
\begin{equation}
\label{eqn:dim4}
\ell \ell \bar{e}\;, q \ell \bar{d}\;\text{ and in addition }\bar{u}\bar{d}\bar{d}
\end{equation}
at the renormalizable level. These dimension 4 operators lead in general to rapid proton decay\index{proton decay}. Therefore, they have to be suppressed. One solution to this problem is $\U{1}_{B-L}$\index{B-L}, as it distinguishes between lepton doublets $\ell$ and Higgs doublets $\phi$. One can easily see that a $\U{1}_{B-L}$ symmetry forbids these dangerous operators.

It is possible to identify a non-anomalous $B-L$ generator for the benchmark model. It reads
\begin{equation}
t_{B-L} ~=~ \left(0 , 0 , 0 , 0 , 0 , -\frac{2}{3} ,-\frac{2}{3} , -\frac{2}{3}\right)\, \left( 0 , 0 , 0 , 0 , 0 , 2 ,0 , 0\right)\;,
\end{equation}
where the first half of $t_{B-L}$ which acts on the first $\E{8}$ has the standard $\SO{10}$ GUT form, but the second part is different. Nevertheless, we have chosen it due to the following two essential properties:
\begin{itemize}
\item the spectrum includes 3 generations of quarks and leptons plus vector-like exotics with respect to $G_\mathrm{SM}\times\U1_{B-L}\;,$ and
\item there are SM singlets with $B-L$ charge $\pm 2$, labeled by $\chi$. Therefor it is important that $t_{B-L}$ acts non-trivially in the second $\E{8}$.
\end{itemize}
Using this definition for $\U{1}_{B-L}$ it turns out that there is only one pair of Higgs candidates $\phi_1$ and $\bar\phi_1$ and $4-1$ lepton doublets. The spectrum of charged matter is summarized in table~(\ref{tab:naming3}). Comparing to eqn.~(\ref{eqn:vacuumconf1b}) we see that in the current vacuum configuration only one SM singlet vev is set to zero, i.e.  $\langle s^0_{19} \rangle = 0$. 

\begin{table}[t]
\centerline{
\begin{tabular}{|c|l|l|c|c|l|l|}
\hline
\# & representation & label & & \# & representation & label\\
\hline
\hline
 3 & $\left( \rep{3},\rep{2};\rep{1},\rep{1}\right)_{(1/6,1/3)}$   & $q_i$ & &
 3 & $\left(\crep{3},\rep{1};\rep{1},\rep{1}\right)_{(-2/3,-1/3)}$ & $\bar u_i$ \\
 3 & $\left(\rep{1},\rep{1};\rep{1},\rep{1}\right)_{(1,1)}$        & $\bar e_i$ & &
 8 & $\left(\rep{1},\rep{2};\rep{1},\rep{1}\right)_{(0,*)}$        & $m_i$ \\
 4 & $\left(\crep{3},\rep{1};\rep{1},\rep{1}\right)_{(1/3,-1/3)}$  & $\bar d_i$ & &
 1 & $\left(\rep{3},\rep{1};\rep{1},\rep{1}\right)_{(-1/3,1/3)}$   & $d_i$ \\
 4 & $\left(\rep{1},\rep{2};\rep{1},\rep{1}\right)_{(-1/2,-1)}$    & $\ell_i$ & &
 1 & $\left(\rep{1},\rep{2};\rep{1},\rep{1}\right)_{(1/2,1)}$      & $\bar \ell_i$ \\
 1 & $\left(\rep{1},\rep{2};\rep{1},\rep{1}\right)_{(-1/2,0)}$     & $\phi_i$ & &
 1 & $\left(\rep{1},\rep{2};\rep{1},\rep{1}\right)_{(1/2,0)}$      & $\bar \phi_i$ \\
 6 & $\left(\crep{3},\rep{1};\rep{1},\rep{1}\right)_{(1/3,2/3)}$   & $\bar\delta_i$ & &
 6 & $\left(\rep{3},\rep{1};\rep{1},\rep{1}\right)_{(-1/3,-2/3)}$  & $\delta_i$ \\
14 & $\left(\rep{1},\rep{1};\rep{1},\rep{1}\right)_{(1/2,*)}$      & $s^+_i$ & &
14 & $\left(\rep{1},\rep{1};\rep{1},\rep{1}\right)_{(-1/2,*)}$     & $s^-_i$ \\
16 & $\left(\rep{1},\rep{1};\rep{1},\rep{1}\right)_{(0,1)}$        & $\bar n_i$ & &
13 & $\left(\rep{1},\rep{1};\rep{1},\rep{1}\right)_{(0,-1)}$       & $n_i$ \\
 5 & $\left(\rep{1},\rep{1};\rep{1},\rep{2}\right)_{(0,1)}$        & $\bar \eta_i$ & &
 5 & $\left(\rep{1},\rep{1};\rep{1},\rep{2}\right)_{(0,-1)}$       & $\eta_i$ \\
10 & $\left(\rep{1},\rep{1};\rep{1},\rep{2}\right)_{(0,0)}$        & $h_i$ & &
 2 & $\left(\rep{1},\rep{2};\rep{1},\rep{2}\right)_{(0,0)}$        & $y_i$ \\
 6 & $\left(\rep{1},\rep{1};\rep{4},\rep{1}\right)_{(0,*)}$        & $f_i$ & &
 6 & $\left(\rep{1},\rep{1};\crep{4},\rep{1}\right)_{(0,*)}$       & $\bar f_i$ \\
 2 & $\left(\rep{1},\rep{1};\rep{4},\rep{1}\right)_{(-1/2,-1)}$    & $f_i^-$ & &
 2 & $\left(\rep{1},\rep{1};\crep{4},\rep{1}\right)_{(1/2,1)}$     & $\bar f_i^+$ \\
 4 & $\left(\rep{1},\rep{1};\rep{1},\rep{1}\right)_{(0,\pm2)}$     & $\chi_i$ & &
32 & $\left(\rep{1},\rep{1};\rep{1},\rep{1}\right)_{(0,0)}$        & $s^0_i$ \\
 2 & $\left(\crep{3},\rep{1};\rep{1},\rep{1}\right)_{(-1/6,2/3)}$  & $\bar v_i$ & &
 2 & $\left(\rep{3},\rep{1};\rep{1},\rep{1}\right)_{(1/6,-2/3)}$   & $v_i$ \\
\hline
\end{tabular}
}
\caption{Spectrum. The quantum numbers under $\SU{3}\times\SU{2}\times[\SU{4}\times\SU{2}]$ are shown in boldface; hypercharge and $B-L$ charge appear as subscripts.  Note that the states $s_i^\pm$, $f_i$, $\bar f_i$ and $m_i$ have different $B-L$ charges for different $i$, which we do not list explicitly.}
\label{tab:naming3}
\end{table}

\subsubsection{The $\boldsymbol{\mu}$-Term}\index{$\mu$-term}

In this model we also find an intriguing correlation between the $\mu$--term
\begin{equation}
\mu ~=~ \left.\frac{\partial^2 \mathcal{W}_\text{total}}{\partial\phi_1\,\partial\bar \phi_1} \right|_{\phi_1=\bar{\phi}_1=0}
\end{equation}
and $\mathcal{W}(\chi,h,s^0)$. We note that both Higgs doublets are untwisted states, the combination $\phi_1\,\bar{\phi}_1$ is a gauge singlet with respect to the full orbifold gauge group and in addition neutral with respect to \emph{all} string selection rules, e.g. $R_{\phi_1\bar{\phi}_1}=(0,0,0,-2)\sim(0,0,0,0)$. Consequently, each monomial of fields appearing in $\mathcal{W}(\chi,h,s^0)$ couples also to the Higgs pair $\phi_1\,\bar{\phi}_1$ giving rise to the $\mu$--term. This implies that\footnote{This applies to the untwisted Higgs pairs in many models of the Mini-Landscape, for instance also to the model presented in \cite{Buchmuller:2005jr,Buchmuller:2006ik}.}
\begin{equation}
\mu = 0 \quad\Leftrightarrow\quad \langle\mathcal{W}(\chi,h,s^0) \rangle=0\;.
\end{equation}
Consequently, the $\mu$-term is of the order of the expectation value of $\mathcal{W}$, i.e.\ the gravitino mass. This mechanism has been discussed before, see e.g.~\cite{Casas:1992mk}. However, here it was not put in by hand, but it is a result of the string construction.

Note that there is no doublet-triplet splitting problem\index{$\mu$-term!doublet-triplet splitting} for the benchmark model since the Higgs color-triplets have been removed by the orbifold projection.

\subsubsection{Breaking of $\boldsymbol{\U{1}_{B-L}}$ and Neutrino matrices}\index{R-symmetry!R-parity}\index{matter parity}

As mentioned before, there are SM singlets with $B-L$ charge $\pm 2$, labeled by $\chi$. If they acquire vevs, they break $\U{1}_{B-L}$ to $R$--parity (or matter parity). Thus, the dangerous baryon/ lepton number violating operators of eqn.~(\ref{eqn:dim4}) remain forbidden. Furthermore, we can have a see-saw mechanism\index{see-saw mechanism} for light neutrino masses, as we have Majorana and Dirac masses for the neutrinos,
\begin{equation}
\bar{n}\bar{n}\chi \;\text{ and }\; \ell\bar{\phi}\bar{n}\;.
\end{equation}
For more details, see the web page~\cite{WebTables:2006ml} and~\cite{Buchmuller:2007zd}.

\subsubsection{Charged fermion Yukawa matrices}

The charged fermion Yukawa matrices are
\begin{equation}
Y_u= \left(
\begin{array}{ccc}
 \widetilde{s}^5 & \widetilde{s}^5 & \widetilde{s}^5 \\
 \widetilde{s}^5 & \widetilde{s}^5 & \widetilde{s}^5 \\
 \widetilde{s}^6 & \widetilde{s}^6 & 1
\end{array}
\right)\;,\quad Y_d~=~ \left(
\begin{array}{ccc}
 \widetilde{s}^5 & \widetilde{s}^5 & 0  \\
 \widetilde{s}^5 & \widetilde{s}^5 & 0  \\
 \widetilde{s}^6 & \widetilde{s}^6 & 0
\end{array}
\right)\;,\quad Y_e~=~ \left(
\begin{array}{ccc}
 \widetilde{s}^5 & \widetilde{s}^5 & \widetilde{s}^6 \\
 \widetilde{s}^5 & \widetilde{s}^5 & \widetilde{s}^6 \\
 \widetilde{s}^6 & \widetilde{s}^6 & 0
\end{array}
\right)\;.
\end{equation}
The up-type quark Yukawa matrix is given directly in terms of the coupling of the up-type Higgs to the three $q$ and $\bar u$ fields. The down-type quark and charged lepton Yukawa matrices are obtained by integrating out a pair of vector-like $d$- and $\bar d$-quarks and $\ell$- and $\bar\ell$-fields, respectively. We find that the up quark and the charged lepton Yukawa matrices have rank three, while the down quark Yukawa matrix has only rank two, at this order in $\tilde s$ singlets. In fact, to this order in SM singlet fields, the superpotential does not couple two right-handed down quarks, $\bar d_{3,4}$, to the quark doublets. This is because $\bar d_{3,4}$ are in the $T_4$ twisted sector.  However, we have verified that some of the zeros in $Y_d$ get filled in at higher orders and at order 8 $Y_d$ has rank 3. Note that in this vacuum configuration the Yukawa matrices retain a form consistent with the underlying $D_4$ family symmetry.

\subsubsection{Dimension 5 baryon and lepton number violating operators}

We further analyzed the question of dimension 5 proton decay\index{proton decay} operators. We find that both
\begin{equation}
q\,q\,q\,\ell \;\text{ and }\; \bar{u}\,\bar{u}\,\bar{d}\,\bar{e}
\end{equation}
appear at order $\widetilde{s}^6$. They are also generated by integrating out the heavy exotics.  For example, the following couplings exist
\begin{equation}
  q_1 \,\ell_1\,\bar\delta_4\,,\; q_1 \,\ell_1\,\bar\delta_5\,,\; q_2 \,\ell_2\,\bar\delta_4\,,\;q_2 \,\ell_2\,\bar\delta_5\,,\; q_1 \,q_1\,\delta_4\,,\;  q_1 \,q_1\,\delta_5\,,\;q_2 \,q_2\,\delta_4\,,\;q_2 \,q_2\,\delta_5\;.
\label{eq:trilinear_D5_operators}
\end{equation}
Hence integrating out the states $\bar\delta_i, \ \delta_i$ produces dangerous dimension 5 operators. These must be sufficiently suppressed to be consistent with present bounds on proton decay~\cite{Hinchliffe:1992ad,Dermisek:2000hr}. We have verified that, for some particular $\widetilde s$ vevs, it is possible to suppress the $q\,q\,q\,\ell$ operators induced by the trilinear couplings of eqn.~(\ref{eq:trilinear_D5_operators}). However, higher order couplings also introduce baryon and lepton number violating operators. We have not been able to identify a suppression mechanism for such operators yet.

\chapter{Blow--up of Orbifold Singularities}
\label{sec:blowup}

From the construction of MSSM candidates in the framework of $\Z{6}$-II orbifolds as discussed in chapter~\ref{sec:ML}, we know that twisted fields have to attain vevs due to two main reasons. First of all, the Fayet--Iliopoulos $D$--term forces some fields to get vevs in order to maintain a supersymmetric solution. Secondly, from a phenomenological point of view we need to give large vevs to some Standard Model singlets in order to generate mass terms for the unwanted vector-like exotics. 

On the other hand, it is known from the early beginnings of orbifold construction that a twisted field attaining a vev can be interpreted as a so-called \emph{blow--up mode}~\cite{Hamidi:1986vh,Font:1988tp}\index{blow--up mode}. By turning on its vev the singularity of the corresponding fixed point is smoothed out, yielding the resolution space. In this chapter, we aim at a deeper understanding of this blow--up procedure. For doing so, we concentrate on the easiest orbifold singularities, the ones appearing in $\Z{3}$ orbifolds. Furthermore, we restrict to the $\E{8}\times\E{8}$ heterotic string as we want to connect the blow--up method with early $\Z{3}$ MSSM candidates\footnote{The $\SO{32}$ theory was considered in~\cite{Nibbelink:2007rd,GrootNibbelink:2007ew}.}. The general outline of this chapter reads: First, the local resolution space of a single singularity is constructed explicitly building on the results of~\cite{Nibbelink:2007rd}. Then, the transition between the string model on the singular space and the effective field theory on the resolution space is discussed in detail, focusing on the role of the blow--up mode. Afterwards, it is shown how the singularities of a compact $\Z{3}$ orbifold can be resolved, even in the presence of Wilson lines. Finally, the blow--up procedure is applied to the $\Z{3}$ MSSM candidate of reference~\cite{Ibanez:1987sn}. The results of this chapter have been published in~\cite{Nibbelink:2008tv}.

\section[Blow--up of Local $\mathbbm C^3/\Z{3}$ Orbifold]{Blow--Up of Local $\boldsymbol{\mathbbm C^3/\Z{3}}$ Orbifold} 
\label{sc:C3Z3res}

We consider the heterotic string quantized on the singular non-compact space $M_{3,1}\times \mathbbm C^3/\Z{3}$ and on its resolution denoted by $\mathcal{M}^3$. We start by giving the geometrical details of the $\mathbbm C^3/\Z{3}$ singularity. Then we show how to resolve it and how to construct gauge fluxes on the resolution. After this study of the geometry, we consider the heterotic string on the singular space and on the resolution, leading to 4d heterotic orbifold and resolved models, respectively. 

Finally, we briefly comment on the anomaly cancelation in both cases: in contrast to the orbifold model which allows for at most one anomalous $\U{1}$, the resolution model can have up to two. On the orbifold side, the standard Green--Schwarz mechanism, involving one single universal axion, is combined with a Higgs mechanism giving rise to the blow--up. On the resolution, this combination is mapped into a Green--Schwarz mechanism involving two axions. These axions are mixtures of the orbifold axion and of the blow--up mode. This identification is completed by the observation that the new Fayet--Iliopoulos term produced on the resolution is nothing else than the (tree--level) $D$--term due to the non--vanishing vev of the blow--up mode.

\begin{figure}[t!]
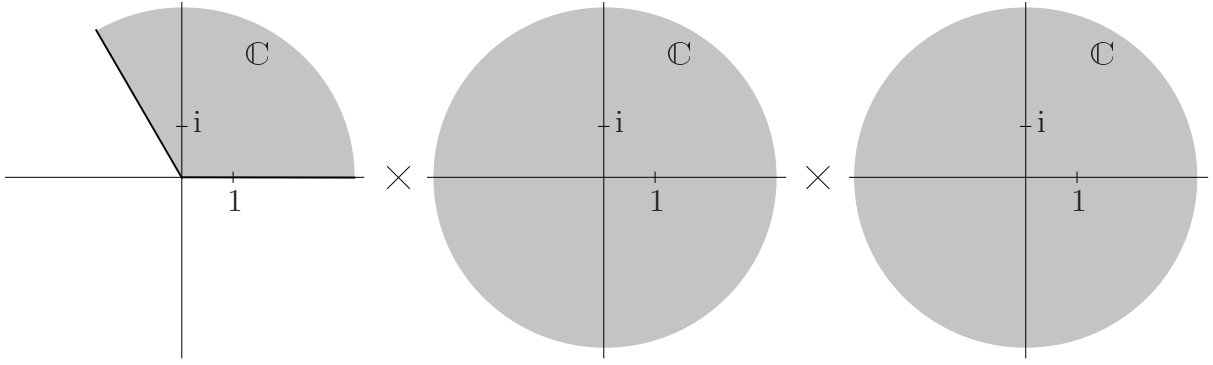

\centerline{\input C3OverZ3FundamentalD.pstex_t}\vspace{-0.2cm}
\caption{The three complex planes ${\mathbbm C}^3$ are modded out by the $\Z{3}$ twist $\theta$. The grey region depicts a fundamental domain of $\mathbbm C^3/\Z{3}$ and corresponds to the patch $U_{(1)}$. The black lines in the first complex plane illustrate that the angle is restricted to be $0<\mbox{arg}(Z^1)<2\pi/3$. In the case of $U_{(2)}$ and $U_{(3)}$ this (deficit) angle lies in the second and third plane, respectively.}
\label{fig:C3OverZ3}
\end{figure}

\subsection{Orbifold and Blow--up Geometry}\index{singularity}
\label{sec:blowupgeometry}

In order to describe a $\Z{3}$ singularity locally, we start from $\mathbbm C^3$ parameterized by the three complex coordinates $Z^i$ ($i = 1,2,3$). Then, the $\Z{3}$ orbifold rotation $\theta$ acts as 
\begin{equation}\label{eq:twist}
\theta:\,\, Z^i\longmapsto e^{2\pi \I v_i/3} Z^i,\quad v = (1,1,-2)\;.
\end{equation}
Note that in this chapter we define the twist vector without the factor $1/3$. The $\Z{3}$ singularity is locally described by the non--compact orbifold 
\begin{equation}
\mathbbm C^3/\Z{3}\;,
\end{equation}
which is obtained by identifying those points in $\mathbbm C^3$ that are mapped into each other by $\theta$. In other words, the equivalence relation $\theta Z \sim Z$ defines the quotient space $\mathbbm C^3/\Z{3}$. Following the holonomy arguments of section~\ref{sec:geometry}, such a space is singular in the fixed point at the origin $\{0\}$. The orbifold space $\mathbbm{C}^3/\Z{3}$-\{0\} is naturally equipped with a \emph{K\"ahler potential}\index{K\"ahler potential}, inherited from $\mathbbm{C}^3$, which reads
\begin{equation}
\label{eq:orbKahler}
\mathcal{K}_{\mathbbm{C}^3/\mathbbm Z_3} = \sum_i\bar{Z}^iZ^i\;,
\end{equation}
such that the metric
\begin{equation}
g_{i\bar{j}} = \frac{\partial}{\partial Z^i}\frac{\partial}{\partial \bar{Z}^j}\mathcal{K}_{\mathbbm{C}^3/\mathbbm Z_3} = \delta_{i\bar{j}}\;
\end{equation}
is Euclidean and the space is flat apart from the origin. We can cover $\mathbbm{C}^3/\Z{3}$-\{0\} by means of three coordinate patches, defined as
\begin{equation}
U_{(i)}\equiv \lbrace Z\in \mathbbm{C}^3|Z^i \neq 0\,,\, 0<\mbox{arg}(Z^i)<2\pi/3 \rbrace\;,\;\;\; i =1,2,3 \;.
\end{equation}
Each of this patches gives a representative of the fundamental domain of $\mathbbm C^3/\Z{3}$. As an example, the coordinate patch $U_{(1)}$ is depicted in figure~(\ref{fig:C3OverZ3}).

\subsubsection{Convenient Coordinates}

It is convenient to choose new coordinates on the orbifold $\mathbbm{C}^3/\Z{3}$, which allow for a systematic construction of a resolution of the singularity as a line bundle over $\mathbbm{CP}^{2}$. In the language of toric geometry~\cite{Lust:2006zh,Nibbelink:2007pn}, the $\mathbbm{CP}^2$ is called an exceptional divisor, and it replaces the singularity in the resolution $\mathcal{M}^3$ of $\mathbbm C^3/\Z{3}$. When its volume shrinks to zero, the singularity is recovered, and the space $\mathcal{M}^3$ approaches $\mathbbm C^3/\Z{3}$ (blow--down). Thus, the blowing--up/down procedure is controlled by the size of the exceptional divisor.  To make this more explicit we consider the patch $U_{(i)}$, where $Z^i\neq 0$, and define $z^j\equiv Z^j/Z^i$ for  $j\neq i$. To remove the deficit angle of $Z^i$ we perform the coordinate transformation $Z^i\mapsto x \equiv (Z^i)^3$, such that the new coordinates are $\theta$-invariant, e.g. $(x,z^2,z^3)$ in the patch $U_{(1)}$. In this way the K\"ahler potential, eqn.~(\ref{eq:orbKahler}), becomes 
\begin{equation}
\label{eq:orbKahler2}
\mathcal{K}_{\mathbbm{C}^3/\mathbbm Z_3} = X^{\frac{1}{3}}\;,
\end{equation}
where $X$ is defined (in the patch $U_{(i)}$) as
\begin{equation}
\label{eqn:DefX}
X \equiv \bar{x}(1 + \bar{z}z)^3 x\quad\Rightarrow\quad X = (\bar{Z}^i)^3 \Big(1 + \sum_{j\neq i}\frac{\bar{Z}^j}{\bar{Z}^i}\frac{Z^j}{Z^i}  \Big)^3  (Z^i)^3 = \Big(\sum_j \bar{Z}^j Z^j\Big)^3\;.
\end{equation}
Thus, we see that the K\"ahler potentials eqns.~(\ref{eq:orbKahler}) and~(\ref{eq:orbKahler2}) are equivalent.

\subsubsection{The Resolution Space}\index{singularity!resolution}

\begin{figure}[t]
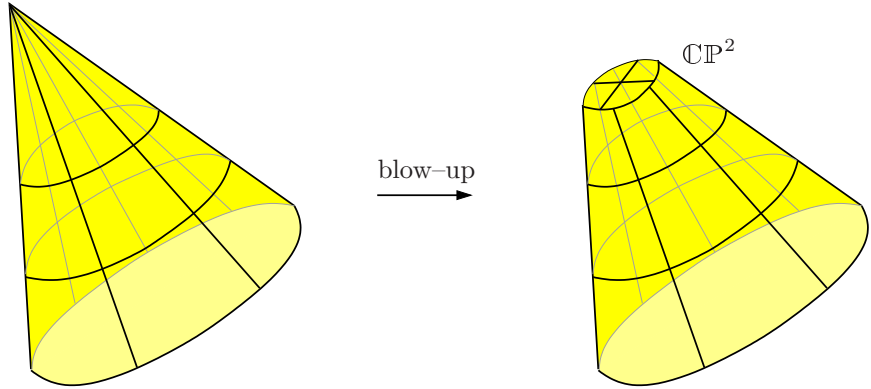

\centerline{\input C3OverZ3Blowup.pstex_t}\vspace{-0.3cm}
\caption{Visualization: the parameter $r$ controls the size of the $\mathbbm{CP}^{2}$. Starting with $r=0$ and the singularity, the orbifold is blown--up by increasing $r$.}
\label{fig:blowup}
\end{figure}

A resolution $\mathcal M^3$ of the orbifold is given by considering the open patches introduced above, equipped with a new K\"ahler potential \cite{Nibbelink:2007rd} 
\begin{equation}
\label{eqn:KahlerResolution}
\mathcal{K}_{\mathcal{M}^3} = \int_{1}^{X}\frac{\mbox{d}X'}{X'} M(X')\;, \quad M(X) = \frac{1}{3}(r + X)^{\frac{1}{3}}\;,
\end{equation}
that is Ricci--flat and matches the orbifold K\"ahler potential eqn.~(\ref{eq:orbKahler2}) in the $r\rightarrow 0$ limit (up to a constant which is irrelevant for a K\"ahler potential). In this limit the curvature vanishes for points $x\neq 0$, whereas for $x=0$ it diverges. Moreover, it vanishes for any value of $r$ when $|x|\rightarrow\infty$. Therefore, blowing up means that the orbifold singularity is replaced by the smooth compact $\mathbbm{CP}^{2}$ that shrinks to zero as $r\rightarrow 0$ (the situation is illustrated in figure~\ref{fig:blowup}).

\subsection{Gauge Fluxes}
\label{sc:gaugefluxes}

Now, we turn on \emph{gauge fluxes}\index{gauge flux}, i.e. non-trivial background values for the internal field strength $\mathcal{F}$, for both the singular orbifold $\mathbbm C^3/\Z{3}$ and its smooth counterpart, the resolution space $\mathcal M^3$.

\subsubsection{Gauge Fluxes Wrapped on the Orbifold}

When defining the heterotic string on $\mathbbm C^3/\Z{3}$, the 10d gauge group $\E{8}\times\E{8}$ is broken by the orbifolding procedure. We can understand this breaking from two perspectives:

On the one hand, following the discussion on compact orbifolds of section~\ref{sec:HetOnOrb}, we know that we have to embed the $\Z{3}$ point group into the gauge degrees of freedom. As before, we do this by a shift embedding $V_\text{orb}$. For later convenience, we define the shift in this chapter as a lattice vector $V_\text{orb} \in \Lambda$. In detail, we absorb the factor $1/3$ into the transformation phase $e^{2\pi\I/3(p_\text{sh} \cdot V_\text{orb} - q_\text{sh}\cdot v)}$ such that the unbroken gauge group is determined by the roots fulfilling
\begin{equation}
p\cdot V_\text{orb} = 0 \mod 3\;.
\end{equation}

On the other hand, we can also understand the gauge symmetry breaking from an effective field theory perspective: let $\I \mathfrak{A}$ be the one--form gauge field, which takes values in the Lie algebra of $\E{8}\times\E{8}$. $\I \mathfrak{F}$ denotes its field strength. Moreover, define $H_I$, for $I=1,\ldots,16$, as the basis elements of the Cartan subalgebra of $\E{8}\times\E{8}$. In a given coordinate patch with local coordinates $z, x=|x|e^{\I\phi}$, the orbifold action $\theta$ is realized as $\phi\rightarrow\phi+2\pi$. On the orbifold there can be non--trivial orbifold boundary conditions for $\mathfrak{A}$ 
\begin{equation}
\label{orbbreak}
\I \mathfrak{A}(\theta\, Z) = \I \mathfrak{A}(z,|x|,\phi+2\pi) = U \I \mathfrak{A}(z,|x|,\phi)U^{-1}\;, 
\end{equation} 
where $U=e^{2\pi\I (V^I_\text{orb} H_I)/ 3}$ and $V_\text{orb}$ is a vector in the $\E{8}\times\E{8}$ root lattice as before. These boundary conditions correspond to a constant Wilson line background (see eqn.~(22) of~\cite{Nibbelink:2007rd}) and hence induce a gauge symmetry breaking, precisely to those $\E{8}\times\E{8}$ algebra elements with root vectors $p$ such that $p \cdot V_\text{orb} = 0\text{ mod } 3$. 

\subsubsection{Gauge Fluxes Wrapped on the Resolution}

The non--trivial orbifold boundary conditions of eqn.~(\ref{orbbreak}) can be reformulated in terms of fields with trivial ones, but having a non--zero constant gauge background. The existence of this non--vanishing gauge flux, localized at the singularity, should become ``visible'' on the resolution. To obtain a matching of orbifold models with models built on the resolved space, we consider the possibility of a gauge bundle wrapped around the resolution. In general such a bundle has structure group $\mathcal{J}$ embedded into $\E{8}\times\E{8}$. This embedding breaks the 10d gauge group $\E{8}\times\E{8}$ to the maximal subgroup $H \subset \E{8}\times\E{8}$ that commutes with $\mathcal{J}$. We therefore expand the 10d field strength two--form
\begin{equation}
\I\mathfrak{F}=\I\mathcal{F}+\I F
\end{equation}
around the internal background $\I\mathcal{F}$, living in the algebra of $\mathcal{J}$, in terms of the 4d field strength $\I F$, taking values in the algebra of $H$. To preserve $\mathcal{N}=1$ supersymmetry in four dimensions, the bundle field strength has to satisfy the Hermitian Yang--Mills equations~\cite{Candelas:1985en}\footnote{Here we ignore loop corrections to these equations, discussed in~\cite{Blumenhagen:2005ga}. We will return to this point later. }   
\begin{equation}
\label{eq:YM}
\mathcal{F}_{ij}=0\;,\quad \mathcal{F}_{\bar{i}\bar{j}}=0\;, \quad G^{i\bar{i}}\mathcal{F}_{i\bar{i}}=0\;,
\end{equation}
where $G^{i\bar{i}}$ denotes the inverse Hermitian metric of $\mathcal{M}^3$, computed from the K\"ahler potential eqn.~(\ref{eqn:KahlerResolution}). One further (topological) consistency requirement follows from the integrated Bianchi--identity of the two--form $B$ of the supergravity multiplet: \index{Bianchi--identity}
\begin{equation}
\label{eq:Bianchi1}
\int_{\mathcal{C}^4}\left(\text{tr}\mathcal{R}^2 - \text{tr}(\I\mathcal{F})^2 \right) = 0\;,
\end{equation}
for all compact four--cycles $\mathcal{C}^4$ of the resolution and $\mathcal{R}$ denotes the curvature of the internal space $\mathcal{M}^3$. This condition is crucial to ensure that the effective four dimensional theory is free of non-Abelian anomalies~\cite{Witten:1984dg}. The resolution space $\mathcal{M}^3$ only contains a single compact four--cycle, the $\mathbbm{CP}^2$ at the resolved singularity, leading to a single consistency condition.

\subsubsection{Examples}

We give two examples of gauge fluxes on the resolution that satisfy eqns.~(\ref{eq:YM}) and~(\ref{eq:Bianchi1}). The simplest construction of such a bundle is the \textit{standard embedding}\index{strings on orbifolds!standard embedding} (to which we refer as ``AS'') with the gauge connection taken to be equal to the spin connection~\cite{Candelas:1985en}. In terms of the curvature this means $\I \mathcal{F}=\mathcal{R}$. Since $\mathcal{R}\in\SU{3}$, this describes an $\SU{3}$ bundle, embedded into $\E{8}\times\E{8}$, leading to the 4d gauge group $H=\E{6}\times\E{8}$. However, we will focus on the case of a $\U{1}$ gauge bundle with a background field strength two--form
\begin{equation}
\label{eq:Uonebundles1}
\I \mathcal{F} =  \Big(\frac r{r + X} \Big)^{1-\frac 1n} \Big( \bar e e - \frac {n-1}{n^2} \, \frac 1{r+X}\, \bar \epsilon \epsilon \Big)\;, 
\end{equation}
where $n=3$ is the order of the orbifold, $X$ is defined in equation~(\ref{eqn:DefX}), $r$ is the blow--up parameter and $e$, $\epsilon$ denote the holomorphic vielbein one--forms. Note that this choice for the fields strength satisfies the Hermitian Yang--Mills equations~(\ref{eq:YM}). For more detail see~\cite{Nibbelink:2007rd}. Such a $\U{1}$ bundle can be embedded into $\E{8}\times\E{8}$ as 
\begin{equation}
\label{eq:Uonebundles2}
\I \mathcal{F}_V = \I \mathcal{F}\, H_V\;,
\end{equation}
where we use the notation $H_V \equiv V^I H_I$. In other words, we have chosen a non-trivial gauge background for the field strength of a single $\U{1}$ direction of the Cartan subalgebra of $\E{8}\times\E{8}$ and the background values of the field strengths corresponding to the other generators of $\E{8}\times\E{8}$ vanish. Since the bundle is only well--defined if its first Chern class, integrated over all compact two--cycles, is integral, an extra consistency requirement arises for the vector $V^I$. For the two--cycle $\mathbbm{CP}^1$ at $x=0$, 
\begin{equation}
\label{eqn:intF}
\frac{1}{2\pi \I} \int_{\mathbbm{CP}^1} \I \mathcal{F}_V = V^I H_I
\end{equation}
must be integral for all $\E{8}\times\E{8}$ roots. This implies that $V$ has to be an $\E{8}\times\E{8}$ root lattice vector itself. The two--form $\mathcal{F}$ as given in eqn.~(\ref{eq:Uonebundles1}) is regular everywhere for $r\neq 0$. In the blow--down limit $r\rightarrow 0$, it is zero for $x\neq 0$ and it diverges for $x=0$, in such a way that the integral eqn.~(\ref{eqn:intF}) remains constant. This means that the bundle is ``visible'' as a two--form only in the blow--up, but in the blow--down it localizes in the singularity to a delta--peak. Thus, its physical effect is not lost. In this sense, this bundle is exactly the counterpart of the orbifold boundary conditions discussed above.

\subsection{Classifying Orbifold and Resolution Models}
\label{sec:classification}

\subsubsection{The Orbifold Models}

The heterotic string on the $\mathbbm C^3/\Z{3}$ is specified by the orbifold gauge shift vector $V_\text{orb}$ defined in eqn.~(\ref{orbbreak}). The freedom in the choice of $V_\text{orb}$ is constrained by modular invariance of the string partition function. In the case of the $\Z{3}$ twist eqn.~(\ref{eq:twist}) the modular invariance constraint eqn.~(\ref{eq:znmodularinv}) reads
\begin{equation}
V_\text{orb}^2 = 0 ~\text{mod}~6\;.
\end{equation}
There are only five inequivalent shift vectors~\cite{Dixon:1986jc,Ibanez:1986tp}, each of them giving rise to a different orbifold model. In table~(\ref{orbshift}) we list the possible $V_\text{orb}$ together with the gauge groups surviving the orbifold projection. Using the standard CFT procedure as discussed in section~\ref{sec:HetOnOrb}, it is possible to compute the spectra of these models. They are listed in the second column of table~(\ref{tab:match}). The spectra are given with the multiplicity numbers with which the various states contribute to the 4d anomaly polynomial localized in the singularity. Thus, these numbers can be fractional if the corresponding states are not localized in the $\mathbbm C^3/\Z{3}$ singularity. The untwisted states have multiplicities that are multiples of $3/27$, because the compact orbifold $T^6/\Z{3}$ has 27 singularities and untwisted states come with multiplicity three. On the other hand, these multiplicities are integers for localized (i.e. twisted) states.

\begin{table}[ht]
\begin{center}
{\small
\begin{tabular}{|ccc|c|c|}
\hline
 & bundle vector  & & gauge group & label \\
 & $V = (V_1)(V_2)$  & & $V_1^2 + V_2^2 = V^2 = 12$       &   \phantom{$I^{I^I}$}     \\
\hline
\hline
$\left(3,1^3,0^4\right)\left(0^8\right)$                                   & $\left(2^3,0^5\right)\left(0^8\right)$                                          & $\left(2^2,1^4,0^2\right)\left(0^8\right)$                                         & $\text{SO}(10)\times \text{U}(3) \times \text{E}'_8$      & \phantom{$I^{I^I}$}\hspace{-0.5cm}AI \\
$\left(\frac{5}{2}, \frac{3}{2}^2,\frac{1}{2}^5\right)\left(0^8\right)$    & $\left(\frac{3}{2}^4,-\frac{3}{2},\frac{1}{2}^3\right)\left(0^8\right)$         &                                                                                    & $12 + 0$                                                  &    \\
\hline
$\left(2,1^2,0^5\right)\left(2,1^2,0^5\right)$                             & $\left(2,1^2,0^5\right)\left(1^6,0^2\right)$                                    & $\left(2,1^2,0^5\right)\left(\frac{3}{2}^2,\frac{1}{2}^6\right)$                   & $\left(\text{E}_6 \times \text{U}(2)\right)^2$            & \phantom{$I^{I^I}$}\hspace{-0.5cm}BI \\
$\left(1^6,0^2\right)\left(1^6,0^2\right)$                                 & $\left(1^6,0^2\right)\left(\frac{3}{2}^2,\frac{1}{2}^6\right)$                  & $\left(\frac{3}{2}^2,\frac{1}{2}^6\right)\left(\frac{3}{2}^2,\frac{1}{2}^6\right)$ & $6 + 6$                                                   &    \\
\hline
$\left(2^2,0^6\right)\left(2,0^7\right)$                                   & $\left(2^2,0^6\right)\left(1^4,0^4\right)$                                      & $\left(2^2,0^6\right)\left(-\frac{3}{2},\frac{1}{2}^7\right)$                      & $\text{E}_7 \times \text{SO}(14)' \times \text{U}(1)^2$   & \phantom{$I^{I^I}$}\hspace{-0.5cm}CI \\
$\left(1^8\right)\left(2,0^7\right)$                                       & $\left(1^8\right)\left(1^4,0^4\right)$                                          & $\left(1^8\right)\left(-\frac{3}{2},\frac{1}{2}^7\right)$                          & $8 + 4$                                                   &    \\
\hline
$\left(1^2, 0^6\right)\left(3,1,0^6\right)$                                & $\left(1^2, 0^6\right)\left(2^2,1^2,0^4\right)$                                 & $\left(1^2, 0^6\right)\left(2,1^6,0\right)$                                        & $\text{E}_7 \times \text{SO}(12)' \times \text{U}(1)^3$   & \phantom{$I^{I^I}$}\hspace{-0.5cm}CII \\
$\left(1^2, 0^6\right)\left(\frac{5}{2},-\frac{3}{2},\frac{1}{2}^6\right)$ & $\left(1^2,0^6\right)\left(\frac{3}{2}^4,\frac{1}{2}^4\right)$                  & $\left(\frac{1}{2}^8\right)\left(3,1,0^6\right)$                                   & $2 + 10$                                                  &     \\

$\left(\frac{1}{2}^8\right)\left(2^2,1^2,0^4\right)$                       & $\left(\frac{1}{2}^8\right)\left(2,1^6,0\right)$                                & $\left(\frac{1}{2}^8\right)\left(\frac{5}{2},-\frac{3}{2},\frac{1}{2}^6\right)$    &                                                           &     \\
$\left(\frac{1}{2}^8\right)\left(\frac{3}{2}^4,\frac{1}{2}^4\right)$       &                                                                                 &                                                                                    &                                                           &     \\
\hline
$\left(2,1^4,0^3\right)\left(2,0^7\right)$                                 & $\left(2,1^4,0^3\right)\left(1^4,0^4\right)$                                    & $\left(2,1^4,0^3\right)\left(-\frac{3}{2},\frac{1}{2}^7\right)$                    & $\text{SU}(8) \times \text{SO}(14)' \times \text{U}(1)^2$ & \phantom{$I^{I^I}$}\hspace{-0.5cm}DI \\
$\left(-\frac{3}{2}^3,\frac{1}{2}^5\right)\left(2,0^7\right)$              & $\left(-\frac{3}{2}^3,\frac{1}{2}^5\right)\left(1^4,0^4\right)$                 & $\left(-\frac{3}{2}^3,\frac{1}{2}^5\right)\left(-\frac{3}{2},\frac{1}{2}^7\right)$ & $8 + 4$                                                   &    \\
$\left(\frac{5}{2},\frac{1}{2}^7\right)\left(2,0^7\right)$                 & $\left(\frac{5}{2},\frac{1}{2}^7\right)\left(1^4,0^4\right)$                    & $\left(\frac{5}{2},\frac{1}{2}^7\right)\left(-\frac{3}{2},\frac{1}{2}^7\right)$    &                                                           &    \\
$\left(-1,1^7\right)\left(2,0^7\right)$                                    & $\left(-1,1^7\right)\left(1^4,0^4\right)$                                       & $\left(-1,1^7\right)\left(-\frac{3}{2},\frac{1}{2}^7\right)$                       &                                                           &    \\
\hline
\end{tabular} 
}\vspace{-0.2cm}
\caption{\label{tab:E8shifts}
This table lists all consistent U(1) bundles embedded into $\E{8}\times\E{8}$. Those bundle vectors $V$ that produce the same gauge symmetry breaking and localized spectrum are grouped together. Each group corresponds to a distinct blow--up of the orbifold models. The bundle vector $V$ contains two parts corresponding to both $\E{8}$'s. Most models are characterized by the values of $V_1^2$ and $V_2^2$ given in the 2nd column; only the splitting $8+4$ has two realizations.}
\end{center}
\end{table}

\subsubsection{The Resolution Models}

The resolution model is completely specified by the way how the gauge flux is embedded in $\E{8}\times\E{8}$, i.e. by the vector $V$. The Bianchi identity eqn.~(\ref{eq:Bianchi1}) integrated over $\mathbbm C \mathbbm P^2$ yields the consistency condition
\begin{equation}
\label{eq:Bianchi2}
V^2 = 12
\end{equation}  
and enormously constrains the number of possible models to a finite number. Having specified the gauge background by $V$, we can analyze the ``compactification'' of the $\E{8}\times\E{8}$ vector multiplet on $\mathcal{M}^3$. Generically, it decomposes into 4d vector multiplets of the unbroken gauge group and 4d chiral multiplets giving rise to charged matter. In detail, the unbroken gauge group in 4d is given by the roots of $\E{8}\times\E{8}$ which fulfill $p\cdot V = 0$ (no mod condition). The chiral matter content, on the other hand, originates from those roots fulfilling $p\cdot V \neq 0$. Their multiplicity is determined by the Dirac index theorem that for $\U{1}$ bundles takes the form
\begin{equation}
\label{eq:index2}
N_V = \frac{1}{18} (H_V)^3 - \frac{1}{6} H_V\;,
\end{equation}
see \cite{Nibbelink:2007rd} for details. In practice, the eigenvalues of $H_V$ and of the multiplicity operator $N_V$ are restricted to five inequivalent cases, listed in table~(\ref{tab:NV}). Let us consider one example in detail: Starting from $V=\left(3,1^3,0^4\right)\left(0^8\right)$, the 480 charged roots of $\E{8}\times\E{8}$ split into:
\begin{itemize}
\item 286 roots with $p\cdot V = 0$ plus Cartans transforming in the adjoint of $\SO{10}\times\U{3}\times\E{8}$
\item  48 roots with $p\cdot V = -1$ transforming as $( \rep{16}, \rep{3},\rep{1})_{-1}$
\item  30 roots with $p\cdot V =  2$ transforming as $( \rep{10}, \rep{3},\rep{1})_{ 2}$
\item  16 roots with $p\cdot V =  3$ transforming as $( \rep{16}, \rep{1},\rep{1})_{ 3}$
\item   3 roots with $p\cdot V =  4$ transforming as $( \rep{ 1},\crep{3},\rep{1})_{ 4}$
\end{itemize}
Note that for the matter representations, we only display one chirality (with positive multiplicity operator $N_V$). The other chirality corresponds to $p\cdot V = 1,-2,-3,-4$. The total number of states is $286 + 2(48+30+16+3) = 480$.

All solutions to eqn.~(\ref{eq:Bianchi2}) together with the corresponding unbroken gauge groups are given in table~(\ref{tab:E8shifts}). The computation of the spectra for each of the $\U{1}$ embeddings shows that there are in fact only five inequivalent models amongst them. We distinguish them by their chiral spectra, which are given in the third column of table~(\ref{tab:match}). 

\begin{table}[t]
\begin{center}
\begin{tabular}{|c|c|c|}
\hline
$H_V|p\rangle = p\cdot V |p\rangle$ & $N_V$ & interpretation$\phantom{I^{I^I}}$\\[0.1cm]
\hline
\hline
 -1 & $\frac{1}{9}$   &  ``untwisted'' matter$\phantom{I^{I^I}}$\\[0.1cm]
\hline
  0 &                 & unbroken gauge group $\phantom{I^{I^I}}$\\[0.1cm]
\hline
  2 & $\frac{1}{9}$   & ``untwisted '' matter$\phantom{I^{I^I}}$\\[0.1cm]
\hline
  3 & $1$             & ``twisted'' matter$\phantom{I^{I^I}}$\\[0.1cm]
\hline
 4 & $3-\frac{1}{9}$ & 3 ``twisted'' matter$\phantom{I^{I^I}}$\\[0.1cm]
   &                 &+ $\frac{1}{9}$ complex conjugate ``untwisted'' matter$\phantom{I^{I^I}}$\\[0.1cm]
\hline
\end{tabular} \vspace{-0.2cm}
\caption{The eigenvalues of the multiplicity operator $N_V$ and their interpretation in the case of the $\Z{3}$ resolution.}
\label{tab:NV}
\end{center}
\end{table}

\subsection{Matching Orbifold and Resolution Models}
\label{sec:localmatching}

Now, we want to investigate the matching between the heterotic orbifold models and the resolution models discussed in the previous section. This matching can be considered at various levels and we begin with some simple observations before entering more subtle issues. 

The first basic observation was made in eqn.~(\ref{orbbreak}): the embedding of the orbifold rotation $\theta$ into the gauge degrees of freedom (via the shift $V_\text{orb}$) can be seen as the presence of a Wilson line, i.e. a gauge flux localized in the singularity as a delta-peak. On the resolution, this gauge flux spreads out on the smooth space, still localized in the region of the former singularity. Going once around the resolved singularity on a circle with ``infinite'' radius in this non-trivial background yields a Wilson line phase
\begin{equation}
\label{matchV}
\frac{1}{2\pi \I}\int_c \I\cA_V = \frac{1}{2\pi \I}\int_\mathbbm{C} \I\cF_V \; \stackrel{r\rightarrow 0}{\longrightarrow}\; \frac{1}{3}V^I H_I\;.
\end{equation}
In detail, the first integration is a contour integral of the gauge potential one--form $\cA_V$ along the contour $c$. This contour in turn is described by the phase $0< \phi < 2\pi$ for large $|x|$, i.e. $x = |x|e^{\I\phi}$ as defined in eqn.~(\ref{orbbreak}). Then, by using Stokes' theorem, the first integration can be reexpressed as an integration of the field strength two--form $\cF_V$ over the variable $x$. Using the integrals of~\cite{Nibbelink:2007rd} yields the behavior of the Wilson line phase eqn.~(\ref{matchV}) in the blow--down limit of the resolved space $r\rightarrow 0$.

In summary, on the singular orbifold $\mathbbm C^3/\Z{3}$ the boundary conditions eqn.~(\ref{orbbreak}) correspond to a Wilson line $V^I_\text{orb} H_I$. Furthermore, in the blow--down limit of the resolution $\mathcal{M}^3$ the gauge flux can be interpreted as a constant Wilson line $V^I H_I$. Obviously these two Wilson line phases have to match in order for the corresponding models to be equal. This is the case if they are related as follows
\begin{equation}
V = V_\text{orb} + 3\lambda \qquad\text{for }\lambda\in\Lambda\;.
\end{equation}

\subsubsection{Connection between $\boldsymbol{V}$ and $\boldsymbol{V_\text{orb}}$}

This basic observation is supported by the fact that any resolution shift $V$ is automatically modular invariant, i.e.
\begin{equation}
V^2 = 12 = 0 \mod 6\;.
\end{equation}
Thus, any resolution shift $V$ can be used as an orbifold shift.

At first sight, the converse, i.e. that any orbifold shift $V_\text{orb}$, classified in table~(\ref{orbshift}) corresponds to a resolution, does not seem to be true. For example, the standard embedding (shift A) has length $V^2 = 6$. However, we should take into account that two $\Z{3}$ orbifold shift vectors are equivalent, i.e.\ lead to the same model if: i) they differ by $3\lambda$ where $\lambda$ is any element of the root lattice of $\E{8}\times\E{8}$, ii) they differ by sign flips of an even number of  entries, or iii) are related by Weyl reflections. By properly combining these operations one can show that all blow--up vectors %
\begin{landscape}
\begin{table}[ht]
\begin{center}
\begin{tabular}{|c||c|c|c|}
\hline&&&\\[-2ex]
 & orbifold model & resolution model & field redefinitions\\[.5ex]
\hline\hline&&&\\[-2ex]
& $\text{E}_6\times \text{SU(3)} \times \text{E}_8'$ & 
$\text{SO(10)}\times \text{U(3)} \times \text{E}_8'$ &
\\[.5ex]\cline{2-3}&&&\\[-2ex]
\hspace{-10pt} 
\begin{tabular}{c} 
A \\ $\downarrow$ \\ AI 
\end{tabular} 
\hspace{-11pt}
&
$\begin{array}{c}
\frac{1}{9} (\rep{27},\rep{3};\rep{1})\\+
{(\rep{27},\rep{1};\rep{1})}+
3(\rep{1},\crep{3};\rep{1})
\end{array}$
&\mbox{}\hspace{-12pt}
$\begin{array}{c}
\frac{1}{9}\left[ (\rep{16},\rep{3};\rep{1})_{\text{-}1}+
(\rep{10},\rep{3};\rep{1})_{2}+
(\rep{1},\rep{3};\rep{1})_{\text{-}4}\right]\\
+(\rep{16},\rep{1};\rep{1})_{3}
+3 (\rep{1},\crep{3};\rep{1})_{4}
\end{array}$\hspace{-8pt}
&
$\begin{array}{lll}\vspace{-36pt}
\as\as\\
(\rep{27},\rep{1};\rep{1})
\as  \hspace{-5pt} \rightarrow \as 
\hspace{-5pt}
\left\{ \hspace{-6pt} 
\begin{array}{lcl}
(\rep{1},\rep{1};\rep{1})_{\text{-}4}\ass=\as e^T v\\
(\rep{16},\rep{1};\rep{1})_{\text{-}1}\ass=\as e^T (\rep{16},\rep{1};\rep{1})_{3}\\
(\rep{10},\rep{1};\rep{1})_{2}\ass=\as e^{-T/2} (\rep{10},\rep{1};\rep{1})^{m}_{0}
\end{array}\right. 
\\[4ex] 
(\rep{1},\crep{3};\rep{1})
\as  \hspace{-5pt} \rightarrow \as  
\hspace{6pt} (\rep{1},\crep{3};\rep{1})_{0} = e^T
(\rep{1},\crep{3};\rep{1})_{4}
\end{array} 
$\\[2.5ex]
\hline\hline&&&\\[-2ex]
& 
$\left[\text{E}_6\times \text{SU(3)}\right]^2$ & 
$\left[\text{E}_6\times \text{U(2)}\right]^2$ &
\\[.5ex]\cline{2-3}&&&\\[-2ex]
\hspace{-10pt}
\begin{tabular}{c} 
B \\ $\downarrow$ \\ BI 
\end{tabular} 
\hspace{-11pt}
&
$\begin{array}{c}
\frac{1}{9} \left[(\rep{27},\crep{3};\rep{1},\rep{1})\right.\\\left.+
(\rep{1},\rep{1};\rep{27},\crep{3})
\right]\\
+{(\rep{1},\rep{3};\rep{1},\rep{3})}
\end{array}$
&
$\begin{array}{c}
\frac{1}{9}\left[ (\rep{27},\rep{2};\rep{1},\rep{1})_{\text{-}1,\text{-}1}+
 (\rep{27},\rep{1};\rep{1},\rep{1})_{2,2}\right.\\
\left.+(\rep{1},\rep{1};\rep{27},\rep{2})_{\text{-}1,1}+
 (\rep{1},\rep{1};\rep{27},\rep{1})_{2,\text{-}2}
\right]\\
+(\rep{1},\rep{2};\rep{1},\rep{1})_{3,3}+
(\rep{1},\rep{1};\rep{1},\rep{2})_{3,\text{-}3}
\end{array}$
&\mbox{}\hspace{-12pt}
$\begin{array}{lll}
\vspace{-36pt}\as\as\\
(\rep{1},\rep{3};\rep{1},\rep{3})
\as  \hspace{-5pt} \rightarrow \as 
\hspace{-5pt}
\left\{ \hspace{-6pt} 
\begin{array}{lcl}
(\rep{1},\rep{1};\rep{1},\rep{1})_{\text{-}4,0}\ass=\as e^T v\\
(\rep{1},\rep{2};\rep{1},\rep{1})_{\text{-}1,3}\ass=
\as e^T (\rep{1},\rep{2};\rep{1},\rep{1})_{3,3}\\
(\rep{1},\rep{1};\rep{1},\rep{2})_{\text{-}1,\text{-}3}\ass=
\as e^T (\rep{1},\rep{1};\rep{1},\rep{2})_{3,\text{-}3}\\
(\rep{1},\rep{2};\rep{1},\rep{2})_{2,0}\ass=
\as e^{-T/2} (\rep{1},\rep{2};\rep{1},\rep{2})^{m}_{0,0}
\end{array}
\right. 
\end{array} 
$\hspace{-14pt}
\\[3ex]
\hline\hline&&&\\[-2ex]
& 
$\text{E}_7\times \text{SO(14)}\times \text{U(1)}^2$ & 
$\text{E}_7\times \text{SO(14)}\times \text{U(1)}^2$ &
\\[.5ex]\cline{2-3}&&&\\[-2ex]
\hspace{-10pt}
\begin{tabular}{c} 
C \\ $\downarrow$ \\ CI 
\end{tabular} 
\hspace{-11pt}
&
\mbox{}\hspace{-12pt}
$\begin{array}{c}
\frac{1}{9} \left[(\rep{56};\rep{1})_{2,2}
+(\rep{1};\rep{1})_{\text{-}4,\text{-}4} \right.\\\left.
+(\rep{1};\rep{64})_{\text{-}1,2}
+(\rep{1};\rep{14})_{2,\text{-}4}
\right]+ \\
(\rep{1};\rep{14})_{2,0}\hspace{-4pt}
+{(\rep{1};\rep{1})}_{\text{-}4,0}\hspace{-4pt}
+3 (\rep{1};\rep{1})_{0,4}
\end{array}$\hspace{-8pt}
&
$\begin{array}{c}
\frac{1}{9} \left[(\rep{56};\rep{1})_{2,2}
+(\rep{1};\rep{1})_{\text{-}4,\text{-}4} \right.\\\left.
+(\rep{1};\rep{64})_{\text{-}1,2}
+(\rep{1};\rep{14})_{2,\text{-}4}
\right]\\
+3 (\rep{1};\rep{1})_{4,4}
\end{array}$
&
$\begin{array}{lcl}
\vspace{-36pt}\as\as\\
(\rep{1};\rep{1})_{\text{-}4,0} \ass = \as e^T v
\\[1ex] 
 (\rep{1};\rep{1})_{0,4}
 \ass = \as e^{T} (\rep{1};\rep{1})_{4,4} 
\\[1ex] 
(\rep{1};\rep{14})_{2,0}
\ass = \as e^{-T/2}(\rep{1};\rep{14})_{0,0}^{m}
\end{array}$
\\[3ex]
\hline\hline&&&\\[-2ex]
& 
$\text{E}_7\times \text{SO(14)}\times \text{U(1)}^2$ & 
$\text{E}_7\times \text{SO(12)}\times \text{U(1)}^3$ &
\\[.5ex]\cline{2-3}&&&\\[-2ex]
\hspace{-10pt}
\begin{tabular}{c} 
C \\ $\downarrow$ \\ CII 
\end{tabular} 
\hspace{-11pt}
&
\mbox{}\hspace{-12pt}
$\begin{array}{c}
\frac{1}{9} \left[(\rep{56};\rep{1})_{2,2}
+(\rep{1};\rep{64})_{\text{-}1,2} \right.\\\left.
+(\rep{1};\rep{1})_{\text{-}4,\text{-}4} 
+(\rep{1};\rep{14})_{2,\text{-}4}
\right]+\\
{(\rep{1};\rep{14})}_{2,0}\hspace{-4pt}
+(\rep{1};\rep{1})_{\text{-}4,0}\hspace{-4pt}
+3 (\rep{1};\rep{1})_{0,4}
\end{array}$\hspace{-8pt}
&
\mbox{}\hspace{-12pt}
$\begin{array}{c}
\frac{1}{9} [(\rep{56};\rep{1})_{\text{-}1,2
,\text{-}2}\hspace{-3pt}
+(\rep{1};\rep{32})_{\text{-}1,2
,2}+
(\rep{1};\crep{32})_{2,2
,0}
\\
+(\rep{1};\rep{1})_{\text{-}4,\text{-}4
,0}
\!+\! (\rep{1};\rep{12})_{\text{-}1,\text{-}4
,\text{-}2} \!+
(\rep{1};\rep{1})_{2,\text{-}4
,\pm 4}]+
\\ 
(\rep{1};\rep{12})_{3,0,\text{-}2}
\!+\! 3 (\rep{1};\rep{1})_{4,4
,0}
\end{array}$\hspace{-8pt}
&
$
\begin{array}{lll}
\vspace{-35pt}\as\as\\
(\rep{1};\rep{14})_{2,0}
\as  \hspace{-5pt} \rightarrow \as 
\hspace{-5pt}
\left\{ \hspace{-6pt}
\begin{array}{lcl}
(\rep{1};\rep{1})_{\text{-}4,0,0}\ass=\as e^T v\\ 
(\rep{1};\rep{12})_{\text{-}1,0,\text{-}2}\ass=
\as e^T (\rep{1};\rep{12})_{3,0,\text{-}2}\\
(\rep{1};\rep{1})_{2,0,\text{-} 4}\ass=
\as e^{-T/2} (\rep{1};\rep{1})^{m}_{0,0,\text{-}4}
\end{array}
\right. 
\\[4ex]  
(\rep{1};\rep{1})_{\text{-}4,0}\hspace{-4pt}
\as  \hspace{-5pt} \rightarrow \as 
\begin{array}{lcl}
(\rep{1};\rep{1})_{2,0,4}\ass=
\as e^{-T/2} (\rep{1};\rep{1})^{m}_{0,0,4}
\end{array}
\\[1ex]  
(\rep{1};\rep{1})_{0,4}
\as \hspace{-5pt} \rightarrow \as 
\begin{array}{lcl}
(\rep{1};\rep{1})_{0,4
,0}\ass=
\as e^T (\rep{1};\rep{1})_{4,4
,0}\\
\end{array}
\end{array}
$
\\\hline\hline&&&\\[-2ex]
&
$\text{SU(9)}\times \text{SO(14)}\times \text{U(1)}$ & 
$\text{SU(8)}\times \text{SO(14)}\times \text{U(1)}^2$ &
\\[.5ex]\cline{2-3}&&&\\[-2ex]
\hspace{-10pt}
\begin{tabular}{c} 
D \\ $\downarrow$ \\ DI 
\end{tabular}
\hspace{-11pt}  
&
\mbox{}\hspace{-12pt}
$\begin{array}{c}
\frac{1}{9} \left[(\rep{84};\rep{1})_{0}\hspace{-4pt}
+(\rep{1};\rep{64})_{\text{-}1}\hspace{-4pt}
+(\rep{1};\rep{14})_{2}
\right]\\
+{(\crep{9};\rep{1})}_{\text{-}4/3}
\end{array}$\hspace{-8pt}
&
$\begin{array}{c}
\frac{1}{9} \left[(\rep{56};\rep{1})_{\text{-}1,\text{-}1}\hspace{-4pt}+
(\rep{28};\rep{1})_{2,2}\hspace{-4pt}
+(\rep{1};\rep{64})_{\text{-}1,2}\right.\\\left.
+(\rep{1};\rep{14})_{2,\text{-}4}
\right]
+(\crep{8};\rep{1})_{3,3}
\end{array}$\hspace{-8pt}
&
$ \begin{array}{l}
\vspace{-40pt} \\
(\crep{9};\rep{1})_{\text{-}4/3} \rightarrow 
\left\{ \begin{array}{l}
(\rep{1};\rep{1})_{\text{-}4,0}  =  e^T v
\\[1ex] 
 (\crep{8};\rep{1})_{\text{-}1,3}
 =  e^{T} (\crep{8};\rep{1})_{3,3} 
\end{array}
\right. 
\end{array}
$
\\[2ex]\hline
\end{tabular} \vspace{-0.2cm}
\caption{\label{tab:match}
We define the matching orbifold and blow--up models in the first
column. The second and third columns give their orbifold and resolution spectra,
respectively. The final column gives the field redefinitions necessary
to match the two spectra.  For blow--up models CII and DI a change of
U(1) basis accompanies the branching (indicated by $\rightarrow$) to
ensure that the state getting the vev is charged under the first blow--up
U(1) only.  
The superscript $m$ indicates non--chiral states that get a mass in
blow--up, and therefore decouple from the massless spectrum.}
\end{center}
\end{table}
\end{landscape}%
\hspace{-0.5cm}of table~(\ref{tab:E8shifts}) can be obtained from the orbifold shifts in table~(\ref{orbshift}). (Only the first model in table~(\ref{orbshift}), characterized by the zero vector $(0^8)(0^8)$, does not have a resolution counterpart in table~(\ref{tab:E8shifts}).)  This leads to a direct matching between orbifold and resolution models. Using the notation from the tables~(\ref{orbshift}) and~(\ref{tab:E8shifts}), we match model B with BI, model D with DI. We also see that even though CI and CII are different resolution models, they correspond to the same orbifold theory C. The same applies to the $\U{1}$ bundle model AI and the standard embedding model AS (introduced in section~\ref{sc:gaugefluxes}): they are both related to the orbifold model A.

\subsubsection{Matching of Gauge Groups and Spectra}

Given the matching at the level of the gauge bundles, we can pass to checks at the level of the 4d gauge groups. A quick glance over the tables~(\ref{orbshift}) and~(\ref{tab:E8shifts}) shows that their gauge groups are never the same. This is easily explained from the orbifold perspective: the blow--up is generated by a non--vanishing vev of some twisted state, the so--called blow--up mode\index{blow--up mode}.  As all twisted states are charged, this vev induces a Higgs mechanism accompanied with gauge symmetry breaking and mass terms. It is not difficult to see from these tables that all non-Abelian resolution gauge groups can be obtained from the orbifold gauge groups by switching on suitable vevs of twisted states. 

Even after taking symmetry breaking, i.e. the branching of the representations of the orbifold state, into account the spectra of the orbifold models still do not agree with the ones of the resolved models: singlets w.r.t.\ non-Abelian groups, and some vector--like states are missing on the resolution. Moreover, the $\U{1}$ charges of localized states do not coincide with the ones expected from the branchings. This can be confirmed from table~(\ref{tab:match}): for each model we give the orbifold spectrum (second column) and the resolution spectrum (third column).  

All these differences can be understood by taking into account more carefully the possible consequences of a twisted state's vev $v$. After branching, this field is a singlet of the non-Abelian gauge group. In the quantum theory this means that the corresponding chiral superfield $\Psi_q$ with charge $q$ under the broken $\U{1}$ never vanishes. Hence, it can be redefined as
\begin{equation}
\Psi_q = v e^T\;,
\end{equation}
where $T$ is a new chiral superfield taking unconstraint values. As it transforms like an axion 
\begin{equation}\label{eq:axiontransf}
T \longrightarrow T + \I q \phi\;,
\end{equation}
under a $\U{1}$ transformation with parameter $\phi$, it is neutral. Hence, it is not part of the charged chiral spectrum computed using the Dirac index eqn.~\eqref{eq:index2} on the resolution. In addition, we can use this axion chiral superfield $T$ to redefine the charges of other twisted states (see the last column of table~(\ref{tab:match})) so that all $\U{1}$ charges of the twisted states agree with the ones of the localized resolution fields. For models CII and DI one needs in addition to change the $\U{1}$ basis when identifying the orbifold and resolution states if one enforces that the field getting a vev is only charged under the first $\U{1}$ factor.

Finally, the remaining states that are missing on the resolution (denoted by a superscript $m$ in table~(\ref{tab:match})) have Yukawa couplings with the blow--up mode, so that they get a mass term in the blow--up. Taking all these blow--up effects into account shows that the spectra of the blown--up orbifold and resolution models become perfectly identical.

\subsection[$F$-- and $D$--flatness of the Blow--up Mode]{$\boldsymbol{F}$-- and $\boldsymbol{D}$--flatness of the Blow--up Mode}
\label{sec:Fflatness}

In the matching of heterotic orbifold models with their resolved counterparts we assumed that a single twisted field of the orbifold model was responsible for generating the blow--up. No other twisted or untwisted states attained non--vanishing vevs. However, in order to obtain a supersymmetric configuration, we have to pay attention to possible non--vanishing $D$-- and $F$--terms arising from the non--zero vev. 

\subsubsection[$F$--Terms]{$\boldsymbol{F}$--Terms}\index{F--flatness}

The analysis of the $F$--flatness for a superpotential $\mathcal W$ is rather involved in the context of heterotic orbifold model building because, in principle, it contains an infinite set of terms with coefficients determined by complicated string amplitudes. In practice, string selection rules can be used to argue that a large class of these coefficients vanishes identically, while the others are taken to be arbitrary, see section~\ref{sec:Yukawa}.

Our assumption above that only a single twisted superfield $\Psi$ has a non--vanishing vev greatly simplifies the $F$--flatness analysis: non--vanishing $F$--terms can only arise from terms in the superpotential that are at most linear in fields $\xi$ having zero vevs
\begin{equation}
\mathcal{W}_\text{relevant} = c_1 \Psi^a + c_2 \Psi^b \xi + \ldots\;,
\end{equation}
where $a,b \in \mathbbm{N}$ and $c_i$ denote the coupling strengths. As in most of the cases the vanishing vev fields $\xi$ form non-Abelian representations, gauge invariance of the superpotential implies that they cannot appear linearly, thus $c_2 = 0$. This means that the complicated analysis of the superpotential involving many superfields often reduces to the analysis of a complex function of a single variable. In reference~\cite{Nibbelink:2008tv} it is shown that all the blow--ups described previously are $F$--flat and therefore constitute viable resolutions of orbifold models.  

\subsubsection{$\boldsymbol{D}$--Terms}\index{D--flatness}

Non--vanishing $D$--terms can only arise under the following conditions~\cite{Buccella:1982nx}: let $\varphi_q$ be the scalar component of the only superfield $\Psi_q$ that acquires a vev $\langle\varphi_q\rangle \neq 0$. In general, the $D$--terms are proportional to
\begin{equation}
D^a \sim \bar{\varphi}_qT^a \varphi_q\;,
\end{equation} 
where $T_a$ are the generators of the orbifold gauge group $G$. Therefore, certainly all $D$--terms corresponding to the generators $T_a$ that annihilate $\langle\varphi_q\rangle$ vanish. They generate the {\em little group} $H$ of gauge symmetries unbroken by the vev. Consequently, non--vanishing $D$--terms are only possible for the generators $T^a$ of the coset $G/H$. Under an infinitesimal gauge transformation with parameter $\epsilon$ the $D$--terms transform as $D^a \rightarrow D^a + \bar{\varphi}_q[\epsilon, T^a] \varphi_q$. This means that for all generators $T^a$ which do not commute with all generators of $G/H$, we can find a gauge such that the $D^a$'s associated to them vanish. But since $(D^a)^2$ defines a gauge invariant object, all these $D^a$ have to vanish in any gauge. The only possibly non--vanishing $D$--terms correspond to the Abelian subgroup of the coset $G/H$. As we explain in the next subsection, precisely those $D$--terms, which are associated with anomalous U(1)'s on the resolution $\mathcal{M}^3$, are non--vanishing. Apart from this subtle issue, $D$--flatness is automatically guaranteed.

\subsection{Multiple Anomalous U(1)'s on the Blow--up}\index{anomaly!multiple anomalous $\U{1}$'s}
\label{sec:anomalies}

In~\cite{GrootNibbelink:2007ew,Nibbelink:2008tv} it is shown that there can be at most two anomalous $\U{1}$'s on the local resolution $\mathcal{M}^3$. Their cancelation involves two axions~\cite{Peccei:1977hh,Peccei:1977ur}, the \textit{model--independent} and a \textit{model--dependent} one, denoted by $a^\text{mi}$ and $a^\text{md}$, respectively. From the singular orbifold perspective, the counterpart of such an anomaly cancelation is a mixture of the standard orbifold Green--Schwarz mechanism (involving the axion $a^\text{het}$) and the Higgs mechanism related to the blow--up mode. As we have seen in eqn.~(\ref{eq:axiontransf}) the imaginary part of $T$ transforms like an axion, denoted by $a^T$. By comparing the anomaly polynomials on the resolution and on the orbifold, the corresponding axions can be matched precisely, $a^\text{mi} = a^\text{het} + \alpha a^T$ and $a^\text{md} = \beta a^T$, where $\alpha$ and $\beta$ are some model dependent constants~\cite{Nibbelink:2008tv,Klevers:2007dp}.

\subsection[$D$--terms in Directions of Anomalous U(1)'s]{$\boldsymbol{D}$--terms in Directions of Anomalous U(1)'s}
\label{sec:Dflatness}

The only non--vanishing $D$--term corresponds to the broken $\U{1}$ generator because only a single twisted chiral superfield developed a vev. The presence of such a $D$--term is consistent: the non--vanishing $D$--term on the blown--up orbifold corresponds to an FI--term on the resolution. In spite of the original orbifold having at most a single anomalous $\U{1}$ and thus a single FI--term, the resolved models can have two. The second one is just the counterpart of the $D$--term generated by the vev. Hence, we conclude that $D$--flatness is guaranteed for all generators except the one corresponding to the broken $\U{1}$. But this non--vanishing $D$--term is required to make the FI--terms coincide: one might interpret this as a matching of two dynamically unstable models. However, we will see at the end of the next section, section~\ref{sec:DflatnessExamples}, that $D$--flatness can be ensured in the compact case. There, we will use the local models (with $D\neq0$) as building blocks for the construction of compact ones and present various methods to obtain $D$--flatness afterwards.

\section[Blow--up of Compact $\Z{3}$ Orbifold]{Blow--up of Compact $\boldsymbol{\Z{3}}$ Orbifold}
\label{sec:compactorbifold}

The local study of orbifold singularities captures a lot of the physics of compact orbifolds. The compact case has some important new aspects as we demonstrate by studying the blow--up of the $T^6/\Z{3}$ orbifold, see section~\ref{sec:Z3Orbifold}. The latter is a space which is flat everywhere except at the 27 fixed points. For later use we enumerate the fixed points as $f = (f_1, f_2, f_3)$ with $f_i = 0,1,2$. The fixed point $0=(0,0,0)$ is obviously localized at the origin. The index $i$ labels the three complex planes. The fixed points are singular and the singularity is identical to the $\mathbbm C^3/\Z{3}$ singularity studied in the previous section. Thus, a sensible resolution of $T^6/\Z{3}$ can be constructed by cutting an open patch around each singularity and replacing it with the smooth space studied above.  

To perform this procedure in detail one has to face the following complicating issues: first of all one has to worry whether the gluing process can be carried out properly. Constructing the blow--up of $T^6/\Z{3}$ by naively joining 27 resolutions of $\mathbbm{C}^3/\Z{3}$ with finite volume seems to lead to a space that is not completely smooth. We ignore this complication by assuming that a more complicated smooth gluing procedure exists, and that for essentially topological questions (e.g. what models do exist and what are their spectra?) this procedure can be trusted. As we are not only gluing together the $\mathbbm C^3/\Z{3}$ blow--ups but also the bundles on them, we have to confirm that the resulting bundle on the resolution of $T^6/\Z{3}$ actually exists. There are two different ways of analyzing this: we can check various consistency conditions ensuring the existence or, from the orbifold point of view, we have to show that $F$-- and $D$--flat directions are allowed by the (super)potential of the compact orbifold theory. 

In this section, we start wit a  brief discussion on resolutions of compact orbifold models without Wilson lines. Next, we review properties of $\Z{3}$ orbifold models with Wilson lines and their resolutions. We finish this section by two examples: the first example considers the blow--up of an orbifold with a single Wilson line, illustrating the gluing procedure of the gauge bundle. The second one examines an orbifold with two Wilson lines and defines an MSSM--like model. Therefore, it is phenomenologically interesting to see whether this model can exist in the blow--up.

\subsection[Resolution of the $\Z{3}$ Orbifolds without Wilson Lines]{Resolution of the $\boldsymbol{\Z{3}}$ Orbifolds without Wilson Lines} 
\label{sec:withoutWilson}

The easiest possibility to construct a smooth compact $\Z{3}$ resolution is to choose the same $\U{1}$ bundle at each fixed point. In such a case, the local consistency conditions are enough to guarantee the existence of the bundle. Indeed, the only extra conditions on the bundle would come from Bianchi identities integrated over the new compact 4--cycles, which are generated by the gluing and thus ``inherited'' from the torus $T^6$. On the other hand, these new 4--cycles are obtained by combining the non--compact 4--cycles of the resolved $\mathbbm C^3/\Z{3}$ singularities. However, for the resolution $\mathcal{M}^3$ the local Bianchi identity on $\mathbbm C^3/\Z{3}$ implies the Bianchi identity on these non--compact 4--cycles, see~\cite{Nibbelink:2007rd,Nibbelink:2007pn}. Thus, the local consistency conditions ensure that the new consistency conditions, due to the gluing, are satisfied. Therefore, all local models can be naturally extended to global ones, with spectra given by 27 copies of the local spectra. 

This compact resolution corresponds to a blown--up orbifold without Wilson lines such that all 27 local spectra are identical. When identical twisted states at all fixed points acquire non--vanishing vevs of the same magnitude and identical orientation, they blow-up the associated fixed points. However, from the orbifold perspective, it requires a little more work to show that this blow--up exists: $F$-- and $D$--flatness have to be checked again. $D$--flatness does not constitute a problem: the total $D$-term $D^a$ entering the scalar potential is simply the sum of the local fixed point contributions $D^{(f)a}$. Since at all fixed points identical twisted states, the blow--up modes, attain exactly the same vev, the individual $D$--terms $D^{(f)a}$ are all the same. For the compact models investigated here all $D$--terms vanish, except possibly the ones associated with the local anomalous $\U{1}$'s, analogously to the non--compact models studied before. The non-vanishing $D$-term corresponding to the anomalous $\U{1}$ can be interpreted as FI term on the resolution, like in the non--compact situation of section~\ref{sec:Dflatness}. On the other hand, $F$--flatness of the compact blow--up does not automatically follow from $F$--flatness of the local $\mathbbm{C}^3/\Z{3}$ blow--ups, because the superpotential of the compact orbifold is much richer than its non--compact counterpart. Of course, all local fixed point couplings that were allowed on $\mathbbm C^3/\Z{3}$ are still allowed. However, new non--local interactions between states from different fixed points are present on the compact orbifold. Furthermore, the $R$--symmetry group of the compact orbifold can be different to the one of the local orbifold singularity. Taking this into account it is shown in~\cite{Nibbelink:2008tv} that a simultaneous blow--up of all 27 fixed points allows for $D$-- and $F$--flatness, when the same blow--up mode at each fixed point acquires the same non--vanishing vev.

\subsection{Orbifolds with Wilson lines}
\label{sec:withWilson}

Orbifolds with Wilson lines have been discussed in section~\ref{sec:HetOnOrb}. We will summarize the main aspects relevant for the $\Z{3}$ case here.

Due to the presence of Wilson lines, there can be different shift-embeddings $V_\text{orb}^{(f)}$ for the fixed points labeled by $f$. Each shift has to fulfill the local version of the modular invariance condition,
\begin{equation}\label{modinv}
(V_\text{orb}^{(f)})^2=0\text{ mod }6\;.
\end{equation}
Furthermore, in the case of $T^6/\mathbbm{Z}_3$, the model is locally completely determined by the gauge groups and spectra listed in table~(\ref{tab:match}). 

In terms of the local shifts $V_\text{orb}^{(g)}$ and $V_\text{orb}^{(f)}$ at the fixed points $g$ and $f$, the Wilson line connecting these points is given by
\begin{equation}
A_\text{orb}^{(fg)}=V_\text{orb}^{(g)}-V_\text{orb}^{(f)}\;.
\end{equation}
Conversely, one can start with a global shift $V_\text{orb}$ and three Wilson lines $A^{(i)}_\text{orb}$. Then, the local shifts are given by $V_\text{orb}^{(f)} \equiv V_\text{orb} + f_i A^{(i)}_\text{orb}$.~\footnote{The three Wilson lines $A^{(i)}_\text{orb}$ are given by $A^{(i)}_\text{orb}=V_\text{orb}^{(i)}-V_\text{orb}^{(0)}$, where $i=(\delta_{1i},\delta_{2i},\delta_{3i})$ indicates the fixed point which lies in the $i$th complex plane.} The $\equiv$ symbol means that the two sides of the equation are equal up to $3\lambda$, where $\lambda\in\Lambda$. Shift and Wilson lines are of order 3, i.e.
\begin{equation}
3 V_\text{orb} \equiv 3 A^{(i)}_\text{orb} \equiv 0\;.
\end{equation} 

The main observation which we will need later for the compact resolution model is the following
\begin{equation}\label{geocondition}
V_\text{orb}^{(f_1,f_2, 0)}+V_\text{orb}^{(f_1,f_2,1)}+V_\text{orb}^{(f_1,f_2,2)}\equiv 0\;.
\end{equation}
One can interpret this equation as follows: sitting at a fixed point $(f_1,f_2)$ of the first two tori but moving in the third one from one fixed point to the other, the total shift has to be trivial after a closed loop. Similar conditions have to be imposed for the other choices of tori. 

At each of the 27 fixed points $f$, the local shift $V_\text{orb}^{(f)}$ induces a (different) gauge symmetry breaking, see the discussion on local GUTs in section~\ref{sec:localGUT}. The resulting 4d gauge group is the common intersection of the local ones, i.e. it survives all local projections simultaneously. In terms of the roots $p$, the unbroken gauge group is determined by
\begin{equation}
\label{orbproj}
V_\text{orb}\cdot p = 0\text{ mod }3\;,\,\,\,\,\,\, A_\text{orb}^{(i)}\cdot p = 0\text{ mod }3\;, \,\,\,\,\text{for}\,\,i=1,\,2,\,3\;.
\end{equation}
The spectrum on $T^6/\mathbbm{Z}_3$ is given by localized and delocalized matter corresponding to the twisted and the untwisted sector. The twisted states localized in the fixed points are organized into representations of the larger GUT gauge group at the respective fixed point, determined by $V_\text{orb}^{(f)}$ {\em only}. They are listed in table~(\ref{tab:match}). Since $T^6/\mathbbm{Z}_3$ has no fixed tori, only the untwisted matter is delocalized. Thus, only the untwisted matter feels the action of {\it all} local projections.

\subsection[The Resolution of $\Z{3}$ Models with Wilson Lines]{The Resolution of $\boldsymbol{\Z{3}}$ Models with Wilson Lines}

Now, we allow for different $\U{1}$ backgrounds in different regions of the smooth compact resolution space. In detail, we have 27 local resolution spaces $\mathcal{M}^3$ combined together to one compact smooth space. Each of the local resolutions is equipped with a $\U{1}$ bundle corresponding to the shifts $V^f$. They are constraints by the local Bianchi identities, related to the localized 4--cycles: for each fixed point $f$ we have a condition\footnote{As explained in the previous section, the new conditions due to the presence of new compact 4--cycles are automatically satisfied once the local conditions are.}
\begin{equation}
(V^{(f)})^2=12\;.
\end{equation}
Moreover, new conditions arise due to the fact that the gauge bundles are not localized, but rather extend over the whole space. Hence, the gluing of different patches requires the various gauge backgrounds to be related in a consistent way on non--trivial overlaps. Therefore, we consider open patches $U^{(f)}$ and $U^{(g)}$ around the resolutions of orbifold singularities labeled by $f$ and $g$ with gauge configurations $\fA_1^{(f)}$ and $\fA_1^{(g)}$, respectively. The transition function $g^{(fg)} = (g^{(gf}){}^{-1}$ describes the relation between the two gauge one--form potentials on the intersection of the two patches:  
\begin{equation}
\fA_1^{(g)} ~=~ g^{(gf)} (\fA_1^{(f)} + \text{d} )g^{(fg)}~.
\end{equation}
Given a point where (any) three patches $f$, $g$ and $h$ overlap, we need to impose the condition $g^{(fg)} g^{(gh)} g^{(hf)} =1$. Moreover, we can identify the transition function $g^{(fg)}$ in the case of a $\text{U}(1)$ gauge bundle with a function $A^{(fg)}$ between the two fixed points $f$ and $g$ as
\begin{equation}
g^{(fg)} ~=~ e^{2\pi\I\, A^{(fg)I} H_I/3}~.
\end{equation}
The function $A^{(fg)}$ is generically not constant. However, in the blow--down limit it becomes constant and can be identified with a discrete Wilson line $A^{(fg)}_\text{orb}$ on the singular orbifold between the fixed points $f$ and $g$. In this limit, we have
\begin{equation}
A^{(fg)} \equiv A^{(fg)}_\text{orb}
\end{equation}
and for that reason we may refer to the function $A^{(fg)}$ as a Wilson line on the resolved space. The co--cycle condition $g^{(fg)} g^{(gh)} g^{(hf)} =1$ can be expressed in terms of these Wilson lines as 
\begin{equation}
A^{(fg)}+A^{(gh)}+A^{(hf)}\equiv 0\;.
\end{equation}
This condition applies to any manifold. It states conditions for the existence of a global flux in the case a space cannot be covered with a single open patch.  

As we are interested in the connection to the singular $\Z{3}$ orbifold with Wilson lines, we require the local identification
\begin{equation}
\label{matchVcmpct} 
V^f \equiv V^f_\text{orb} 
\end{equation} 
at each fixed point, as explained in eqn.~(\ref{matchV}). Thus, the condition eqn.~(\ref{geocondition}) has to be imposed on the resolution shifts, too, and we have
\begin{equation}
\label{res-iden}
V^{(f_1,f_2,0)}+V^{(f_1,f_2,1)}+V^{(f_1,f_2,2)}\equiv 0 
\end{equation}
and corresponding expressions for the other tori. 

In~\cite{Nibbelink:2008tv} it is shown that the unbroken gauge group on the compact resolution is determined by projection conditions in terms of bundle shift $V$ and Wilson lines $A^{(f)} = A^{(f0)}$, i.e.
\begin{equation}
\label{rescond}
V\cdot p =0\;,\,\,\,\,\,A^{(f)}\cdot p =0\;,\,\,\,\text{for}\,\,f \neq 0\;.
\end{equation}
As compared to the maximally four projection conditions for the effective 4d gauge group on the orbifold, we see that there are generically more and stronger conditions on the surviving 4d gauge group on the resolution. 

The main reason for the additional gauge symmetry breaking on the resolution is that the conditions of eqn.~(\ref{rescond}) are not ``$\text{mod}\,\,3$", as they were in the orbifold case. This means that we cannot neglect (triple multiples of) $\E{8}\times\E{8}$ lattice vectors and reduce to four projections at most. In particular, this implies that an orbifold {\em irrelevant Wilson line}, i.e.\ just being three times an $\E{8}\times\E{8}$ lattice vector, can have a non--trivial effect on the resolution gauge group. In this case the same $\U{1}$ bundle is chosen at each fixed point, but they are differently aligned in the $\E{8}\times\E{8}$. From the orbifold point of view this choice corresponds to identical twisted states at all fixed points acquiring non--vanishing vevs of the same magnitude, but different orientation. It will be shown later that this can help to ensure $D$--flatness for all $\U{1}$'s in the compact resolution.

Finally, we describe the consequences of this for the matter states of the various resolved fixed points of the compact smooth space. Locally, the delocalized matter was identified by the fact that it has a fractional multiplicity factor, $\frac{1}{9}$ (or multiples), see section~\ref{sec:localmatching}. Because it is distributed over all patches, it feels projection conditions due to the transition functions between the patches. Thus, given a resolved singularity, say 0, we have to impose 
\begin{equation}
\label{eq:matterprojections}
A^{(f)}\cdot p =0 \text{ mod }3,\,\,\,\text{for}\,\,f \neq 0
\end{equation}
on its delocalized matter. The localized matter, with integral multiplicity, does not reach the overlap regions with the other patches and therefore feels no further projection conditions. Hence, the matter representations of the localized matter just branch with respect to the global unbroken 4d gauge group.

\subsection{One Wilson Line Model with three Anomalous U(1)'s}\index{orbifold!$\Z{3}$}
\label{sec:wilsonex}

\begin{table}[t]\small
\begin{center}
\begin{tabular}{|c||c|>{\!\!}l<{\!\!}|>{\!\!}l<{\!\!}|>{\!\!}ll<{\!\!}|}
\hline
fixed point        & \multicolumn{3}{c|}{matter}      & decomposition    & \mbox{}\hspace{-6pt}field  \\
\cline{2-4}
(loc.) gauge group & \mbox{}\hspace{-4pt}mult.\hspace{-6pt} & local matt. & 4d matt. & to blow--up group & \mbox{}\hspace{-6pt}redefinition \\
\hline\hline&&&&&\\[-2ex]
U Sector                         &3&                             &$(\rep{27},\rep{1})_{(2,2,0)}$ &$(\rep{16},\rep{1})_{(2,2,0,-1)}$ &\\[.5ex]
$\E{6}\times\SO{14}\times\U{1}^3$& &                             &                               &$(\rep{10},\rep{1})_{(2,2,0,2)}$  &\\[.5ex]
                                 & &                             &                               &$(\rep{1},\rep{1})_{(2,2,0,-4)}$  &\\[.5ex]
\hline\hline&&&&&\\[-2ex]
$g_1=\left(\theta,0\right)$      &1&$(\rep{1},\rep{14})_{(2,0)}$ &$(\rep{1},\rep{14})_{(2,0,0)}$ &$(\rep{1},\rep{14})_{(2,0,0,0)}$  &$=e^{-\frac{1}{2}T_1}(\rep{1},\rep{14})^m_{(0,0,0,0)}$\\[.5ex]
\cline{2-6}&&&&&\\[-2ex]
$\E{7}\times\SO{14}\times\U{1}^2$&1&$(\rep{1},\rep{1})_{(-4,0)}$ &$(\rep{1},\rep{1})_{(-4,0,0)}$ &$(\rep{1},\rep{1})_{(-4,0,0,0)}$  &$=v_1 e^{T_1}$\\[.5ex]
\cline{2-6}&&&&&\\[-2ex]
                                 &3&$(\rep{1},\rep{1})_{(0,4)}$  &$(\rep{1},\rep{1})_{(0,2,-2)}$ &$(\rep{1},\rep{1})_{(0,2,-2,0)}$  &$=e^{T_1}(\rep{1},\rep{1})_{(4,2,-2,0)}$\\[.5ex]
\cline{2-6}
local blow--up at $g_1$           & \multicolumn{5}{|l|}{CI}\\
\hline\hline&&&&&\\[-2ex]
$g_2=\left(\theta,e_1\right)$    &1&$(\rep{27},\rep{1}, \rep{1})$&$(\rep{27},\rep{1})_{(0,0,0)}$ &$(\rep{16},\rep{1})_{(0,0,0,-1)}$ &$=e^{T_2}(\rep{16},\rep{1})_{(0,0,0,3)}$\\
$\E{6}\times\SU{3}\times\E{8}$   & &                             &                               &$(\rep{10},\rep{1})_{(0,0,0,2)}$  &$=e^{-\frac{1}{2}T_2}(\rep{10},\rep{1})^m_{(0,0,0,0)}$\\ 
                                 & &                             &                               &$(\rep{1},\rep{1})_{(0,0,0,-4)}$  &$=v_2 e^{T_2}$\\
\cline{2-6} &&&&&\\[-2ex]
                                 &3&$(\rep{1},\rep{3}, \rep{1})$ &$(\rep{1},\rep{1})_{(-2,-2,0)}$&$(\rep{1},\rep{1})_{(-2,-2,0,0)}$ &$=e^{T_2}(\rep{1},\rep{1})_{(-2,-2,0,4)}$\\ 
                                 & &                             &$(\rep{1},\rep{1})_{(0,2,2)}$  &$(\rep{1},\rep{1})_{(0,2,2,0)}$   &$=e^{T_2}(\rep{1},\rep{1})_{(0,2,2,4)}$\\ 
                                 & &                             &$(\rep{1},\rep{1})_{(2,0,-2)}$ &$(\rep{1},\rep{1})_{(2,0,-2,0)}$  &$=e^{T_2}(\rep{1},\rep{1})_{(2,0,-2,4)}$\\
\cline{2-6}
local blow--up at $g_2$           & \multicolumn{5}{|l|}{AI}\\
\hline\hline&&&&&\\[-2ex]
$g_3=\left(\theta,e_1+e_2\right)$&1&$(\rep{1},\rep{14})_{(0,2)}$ &$(\rep{1},\rep{14})_{(0,2,0)}$ &$(\rep{1},\rep{14})_{(0,2,0,0)}$  &$=e^{-\frac{1}{2}T_3}(\rep{1},\rep{14})^m_{(0,0,0,0)}$\\
\cline{2-6}&&&&&\\[-2ex]
$\E{7}\times\SO{14}\times\U{1}^2$&1&$(\rep{1},\rep{1})_{(0,-4)}$ &$(\rep{1},\rep{1})_{(0,-4,0)}$ &$(\rep{1},\rep{1})_{(0,-4,0,0)}$  &$=v_3 e^{T_3}$\\
\cline{2-6}&&&&&\\[-2ex]
                                 &3&$(\rep{1},\rep{1})_{(4,0)}$  &$(\rep{1},\rep{1})_{(2,0,2)}$  &$(\rep{1},\rep{1})_{(2,0,2,0)}$   &$=e^{T_3}(\rep{1},\rep{1})_{(2,4,2,0)}$\\
\cline{2-6} 
local blow--up at $g_3$           & \multicolumn{5}{|l|}{CI}\\
\hline
\end{tabular} 
\caption{\label{tab:table2}
This table gives an overview of the complete global 4d spectrum of the blown up orbifold theory. The field redefinitions that are necessary for the matching between the orbifold blow--up theory and the resolution model are indicated. The $\U{1}^4$-generators of the 4d gauge group in blow--up are $Q_1 = (2,2,0^6)(2,0^7)$, $Q_2 = (2,0,-2,0^5)(-2,0^7)$, $Q_3 = (0,-2,-2,0^5)(2,0^7)$ and $Q_4 = (2,-2,2,0^5)(0^8)$. There are two anomalous combinations: $Q_1^{an} = Q_1+Q_2$ and $Q_2^{an} = Q_4$.}
\end{center}
\end{table}

In the following we give a specific example of an orbifold model in the presence of a discrete Wilson line, and study one of its blown up versions. On the resolution, the model has three anomalous U(1)'s. The bulk universal and the local model--dependent axions are all involved in the anomaly cancelation.

To make the general discussion more explicit, we consider the model obtained from the $T^6/\mathbbm{Z}_3$ orbifold with torus lattice $\SU{3}^3$ and gauge-embedding
\begin{equation}
V_\text{orb}=\left(2,2,0^6\right)\left(2,0^7\right)\;.
\end{equation}
This shift is equivalent to shift C of table~(\ref{orbshift}). Furthermore, we turn on a Wilson line
\begin{equation}
A_\text{orb}=\left(0,-4,2,0^5\right)\left(-2,0^7\right)
\end{equation}
in the directions $e_1$ and $e_2$ of the first complex plane. First, we look at the orbifold and then investigate its resolution. Due to the Wilson line on the orbifold, the $27$ fixed points are grouped together in three sets of nine fixed points each. The three sets are characterized by the local shift vectors $V_\text{orb}$, $V_\text{orb}+A_\text{orb}$ and $V_\text{orb}+2A_\text{orb}$ respectively. The same local gauge group and charged matter is present at all nine fixed points of each set. Details are given in table~(\ref{tab:table2}), where representatives of the three sets of fixed points are identified by their space group representatives $g_1, g_2$ and $g_3$, respectively.

The next task is to find a resolution model that reduces to this orbifold model in the blow down limit. We find that at the $g_1$ singularities we have to choose the CI resolution with a gauge bundle defined by the blow--up shift $V_1 = V_\text{orb}$. At the $g_2$ singularities the AI resolution with $V_2=V_\text{orb}+A_\text{orb}$ is chosen. Finally, at the $g_3$ singularities we have to choose the resolution CI again, but with a different shift $V_3 = V_\text{orb} + 2A_\text{orb} + 3\lambda$. Note that
\begin{equation}
3\lambda=\left(0,6,-6,0^5\right)\left(0^8\right)
\end{equation}
represents, from the orbifold perspective, an irrelevant Wilson line (even if one takes the concept of brother models into account). Nevertheless, it is crucial to ensure that $V_3$ satisfies the local Bianchi identity. This ``irrelevant'' Wilson line leads to additional gauge symmetry breaking on the resolution. The local gauge group and the chiral matter on each of the three sets of nine patches can be found in table~(\ref{tab:match}). The different bundle vectors $V_1$, $V_2$ and $V_3$ combined lead to further symmetry breaking of the local gauge groups at the 27 resolved fixed points to the global 4d gauge group: 
\begin{equation}
\SO{10}\times\SO{14}\times\U{1}^4\;.
\label{G_ex1}
\end{equation}
Consequently, the representations of the local spectra on each of the $3\times 9$ fixed point resolutions become 
\begin{equation}
\begin{array}{ll}
g_1:~ CI: ~ & \frac{1}{9}\left[(\rep{16};\rep{1})_{(2,2,0,\text{-}1)} + (\rep{10};\rep{1})_{(2,2,0,2)} + (\rep{1};\rep{1})_{(2,2,0,\text{-}4)} \right] + 3\,(\rep{1};\rep{1})_{(4,2,\text{-}2,0)}\;, \\[2ex] 
g_2:~ AI: ~ & \frac{1}{9}\left[(\rep{16};\rep{1})_{(2,2,0,\text{-}1)} + (\rep{10};\rep{1})_{(2,2,0,2)} + (\rep{1};\rep{1})_{(2,2,0,\text{-}4)} \right] + (\rep{16};\rep{1})_{(0,0,0,3)} \\[2ex]
            & + \, 3\left[ \,(\rep{1};\rep{1})_{(\text{-}2,\text{-}2,0,4)} + (\rep{1};\rep{1})_{(0,2,2,4)} + (\rep{1};\rep{1})_{(2,0,\text{-}2,4)} \right]\;,\\[2ex]
g_3:~ CI: ~ & \frac{1}{9}\left[(\rep{16};\rep{1})_{(2,2,0,\text{-}1)} + (\rep{10};\rep{1})_{(2,2,0,2)} + (\rep{1};\rep{1})_{(2,2,0,\text{-}4)} \right] + 3\,(\rep{1};\rep{1})_{(2,4,2,0)}\;.
\end{array} 
\label{localResSpec}
\end{equation}
Comparing this with table~(\ref{tab:match}), the localized states (with integral multiplicities) are simply branched to representations of the unbroken 4d gauge group, while some delocalized states (with multiplicity 1/9) are projected out. Because these delocalized states live everywhere on the compact resolution, their spectra at the three types of patches are all the same. The complete resolution spectrum is obtained by multiplying each line of~\eqref{localResSpec} by nine.

We can also study this resolved model from the orbifold blow--up perspective: we select a single twisted field per fixed point that attains a vev chosen along a $F$--flat direction, but some $D$--terms are induced in order to match the FI--terms of the resolution.\footnote{For complete $F$-- and $D$--flatness, we can choose another vacuum configuration, defined by the monomial $(\rep{27}, \rep{1})_{(2,2,0)}^2 (\rep{1}, \rep{1})_{(-4,0,0)} (\rep{27}, \rep{1})_{(0,0,0)} (\rep{1}, \rep{1})_{(0,-4,0)}$. This means that the additional untwisted field  $(\rep{27}, \rep{1})_{(2,2,0)}$ gets a vev leading to a further gauge symmetry break down.} We determine the gauge symmetry breaking induced by this. Each set of singularities $g_i$ has a different blow--up mode and hence a different gauge symmetry breaking:  
\begin{equation} 
\begin{array}{lrccl}
g_1:~ & \langle (\rep{1};\rep{1})_{(-4,0,0)}  \rangle\neq 0\;:~ & \E{7}\times\SO{14}\times\U{1}^2 &\rightarrow& \E{7}  \times\SO{14}\times\U{1}\;, \\[2ex]  
g_2:~ & \langle (\rep{27};\rep{1})_{(0,0,0)}  \rangle\neq 0\;:~ & \E{6}\times\SU{3} \times\E{8}   &\rightarrow& \SO{10}\times\U{3}  \times\E{8}\;, \\[2ex]  
g_3:~ & \langle (\rep{1}; \rep{1})_{(0,-4,0)} \rangle\neq 0\;:~ & \E{7}\times\SO{14}\times\U{1}^2 &\rightarrow& \E{7}  \times\SO{14}\times\U{1}\;. 
\end{array}
\end{equation} 
The global 4d gauge group can be obtained as the intersection of the three local ones, and coincides with the one given in \eqref{G_ex1}. By performing the appropriate field redefinitions on the orbifold, given in table~(\ref{tab:table2}), the blown--up orbifold and the smooth resolution model match perfectly.

By considering the anomaly polynomial of the resolution model, one can easily identify two anomalous $\U{1}$'s. However, if we more physically define the number of anomalous $\U{1}$'s as the number of independent massive $\U{1}$ gauge fields, the number is three: three different vevs $v_1$, $v_2$ and $v_3$ break the $\U{1}$ symmetries $Q_1$, $Q_4$ and $Q_2$, respectively. The three axions $T_1, T_2$ and $T_3$ that do transform under three different combinations of the $\U{1}$'s couple to the corresponding gauge field strengths, leading to three massive gauge fields. More details on this can be found in~\cite{Nibbelink:2008tv}.

\subsection[Blow--up of a $\Z{3}$ MSSM]{Blow--up of a $\boldsymbol{\Z{3}}$ MSSM}\index{orbifold!$\Z{3}$}\index{MSSM!candidates from orbifolds}

We consider the $\Z{3}$ orbifold model with two Wilson lines initially introduced in~\cite{Ibanez:1987sn}. This model is interesting because it was one of the first string models with Standard Model gauge group and three generations of quarks and leptons. A potential problem of this model is the set of vector--like exotics in the spectrum. Only if these exotic states can all be made heavy, the effective low energy spectrum will be identical to that of the MSSM. The way this may happen is by turning on appropriate vevs. As vevs of twisted states lead to blow--ups of the singularities on which they are localized, it is interesting to investigate blow--up versions of this model. Therefore, we assume that the blow--up of this model is generated by single vevs of twisted states at each of the 27 fixed points. This assumption guarantees that we can rely on the Abelian bundles, constructed in section~\ref{sec:classification}, only. We focus on the question whether crucial properties of the MSSM are maintained in blow--up.

The work of~\cite{Ibanez:1987sn,Casas:1987us,Casas:1988se} revealed the possibility to choose one of two hypercharge candidates from the eight $\U{1}$ factors of the model. Each choice corresponds to an ambiguity of identifying the MSSM particle spectrum. However, for either choice the orbifold theory cannot be completely blown up without breaking hypercharge. To resolve all singularities simultaneously, one blow--up mode has to be chosen per fixed point. Table~(1) of~\cite{Casas:1988se} implies that all the states at the fixed point $(n_1,n_3) = (-1,-1)$ carry the same charge under both hypercharge candidates. Hence, by blowing up this singularity, we inevitably break hypercharge. There is only one way to avoid the end of any phenomenology in this orbifold model in full blow--up: the Higgs doublet $H_1$ of the MSSM at $(-1,-1)$ has to obtain a vev. Hence, the blow--up procedure has the interpretation of electroweak symmetry breaking. As far as we have been able to confirm, such a scenario still does not lead to a phenomenologically acceptable situation, because the vanishing of all the $D$--terms requires the vev of $H_1$ to be of the order of the compactification scale, i.e.\ far too large. 

\begin{table}[t]
\begin{center}
\begin{tabular}{|c|rr|rrrrrrrr|c|c|}
\hline
state & \multicolumn{2}{c}{fixed point} & \multicolumn{8}{|c|}{$\U{1}$ charges}                         & \multicolumn{1}{c|}{hyper} & local\\
label & $~~~n_1$ & $n_3$                & $Q_1$ & $Q_2$ & $Q_3$ & $Q_4$ & $Q_5$ & $Q_6$ & $Q_7$ & $Q_8$ &  charge $Y$                & blow--up\\
\hline
\hline
$h_2$    & 0 & 0 & -3 & -2 &  3 &  3 & -3 &  4 &  0 &  0 &   0 & DI\\ 
$h_{10}$ & 1 & 0 & -3 & -2 &  3 &  3 &  1 & -2 &  2 & -4 &   0 & BI\\ 
$h_{14}$ &-1 & 0 &  6 &  4 &  0 &  0 &  2 &  4 & -2 & -2 &   0 & BI\\ 
$h_{15}$ & 0 & 1 & -6 &  0 &  0 &  2 & -4 &  0 & -4 &  0 &   0 & DI\\ 
$h_{17}$ & 0 &-1 &  0 & -4 &  0 & -2 & -2 & -4 &  4 &  0 &   0 & CI\\ 
$h_{21}$ & 1 & 1 & -6 &  0 &  0 &  2 &  0 &  0 &  4 & -4 &   0 & CI\\ 
$h_{23}$ &-1 & 1 &  3 &  6 & -3 & -1 &  1 &  0 &  0 &  4 &   0 & DI\\ 
$h_{24}$ & 1 &-1 &  0 & -4 &  0 & -2 &  2 & -4 &  0 & -4 &   0 & CI\\
\hline
\end{tabular}
\end{center} \vspace{-0.5cm}
\caption{\label{tab:Z3-2WLblowupmodes}
The eight blow--up modes (one per resolved fixed point) are chosen to be singlets with respect to $\SU{3}\times\SU{2}_L\times\U{1}_Y$. The notation used here follows~\cite{Casas:1988se}.}
\end{table}

For this reason, we explore a second possibility and resolve all singularities except the one at $(n_1,n_3) = (-1,-1)$. This partial resolution can be performed in an entirely  $F$-- and $D$--flat way, in all $\U{1}$ directions including the anomalous one and without breaking the hypercharge. For $F$--flatness\index{F--flatness}, we need higher orders in the superpotential to guarantee that the derivative of the superpotential has a zero. For concreteness, consider the situation in which the fields listed in table~(\ref{tab:Z3-2WLblowupmodes}) all have non--vanishing vevs. Their gauge invariant monomial 
\begin{equation}
h_2\, (h_{10})^2\, (h_{14})^2\, h_{15}\, (h_{17})^3\, h_{21}\, (h_{23})^3\, (h_{24})^2
\end{equation}
corresponds to the following relation between the vevs~\cite{Buccella:1982nx, Buccella:1981ib} 
\begin{equation}
\sqrt{6} h_2 = \sqrt{3}h_{10} = \sqrt{3}h_{14} = \sqrt{6}h_{15} = \sqrt{2}h_{17} = \sqrt{6} h_{21} = \sqrt{2} h_{23} = \sqrt{3}h_{24}\;,
\end{equation}
which ensures $D$--flatness\index{D--flatness} as discussed in section~\ref{sec:PropMSSM}. In this configuration, the hypercharge is identified to be
\begin{equation}
Y =\frac 16 \big( \frac{1}{3} Q_1 - \frac{1}{2}Q_2 - Q_3 + Q_4 \big)\;\Leftrightarrow\; t_Y = \left(-\tfrac{1}{3},-\tfrac{1}{3},-\tfrac{1}{3},\tfrac{1}{2},-\tfrac{1}{2},1,-1,0\right)\left(0^8\right)\;,
\end{equation}
\index{hypercharge}
so that none of the blow--up modes is charged under it. Since $H_1$ is massless but does not constitute a flat direction of the effective scalar potential away from this point (i.e.\ at least as long as supersymmetry is not broken), the Higgs cannot acquire a vev. Consequently, electroweak symmetry breaking can only occur at low energies. Furthermore, in this vev configuration all extra $\U{1}$'s are broken and all extra color triplets acquire high masses from trilinear couplings. However, some of the other vector--like exotics stay massless at this order in the superpotential. Thus finally, neither the singular orbifold nor the everywhere smooth resolution of all the fixed points, but the partial blow--up to this \textit{hybrid model} can potentially save phenomenology.

\subsection[$F$-- and $D$--terms for Compact Blow--ups]{$\boldsymbol{F}$-- and $\boldsymbol{D}$--terms for Compact Blow--ups}
\label{sec:DflatnessExamples}

We have mainly focused on compact resolutions with multiple anomalous $\U{1}$'s and corresponding FI--terms. From the orbifold perspective, we have seen that these terms can be interpreted as non--vanishing $D$--terms induced by vevs of the blow--up modes. This situation is exactly the same as explained in section~\ref{sec:Dflatness}. In the following, we will discuss various possibilities to obtain  stable resolutions by finding orbifold blow--ups corresponding to vacua with $F=D=0$.

The first method was discussed in the previous section, where it was necessary to blow--up the orbifold only partially in order to obtain $F=D=0$. This may seem a rather easy way out. A more interesting possibility is that some additional matter fields, either twisted or untwisted, take non--vanishing vevs. When more than one twisted state develop vevs at a single fixed point, a non-Abelian gauge background is expected to be generated on the resolution. On the other hand, a vev for an untwisted state leads to a {\em continuous} Wilson line. An example of the latter case was briefly mentioned in section~\ref{sec:wilsonex}, where the vev of the untwisted state $(\rep{27}, \rep{1})_{(2,2,0)}$ yielded a stable vacuum. 

The general idea of a third method is to perform different blow--ups of degenerate fixed points, i.e. of fixed points not distinguished by Wilson lines from the orbifold perspective. This can be achieved by choosing different blow--up modes at the various fixed points. They may be either contained in different types of non-Abelian representations or in the same ones, but in different components. This allows for choosing the vevs at the different fixed points such that all $D$--terms vanish globally. 

We can exemplify the latter possibility by considering the blow--up of the compact orbifold B without Wilson lines, compare to section~\ref{sec:withoutWilson}. Here, the blow--up mode is contained in the representation $(\rep{1},\rep{3};\rep{1},\rep{3})$, denoted by the matrix $C$. $D$--flatness can be guaranteed by assigning a vev of the same magnitude, but {\it different orientation} to each of the fields $C_i$, localized at one of the 27 fixed points $i=1,\ldots,27$. This corresponds to a gauge invariant monomial of the form
\begin{equation}
\prod_{i=1}^{27} C_i\,,
\end{equation}
breaking the $\SU{3}^2$ factors of the 4D orbifold gauge group to $\U{1}^4$.  Furthermore, $F$--flatness $F=0$ can be achieved at isolated points using higher order couplings in the superpotential yielding stable SUSY preserving vacua.

\chapter{Conclusion}
\label{sec:conclusion}

We gave a precise description of the construction of Abelian heterotic orbifolds. This was done in a very detailed way, on the one hand in order to serve as an introduction to this field. On the other hand, it should provide an intuitive understanding of strings on orbifolds. 

All parameters of this compactification scheme, being the point group $P$, the torus lattice $\Gamma$, the shift-embedding(s) $V$ (or $V_1$ and $V_2$ in the case of $\Z{N}\times\Z{M}$), the Wilson lines $A_\alpha$ and the (new) discrete torsion parameters $a$, $b_\alpha$, $c_\alpha$ and $d_{\alpha\beta}$, are presented in detail. Special focus lies on the conditions one has to impose on the shift(s) and the Wilson lines due to modular invariance. The corresponding conditions one could find in the literature before were incomplete. Thus, this is an important result of this thesis.

Beside a theoretical derivation of these conditions, their correctness could be verified ``experimentally'' by the explicit construction of several million inequivalent orbifold models based on various point groups and torus lattices. By checking the non-trivial anomaly freedom of the resulting massless spectra, we become confident on their properness. This was done using a newly developed c++ computer program, named the \emph{c++ orbifolder}\index{orbifolder (c++ program)}. Given the input parameters as described above, it computes the massless spectrum in less than a second. In addition, one can automatically analyze the properties of the models in order to identify MSSM candidates. This analysis includes for example the number of families of quarks and leptons, the identification of hypercharge generators and the Yukawa couplings that give mass terms for vector-like exotics. Furthermore, a user-friendly prompt was incorporated that allows the usage of the program without any c++ knowledge. The orbifolder was written in collaboration with Dr.~Sa\'ul Ramos-S\'anchez and Dr.~Ak\i n Wingerter and represents by itself another significant result of this work. As this tool might be of great interest to the community, we are planing to publish it later.

We introduced new (generalized) discrete torsion parameters in the context of heterotic orbifolds. Their existence was revealed by the observation that orbifold models whose shift-embeddings differ by lattice vectors are not necessarily identical, yielding the concept of brother-models. This is in contrast to many statements in the literature. The reason for this misunderstanding was the fact that the lattice-symmetry was proven in the $\Z{3}$ case only, where it is indeed a symmetry. Then, it was conjectured to be valid for all orbifolds. However, we have shown that this is not true in general. The structure of the brother models immediately suggested a relation to discrete torsion. An obvious generalization was the introduction of generalized brother models, where shifts and Wilson lines differ by lattice vectors. This construction could then be mapped to generalized discrete torsion, which seems to be the most general solution to the conditions of modular invariance of the partition function. As a consequence, discrete torsion can likewise appear in $\Z{N}$ orbifolds. In the general case we found equivalence of generalized discrete torsion and generalized brothers. However, we also found an exception in the case of $\Z{3}\times\Z{3}$, whose origin remains unsolved. Another important observation was the connection between generalized discrete torsion and orbifold compactifications on non-factorizable lattices. Orbifolds on factorizable torus lattices with generalized discrete torsion turned on can be equivalent to torsionless orbifolds on non-factorizable torus lattices. The latter one generically have less fixed points. Exactly those fixed points that are too much in the factorizable case can be empty (i.e. have no massless twisted strings) if an appropriate choice of generalized torsion is made. Finally, the discovery of generalized discrete torsion, generalized brother models and their relation to non-factorizable torus lattices seems very surprising. Even though many people might think that the orbifold construction is understood completely, there are still important issues to clarify. Their relevance might be even greater once one goes from the singular orbifold to the smooth Calabi-Yau.

The orbifolder was also used to classify all shift-embeddings for heterotic $\SO{32}$ on $\Z{N}$ orbifolds. This classification completed an analogous task that started almost 20 years ago with the case of $\E{8}\times\E{8}$~\cite{Katsuki:1989cs}. It revealed some interesting aspects of the $\SO{32}$ string, like the frequent presence of spinor representations in twisted sectors, which is relevant for the duality between heterotic $\SO{32}$ and type I. However, a systematic analysis of the $\SO{32}$ heterotic string including Wilson lines remains an open task.

One of the major results of this thesis is the construction of about 200 MSSM candidates from the $\Z{6}$-II orbifold. The main task for this project was to invent a suitable search strategy based on the concept of local GUTs. Roughly speaking, the search strategy of local GUTs suggests not to look arbitrarily for MSSM candidates but to restrict to the promising regions of the string landscape, being the ones equipped with a GUT structure of $\SO{10}$ or $\E{6}$. Out of 30.000 inequivalent models from the local GUT regions of the Mini-Landscape, 200 can serve as MSSM candidates. This is a huge fraction compared to other constructions and underlines the success of the local GUT strategy. Furthermore, we discussed gaugino condensation and SUSY breaking for these models with the result that intermediate SUSY breaking by gaugino condensation in the hidden sector is preferred. The reason being that the good models require complicated Wilson lines that break the hidden sector $\E{8}$. It turns out that the remnants of the hidden $\E{8}$ are not too big, nor too small: $\SU{4}$ and $\SO{8}$ are the most common cases. Finally, we analyzed one model, the so-called benchmark model, in detail. Questions about supersymmetric vacuum configurations, the $\mu$-term, $\U{1}_{B-L}$ and $R$-parity, neutrino masses, Yukawa matrices of charged quarks and leptons and finally proton decay were addressed. This demonstrates how far phenomenological aspects of heterotic models can be analyzed today. 

We clarified a serious error concerning the selection rules for allowed string Yukawa couplings. In the literature, there existed a so-called $\gamma$ selection rule stating that the quantum number $\gamma$ should be conserved. Unfortunately, these $\gamma$'s were computed inaccurately. Consequently, the $\gamma$ rule yielded wrong results. We have shown that, using a fully consistent approach for the $\gamma$'s, the $\gamma$ selection rule is indeed trivial, i.e. once the other string selection rules are satisfied the $\gamma$ rule is automatically fulfilled. Therefore, it could be shown that there is no need for an additional $\gamma$ rule. In this context, the string selection rules can be interpreted as discrete symmetries of the superpotential. We analyzed anomalies of these symmetries and revealed intrinsic relations between them. For example, the discrete anomalies corresponding to the space group symmetry are associated to the generator of the anomalous $\U{1}$. 

Furthermore, we performed a detailed analysis of the blow--up of $\Z{3}$ singularities, both in the local and the global orbifold case, with and without Wilson lines. The blown-up orbifold could be matched to smooth resolution models with $\U{1}$ bundles. We briefly addressed the issue of multiple anomalous $\U{1}$'s on the resolution including the identification of the corresponding axions. Finally, we applied the blow--up procedure to a compact $\Z{3}$ orbifold model that is one of the earliest string MSSM candidates. We see that only a partial blow--up of some of its fixed points can retain the nice properties of this model.

\appendix

\chapter{Details}
\label{app:comput}

\section[Continuous Deformations of the $\Z{3}$ Orbifold]{Continuous Deformations of the $\boldsymbol{\Z{3}}$ Orbifold}\index{orbifold!$\Z{3}$}
\label{sec:Z3Deform}
The $\mathbbm{Z}_3$ orbifold is defined by the $\Z{3}$ point group specified by the twist vector $v=(0,\tfrac{1}{3},\tfrac{1}{3},-\tfrac{2}{3})$ and some torus lattice $\Gamma$ that obeys this symmetry. We have chosen  $\Gamma = \SU{3}\times\SU{3}\times\SU{3}$. However, this is not the most general lattice $\Gamma$.

First of all, $\Gamma = \SU{3}\times\SU{3}\times\SU{3}$ can be deformed continuously. Following~\cite{Casas:1991ac}, these deformations can be parameterized by three radii\index{modulus}
\begin{equation}
R_1 \equiv |e_1| = |e_2|,\quad R_3 \equiv |e_3| = |e_4|,\quad R_5 \equiv |e_5| = |e_6|\;,
\end{equation}
and six angles associated to $\alpha_{13},\alpha_{15},\alpha_{35},\alpha_{14},\alpha_{16},\alpha_{36}$. In general, the $\alpha_{ij}$'s correspond to the 15 angles between $e_i$ and $e_j$, i.e.
\begin{equation}
\alpha_{ij} \equiv \cos(\phi_{ij}) = \frac{e_i \cdot e_j}{|e_i| |e_j|}\;,\quad i<j\text{ and }i,j=1,\ldots,6\;.
\end{equation}
Six of them are free parameters, and the other nine angles are fixed in order to preserve the $\Z{3}$ symmetry. To see that the nine continuous parameters $R_1, R_3, R_5, \alpha_{13},\alpha_{15},\alpha_{35},\alpha_{14},\alpha_{16}$ and $\alpha_{36}$ do not affect the $\Z{3}$ point group, we compute the torus metric $g$. Using the real basis vectors $e_i$ the deformed metric $g_{ij} = e_i \cdot e_j$ is given by
\begin{equation}
\label{eqn:generalZ3torusmetric}
g = \left(
\begin{array}{cccccc}
R_1^2               & -\frac{1}{2}R_1^2   & R_1 R_3 \alpha_{13}  & R_1 R_3 \alpha_{14} & R_1 R_5 \alpha_{15} & R_1 R_5 \alpha_{16} \\
-\frac{1}{2}R_1^2   & R_1^2               & R_1 R_3 \alpha_{23}  & R_1 R_3 \alpha_{13} & R_1 R_5 \alpha_{25} & R_1 R_5 \alpha_{15} \\
R_1 R_3 \alpha_{13} & R_1 R_3 \alpha_{23} & R_3^2                & -\frac{1}{2}R_3^2   & R_3 R_5 \alpha_{35} & R_3 R_5 \alpha_{36} \\
R_1 R_3 \alpha_{14} & R_1 R_3 \alpha_{13} & -\frac{1}{2}R_3^2    & R_3^2               & R_3 R_5 \alpha_{45} & R_3 R_5 \alpha_{35} \\
R_1 R_5 \alpha_{15} & R_1 R_5 \alpha_{25} & R_3 R_5 \alpha_{35}  & R_3 R_5 \alpha_{45} & R_5^2               & -\frac{1}{2}R_5^2   \\
R_1 R_5 \alpha_{16} & R_1 R_5 \alpha_{15} & R_3 R_5 \alpha_{36}  & R_3 R_5 \alpha_{35} & -\frac{1}{2}R_5^2   & R_5^2               \\
\end{array}
\right)\;,
\end{equation}
where $\alpha_{23}$, $\alpha_{25}$ and $\alpha_{45}$ are expressed in terms of the free parameters as
\begin{equation}
 \alpha_{i+1,j} = -\left(\alpha_{ij} + \alpha_{i,j+1}\right) \quad i,j=1,3,5\;,\; i<j\;.
\end{equation}
The Coxeter element corresponding to the twist vector $v$ reads
\begin{equation}
\label{eqn:Z3Coxeter}
\hat\theta = \left(
\begin{array}{rrrrrr}
 0 &-1 & 0 & 0 & 0 & 0 \\
 1 &-1 & 0 & 0 & 0 & 0 \\
 0 & 0 & 0 &-1 & 0 & 0 \\
 0 & 0 & 1 &-1 & 0 & 0 \\
 0 & 0 & 0 & 0 & 0 &-1 \\
 0 & 0 & 0 & 0 & 1 &-1 \\
\end{array}
\right)\;,
\end{equation}
such that for example $e_1 \mapsto e_2$ and $e_2 \mapsto -e_1 - e_2$. Then, one can check that the torus metric $g$ is invariant, i.e.
\begin{equation}
\hat\theta^T g \hat\theta = g\;,
\end{equation}
for any value of the deformation parameters.

It is easy to see that the metric eqn.~(\ref{eqn:generalZ3torusmetric}) coincides with the Cartan matrix of $\SU{3}\times\SU{3}\times\SU{3}$ in the special case of $R_1 = R_3 = R_5 = \sqrt{2}$ and $\alpha_{ij} = 0$ for the six free angles. However, the metric  eqn.~(\ref{eqn:generalZ3torusmetric}) cannot be deformed such that it matches to the one of an $\E{6}$ lattice. This shows explicitly that the moduli space is in general disconnected, i.e. it is not possible to deform one allowed torus lattice continuously into any other allowed one while keeping the point group symmetry.

\section{Symmetries under Particle Exchange}
\label{app:gaugeinvariancesymmetry}

As discussed in section~\ref{sec:Yukawa} gauge invariant couplings might vanish because of an antisymmetry of the coupling under particle exchange. In this appendix, we discuss one further example explicitly and list many more in table~(\ref{tab:Symm}). For the example, we assume a gauge group that contains an $\SU{4}$ factor. Furthermore, the coupling $A_i A_j B^{ij} ~=~ (\crep{4})_a (\crep{4})_a ( \rep{6})_b$ is assumed to be invariant under all string selection rules, where the summation over the $\SU{4}$ indices $i,j = 1,\ldots,4$ is implicit. However the two $\crep{4}$-plets are identical. Consequently, the coupling vanishes since $A_i A_j B^{ij} ~=~ -A_i A_j B^{ji} ~=~ 0$, because $B^{ij}$ is antisymmetric in $i$ and $j$, i.e. the $\rep{6}$ is the two index \emph{antisymmetric} tensor of $\SU{4}$ with $4\times 3/2 = 6$ components.

\begin{table}[!ht]
\begin{center}
\begin{tabular}{|c|l|}
\hline
gauge group               & vanishing coupling \\
\hline
\hline
$\SU{2}$                  & $( \rep{2})_a ( \rep{2})_a$ \\
\hline
$\SU{3}$                  & $(\crep{3})_a (\crep{3})_a (\crep{3})_b$ \\
                          & $(\crep{3})_a (\crep{3})_a (\crep{3})_a$ \\
\hline
$\SU{4}$                  & $(\crep{4})_a (\crep{4})_a ( \rep{6})_b $ \\
                          & $(\crep{4})_a (\crep{4})_a (\crep{4})_b (\crep{4})_c$ \\
                          & $(\crep{4})_a (\crep{4})_a (\crep{4})_b (\crep{4})_b$ \\
\hline
$\SO{12}$                 & $(\rep{32})_a (\rep{32})_a$ \\
\hline
$\SU{3}\times\SU{2}$      & $( \rep{3}, \rep{1})_a ( \rep{3}, \rep{1})_a ( \rep{3}, \rep{2})_b ( \rep{1}, \rep{2})_c$  \\
                          & $( \rep{3}, \rep{2})_a ( \rep{3}, \rep{2})_a (\crep{3}, \rep{1})_b (\crep{3}, \rep{1})_b$  \\
                          & $( \rep{3}, \rep{2})_a ( \rep{3}, \rep{2})_a ( \rep{3}, \rep{2})_a ( \rep{1}, \rep{2})_b$  \\
                          & $(\crep{3}, \rep{1})_a (\crep{3}, \rep{1})_a ( \rep{1}, \rep{2})_b (\crep{3}, \rep{2})_c$  \\
\hline
$\SU{3}^2$                & $( \rep{1}, \rep{3})_a ( \rep{1}, \rep{3})_a ( \rep{3}, \rep{1})_b (\crep{3}, \rep{3})_c$ \\
\hline
$\SU{4}\times\SU{2}$      & $(\rep{6},\rep{2})_a (\rep{6},\rep{2})_a$ \\
                          & $(\rep{4},\rep{1})_a (\rep{4},\rep{1})_a (\rep{4},\rep{2})_b (\rep{1},\rep{2})_c (\rep{4},\rep{1})_d$ \\
\hline
$\SO{14}\times\SU{2}$     & $(\rep{14},\rep{2})_a (\rep{14},\rep{2})_a$ \\
\hline
$\SU{2}^3$                & $(\rep{2},\rep{2},\rep{2})_a (\rep{2},\rep{2},\rep{2})_a$ \\
\hline
$\SU{4}\times\SU{2}^2$    & $(\crep{4},\rep{1},\rep{1})_a (\crep{4},\rep{1},\rep{1})_a ( \rep{1},\rep{2},\rep{2})_b ( \rep{1},\rep{2},\rep{1})_c ( \rep{6},\rep{1},\rep{2})_d$ \\
\hline
\end{tabular}
\caption{Examples for gauge invariant couplings that nevertheless vanish because of antisymmetry under particle exchange. For each coupling a subindex $a,b,c,\ldots$ labels the different states, e.g. for $(\crep{3})_a (\crep{3})_a (\crep{3})_b$ two $\crep{3}$-plets are identical and the third is different. Note that the states are supposed to be singlets with respect to further non-Abelian gauge group factors.}
\label{tab:Symm}
\end{center}\index{orbifold!$\Z{3}\times\Z{3}$}
\end{table}

\section{Detailed Discussion on the Gamma--Rule}\index{Yukawa couplings!no need for a $\gamma$ rule}
\label{app:gamma}\index{strings on orbifolds!gamma-phase}

In section~\ref{sec:GammaShort} it was shown that, under the assumption of a trivial vacuum phase\index{strings on orbifolds!vacuum phase} $\Phi_\text{vac}=1$, the gamma selection rule for a coupling of the fields $\Psi_i$, $i=1,\ldots,n$
\begin{equation}
W \supset \Psi_1 \ldots \Psi_n
\end{equation}
is satisfied automatically if gauge invariance and $R$--charge conservation are fulfilled. Now, allowing for arbitrary $\Phi_\text{vac}\neq 1$, it is emphasized that the gamma-rule remains trivial under the additional assumption that the coupling fulfills the space group selection rule and by using the modular invariance conditions for shift and Wilson lines.

The constructing elements corresponding to the fields $\Psi_i$ are labeled by $g_i = \left(\theta^{k_i}, n_{i\alpha} e_\alpha \right)$ in the case of a $\Z{N}$ point group. The gamma--rule is now checked with respect to a transformation $h=\left(\theta^t, m_\alpha e_\alpha \right)$. Using the definition of the gamma--phase eqn.~(\ref{eqn:projectionwithgamma}) yields
\begin{eqnarray}
\sum_i \gamma_i & = & \sum_i \left(R_i\cdot v_h - p_{\text{sh},i}\cdot V_h + \frac{1}{2}\left(V_{g_i}\cdot V_h - v_{g_i}\cdot v_h \right)\right)\\
                & = & \left(\sum_i R_i\right) \cdot v_h + \left(\sum_i p_{\text{sh},i}\right) \cdot V_h + \frac{1}{2}\left(\left(\sum_i V_{g_i}\right)\cdot V_h - \left(\sum_i v_{g_i}\right)\cdot v_h \right) \nonumber \\
                & = & \frac{1}{2}\sum_i \left(V_{g_i}\cdot V_h - v_{g_i}\cdot v_h \right) + \text{integer}\nonumber
\end{eqnarray}
where in the last line gauge invariance and R--charge conservation have been used, as in eqn.~(\ref{eqn:GammaShort}). The remaining part originating from the vacuum phase $\Phi_\text{vac}$ is now split into its pieces, i.e.
\begin{eqnarray}
\sum_i \gamma_i & = & \frac{1}{2}\left(\left(\sum_i k_i V +  n_{i\alpha} A_\alpha \right)\cdot \left(t V + m_\beta A_\beta\right) - \left(\sum_i k_i v\right)\cdot\left(t v\right)\right)  + \text{integer}\\
                & = & \frac{1}{2}\left(t\left(\sum_i k_i\right) \left(V^2 -v^2\right) + m_\beta \left(\sum_i k_i\right) V\cdot A_\beta + \left(\sum_i n_{i\alpha} A_\alpha \right)\cdot\left(tV +  m_\beta A_\beta\right) \right) \nonumber \\
                &   & + \text{integer} \nonumber \\
                & = & \frac{Nat}{2}\left(V^2 - v^2\right) + \frac{Nam_\beta}{2} V\cdot A_\beta + \frac{1}{2}\left(\sum_i n_{i\alpha} A_\alpha \right)\cdot\left(tV + m_\alpha A_\alpha\right) + \text{integer}\nonumber
\end{eqnarray}
where we used point group invariance, i.e. $\sum_i k_i = aN$ with $a \in {\mathbb Z}$ and $N$ being the order of the orbifold. The first two terms are integer due to the modular invariance conditions eqns.~(\ref{eq:znmodularinv}) and~(\ref{eq:fsmi3}), i.e.
\begin{equation}
N\left(V^2 - v^2\right) = 0 \text{ mod } 2 \quad\text{ and }\quad N_\alpha \left( A_\alpha\cdot V\right) = 0 \text{ mod } 2
\end{equation}
and we are left with the expression
\begin{equation}
\label{eqn:GammaBeforeSG}
\sum_i \gamma_i = \frac{1}{2}\left(\sum_i n_{i\alpha} A_\alpha \right)\cdot\left(tV + m_\beta A_\beta\right) + \text{integer}\;.
\end{equation}

For the evaluation of this remaining term, we need to consider some consequences arising from the space group selection rule. We assume that the coupling fulfills the space group selection rule, that is
\begin{equation}
\label{eqn:sgsrc}
\prod_i f_i g_i f_i^{-1} = \left(\mathbbm{1},0\right)\;,
\end{equation}
for some choice of the conjugation elements $f_i = \left(\theta^{s_i}, l_i \right)$. Straightforward evaluation of eqn.~(\ref{eqn:sgsrc}) and using the fact that the action of any power of$\theta$ on an arbitrary lattice vector $l_i$ yields some other lattice vector (denoted by $l'_i$), we get
\begin{equation}
\label{eqn:sgsr}
\sum_i \theta^{\left(\sum_{j=1}^{i-1}k_j + s_i\right)} n_{i\alpha} e_\alpha = \sum_i \left(\mathbbm{1} - \theta^{k_i} \right) l'_i\;.
\end{equation}
As explained in the context of the order of Wilson lines in section~\ref{sec:ConditionsGaugeEmbedding}, we embed this equation into the gauge degrees of freedom. Thus, we use
\begin{eqnarray}
e_\alpha        & \hookrightarrow & A_\alpha \\
\theta e_\alpha & \hookrightarrow & A_\alpha + a_\alpha N_\alpha A_\alpha\;,
\end{eqnarray}
where $a_\alpha \in \mathbbm{Z}$ and $N_\alpha$ is the order of the Wilson line $A_\alpha$, such that $a_\alpha N_\alpha A_\alpha \in \Lambda$ is a lattice vector in the direction of $A_\alpha$ and therefore trivial. The reason for this lattice vector becomes clear by considering an example: for $\Z{3}$ we know that $\theta e_1 = e_2$ and $\theta e_2 = -e_1-e_2$; thus, $A_1 = A_2$ and $A_2 = -A_1-A_2 =-2A_2=A_2 - 3A_2$ and $a_2 = -1$ and $N_1 = N_2 = 3$, in this case. Consequently the embedding of eqn.~(\ref{eqn:sgsr}) reads
\begin{equation}
\label{eqn:FromSpaceGroup}
\sum_i n_{i\alpha} A_\alpha = a_\alpha N_\alpha A_\alpha\;.
\end{equation}
Note the summation over $\alpha$ on both sides of the equation.

Now, we can insert the result obtained from the space group selection rule eqn.~(\ref{eqn:FromSpaceGroup}) into the gamma--rule eqn.~(\ref{eqn:GammaBeforeSG}) and obtain
\begin{equation}
\sum_i \gamma_i = \frac{1}{2}\left(\sum_\alpha a_\alpha N_\alpha A_\alpha \right)\cdot\left(tV + m_\beta A_\beta\right) + \text{integer}
\end{equation}
Using the conditions from modular invariance eqns.~(\ref{eq:fsmi3}), (\ref{eq:fsmi4}) and~(\ref{eq:fsmi5}) it is easy to see that the gamma--rule is satisfied automatically,
\begin{equation}
\sum_i \gamma_i = \text{integer}\;,
\end{equation}
after imposing the other string selection rules.

\section[Geometrical Interpretation of $\ensuremath{[g,h]=0}$]{Geometrical Interpretation of $\boldsymbol{[g,h] = 0}$}
\label{app:commuting}

In this section, we want to give an easy geometrical interpretation of commuting space group elements, either as being associated to the same fixed point or as acting in orthogonal directions. Consider some general orbifold with space group $S$. Its point group might be $\Z{N}$ or $\Z{N}\times\Z{M}$. Take some constructing element
\begin{equation}
g~=~\left(\vartheta, n_\alpha e_\alpha\right) \in S
\end{equation}
which shall correspond to the boundary condition of a massless string. In the case when $\vartheta$ has a fixed torus (i.e. some component of the associated twist vector is zero: $v^i(\vartheta) = 0$ for an index $i \neq 0$), this gives a condition on the constructing element $g$: in order to be massless, the string is not allowed to have windings in the direction of the fixed torus. Technically, this is equivalent to demanding that $n_\alpha = 0$ if $e_\alpha$ lies in the fixed torus. Furthermore, assume that $h \in S$ is some commuting element
\begin{equation}
[g,h] = 0\;.
\end{equation}
Now, it is convenient to distinguish between the following two cases:

\subsubsection{Case 1:  $\vartheta$ has no Fixed Torus}

\begin{figure}[t]
\centerline{
\input{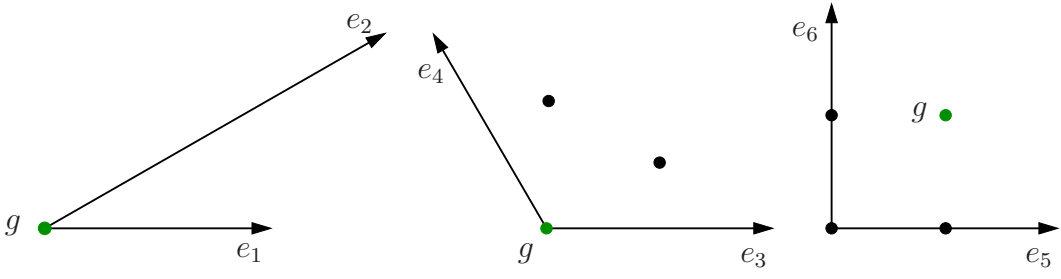}}
\caption{A fixed point $g = \left(\theta,e_5+e_6\right)$ from the first twisted sector is highlighted (in green). All its commuting elements can be written as $h = g^p$ for $p=0,\ldots,5$ such that they are associated to $g$'s fixed point coordinate $z_f$.}
\label{fig:Z6IICommuting1}
\end{figure}

In this case, $g$ corresponds to a (6d) fixed point with coordinates denoted by $z_f$. We easily see that by multiplying $h$ to the fixed point equation
\begin{equation}
g z_f = z_f \quad\Rightarrow\quad hg z_f = hz_f
\end{equation}
and using the assumption that $g$ and $h$ commute, we get $g\left(h z_f\right) = \left(h z_f\right)$. Since the twist $\vartheta$ of $g$ has no fixed torus, the coordinates $z_f$ of the fixed point associated to $g$ are uniquely determined by $g$ (cf. eqn.~(\ref{eqn:FixedPointCoordinate})). Hence, this yields
\begin{equation}
h z_f = z_f\;.
\end{equation}
Thus, we have proven that $g$ and $h$ correspond to the same fixed point. Note that, in contrast to $g$, $h$ might have a fixed torus. For an example in the case of the first twisted sector of the $\Z{6}$-II orbifold\index{orbifold!$\Z{6}$-II} with $\G{2}\times\SU{3}\times\SU{2}^2$ torus lattice see figure~(\ref{fig:Z6IICommuting1}). 

\subsubsection{Case 2: $\vartheta$ has a Fixed Torus}

\begin{figure}[t]
\centerline{
\input{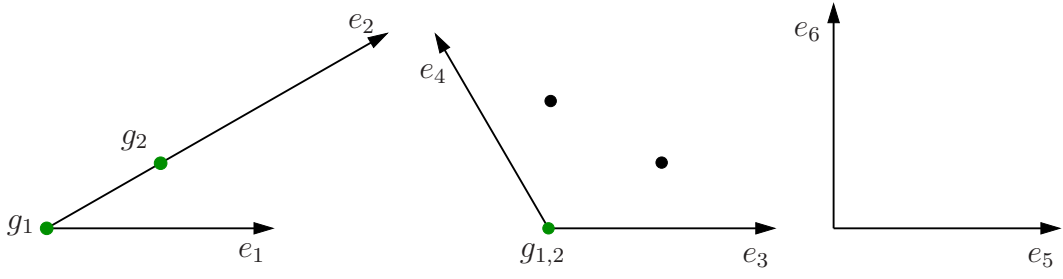}}
\caption{Two fixed points from the $\theta^2$ sector are highlighted (in green), the first being associated to $g_1 = \left(\theta^2,0\right)$ and the second to $g_2 = \left(\theta^2,e_1\right)$. Their corresponding strings are localized at the origin in the $\SU{3}$ lattice and are free to move in the fixed torus $\SU{2}^2$. Their commuting elements are: First of all, $h = \left(\mathbbm{1}, n_5 e_5 + n_6 e_6\right) \in S$ commutes with both, since it acts in orthogonal directions. For $g_1$ there are additional commuting elements from \emph{any} twisted sector $h=\left(\theta^t,n_5 e_5 + n_6 e_6\right)$, since $h$ and $g$ are associated to the same fixed point in the first and second plane. For $g_2$ only elements from the second and forth twisted sector can correspond to $g_2$'s coordinate in the $\G{2}$ plane. Thus, only $h=\left(\mathbbm{1},n_5 e_5 + n_6 e_6\right)\cdot g_2^p$ with $t=0,1,2$ commutes with $g_2$.}
\label{fig:Z6IICommuting2}
\end{figure}

Denote the torus directions in the fixed torus of $\vartheta$ by $t_1$ and $t_2$. Then, by assumption
\begin{equation}
\vartheta \left( a_1 t_1 + a_2 t_2\right) = a_1 t_1 + a_2 t_2 \qquad\text{for }a_1,a_2 \in\mathbbm{R}\;.
\end{equation}
Consequently, we can add this direction to the fixed point equation $g z_f = z_f$ and we get
\begin{equation}
g\left(z_f + a_1 t_1 + a_2 t_2\right) = z_f + a_1 t_1 + a_2 t_2
\end{equation}
Multiplying with $h$ and using the commutator results in $gh\left(z_f + a_1 t_1 + a_2 t_2\right) = h\left(z_f + a_1 t_1 + a_2 t_2\right)$. This time the fixed point of $g$ is not uniquely determined, but only up to some vector in the fixed torus. Therefore
\begin{equation}
\label{eqn:case2step}
h\left(z_f + a_1 t_1 + a_2 t_2\right) = z_f + b_1 t_1 + b_2 t_2\qquad\text{for some }b_1,b_2 \in\mathbbm{R}\;.
\end{equation}
Since $h\in S$, we know that $b_1 t_1 + b_2 t_2 - h( a_1 t_1 + a_2 t_2) = m_1 t_1 + m_2 t_2\in\Gamma$, thus $m_1, m_2 \in \Z{}$. Finally, we see from eqn.~(\ref{eqn:case2step}) that any commuting element $h$ of $g$ has to fulfill the equation
\begin{equation}
hz_f = z_f + m_1 t_1 + m_2 t_2\;.
\end{equation}
In other words, if $h$ is non-trivial, it must be associated to the same fixed point as $g$. In addition, it might act in the fixed torus directions of $g$. Examples are $h=\left(\mathbbm{1}, m_1 t_1 + m_2 t_2\right)$ which purely acts in the fixed torus of $g$ or $h=\left(\mathbbm{1}, m_1 t_1 + m_2 t_2\right)\cdot g^p$ for any $p \in\mathbbm{N}$. For an example in the case of the second twisted sector of the $\Z{6}$-II orbifold\index{orbifold!$\Z{6}$-II} see figure~(\ref{fig:Z6IICommuting2}). 

\section[$g$ level-matching $\Rightarrow$ $g$ invariant]{$\boldsymbol{g}$ level-matching $\boldsymbol{\Rightarrow}$ $\boldsymbol{g}$ invariant}
\label{app:invariant}
In this section, we briefly show that if a string with boundary condition $g$ fulfills the level-matching\index{level-matching} condition $M_\text{L}^2 = M_\text{R}^2$, then it is automatically invariant under transformations with respect to this element $g$.

We denote the local shift and the local twist associated to the constructing element $g$ by $V_g$ and $v_g$, respectively. Then, the mass equations~(\ref{eqn:masslesstwistedleftmover}) and~(\ref{eqn:masslesstwistedrightmover}) for twisted left- and right-movers read
\begin{equation}
\frac{M_\text{L}^2}{8} = \frac{\left(p + V_g\right)^2}{2} - 1 + \tilde{N} + \delta c \quad\text{and}\quad \frac{M_\text{R}^2}{8} = \frac{\left(q + v_g\right)^2}{2} - \frac{1}{2} + \delta c\;.
\end{equation}
Taking their difference\footnote{This corresponds to $L_0 - \tilde{L}_0$, the generator of translations in the $\sigma$ direction of the strings world-sheet. Note that invariance under the action of $L_0 - \tilde{L}_0$ is the origin of the level-matching condition eqn.~(\ref{eqn:levelmatching}).} yields
\begin{equation}
\left(p + V_g\right)^2 - \left(q + v_g\right)^2 - 1 + 2\tilde{N} = 0\;.
\end{equation}
Since $p^2$ is even and $q^2$ is odd, we find
\begin{eqnarray}
2p\cdot V_g + V_g^2 - 2q\cdot v_g - v_g^2 + 2\tilde{N}                                   & = & \text{even}\;,\\
2\left(p+V_g\right)\cdot V_g - V_g^2 - 2\left(q+v_g\right)\cdot v_g + v_g^2 + 2\tilde{N} & = & \text{even}\;,\\
\left(p+V_g\right)\cdot V_g - \left(q+v_g\right)\cdot v_g + \tilde{N} - \frac{1}{2}\big[V_g^2 - v_g^2\big]  & = & \text{int} \;.
\end{eqnarray}
Finally, we insert the shifted momenta $p_\text{sh}$ and $q_\text{sh}$ and obtain
\begin{equation}
\label{eqn:trafofromlevel}
p_\text{sh}\cdot V_g  - q_\text{sh}\cdot v_g + \tilde{N} - \frac{1}{2}\big[V_g^2 - v_g^2\big] = \text{int} \;.
\end{equation}

If there is no left-moving oscillator excitation, i.e. $\tilde{N} = 0$, we easily see that the transformation phase under $g$ eqn.~(\ref{eq:transformationphase}) is fulfilled automatically. In the oscillator case $\tilde{N}\neq 0$, we can split the (eigenvalue of the) oscillator number according to
\begin{equation}
\tilde{N} = \omega_i \tilde{N}^i + \bar{\omega}_i \tilde{N}^{*i}\;,
\end{equation}
as in equation~(\ref{eqn:oscillatornumbersplit}). Using $\omega_i \tilde{N}^i + \bar{\omega}_i \tilde{N}^{*i} = (v_g)_i (\tilde{N}^i - \tilde{N}^{*i}) + \text{ int}$, we can combine this with $q_\text{sh}$ resulting in the $R$ charge $R = q_\text{sh} - \tilde{N} + \tilde{N}^*$ and thus
\begin{equation}
p_\text{sh}\cdot V_g  - R\cdot v_g - \frac{1}{2}\big[V_g^2 - v_g^2\big] = \text{int} \;.
\end{equation}

This proves the general case. If a string with boundary condition $g$ fulfills the level-matching condition (for massless or massive strings), it is automatically invariant under transformations with respect to itself. Note that this necessarily requires the presence of the vacuum phase $-\tfrac{1}{2}[V_g^2 - v_g^2]$.\index{strings on orbifolds!vacuum phase}

\newpage
\section[First Twisted Sector of $\Z{N}$ is Invariant]{First Twisted Sector of $\boldsymbol{\Z{N}}$ is Invariant}
\label{app:T1IsInvariant}

We take some constructing element $g\in S$ from the first twisted sector of some $\Z{N}$ orbifold, i.e. $g=\left(\theta, n_\alpha e_\alpha\right)$. Due to the requirement of $\mathcal{N} = 1$ supersymmetry, we know that $g$ is not associated to a fixed torus, but to a (6d) fixed point. Assume that we have constructed the (massless) spectrum corresponding to the boundary condition $g$. Now, we have to ensure invariance under all commuting elements $h\in S$. Following the discussion of appendix~\ref{app:commuting}, we see that the only commuting elements of $g$ are $h=g^p$ for $p\in\Z{}$. Since $g$ is invariant under the transformation with respect to itself as discussed in appendix~\ref{app:invariant}, it is clearly invariant under all powers of itself.

In summary, states from the first twisted sector are automatically invariant on the $\Z{N}$ orbifold and are not subject to further projection conditions.

\section{Modular Invariance of Shift and Wilson Lines}
\label{app:ModularInvariance}

In this section, we briefly derive the modular invariance conditions on shifts and Wilson lines as listed in section~\ref{sec:SummaryConditions}. We start from the transformation phase of a string with constructing element $g$ and projecting element $h$. The full phase reads
\begin{equation}
\Phi ~\equiv~ e^{2\pi\I\,[p_\text{sh}\cdot V_h - R \cdot v_h ]}\, \Phi_\mathrm{vac}\;,
\end{equation}
where the vacuum phase is defined by
\begin{equation}
\Phi_\mathrm{vac} ~=~ e^{2\pi\I\,[-\frac{1}{2}(V_g\cdot V_h - v_g\cdot v_h)]}\;,
\end{equation}
as stated in section~\ref{sec:HetOnOrb}.~\footnote{Note that the vacuum phases are contained in the corresponding terms $\mathcal{Z}(g,h)$ of the partition function $\mathcal{Z}$. By applying $T$ transformations one can easily reproduce the modular invariance condition on the shift $N(V^2-v^2) = 0\mod 2$, compare to the end of section 3 of~\cite{Ibanez:1987pj} and to section 7.1.2 of~\cite{Choi:2006qh}.}

Now consider two space group elements $g,h \in S$ of order $n$ and $s$, i.e. $g^n = h^s = \mathbbm{1}$. It seems reasonable to demand that the transformation phases for twisted strings with either constructing element $g$ or $g^{n+1}$ are identical, i.e.
\begin{equation}
\label{eq:TransfModified}
\Phi(g,h) ~\stackrel{!}{=}~ \Phi(g^{n+1},h)\;.
\end{equation}
Since $g$ is embedded as shift $V_g$ into the gauge degrees of freedom, the element $g^{n+1}$ will be embedded as $V_{g^{n+1}} = (n+1)\,V_g$. In addition, the local twist of $g^{n+1}$ is $(n+1)\,v_g$. Consequently, eqn.~(\ref{eq:TransfModified}) yields
\begin{equation}
\label{eq:TransfModified2}
\Phi(g^{n+1},h) ~=~ \Phi(g,h)\,\Phi_\mathrm{vac}(g,h)^n \,.
\end{equation}
Thus, we find the condition $\Phi_\mathrm{vac}(g,h)^{n}\stackrel{!}{=}1$ on the vacuum phase\index{strings on orbifolds!vacuum phase}, which is equivalent to the condition
\begin{equation}
n\left(V_g\cdot V_h - v_g\cdot v_h \right)~=~ 0 \mod 2
\end{equation}
An analogous reasoning starting with $\Phi(g,h^{s+1})$ leads to $\Phi_\mathrm{vac}(g,h)^{s}~\stackrel{!}{=}~1$ and thus finally to 
\begin{equation}\label{eq:consistency}
\Phi_\mathrm{vac}(g,h)^{\mathrm{gcd}(n,s)}~\stackrel{!}{=}~1\;.
\end{equation}
This can be rewritten into the modular invariance condition on shifts and Wilson lines\index{modular invariance!of shifts and Wilson lines}
\begin{equation}
\text{gcd}(n,s)\left(V_g\cdot V_h - v_g\cdot v_h \right)~=~ 0 \mod 2\;,
\end{equation}
for arbitrary elements $g,h \in S$ of order $n$ and $s$.

\chapter{Tables}
\label{app:tables}

\begin{table}[!ht]
\begin{center}
\begin{tabular}{|c|rl|rl|}
\hline
Sector      & \multicolumn{2}{c|}{Spectrum M}          & \multicolumn{2}{c|}{Spectrum M$'$}\\
\hline
\hline
$\phantom{I^{I^I}}U\phantom{I^{I^I}}$ & $  1 $&$ (\crep{27},\rep{1})_{(-6, 0)} $ & $  1 $&$ (\crep{27},\rep{1})_{(-6, 0)} $ \\
            & $  1 $&$ (\crep{27},\rep{1})_{( 3,-3)} $ & $  1 $&$ (\crep{27},\rep{1})_{( 3,-3)} $ \\
            & $  1 $&$ (\crep{27},\rep{1})_{( 3, 3)} $ & $  1 $&$ (\crep{27},\rep{1})_{( 3, 3)} $ \\
\hline
$T_{(0,1)}$ & $ 18 $&$ (  \rep{1},\rep{1})_{( 0, 2)} $ & $  9 $&$ (  \rep{1},\rep{1})_{( 0, 2)} $ \\
            & $  9 $&$ (  \rep{1},\rep{1})_{( 0,-4)} $ & $  9 $&$ (  \rep{1},\rep{1})_{(-9,-1)} $ \\
            & $  9 $&$ (\crep{27},\rep{1})_{( 3,-1)} $ & $  9 $&$ (  \rep{1},\rep{1})_{( 9,-1)} $ \\
\hline
$T_{(0,2)}$ & $ 18 $&$ (  \rep{1},\rep{1})_{( 0,-2)} $ & $ 18 $&$ (  \rep{1},\rep{1})_{( 0,-2)} $ \\
            & $  9 $&$ (  \rep{1},\rep{1})_{( 0, 4)} $ & $  9 $&$ (  \rep{1},\rep{1})_{( 0, 4)} $ \\
            & $  9 $&$ (\crep{27},\rep{1})_{( 3, 1)} $ & $  9 $&$ ( \rep{27},\rep{1})_{(-3, 1)} $ \\
\hline
$T_{(1,0)}$ & $ 18 $&$ (  \rep{1},\rep{1})_{( 3, 1)} $ & $ 18 $&$ (  \rep{1},\rep{1})_{( 3, 1)} $ \\
            & $  9 $&$ (  \rep{1},\rep{1})_{(-6,-2)} $ & $  9 $&$ (  \rep{1},\rep{1})_{(-6,-2)} $ \\
            & $  9 $&$ (\crep{27},\rep{1})_{(-3, 1)} $ & $  9 $&$ ( \rep{27},\rep{1})_{( 0,-2)} $ \\
\hline
$T_{(1,1)}$ & \multicolumn{2}{c|}{-}                   & \multicolumn{2}{c|}{-}                   \\
\hline
$T_{(1,2)}$ & $ 18 $&$ (  \rep{1},\rep{1})_{( 3,-1)} $ & $  9 $&$ (  \rep{1},\rep{1})_{( 3,-1)} $ \\
            & $  9 $&$ (  \rep{1},\rep{1})_{(-6, 2)} $ & $  9 $&$ (  \rep{1},\rep{1})_{(-6,-4)} $ \\
            & $  9 $&$ (\crep{27},\rep{1})_{(-3,-1)} $ & $  9 $&$ (  \rep{1},\rep{1})_{( 3, 5)} $ \\
\hline
$T_{(2,0)}$ & $ 18 $&$ (  \rep{1},\rep{1})_{(-3,-1)} $ & $  9 $&$ (  \rep{1},\rep{1})_{(-3,-1)} $ \\
            & $  9 $&$ (  \rep{1},\rep{1})_{( 6, 2)} $ & $  9 $&$ (  \rep{1},\rep{1})_{(-3, 5)} $  \\
            & $  9 $&$ (\crep{27},\rep{1})_{( 0, 2)} $ & $  9 $&$ (  \rep{1},\rep{1})_{( 6,-4)} $  \\
\hline
$T_{(2,1)}$ & $ 18 $&$ (  \rep{1},\rep{1})_{(-3, 1)} $ & $ 18 $&$ (  \rep{1},\rep{1})_{(-3, 1)} $ \\
            & $  9 $&$ (  \rep{1},\rep{1})_{( 6,-2)} $ & $  9 $&$ (  \rep{1},\rep{1})_{( 6,-2)} $ \\
            & $  9 $&$ (\crep{27},\rep{1})_{( 0,-2)} $ & $  9 $&$ ( \rep{27},\rep{1})_{( 3, 1)} $ \\
\hline
$T_{(2,2)}$ & $ 27 $&$ (  \rep{1},\rep{1})_{(-3,-3)} $ & $ 27 $&$ (  \rep{1},\rep{1})_{(-3,-3)} $ \\
            & $ 27 $&$ (  \rep{1},\rep{1})_{( 6, 0)} $ & $ 27 $&$ (  \rep{1},\rep{1})_{( 6, 0)} $ \\
            & $ 27 $&$ (  \rep{1},\rep{1})_{(-3, 3)} $ & $ 27 $&$ (  \rep{1},\rep{1})_{(-3, 3)} $ \\
            & $ 27 $&$ (\crep{27},\rep{1})_{( 0, 0)} $ &       &                                  \\
\hline
\end{tabular}
\caption{The charged spectra of the $\Z{3}\times\Z{3}$ brother models M and M$'$ separated by untwisted and twisted sectors. The 4d gauge group is $\E{6}\times\U{1}^2\times\E{8}$ where the $\U{1}$ generators are chosen to be $t_1 = \left(6,-3,-3,0^5\right)\left(0^8\right)$ and $t_2 = \left(0,3,-3,0^5\right)\left(0^8\right)$.}
\label{tab:Z3xZ3Example}
\end{center}\index{orbifold!$\Z{3}\times\Z{3}$}
\end{table}

\begin{table}[!ht]
\begin{center}
\begin{tabular}{|c|rl|rl|}
\hline
Sector  & \multicolumn{2}{c|}{Spectrum M}             & \multicolumn{2}{c|}{Spectrum M$'$}\\
\hline
\hline
$\phantom{I^{I^I}}U\phantom{I^{I^I}}$ & $  2 $&$ (\crep{27},\rep{2},\rep{1})_{ 2} $ & $  2 $&$ (\crep{27},\rep{2},\rep{1})_{ 2} $ \\
        & $  2 $&$ (  \rep{1},\rep{2},\rep{1})_{-6} $ & $  2 $&$ (  \rep{1},\rep{2},\rep{1})_{-6} $ \\
        & $  1 $&$ (\crep{27},\rep{1},\rep{1})_{-4} $ & $  1 $&$ (\crep{27},\rep{1},\rep{1})_{-4} $ \\
        & $  1 $&$ ( \rep{27},\rep{1},\rep{1})_{ 4} $ & $  1 $&$ ( \rep{27},\rep{1},\rep{1})_{ 4} $ \\
\hline
$T_{1}$ & \multicolumn{2}{c|}{-}                 & \multicolumn{2}{c|}{-}              \\
\hline
$T_{2}$ & $ 32 $&$ (  \rep{1},\rep{2},\rep{1})_{ 0} $ & $ 16 $&$ (  \rep{1},\rep{2},\rep{1})_{ 0} $ \\
        & $ 10 $&$ (  \rep{1},\rep{1},\rep{1})_{-6} $ & $  6 $&$ (  \rep{1},\rep{1},\rep{1})_{-6} $ \\
        & $ 10 $&$ (\crep{27},\rep{1},\rep{1})_{ 2} $ & $  6 $&$ (\crep{27},\rep{1},\rep{1})_{ 2} $ \\
        & $  6 $&$ ( \rep{27},\rep{1},\rep{1})_{-2} $ & $  2 $&$ ( \rep{27},\rep{1},\rep{1})_{-2} $ \\
        & $  6 $&$ (  \rep{1},\rep{1},\rep{1})_{ 6} $ & $  2 $&$ (  \rep{1},\rep{1},\rep{1})_{ 6} $ \\
\hline
$T_{3}$ & $ 80 $&$ (  \rep{1},\rep{1},\rep{1})_{ 3} $ & $ 80 $&$ (  \rep{1},\rep{1},\rep{1})_{ 3} $ \\
        & $ 32 $&$ (  \rep{1},\rep{2},\rep{1})_{-3} $ & $ 32 $&$ (  \rep{1},\rep{2},\rep{1})_{-3} $ \\
        & $ 16 $&$ (\crep{27},\rep{1},\rep{1})_{-1} $ & $ 16 $&$ (\crep{27},\rep{1},\rep{1})_{-1} $ \\
\hline
\end{tabular}
\caption{The charged spectra of the $\Z{4}$ brother models M and M$'$ separated by untwisted and twisted sectors. Both models have the standard embedding shift $V = \left(-\tfrac{1}{2},\tfrac{1}{4},\tfrac{1}{4},0^5\right)\left(0^8\right)$. In addition, model M$'$ has lattice-valued Wilson lines in the $e_1$, $e_5$ and $e_6$ direction, see eqn.~(\ref{eqn:Z4BrotherModelWL}).  The 4d gauge group is $\E{6}\times\SU{2}\times\U{1}\times\E{8}$ where the $\U{1}$ generator is chosen to be $t_1 = \left(4,-2,-2,0^5\right)\left(0^8\right)$.}
\label{tab:Z4LatticeWL}
\end{center}\index{orbifold!$\Z{4}$}
\end{table}

\begin{table}[!ht]
\begin{center}
\begin{tabular}{|c|rl|c|c|rl|}
\hline
Sector      & \multicolumn{2}{c|}{Spectrum} & & Sector      & \multicolumn{2}{c|}{Spectrum}\\
\hline
\hline
$\phantom{I^{I^I}}U\phantom{I^{I^I}}$ & $  1 $&$ (\crep{27},\rep{1})_{(-6, 0)} $ && $T_{(1,2)}$ & $ 15 $&$ (  \rep{1},\rep{1})_{( 3,-1)} $ \\
            & $  1 $&$ (\crep{27},\rep{1})_{( 3,-3)} $ &&             & $  6 $&$ (  \rep{1},\rep{1})_{(-6, 2)} $ \\
            & $  1 $&$ (\crep{27},\rep{1})_{( 3, 3)} $ &&             & $  3 $&$ (  \rep{1},\rep{1})_{(-6,-4)} $ \\
\cline{1-3}
$T_{(0,1)}$ & $ 15 $&$ (  \rep{1},\rep{1})_{( 0, 2)} $ &&             & $  3 $&$ (  \rep{1},\rep{1})_{( 3, 5)} $ \\
            & $  6 $&$ (  \rep{1},\rep{1})_{( 0,-4)} $ &&             & $  3 $&$ (\crep{27},\rep{1})_{(-3,-1)} $ \\
            & $  3 $&$ (  \rep{1},\rep{1})_{(-9,-1)} $ &&             & $  3 $&$ ( \rep{27},\rep{1})_{( 0, 2)} $ \\
\cline{5-7}
            & $  3 $&$ (  \rep{1},\rep{1})_{( 9,-1)} $ && $T_{(2,0)}$ & \multicolumn{2}{c|}{-}                   \\
\cline{5-7}
            & $  3 $&$ (\crep{27},\rep{1})_{( 3,-1)} $ && $T_{(2,1)}$ & $ 15 $&$ (  \rep{1},\rep{1})_{(-3, 1)} $ \\
            & $  3 $&$ ( \rep{27},\rep{1})_{(-3,-1)} $ &&             & $  6 $&$ (  \rep{1},\rep{1})_{( 6,-2)} $ \\
\cline{1-3}
$T_{(0,2)}$ & $ 15 $&$ (  \rep{1},\rep{1})_{( 0,-2)} $ &&             & $  3 $&$ (  \rep{1},\rep{1})_{(-3,-5)} $ \\
            & $  6 $&$ (  \rep{1},\rep{1})_{( 0, 4)} $ &&             & $  3 $&$ (  \rep{1},\rep{1})_{( 6, 4)} $ \\
            & $  3 $&$ (  \rep{1},\rep{1})_{(-9, 1)} $ &&             & $  3 $&$ (\crep{27},\rep{1})_{( 0,-2)} $ \\
            & $  3 $&$ (  \rep{1},\rep{1})_{( 9, 1)} $ &&             & $  3 $&$ ( \rep{27},\rep{1})_{( 3, 1)} $ \\
\cline{5-7}
            & $  3 $&$ (\crep{27},\rep{1})_{( 3, 1)} $ && $T_{(2,2)}$ & $ 27 $&$ (  \rep{1},\rep{1})_{(-3,-3)} $ \\
            & $  3 $&$ ( \rep{27},\rep{1})_{(-3, 1)} $ &&             & $ 27 $&$ (  \rep{1},\rep{1})_{( 6, 0)} $ \\
\cline{1-3}
$T_{(1,0)}$ & \multicolumn{2}{c|}{-}                   &&             & $ 27 $&$ (  \rep{1},\rep{1})_{(-3, 3)} $ \\
\cline{1-3}
$T_{(1,1)}$ & \multicolumn{2}{c|}{-}                   &&             & $  9 $&$ (\crep{27},\rep{1})_{( 0, 0)} $ \\
\hline
\end{tabular}
\caption{The charged spectrum of the $\Z{3}\times\Z{3}$ model with standard embedding and generalized discrete torsion $b_3 = b_4 = 1/3$. The 4d gauge group is $\E{6}\times\U{1}^2\times\E{8}$ where the $\U{1}$ generators are chosen to be $t_1 = \left(6,-3,-3,0^5\right)\left(0^8\right)$ and $t_2 = \left(0,3,-3,0^5\right)\left(0^8\right)$.}
\label{tab:Z3xZ3ExampleBNotZero}
\end{center}\index{orbifold!$\Z{3}\times\Z{3}$}
\end{table}

\begin{table}[ht!]
\begin{center}
\begin{tabular}{|l|r|l|l|}
\hline
orbifold & torsion $\varepsilon$ & shift $V_1$ & shift $V_2$\\
\hline\hline
$\Z2\times\Z2$  & $1$  & $\displaystyle \left(\tfrac{1}{2},0,-\tfrac{1}{2},0,0,0,0,0\right)$   & $\displaystyle \left(0,\tfrac{1}{2},-\tfrac{1}{2},0 ,0, 0, 0, 0\right)$\\
                & $-1$ & $\displaystyle \left(\tfrac{1}{2}, -1,-\tfrac{1}{2},1,0,0,0,0\right)$ & $\displaystyle \left(1,\tfrac{1}{2},-\tfrac{1}{2},0,1,0,0,0\right)$\\
\hline
$\Z4\times\Z2$  & $1$  & $\displaystyle \left(\tfrac{1}{4},0,-\tfrac{1}{4},0,0,0,0,0\right)$   & $\displaystyle \left(0,\tfrac{1}{2},-\tfrac{1}{2},0,0,0,0,0\right)$ \\
                & $-1$ & $\displaystyle \left(\tfrac{1}{4},-1,-\tfrac{1}{4},1,0,0,0,0\right)$  & $\displaystyle \left(2,\tfrac{1}{2},-\tfrac{1}{2},0,0,0,0,0\right)$ \\
\hline
$\Z6\times\Z2$  & $1$  & $\displaystyle \left(\tfrac{1}{6},0,-\tfrac{1}{6},0,0,0,0,0\right)$   & $\displaystyle \left(0,\tfrac{1}{2},-\tfrac{1}{2},0 ,0, 0, 0, 0\right)$\\
                & $-1$ & $\displaystyle \left(\tfrac{1}{6},-1,-\tfrac{1}{6},1,0,0,0,0\right)$  & $\displaystyle \left(3,\tfrac{1}{2},-\tfrac{1}{2},0,1,0,0,0\right)$ \\
\hline
$\Z6'\times\Z2$ & $1$  & $\displaystyle \left(\tfrac{1}{6},\tfrac{1}{6},-\tfrac{1}{3},0,0,0,0,0\right)$  & $\displaystyle \left(\tfrac{1}{2},0,-\tfrac{1}{2},0,0,0,0,0\right)$\\
                & $-1$ & $\displaystyle \left(-\tfrac{5}{6},\tfrac{7}{6},-\tfrac{1}{3},1,1,0,0,0\right)$ & $\displaystyle \left(\tfrac{1}{2},3,-\tfrac{1}{2},1,0,0,0,0\right)$\\
\hline
$\Z3\times\Z3$      & $1$ & $\displaystyle \left(\tfrac{1}{3},0,-\tfrac{1}{3},0,0,0,0,0\right)$  & $\displaystyle \left(0,\tfrac{1}{3},-\tfrac{1}{3},0,0,0,0,0\right)$\\
&$e^{2\pi\I\tfrac{1}{3}}$ & $\displaystyle \left(\tfrac{1}{3},-1,-\tfrac{1}{3},1,0,0,0,0\right)$ & $\displaystyle \left(1,\tfrac{1}{3},-\tfrac{1}{3},0,1,0,0,0\right)$\\
&$e^{2\pi\I\tfrac{2}{3}}$ & $\displaystyle \left(\tfrac{1}{3},-2,-\tfrac{1}{3},0,0,0,0,0\right)$ & $\displaystyle \left(2,\tfrac{1}{3},-\tfrac{1}{3},0,0,0,0,0\right)$\\
\hline
$\Z6\times\Z3$ & $1$ & $\displaystyle \left(\tfrac{1}{6},0,-\tfrac{1}{6},0,0,0,0,0\right)$ & $\displaystyle \left(0,\tfrac{1}{3},-\tfrac{1}{3},0 ,0, 0, 0, 0\right)$\\
&$e^{2\pi\I\tfrac{1}{3}}$ & $\displaystyle \left(\tfrac{1}{6},-1,-\tfrac{1}{6},1,0,0,0,0\right)$ & $\displaystyle \left(2,\tfrac{1}{3},-\tfrac{1}{3},0,0,0,0,0\right)$\\
&$e^{2\pi\I\tfrac{2}{3}}$ & $\displaystyle \left(\tfrac{1}{6},-2,-\tfrac{1}{6},0,0,0,0,0\right)$ & $\displaystyle \left(4,\tfrac{1}{3},-\tfrac{1}{3},0,0,0,0,0\right)$\\
\hline
$\Z4\times\Z4$ & $1$   & $\displaystyle \left(\tfrac{1}{4},0,-\tfrac{1}{4},0,0,0,0,0\right)$  & $\displaystyle \left(0,\tfrac{1}{4},-\tfrac{1}{4},0,0,0,0,0\right)$\\
               & $\I$  & $\displaystyle \left(\tfrac{1}{4},-1,-\tfrac{1}{4},1,0,0,0,0\right)$ & $\displaystyle \left(1,\tfrac{1}{4},-\tfrac{1}{4},0,1,0,0,0\right)$\\ 
               & $-1$  & $\displaystyle \left(\tfrac{1}{4},-2,-\tfrac{1}{4},0,0,0,0,0\right)$ & $\displaystyle \left(2,\tfrac{1}{4},-\tfrac{1}{4},0,0,0,0,0\right)$\\ 
               & $-\I$ & $\displaystyle \left(\tfrac{1}{4},-3,-\tfrac{1}{4},1,0,0,0,0\right)$ & $\displaystyle \left(3,\tfrac{1}{4},-\tfrac{1}{4},0,1,0,0,0\right)$\\ 
\hline
$\Z6\times\Z6$ & $1$ & $\displaystyle \left(\tfrac{1}{6},0,-\tfrac{1}{6},0,0,0,0,0\right)$    & $\displaystyle \left(0,\tfrac{1}{6},-\tfrac{1}{6},0,0,0,0,0\right)$\\
&$e^{2\pi\I\tfrac{1}{6}}$ & $\displaystyle \left(\tfrac{1}{6},-1,-\tfrac{1}{6},1,0,0,0,0\right)$ & $\displaystyle \left(1,\tfrac{1}{6},-\tfrac{1}{6},0,1,0,0,0\right)$\\
&$e^{2\pi\I\tfrac{1}{3}}$ & $\displaystyle \left(\tfrac{1}{6},-2,-\tfrac{1}{6},0,0,0,0,0\right)$ & $\displaystyle \left(2,\tfrac{1}{6},-\tfrac{1}{6},0,0,0,0,0\right)$\\
                   & $-1$ & $\displaystyle \left(\tfrac{1}{6},-3,-\tfrac{1}{6},1,0,0,0,0\right)$ & $\displaystyle \left(3,\tfrac{1}{6},-\tfrac{1}{6},0,1,0,0,0\right)$\\
&$e^{2\pi\I\tfrac{2}{3}}$ & $\displaystyle \left(\tfrac{1}{6},-4,-\tfrac{1}{6},0,0,0,0,0\right)$ & $\displaystyle \left(4,\tfrac{1}{6},-\tfrac{1}{6},0,0,0,0,0\right)$\\
&$e^{2\pi\I\tfrac{5}{6}}$ & $\displaystyle \left(\tfrac{1}{6},-5,-\tfrac{1}{6},1,0,0,0,0\right)$ & $\displaystyle \left(5,\tfrac{1}{6},-\tfrac{1}{6},0,1,0,0,0\right)$\\
\hline
\end{tabular}
\caption{$\Z{N}\times\Z{M}$ models with discrete torsion and standard embedding are equivalent to models without discrete torsion and non-standard embedding. We write the torsion phase factor as $\varepsilon=e^{-2\pi\I\, V_2\cdot\Delta V_1}$. The components of the shifts within the second $\E8$ all vanish. This result also applies to orbifold models in $\SO{32}$.}
\label{tab:DiscreteTorsionBrothers}
\end{center}
\end{table}

\begin{table}[!ht]
\begin{center}
\begin{tabular}{|l|rl|rl|c|rl|}
\hline
\multicolumn{1}{|c|}{non-vanishing} &\multicolumn{5}{c|}{twisted spectrum}                                               &  \multicolumn{2}{c|}{total}\\
discrete torsion parameter          &\multicolumn{2}{c|}{$T_{(1,0)}$} &  \multicolumn{2}{c|}{$T_{(0,1)}$}  & $T_{(1,1)}$ &  \multicolumn{2}{c|}{spectrum}\\
\hline
\hline
$c_4 = \frac{1}{2}$                 & $ 8$ & $(\crep{27},\rep{1})$ & $ 8$ & $(\crep{27},\rep{1})$ & -           &  $19$ & $(\crep{27},\rep{1})\phantom{I^{I^I}}$ \\
                                    & $ 8$ & $(\rep{27},\rep{1})$  & $ 8$ & $(\rep{27},\rep{1})$  & -           &  $19$ & $(\rep{27},\rep{1})$  \\
                                    & $80$ & $(\rep{1},\rep{1})$   & $80$ & $(\rep{1},\rep{1})$   & -           & $166$ & $(\rep{1},\rep{1})$  \\
\hline
$c_3 = d_{46}= \frac{1}{2}$         & $ 4$ & $(\crep{27},\rep{1})$ & $ 8$ & $(\crep{27},\rep{1})$ & -           &  $15$ & $(\crep{27},\rep{1})\phantom{I^{I^I}}$ \\
                                    & $ 4$ & $(\rep{27},\rep{1})$  & $ 8$ & $(\rep{27},\rep{1})$  & -           &  $15$ & $(\rep{27},\rep{1})$  \\
                                    & $40$ & $(\rep{1},\rep{1})$   & $80$ & $(\rep{1},\rep{1})$   & -           & $126$ & $(\rep{1},\rep{1})$  \\
\hline
$c_2 = c_4= \frac{1}{2}$            & $ 8$ & $(\crep{27},\rep{1})$ & -    &                       & -           &  $11$ & $(\crep{27},\rep{1})\phantom{I^{I^I}}$ \\
                                    & $ 8$ & $(\rep{27},\rep{1})$  & -    &                       & -           &  $11$ & $(\rep{27},\rep{1})$  \\
                                    & $80$ & $(\rep{1},\rep{1})$   & -    &                       & -           &  $86$ & $(\rep{1},\rep{1})$  \\
\hline
$c_3=d_{26}=d_{45}=\frac{1}{2}$     & $ 2$ & $(\crep{27},\rep{1})$ & $ 4$ & $(\crep{27},\rep{1})$ & -           &  $ 9$ & $(\crep{27},\rep{1})\phantom{I^{I^I}}$ \\
                                    & $ 2$ & $(\rep{27},\rep{1})$  & $ 4$ & $(\rep{27},\rep{1})$  & -           &  $ 9$ & $(\rep{27},\rep{1})$  \\
                                    & $20$ & $(\rep{1},\rep{1})$   & $40$ & $(\rep{1},\rep{1})$   & -           &  $66$ & $(\rep{1},\rep{1})$  \\
\hline
$c_2=c_3=d_{46}=\frac{1}{2}$        & $ 4$ & $(\crep{27},\rep{1})$ & -    &                       & -           &  $ 7$ & $(\crep{27},\rep{1})\phantom{I^{I^I}}$ \\
                                    & $ 4$ & $(\rep{27},\rep{1})$  & -    &                       & -           &  $ 7$ & $(\rep{27},\rep{1})$  \\
                                    & $40$ & $(\rep{1},\rep{1})$   & -    &                       & -           &  $46$ & $(\rep{1},\rep{1})$  \\
\hline
$c_1=c_3=d_{26}=d_{45}=\frac{1}{2}$ & $ 2$ & $(\crep{27},\rep{1})$ & -    &                       & -           &  $ 5$ & $(\crep{27},\rep{1})\phantom{I^{I^I}}$ \\
                                    & $ 2$ & $(\rep{27},\rep{1})$  & -    &                       & -           &  $ 5$ & $(\rep{27},\rep{1})$  \\
                                    & $20$ & $(\rep{1},\rep{1})$   & -    &                       & -           &  $26$ & $(\rep{1},\rep{1})$  \\
\hline
$b_6 = c_2= \frac{1}{2}$            & -    &                       & -    &                       & -           &  $ 3$ & $(\crep{27},\rep{1})\phantom{I^{I^I}}$ \\
                                    & -    &                       & -    &                       & -           &  $ 3$ & $(\rep{27},\rep{1})$  \\
                                    & -    &                       & -    &                       & -           &  $ 6$ & $(\rep{1},\rep{1})$  \\
\hline
\end{tabular}
\caption{The table shows the charged matter spectra (omitting the $\U{1}$ charges) of $\Z{2}\times\Z{2}$ standard embedding models with generalized discrete torsion yielding a net number of zero families. The gauge group is $\E{6}\times\U{1}^2\times\E{8}$. The untwisted matter contributes with $3(\crep{27},\rep{1})$ + $3(\rep{27},\rep{1})$ + $6(\rep{1},\rep{1})$ to the charged matter spectrum. Interestingly, the last model with $b_6 = c_2= \tfrac{1}{2}$ has no massless twisted matter. Note that there are more models, which share the total number of charged states but differ in their localization, e.g. there is another model with 15 $(\rep{27},\rep{1})$ and 15 $(\crep{27},\rep{1})$ where each twisted sector contributes 4 $(\rep{27},\rep{1})$ and 4 $(\crep{27},\rep{1})$.}
\label{tab:Z2xZ2ExampleZERO}
\end{center}\index{orbifold!$\Z{2}\times\Z{2}$}
\end{table}

\begin{table}[!ht]
\begin{center}
\begin{tabular}{|l|r|r|r|}
\hline
\multicolumn{1}{|c|}{non-vanishing}            & \multicolumn{1}{c|}{total} & \multicolumn{1}{c|}{total}& \multicolumn{1}{c|}{total} \\
discrete torsion parameter$\phantom{I^{I^I}}$  & $\#(\crep{27},\rep{1})$ & $\#(\rep{27},\rep{1})$ & $\#$ singlets \\
\hline
\hline
-                                              & 61                      & 1                      & 252$\phantom{I^{I^I}}$           \\
$a=\frac{1}{2}$                                & 21                      & 9                      & 220$\phantom{I^{I^I}}$           \\
\hline
$d_{46}=\frac{1}{2}$                           & 51                      & 3                      & 212$\phantom{I^{I^I}}$           \\
$a=d_{46}=\frac{1}{2}$                         & 11                      & 11                     & 180$\phantom{I^{I^I}}$           \\
\hline
$d_{26}=\frac{1}{2}$                           & 39                      & 3                      & 184$\phantom{I^{I^I}}$           \\
$a=d_{26}=\frac{1}{2}$                         & 19                      & 7                      & 168$\phantom{I^{I^I}}$           \\
\hline
$d_{26}=d_{46}=\frac{1}{2}$                    & 37                      & 1                      & 164$\phantom{I^{I^I}}$           \\
$a=d_{26}=d_{46}=\frac{1}{2}$                  & 17                      & 5                      & 148$\phantom{I^{I^I}}$           \\
\hline
$d_{24}=d_{26}=d_{46}=\frac{1}{2}$             & 31                      & 7                      & 164$\phantom{I^{I^I}}$           \\
$a=d_{24}=d_{26}=d_{46}=\frac{1}{2}$           & 11                      & 11                     & 148$\phantom{I^{I^I}}$           \\
\hline
$d_{16}=d_{24}=\frac{1}{2}$                    & 27                      & 3                      & 140$\phantom{I^{I^I}}$           \\
$a=d_{16}=d_{24}=\frac{1}{2}$                  & 17                      & 5                      & 132$\phantom{I^{I^I}}$           \\
\hline
\hline
$c_2=\frac{1}{2}$                              & 37                      & 1                      & 196$\phantom{I^{I^I}}$           \\
$c_2=d_{46}=\frac{1}{2}$                       & 27                      & 3                      & 156$\phantom{I^{I^I}}$           \\
$c_2=d_{16}=d_{46}=\frac{1}{2}$                & 25                      & 1                      & 136$\phantom{I^{I^I}}$           \\
$c_2=d_{14}=d_{16}=d_{46}=\frac{1}{2}$         & 19                      & 7                      & 136$\phantom{I^{I^I}}$           \\
\hline
$b_6=\frac{1}{2}$                              & 25                      & 13                     & 196$\phantom{I^{I^I}}$           \\
$b_6=d_{46}=\frac{1}{2}$                       & 31                      & 7                      & 196$\phantom{I^{I^I}}$           \\
$b_6=d_{24}=d_{46}=\frac{1}{2}$                & 21                      & 9                      & 156$\phantom{I^{I^I}}$           \\
$b_6=c_2=d_{14}=\frac{1}{2}$                   & 17                      & 5                      & 116$\phantom{I^{I^I}}$           \\
\hline
\hline
$b_4=b_6=d_{46}=\frac{1}{2}$                   & 19                      & 19                     & 196$\phantom{I^{I^I}}$           \\
$b_4=b_6=c_2=d_{46}=\frac{1}{2}$               & 15                      & 15                     & 156$\phantom{I^{I^I}}$           \\
\hline
\end{tabular}
\caption{$\Z{2}\times\Z{4}$ orbifolds with twist vectors $v_1 = (0,\tfrac{1}{2},0,-\tfrac{1}{2})$ and $v_2 = (0,0,\tfrac{1}{4},-\tfrac{1}{4})$ on an $\SO{4}\times\SO{5}\times\SO{5}$ torus lattice. A summary of the charged matter spectra of standard embedding models with generalized discrete torsion is listed. The gauge group is $\E{6}\times\U{1}^2\times\E{8}$, the $\U{1}$ charges are omitted. Some of these spectra are known in the context of non-factorizable $\Z{2}\times\Z{4}$ orbifolds, compare to~\cite{Takahashi:2007qc}.}
\label{tab:Z2xZ4ExampleGDT}
\end{center}\index{orbifold!$\Z{2}\times\Z{4}$}
\end{table}


\phantom{oeff oeff}

{
\small
\setlength{\LTcapwidth}{14.5cm}
\begin{longtable}{|l<{\!\!\!\!\!\!\!}|c|>{\!\!\!}r>{\!\!\!}r>{\!\!\!}r|>{\!\!\!}r>{\!\!\!}r>{\!\!\!}r>{\!\!\!}r>{\!\!\!}r>{\!\!\!}r>{\!\!\!}r>{\!\!\!}r>{\!\!\!}r|l<{\!\!}|}
\caption{Charged matter spectrum of the benchmark model. The localization $n_6$ in $T_{(3)}$, $T_{(5)}$ takes values $n_6 = 0,1$ and therefore indicates a $D_4$ doublet.}
\\
\hline
sector\MyQSpc & rep. & $R_1$ & $R_2$ & $R_3$ & $Q_{Y}$ & $Q_{2}$ & $Q_{3}$ & $Q_{4}$ & $Q_{5}$ & $Q_{6}$ & $Q_{7}$ & $Q_{8}$ & $Q_{9}$ & label \\
\hline\hline
\endhead
\hline
\endfoot
\label{tab:benchmark1}
$U_1$ \MyQSpc & $(\rep{1}, \rep{1}, \rep{1}, \rep{1})$ & $ -1$ & $ 0$ & $ 0$ & $ 0$ & $-\frac{1}{2}$ & $\frac{1}{2}$ & $-\frac{1}{2}$ & $-\frac{5}{2}$ & $ 0$ & $ 0$ & $ 0$ & $ 0$ &  $\bar{n}_{3}$\\
\MyQSpc & $(\rep{1}, \rep{1}, \rep{1}, \rep{1})$ & $ -1$ & $ 0$ & $ 0$ & $ 1$ & $\frac{1}{2}$ & $-\frac{1}{2}$ & $\frac{1}{2}$ & $-\frac{1}{2}$ & $ 0$ & $ 0$ & $ 0$ & $ 0$ &  $\bar{e}_{3}$\\
\MyQSpc & $(\crep{3}, \rep{1}, \rep{1}, \rep{1})$ & $ -1$ & $ 0$ & $ 0$ & $-\frac{2}{3}$ & $\frac{1}{2}$ & $-\frac{1}{2}$ & $\frac{1}{2}$ & $-\frac{1}{2}$ & $ 0$ & $ 0$ & $ 0$ & $ 0$ &  $\bar{u}_{3}$\\
\MyQSpc & $(\rep{1}, \rep{1}, \crep{4}, \rep{1})$ & $ -1$ & $ 0$ & $ 0$ & $ 0$ & $ 0$ & $ 0$ & $ 0$ & $ 0$ & $\frac{1}{2}$ & $ -1$ & $-\frac{1}{2}$ & $-\frac{1}{2}$ &  $\bar{f}_{1}$\\
\MyQSpc & $(\rep{1}, \rep{1}, \rep{4}, \rep{1})$ & $ -1$ & $ 0$ & $ 0$ & $ 0$ & $ 0$ & $ 0$ & $ 0$ & $ 0$ & $\frac{1}{2}$ & $ 1$ & $-\frac{1}{2}$ & $\frac{1}{2}$ &  $f_{1}$\\
\cline{2-15}
$U_2$ \MyQSpc & $(\rep{1}, \rep{1}, \rep{1}, \rep{1})$ & $ 0$ & $ -1$ & $ 0$ & $ 0$ & $ 0$ & $ 0$ & $ 0$ & $ 0$ & $ 1$ & $ 0$ & $ 1$ & $ 0$ &  $s^0_{2}$\\
\MyQSpc & $(\rep{1}, \rep{1}, \rep{1}, \rep{1})$ & $ 0$ & $ -1$ & $ 0$ & $ 0$ & $ 0$ & $ 0$ & $ 0$ & $ 0$ & $ 1$ & $ 0$ & $ -1$ & $ 0$ &  $s^0_{1}$\\
\MyQSpc & $(\rep{3}, \rep{2}, \rep{1}, \rep{1})$ & $ 0$ & $ -1$ & $ 0$ & $\frac{1}{6}$ & $-\frac{1}{2}$ & $\frac{1}{2}$ & $\frac{1}{2}$ & $-\frac{1}{2}$ & $ 0$ & $ 0$ & $ 0$ & $ 0$ &  $q_{3}$\\
\cline{2-15}
$U_3$ \MyQSpc & $(\rep{1}, \rep{2}, \rep{1}, \rep{1})$ & $ 0$ & $ 0$ & $ -1$ & $-\frac{1}{2}$ & $ 0$ & $ 0$ & $ 1$ & $ -1$ & $ 0$ & $ 0$ & $ 0$ & $ 0$ &  $\phi_{1}$\\
\MyQSpc & $(\rep{1}, \rep{2}, \rep{1}, \rep{1})$ & $ 0$ & $ 0$ & $ -1$ & $\frac{1}{2}$ & $ 0$ & $ 0$ & $ -1$ & $ 1$ & $ 0$ & $ 0$ & $ 0$ & $ 0$ &  $\bar{\phi}_{1}$\\
\hline
\hline
$T_{2(0, 0, 0, 0, 0, 0)}$ \MyQSpc & $(\rep{1}, \rep{1}, \rep{1}, \rep{1})$ & $-\frac{2}{3}$ & $\frac{2}{3}$ & $ 0$ & $ 0$ & $-\frac{2}{3}$ & $ 0$ & $ 0$ & $ 0$ & $-\frac{2}{3}$ & $ 0$ & $ 0$ & $ 0$ &  $s^0_{26}$\\
\MyQSpc & $(\rep{3}, \rep{1}, \rep{1}, \rep{1})$ & $-\frac{2}{3}$ & $-\frac{1}{3}$ & $ 0$ & $-\frac{1}{3}$ & $\frac{1}{3}$ & $ 0$ & $ 0$ & $ 1$ & $-\frac{2}{3}$ & $ 0$ & $ 0$ & $ 0$ &  $\delta_{4}$\\
\MyQSpc & $(\rep{1}, \rep{1}, \rep{1}, \rep{2})$ & $-\frac{2}{3}$ & $-\frac{1}{3}$ & $ 0$ & $ 0$ & $-\frac{2}{3}$ & $ 0$ & $ 0$ & $ 0$ & $\frac{1}{3}$ & $ 1$ & $ 0$ & $ 0$ &  $h_{8}$\\
\MyQSpc & $(\crep{3}, \rep{1}, \rep{1}, \rep{1})$ & $-\frac{2}{3}$ & $-\frac{1}{3}$ & $ 0$ & $\frac{1}{3}$ & $\frac{1}{3}$ & $ 0$ & $ 0$ & $ -1$ & $-\frac{2}{3}$ & $ 0$ & $ 0$ & $ 0$ &  $\bar{\delta}_{4}$\\
\MyQSpc & $(\rep{1}, \rep{1}, \rep{1}, \rep{2})$ & $-\frac{2}{3}$ & $-\frac{1}{3}$ & $ 0$ & $ 0$ & $-\frac{2}{3}$ & $ 0$ & $ 0$ & $ 0$ & $\frac{1}{3}$ & $ -1$ & $ 0$ & $ 0$ &  $h_{7}$\\
\MyQSpc & $(\rep{1}, \rep{1}, \rep{1}, \rep{1})$ & $-\frac{2}{3}$ & $-\frac{1}{3}$ & $ 0$ & $ 0$ & $-\frac{2}{3}$ & $ 0$ & $ 0$ & $ 0$ & $\frac{1}{3}$ & $ 0$ & $ 1$ & $ 0$ &  $s^0_{25}$\\
\MyQSpc & $(\rep{1}, \rep{1}, \rep{1}, \rep{1})$ & $-\frac{2}{3}$ & $-\frac{1}{3}$ & $ 0$ & $ 0$ & $-\frac{2}{3}$ & $ 0$ & $ 0$ & $ 0$ & $\frac{1}{3}$ & $ 0$ & $ -1$ & $ 0$ &  $s^0_{24}$\\
$T_{2(0, 0, 1, 1, 0, 0)}$ \MyQSpc & $(\rep{1}, \rep{1}, \rep{1}, \rep{1})$ & $-\frac{2}{3}$ & $-\frac{1}{3}$ & $ 0$ & $ 0$ & $\frac{5}{6}$ & $-\frac{1}{2}$ & $\frac{1}{6}$ & $\frac{5}{6}$ & $\frac{1}{3}$ & $-\frac{2}{3}$ & $ 0$ & $-\frac{1}{3}$ &  $n_{12}$\\
\MyQSpc & $(\rep{1}, \rep{1}, \crep{4}, \rep{1})$ & $-\frac{2}{3}$ & $-\frac{1}{3}$ & $ 0$ & $ 0$ & $-\frac{1}{6}$ & $\frac{1}{2}$ & $\frac{1}{6}$ & $\frac{5}{6}$ & $-\frac{1}{6}$ & $\frac{1}{3}$ & $-\frac{1}{2}$ & $\frac{1}{6}$ &  $\bar{f}_{4}$\\
\MyQSpc & $(\crep{3}, \rep{1}, \rep{1}, \rep{1})$ & $-\frac{2}{3}$ & $-\frac{1}{3}$ & $ 0$ & $\frac{1}{3}$ & $-\frac{1}{6}$ & $-\frac{1}{2}$ & $\frac{1}{6}$ & $-\frac{1}{6}$ & $\frac{1}{3}$ & $-\frac{2}{3}$ & $ 0$ & $-\frac{1}{3}$ &  $\bar{d}_{3}$\\
$T_{2(0, 0, 0, 1, 0, 0)}$ \MyQSpc & $(\rep{1}, \rep{1}, \rep{1}, \rep{1})$ & $-\frac{2}{3}$ & $\frac{2}{3}$ & $ 0$ & $ 0$ & $-\frac{1}{6}$ & $\frac{1}{2}$ & $-\frac{1}{6}$ & $-\frac{5}{6}$ & $\frac{1}{3}$ & $\frac{2}{3}$ & $ 0$ & $\frac{1}{3}$ &  $\bar{n}_{12}$\\
\MyQSpc & $(\rep{1}, \rep{1}, \rep{1}, \rep{1})$ & $-\frac{2}{3}$ & $-\frac{1}{3}$ & $ 0$ & $ 0$ & $-\frac{1}{6}$ & $\frac{1}{2}$ & $-\frac{1}{6}$ & $-\frac{5}{6}$ & $\frac{1}{3}$ & $-\frac{4}{3}$ & $ 0$ & $\frac{1}{3}$ &  $\bar{n}_{11}$\\
\MyQSpc & $(\rep{1}, \rep{1}, \rep{1}, \rep{1})$ & $-\frac{2}{3}$ & $-\frac{1}{3}$ & $ 0$ & $ 0$ & $-\frac{1}{6}$ & $-\frac{1}{2}$ & $\frac{5}{6}$ & $-\frac{5}{6}$ & $\frac{1}{3}$ & $\frac{2}{3}$ & $ 0$ & $\frac{1}{3}$ &  $\bar{n}_{10}$\\
\MyQSpc & $(\rep{1}, \rep{1}, \rep{1}, \rep{2})$ & $-\frac{2}{3}$ & $-\frac{1}{3}$ & $ 0$ & $ 0$ & $-\frac{1}{6}$ & $\frac{1}{2}$ & $-\frac{1}{6}$ & $-\frac{5}{6}$ & $-\frac{2}{3}$ & $-\frac{1}{3}$ & $ 0$ & $\frac{1}{3}$ &  $\bar{\eta}_{4}$\\
$T_{2(1, 0, 0, 0, 0, 0)}$ \MyQSpc & $(\rep{1}, \rep{1}, \rep{1}, \rep{1})$ & $-\frac{5}{3}$ & $-\frac{1}{3}$ & $ 0$ & $ 0$ & $-\frac{2}{3}$ & $ 0$ & $ 0$ & $ 0$ & $-\frac{2}{3}$ & $ 0$ & $ 0$ & $ 0$ &  $s^0_{30}$\\
\MyQSpc & $(\rep{1}, \rep{1}, \rep{1}, \rep{1})$ & $-\frac{2}{3}$ & $\frac{2}{3}$ & $ 0$ & $ 0$ & $-\frac{2}{3}$ & $ 0$ & $ 0$ & $ 0$ & $-\frac{2}{3}$ & $ 0$ & $ 0$ & $ 0$ &  $s^0_{29}$\\
\MyQSpc & $(\rep{3}, \rep{1}, \rep{1}, \rep{1})$ & $-\frac{2}{3}$ & $-\frac{1}{3}$ & $ 0$ & $-\frac{1}{3}$ & $\frac{1}{3}$ & $ 0$ & $ 0$ & $ 1$ & $-\frac{2}{3}$ & $ 0$ & $ 0$ & $ 0$ &  $\delta_{5}$\\
\MyQSpc & $(\rep{1}, \rep{1}, \rep{1}, \rep{2})$ & $-\frac{2}{3}$ & $-\frac{1}{3}$ & $ 0$ & $ 0$ & $-\frac{2}{3}$ & $ 0$ & $ 0$ & $ 0$ & $\frac{1}{3}$ & $ 1$ & $ 0$ & $ 0$ &  $h_{10}$\\
\MyQSpc & $(\crep{3}, \rep{1}, \rep{1}, \rep{1})$ & $-\frac{2}{3}$ & $-\frac{1}{3}$ & $ 0$ & $\frac{1}{3}$ & $\frac{1}{3}$ & $ 0$ & $ 0$ & $ -1$ & $-\frac{2}{3}$ & $ 0$ & $ 0$ & $ 0$ &  $\bar{\delta}_{5}$\\
\MyQSpc & $(\rep{1}, \rep{1}, \rep{1}, \rep{2})$ & $-\frac{2}{3}$ & $-\frac{1}{3}$ & $ 0$ & $ 0$ & $-\frac{2}{3}$ & $ 0$ & $ 0$ & $ 0$ & $\frac{1}{3}$ & $ -1$ & $ 0$ & $ 0$ &  $h_{9}$\\
\MyQSpc & $(\rep{1}, \rep{1}, \rep{1}, \rep{1})$ & $-\frac{2}{3}$ & $-\frac{1}{3}$ & $ 0$ & $ 0$ & $-\frac{2}{3}$ & $ 0$ & $ 0$ & $ 0$ & $\frac{1}{3}$ & $ 0$ & $ 1$ & $ 0$ &  $s^0_{28}$\\
\MyQSpc & $(\rep{1}, \rep{1}, \rep{1}, \rep{1})$ & $-\frac{2}{3}$ & $-\frac{1}{3}$ & $ 0$ & $ 0$ & $-\frac{2}{3}$ & $ 0$ & $ 0$ & $ 0$ & $\frac{1}{3}$ & $ 0$ & $ -1$ & $ 0$ &  $s^0_{27}$\\
$T_{2(1, 0, 1, 1, 0, 0)}$ \MyQSpc & $(\rep{1}, \rep{1}, \rep{1}, \rep{1})$ & $-\frac{2}{3}$ & $-\frac{1}{3}$ & $ 0$ & $ 0$ & $\frac{5}{6}$ & $-\frac{1}{2}$ & $\frac{1}{6}$ & $\frac{5}{6}$ & $\frac{1}{3}$ & $-\frac{2}{3}$ & $ 0$ & $-\frac{1}{3}$ &  $n_{13}$\\
\MyQSpc & $(\rep{1}, \rep{1}, \crep{4}, \rep{1})$ & $-\frac{2}{3}$ & $-\frac{1}{3}$ & $ 0$ & $ 0$ & $-\frac{1}{6}$ & $\frac{1}{2}$ & $\frac{1}{6}$ & $\frac{5}{6}$ & $-\frac{1}{6}$ & $\frac{1}{3}$ & $-\frac{1}{2}$ & $\frac{1}{6}$ &  $\bar{f}_{5}$\\
\MyQSpc & $(\rep{3}, \rep{1}, \rep{1}, \rep{1})$ & $-\frac{2}{3}$ & $-\frac{1}{3}$ & $ 0$ & $-\frac{1}{3}$ & $\frac{1}{3}$ & $ 0$ & $-\frac{1}{3}$ & $-\frac{2}{3}$ & $\frac{1}{3}$ & $-\frac{2}{3}$ & $ 0$ & $-\frac{1}{3}$ &  $\delta_{6}$\\
\MyQSpc & $(\rep{1}, \rep{1}, \rep{1}, \rep{1})$ & $-\frac{2}{3}$ & $-\frac{1}{3}$ & $ 0$ & $ 0$ & $-\frac{1}{6}$ & $\frac{1}{2}$ & $\frac{1}{6}$ & $\frac{5}{6}$ & $-\frac{2}{3}$ & $-\frac{2}{3}$ & $ 0$ & $\frac{2}{3}$ &  $\bar{n}_{9}$\\
\MyQSpc & $(\rep{1}, \rep{1}, \rep{1}, \rep{2})$ & $-\frac{2}{3}$ & $-\frac{1}{3}$ & $ 0$ & $ 0$ & $-\frac{1}{6}$ & $\frac{1}{2}$ & $\frac{1}{6}$ & $\frac{5}{6}$ & $\frac{1}{3}$ & $\frac{1}{3}$ & $ 0$ & $\frac{2}{3}$ &  $\bar{\eta}_{3}$\\
\MyQSpc & $(\crep{3}, \rep{1}, \rep{1}, \rep{1})$ & $-\frac{2}{3}$ & $-\frac{1}{3}$ & $ 0$ & $\frac{1}{3}$ & $-\frac{1}{6}$ & $-\frac{1}{2}$ & $\frac{1}{6}$ & $-\frac{1}{6}$ & $\frac{1}{3}$ & $-\frac{2}{3}$ & $ 0$ & $-\frac{1}{3}$ &  $\bar{d}_{4}$\\
\MyQSpc & $(\rep{1}, \rep{1}, \rep{1}, \rep{1})$ & $-\frac{2}{3}$ & $-\frac{1}{3}$ & $ 0$ & $ 0$ & $-\frac{2}{3}$ & $ 0$ & $-\frac{1}{3}$ & $-\frac{5}{3}$ & $\frac{1}{3}$ & $-\frac{2}{3}$ & $ 0$ & $-\frac{1}{3}$ &  $s^0_{31}$\\
$T_{2(1, 0, 0, 1, 0, 0)}$ \MyQSpc & $(\rep{1}, \rep{1}, \rep{1}, \rep{1})$ & $-\frac{5}{3}$ & $-\frac{1}{3}$ & $ 0$ & $ 0$ & $-\frac{1}{6}$ & $\frac{1}{2}$ & $-\frac{1}{6}$ & $-\frac{5}{6}$ & $\frac{1}{3}$ & $\frac{2}{3}$ & $ 0$ & $\frac{1}{3}$ &  $\bar{n}_{16}$\\
\MyQSpc & $(\rep{1}, \rep{1}, \rep{1}, \rep{1})$ & $-\frac{2}{3}$ & $\frac{2}{3}$ & $ 0$ & $ 0$ & $-\frac{1}{6}$ & $\frac{1}{2}$ & $-\frac{1}{6}$ & $-\frac{5}{6}$ & $\frac{1}{3}$ & $\frac{2}{3}$ & $ 0$ & $\frac{1}{3}$ &  $\bar{n}_{15}$\\
\MyQSpc & $(\crep{3}, \rep{1}, \rep{1}, \rep{1})$ & $-\frac{2}{3}$ & $-\frac{1}{3}$ & $ 0$ & $\frac{1}{3}$ & $\frac{1}{3}$ & $ 0$ & $\frac{1}{3}$ & $\frac{2}{3}$ & $\frac{1}{3}$ & $\frac{2}{3}$ & $ 0$ & $\frac{1}{3}$ &  $\bar{\delta}_{6}$\\
\MyQSpc & $(\rep{1}, \rep{1}, \rep{1}, \rep{1})$ & $-\frac{2}{3}$ & $-\frac{1}{3}$ & $ 0$ & $ 0$ & $-\frac{1}{6}$ & $\frac{1}{2}$ & $-\frac{1}{6}$ & $-\frac{5}{6}$ & $\frac{1}{3}$ & $-\frac{4}{3}$ & $ 0$ & $\frac{1}{3}$ &  $\bar{n}_{14}$\\
\MyQSpc & $(\rep{1}, \rep{1}, \rep{1}, \rep{1})$ & $-\frac{2}{3}$ & $-\frac{1}{3}$ & $ 0$ & $ 0$ & $-\frac{1}{6}$ & $-\frac{1}{2}$ & $\frac{5}{6}$ & $-\frac{5}{6}$ & $\frac{1}{3}$ & $\frac{2}{3}$ & $ 0$ & $\frac{1}{3}$ &  $\bar{n}_{13}$\\
\MyQSpc & $(\rep{1}, \rep{1}, \rep{1}, \rep{2})$ & $-\frac{2}{3}$ & $-\frac{1}{3}$ & $ 0$ & $ 0$ & $-\frac{1}{6}$ & $\frac{1}{2}$ & $-\frac{1}{6}$ & $-\frac{5}{6}$ & $-\frac{2}{3}$ & $-\frac{1}{3}$ & $ 0$ & $\frac{1}{3}$ &  $\bar{\eta}_{5}$\\
\MyQSpc & $(\rep{1}, \rep{1}, \crep{4}, \rep{1})$ & $-\frac{2}{3}$ & $-\frac{1}{3}$ & $ 0$ & $ 0$ & $-\frac{1}{6}$ & $\frac{1}{2}$ & $-\frac{1}{6}$ & $-\frac{5}{6}$ & $-\frac{1}{6}$ & $-\frac{1}{3}$ & $\frac{1}{2}$ & $-\frac{1}{6}$ &  $\bar{f}_{6}$\\
\MyQSpc & $(\rep{1}, \rep{1}, \rep{1}, \rep{1})$ & $-\frac{2}{3}$ & $-\frac{1}{3}$ & $ 0$ & $ 0$ & $-\frac{2}{3}$ & $ 0$ & $\frac{1}{3}$ & $\frac{5}{3}$ & $\frac{1}{3}$ & $\frac{2}{3}$ & $ 0$ & $\frac{1}{3}$ &  $s^0_{32}$\\
\MyQSpc & $(\rep{1}, \rep{2}, \rep{1}, \rep{1})$ & $-\frac{2}{3}$ & $-\frac{1}{3}$ & $ 0$ & $\frac{1}{2}$ & $-\frac{1}{6}$ & $-\frac{1}{2}$ & $-\frac{1}{6}$ & $\frac{1}{6}$ & $\frac{1}{3}$ & $\frac{2}{3}$ & $ 0$ & $\frac{1}{3}$ &  $\bar{\ell}_{1}$\\
\hline
\hline
$T_{3(0, 0, 0, 0, 1, n_6)}$ \MyQSpc & $(\rep{1}, \rep{1}, \rep{1}, \rep{1})$ & $-\frac{1}{2}$ & $ 0$ & $-\frac{1}{2}$ & $\frac{1}{2}$ & $ 0$ & $ 0$ & $ 0$ & $ 1$ & $-\frac{1}{2}$ & $ -1$ & $-\frac{1}{2}$ & $ 0$ &  $s^+_{9}$, $s^+_{12}$\\
\MyQSpc & $(\rep{1}, \rep{1}, \rep{1}, \rep{1})$ & $-\frac{1}{2}$ & $ 0$ & $-\frac{1}{2}$ & $-\frac{1}{2}$ & $ 0$ & $ 0$ & $ 0$ & $ -1$ & $-\frac{1}{2}$ & $ 1$ & $-\frac{1}{2}$ & $ 0$ &  $s^-_{9}$, $s^-_{12}$\\
$T_{3(0, 1, 0, 0, 0, n_6)}$ \MyQSpc & $(\rep{1}, \rep{1}, \rep{1}, \rep{2})$ & $-\frac{1}{2}$ & $ 0$ & $-\frac{1}{2}$ & $ 0$ & $ 0$ & $\frac{1}{2}$ & $-\frac{1}{2}$ & $ 0$ & $ 0$ & $ -1$ & $ 0$ & $ 0$ &  $h_{4}$,  $h_{6}$\\
\MyQSpc & $(\rep{1}, \rep{1}, \rep{1}, \rep{2})$ & $-\frac{1}{2}$ & $ 0$ & $-\frac{1}{2}$ & $ 0$ & $ 0$ & $-\frac{1}{2}$ & $\frac{1}{2}$ & $ 0$ & $ 0$ & $ 1$ & $ 0$ & $ 0$ &  $h_{3}$, $h_{5}$\\
\MyQSpc & $(\rep{1}, \rep{1}, \rep{1}, \rep{1})$ & $-\frac{1}{2}$ & $ 0$ & $-\frac{1}{2}$ & $ 0$ & $ 0$ & $\frac{1}{2}$ & $-\frac{1}{2}$ & $ 0$ & $ 0$ & $ 0$ & $ 0$ & $ -1$ &  $\chi_{2}$, $\chi_{4}$\\
\MyQSpc & $(\rep{1}, \rep{1}, \rep{1}, \rep{1})$ & $-\frac{1}{2}$ & $ 0$ & $-\frac{1}{2}$ & $ 0$ & $ 0$ & $-\frac{1}{2}$ & $\frac{1}{2}$ & $ 0$ & $ 0$ & $ 0$ & $ 0$ & $ 1$ &  $\chi_{1}$, $\chi_{3}$\\
$T_{3(0, 1, 0, 0, 1, n_6)}$ \MyQSpc & $(\rep{1}, \rep{1}, \rep{1}, \rep{1})$ & $-\frac{1}{2}$ & $ 0$ & $-\frac{1}{2}$ & $\frac{1}{2}$ & $ 0$ & $ 0$ & $ 0$ & $ 1$ & $\frac{1}{2}$ & $ -1$ & $\frac{1}{2}$ & $ 0$ &  $s^+_{11}$,  $s^+_{14}$\\
\MyQSpc & $(\rep{1}, \rep{1}, \rep{1}, \rep{1})$ & $-\frac{1}{2}$ & $ 0$ & $-\frac{1}{2}$ & $-\frac{1}{2}$ & $ 0$ & $ 0$ & $ 0$ & $ -1$ & $\frac{1}{2}$ & $ 1$ & $\frac{1}{2}$ & $ 0$ &  $s^-_{11}$,  $s^-_{14}$\\
\MyQSpc & $(\rep{1}, \rep{1}, \rep{1}, \rep{1})$ & $-\frac{1}{2}$ & $ 0$ & $-\frac{1}{2}$ & $\frac{1}{2}$ & $ 0$ & $ 0$ & $ 0$ & $ 1$ & $-\frac{1}{2}$ & $ -1$ & $-\frac{1}{2}$ & $ 0$ &  $s^+_{10}$, $s^+_{13}$\\
\MyQSpc & $(\rep{1}, \rep{1}, \rep{1}, \rep{1})$ & $-\frac{1}{2}$ & $ 0$ & $-\frac{1}{2}$ & $-\frac{1}{2}$ & $ 0$ & $ 0$ & $ 0$ & $ -1$ & $-\frac{1}{2}$ & $ 1$ & $-\frac{1}{2}$ & $ 0$ &  $s^-_{10}$, $s^-_{13}$\\
\MyQSpc & $(\rep{1}, \rep{1}, \crep{4}, \rep{1})$ & $-\frac{1}{2}$ & $ 0$ & $-\frac{1}{2}$ & $\frac{1}{2}$ & $ 0$ & $ 0$ & $ 0$ & $ 1$ & $ 0$ & $ 0$ & $ 0$ & $\frac{1}{2}$ &  $\bar{f}^+_{1}$, $\bar{f}^+_{2}$\\
\MyQSpc & $(\rep{1}, \rep{1}, \rep{4}, \rep{1})$ & $-\frac{1}{2}$ & $ 0$ & $-\frac{1}{2}$ & $-\frac{1}{2}$ & $ 0$ & $ 0$ & $ 0$ & $ -1$ & $ 0$ & $ 0$ & $ 0$ & $-\frac{1}{2}$ &  $f^-_{1}$, $f^-_{2}$\\
\hline
\hline
$T_{4(0, 0, 0, 0, 0, 0)}$ \MyQSpc & $(\rep{1}, \rep{1}, \rep{1}, \rep{1})$ & $\frac{2}{3}$ & $-\frac{2}{3}$ & $ 0$ & $ 0$ & $\frac{2}{3}$ & $ 0$ & $ 0$ & $ 0$ & $\frac{2}{3}$ & $ 0$ & $ 0$ & $ 0$ &  $s^0_{15}$\\
$T_{4(0, 0, 1, 0, 0, 0)}$ \MyQSpc & $(\rep{1}, \rep{1}, \rep{1}, \rep{1})$ & $-\frac{1}{3}$ & $-\frac{2}{3}$ & $ 0$ & $ 0$ & $\frac{2}{3}$ & $ 0$ & $\frac{1}{3}$ & $\frac{5}{3}$ & $-\frac{1}{3}$ & $\frac{2}{3}$ & $ 0$ & $\frac{1}{3}$ &  $s^0_{20}$\\
\MyQSpc & $(\rep{1}, \rep{1}, \rep{1}, \rep{2})$ & $-\frac{1}{3}$ & $-\frac{2}{3}$ & $ 0$ & $ 0$ & $\frac{1}{6}$ & $-\frac{1}{2}$ & $-\frac{1}{6}$ & $-\frac{5}{6}$ & $-\frac{1}{3}$ & $-\frac{1}{3}$ & $ 0$ & $-\frac{2}{3}$ &  $\eta_{3}$\\
\MyQSpc & $(\rep{1}, \rep{1}, \rep{1}, \rep{1})$ & $-\frac{1}{3}$ & $-\frac{2}{3}$ & $ 0$ & $ 0$ & $\frac{1}{6}$ & $-\frac{1}{2}$ & $-\frac{1}{6}$ & $-\frac{5}{6}$ & $\frac{2}{3}$ & $\frac{2}{3}$ & $ 0$ & $-\frac{2}{3}$ &  $n_{5}$\\
\MyQSpc & $(\crep{3}, \rep{1}, \rep{1}, \rep{1})$ & $-\frac{1}{3}$ & $-\frac{2}{3}$ & $ 0$ & $\frac{1}{3}$ & $-\frac{1}{3}$ & $ 0$ & $\frac{1}{3}$ & $\frac{2}{3}$ & $-\frac{1}{3}$ & $\frac{2}{3}$ & $ 0$ & $\frac{1}{3}$ &  $\bar{\delta}_{2}$\\
$T_{4(0, 0, 1, 1, 0, 0)}$ \MyQSpc & $(\rep{1}, \rep{1}, \rep{1}, \rep{1})$ & $-\frac{1}{3}$ & $-\frac{2}{3}$ & $ 0$ & $ 0$ & $\frac{2}{3}$ & $ 0$ & $-\frac{1}{3}$ & $-\frac{5}{3}$ & $-\frac{1}{3}$ & $-\frac{2}{3}$ & $ 0$ & $-\frac{1}{3}$ &  $s^0_{22}$\\
\MyQSpc & $(\rep{1}, \rep{1}, \rep{4}, \rep{1})$ & $-\frac{1}{3}$ & $-\frac{2}{3}$ & $ 0$ & $ 0$ & $\frac{1}{6}$ & $-\frac{1}{2}$ & $\frac{1}{6}$ & $\frac{5}{6}$ & $\frac{1}{6}$ & $\frac{1}{3}$ & $-\frac{1}{2}$ & $\frac{1}{6}$ &  $f_{5}$\\
\MyQSpc & $(\rep{1}, \rep{1}, \rep{1}, \rep{1})$ & $\frac{2}{3}$ & $-\frac{2}{3}$ & $ 0$ & $ 0$ & $\frac{1}{6}$ & $-\frac{1}{2}$ & $\frac{1}{6}$ & $\frac{5}{6}$ & $-\frac{1}{3}$ & $-\frac{2}{3}$ & $ 0$ & $-\frac{1}{3}$ &  $n_{7}$\\
\MyQSpc & $(\rep{1}, \rep{2}, \rep{1}, \rep{1})$ & $-\frac{1}{3}$ & $-\frac{2}{3}$ & $ 0$ & $-\frac{1}{2}$ & $\frac{1}{6}$ & $\frac{1}{2}$ & $\frac{1}{6}$ & $-\frac{1}{6}$ & $-\frac{1}{3}$ & $-\frac{2}{3}$ & $ 0$ & $-\frac{1}{3}$ &  $\ell_{3}$\\
\MyQSpc & $(\rep{3}, \rep{1}, \rep{1}, \rep{1})$ & $-\frac{1}{3}$ & $-\frac{2}{3}$ & $ 0$ & $-\frac{1}{3}$ & $-\frac{1}{3}$ & $ 0$ & $-\frac{1}{3}$ & $-\frac{2}{3}$ & $-\frac{1}{3}$ & $-\frac{2}{3}$ & $ 0$ & $-\frac{1}{3}$ &  $\delta_{2}$\\
$T_{4(-1, 1, 0, 0, 0, 0)}$ \MyQSpc & $(\rep{1}, \rep{1}, \rep{1}, \rep{2})$ & $-\frac{1}{3}$ & $-\frac{2}{3}$ & $ 0$ & $ 0$ & $\frac{2}{3}$ & $ 0$ & $ 0$ & $ 0$ & $-\frac{1}{3}$ & $ -1$ & $ 0$ & $ 0$ &  $h_{1}$\\
\MyQSpc & $(\crep{3}, \rep{1}, \rep{1}, \rep{1})$ & $-\frac{1}{3}$ & $-\frac{2}{3}$ & $ 0$ & $\frac{1}{3}$ & $-\frac{1}{3}$ & $ 0$ & $ 0$ & $ -1$ & $\frac{2}{3}$ & $ 0$ & $ 0$ & $ 0$ &  $\bar{\delta}_{1}$\\
\MyQSpc & $(\rep{1}, \rep{1}, \rep{1}, \rep{1})$ & $\frac{2}{3}$ & $-\frac{2}{3}$ & $ 0$ & $ 0$ & $\frac{2}{3}$ & $ 0$ & $ 0$ & $ 0$ & $\frac{2}{3}$ & $ 0$ & $ 0$ & $ 0$ &  $s^0_{16}$\\
\MyQSpc & $(\rep{1}, \rep{1}, \rep{1}, \rep{1})$ & $-\frac{1}{3}$ & $-\frac{5}{3}$ & $ 0$ & $ 0$ & $\frac{2}{3}$ & $ 0$ & $ 0$ & $ 0$ & $\frac{2}{3}$ & $ 0$ & $ 0$ & $ 0$ &  $s^0_{19}$\\
\MyQSpc & $(\rep{1}, \rep{1}, \rep{1}, \rep{2})$ & $-\frac{1}{3}$ & $-\frac{2}{3}$ & $ 0$ & $ 0$ & $\frac{2}{3}$ & $ 0$ & $ 0$ & $ 0$ & $-\frac{1}{3}$ & $ 1$ & $ 0$ & $ 0$ &  $h_{2}$\\
\MyQSpc & $(\rep{3}, \rep{1}, \rep{1}, \rep{1})$ & $-\frac{1}{3}$ & $-\frac{2}{3}$ & $ 0$ & $-\frac{1}{3}$ & $-\frac{1}{3}$ & $ 0$ & $ 0$ & $ 1$ & $\frac{2}{3}$ & $ 0$ & $ 0$ & $ 0$ &  $\delta_{1}$\\
\MyQSpc & $(\rep{1}, \rep{1}, \rep{1}, \rep{1})$ & $-\frac{1}{3}$ & $-\frac{2}{3}$ & $ 0$ & $ 0$ & $\frac{2}{3}$ & $ 0$ & $ 0$ & $ 0$ & $-\frac{1}{3}$ & $ 0$ & $ 1$ & $ 0$ &  $s^0_{18}$\\
\MyQSpc & $(\rep{1}, \rep{1}, \rep{1}, \rep{1})$ & $-\frac{1}{3}$ & $-\frac{2}{3}$ & $ 0$ & $ 0$ & $\frac{2}{3}$ & $ 0$ & $ 0$ & $ 0$ & $-\frac{1}{3}$ & $ 0$ & $ -1$ & $ 0$ &  $s^0_{17}$\\
$T_{4(-1, 1, 1, 0, 0, 0)}$ \MyQSpc & $(\rep{1}, \rep{1}, \rep{1}, \rep{1})$ & $-\frac{1}{3}$ & $-\frac{2}{3}$ & $ 0$ & $ 0$ & $\frac{2}{3}$ & $ 0$ & $\frac{1}{3}$ & $\frac{5}{3}$ & $-\frac{1}{3}$ & $\frac{2}{3}$ & $ 0$ & $\frac{1}{3}$ &  $s^0_{21}$\\
\MyQSpc & $(\rep{3}, \rep{1}, \rep{1}, \rep{1})$ & $-\frac{1}{3}$ & $-\frac{2}{3}$ & $ 0$ & $-\frac{1}{3}$ & $\frac{1}{6}$ & $\frac{1}{2}$ & $-\frac{1}{6}$ & $\frac{1}{6}$ & $-\frac{1}{3}$ & $\frac{2}{3}$ & $ 0$ & $\frac{1}{3}$ &  $d_{1}$\\
\MyQSpc & $(\rep{1}, \rep{1}, \rep{1}, \rep{2})$ & $-\frac{1}{3}$ & $-\frac{2}{3}$ & $ 0$ & $ 0$ & $\frac{1}{6}$ & $-\frac{1}{2}$ & $-\frac{1}{6}$ & $-\frac{5}{6}$ & $-\frac{1}{3}$ & $-\frac{1}{3}$ & $ 0$ & $-\frac{2}{3}$ &  $\eta_{4}$\\
\MyQSpc & $(\rep{1}, \rep{1}, \rep{4}, \rep{1})$ & $-\frac{1}{3}$ & $-\frac{2}{3}$ & $ 0$ & $ 0$ & $\frac{1}{6}$ & $-\frac{1}{2}$ & $-\frac{1}{6}$ & $-\frac{5}{6}$ & $\frac{1}{6}$ & $-\frac{1}{3}$ & $\frac{1}{2}$ & $-\frac{1}{6}$ &  $f_{4}$\\
\MyQSpc & $(\rep{1}, \rep{1}, \rep{1}, \rep{1})$ & $-\frac{1}{3}$ & $-\frac{2}{3}$ & $ 0$ & $ 0$ & $-\frac{5}{6}$ & $\frac{1}{2}$ & $-\frac{1}{6}$ & $-\frac{5}{6}$ & $-\frac{1}{3}$ & $\frac{2}{3}$ & $ 0$ & $\frac{1}{3}$ &  $\bar{n}_{8}$\\
\MyQSpc & $(\rep{1}, \rep{1}, \rep{1}, \rep{1})$ & $-\frac{1}{3}$ & $-\frac{2}{3}$ & $ 0$ & $ 0$ & $\frac{1}{6}$ & $-\frac{1}{2}$ & $-\frac{1}{6}$ & $-\frac{5}{6}$ & $\frac{2}{3}$ & $\frac{2}{3}$ & $ 0$ & $-\frac{2}{3}$ &  $n_{6}$\\
\MyQSpc & $(\crep{3}, \rep{1}, \rep{1}, \rep{1})$ & $-\frac{1}{3}$ & $-\frac{2}{3}$ & $ 0$ & $\frac{1}{3}$ & $-\frac{1}{3}$ & $ 0$ & $\frac{1}{3}$ & $\frac{2}{3}$ & $-\frac{1}{3}$ & $\frac{2}{3}$ & $ 0$ & $\frac{1}{3}$ &  $\bar{\delta}_{3}$\\
$T_{4(-1, 1, 1, 1, 0, 0)}$ \MyQSpc & $(\rep{1}, \rep{1}, \rep{1}, \rep{1})$ & $-\frac{1}{3}$ & $-\frac{2}{3}$ & $ 0$ & $ 0$ & $\frac{2}{3}$ & $ 0$ & $-\frac{1}{3}$ & $-\frac{5}{3}$ & $-\frac{1}{3}$ & $-\frac{2}{3}$ & $ 0$ & $-\frac{1}{3}$ &  $s^0_{23}$\\
\MyQSpc & $(\rep{1}, \rep{1}, \rep{4}, \rep{1})$ & $-\frac{1}{3}$ & $-\frac{2}{3}$ & $ 0$ & $ 0$ & $\frac{1}{6}$ & $-\frac{1}{2}$ & $\frac{1}{6}$ & $\frac{5}{6}$ & $\frac{1}{6}$ & $\frac{1}{3}$ & $-\frac{1}{2}$ & $\frac{1}{6}$ &  $f_{6}$\\
\MyQSpc & $(\rep{1}, \rep{1}, \rep{1}, \rep{1})$ & $\frac{2}{3}$ & $-\frac{2}{3}$ & $ 0$ & $ 0$ & $\frac{1}{6}$ & $-\frac{1}{2}$ & $\frac{1}{6}$ & $\frac{5}{6}$ & $-\frac{1}{3}$ & $-\frac{2}{3}$ & $ 0$ & $-\frac{1}{3}$ &  $n_{8}$\\
\MyQSpc & $(\rep{1}, \rep{1}, \rep{1}, \rep{1})$ & $-\frac{1}{3}$ & $-\frac{5}{3}$ & $ 0$ & $ 0$ & $\frac{1}{6}$ & $-\frac{1}{2}$ & $\frac{1}{6}$ & $\frac{5}{6}$ & $-\frac{1}{3}$ & $-\frac{2}{3}$ & $ 0$ & $-\frac{1}{3}$ &  $n_{11}$\\
\MyQSpc & $(\rep{1}, \rep{2}, \rep{1}, \rep{1})$ & $-\frac{1}{3}$ & $-\frac{2}{3}$ & $ 0$ & $-\frac{1}{2}$ & $\frac{1}{6}$ & $\frac{1}{2}$ & $\frac{1}{6}$ & $-\frac{1}{6}$ & $-\frac{1}{3}$ & $-\frac{2}{3}$ & $ 0$ & $-\frac{1}{3}$ &  $\ell_{4}$\\
\MyQSpc & $(\rep{1}, \rep{1}, \rep{1}, \rep{1})$ & $-\frac{1}{3}$ & $-\frac{2}{3}$ & $ 0$ & $ 0$ & $\frac{1}{6}$ & $\frac{1}{2}$ & $-\frac{5}{6}$ & $\frac{5}{6}$ & $-\frac{1}{3}$ & $-\frac{2}{3}$ & $ 0$ & $-\frac{1}{3}$ &  $n_{10}$\\
\MyQSpc & $(\rep{1}, \rep{1}, \rep{1}, \rep{1})$ & $-\frac{1}{3}$ & $-\frac{2}{3}$ & $ 0$ & $ 0$ & $\frac{1}{6}$ & $-\frac{1}{2}$ & $\frac{1}{6}$ & $\frac{5}{6}$ & $-\frac{1}{3}$ & $\frac{4}{3}$ & $ 0$ & $-\frac{1}{3}$ &  $n_{9}$\\
\MyQSpc & $(\rep{1}, \rep{1}, \rep{1}, \rep{2})$ & $-\frac{1}{3}$ & $-\frac{2}{3}$ & $ 0$ & $ 0$ & $\frac{1}{6}$ & $-\frac{1}{2}$ & $\frac{1}{6}$ & $\frac{5}{6}$ & $\frac{2}{3}$ & $\frac{1}{3}$ & $ 0$ & $-\frac{1}{3}$ &  $\eta_{5}$\\
\MyQSpc & $(\rep{3}, \rep{1}, \rep{1}, \rep{1})$ & $-\frac{1}{3}$ & $-\frac{2}{3}$ & $ 0$ & $-\frac{1}{3}$ & $-\frac{1}{3}$ & $ 0$ & $-\frac{1}{3}$ & $-\frac{2}{3}$ & $-\frac{1}{3}$ & $-\frac{2}{3}$ & $ 0$ & $-\frac{1}{3}$ &  $\delta_{3}$\\
\hline
\hline
$T_{5(0, 0, 0, 0, 0, n_6)}$ \MyQSpc & $(\crep{3}, \rep{1}, \rep{1}, \rep{1})$ & $-\frac{1}{6}$ & $-\frac{1}{3}$ & $-\frac{1}{2}$ & $\frac{1}{3}$ & $-\frac{1}{6}$ & $ 0$ & $ 0$ & $\frac{3}{2}$ & $\frac{1}{3}$ & $ 0$ & $ 0$ & $ 0$ &  $\bar{d}_{1}$, $\bar{d}_{2}$\\
\MyQSpc & $(\rep{1}, \rep{1}, \rep{1}, \rep{1})$ & $\frac{5}{6}$ & $-\frac{1}{3}$ & $-\frac{1}{2}$ & $ 0$ & $-\frac{2}{3}$ & $\frac{1}{2}$ & $-\frac{1}{2}$ & $ 0$ & $\frac{1}{3}$ & $ 0$ & $ 0$ & $ 0$ &  $s^0_{6}$,  $s^0_{12}$\\
\MyQSpc & $(\rep{1}, \rep{1}, \rep{1}, \rep{1})$ & $\frac{11}{6}$ & $-\frac{1}{3}$ & $-\frac{1}{2}$ & $ 0$ & $\frac{1}{3}$ & $\frac{1}{2}$ & $\frac{1}{2}$ & $ 0$ & $\frac{1}{3}$ & $ 0$ & $ 0$ & $ 0$ &  $s^0_{8}$,  $s^0_{14}$\\
\MyQSpc & $(\rep{1}, \rep{1}, \rep{1}, \rep{1})$ & $-\frac{1}{6}$ & $\frac{2}{3}$ & $-\frac{1}{2}$ & $ 0$ & $\frac{1}{3}$ & $\frac{1}{2}$ & $\frac{1}{2}$ & $ 0$ & $\frac{1}{3}$ & $ 0$ & $ 0$ & $ 0$ &  $s^0_{4}$, $s^0_{10}$\\
\MyQSpc & $(\rep{1}, \rep{1}, \rep{1}, \rep{1})$ & $-\frac{1}{6}$ & $-\frac{1}{3}$ & $-\frac{1}{2}$ & $ 1$ & $-\frac{1}{6}$ & $ 0$ & $ 0$ & $-\frac{1}{2}$ & $\frac{1}{3}$ & $ 0$ & $ 0$ & $ 0$ &  $\bar{e}_{1}$, $\bar{e}_{2}$\\
\MyQSpc & $(\crep{3}, \rep{1}, \rep{1}, \rep{1})$ & $-\frac{1}{6}$ & $-\frac{1}{3}$ & $-\frac{1}{2}$ & $-\frac{2}{3}$ & $-\frac{1}{6}$ & $ 0$ & $ 0$ & $-\frac{1}{2}$ & $\frac{1}{3}$ & $ 0$ & $ 0$ & $ 0$ &  $\bar{u}_{1}$, $\bar{u}_{2}$\\
\MyQSpc & $(\rep{1}, \rep{2}, \rep{1}, \rep{1})$ & $-\frac{1}{6}$ & $-\frac{1}{3}$ & $-\frac{1}{2}$ & $-\frac{1}{2}$ & $-\frac{1}{6}$ & $ 0$ & $ 0$ & $\frac{3}{2}$ & $\frac{1}{3}$ & $ 0$ & $ 0$ & $ 0$ &  $\ell_{1}$, $\ell_{2}$\\
\MyQSpc & $(\rep{3}, \rep{2}, \rep{1}, \rep{1})$ & $-\frac{1}{6}$ & $-\frac{1}{3}$ & $-\frac{1}{2}$ & $\frac{1}{6}$ & $-\frac{1}{6}$ & $ 0$ & $ 0$ & $-\frac{1}{2}$ & $\frac{1}{3}$ & $ 0$ & $ 0$ & $ 0$ &  $q_{1}$, $q_{2}$\\
\MyQSpc & $(\rep{1}, \rep{1}, \rep{1}, \rep{1})$ & $-\frac{1}{6}$ & $-\frac{1}{3}$ & $-\frac{1}{2}$ & $ 0$ & $-\frac{1}{6}$ & $ 0$ & $ 0$ & $-\frac{5}{2}$ & $\frac{1}{3}$ & $ 0$ & $ 0$ & $ 0$ &  $\bar{n}_{1}$, $\bar{n}_{2}$\\
\MyQSpc & $(\rep{1}, \rep{1}, \rep{1}, \rep{1})$ & $\frac{5}{6}$ & $-\frac{1}{3}$ & $-\frac{1}{2}$ & $ 0$ & $-\frac{2}{3}$ & $-\frac{1}{2}$ & $\frac{1}{2}$ & $ 0$ & $\frac{1}{3}$ & $ 0$ & $ 0$ & $ 0$ &  $s^0_{5}$, $s^0_{11}$\\
\MyQSpc & $(\rep{1}, \rep{1}, \rep{1}, \rep{1})$ & $\frac{11}{6}$ & $-\frac{1}{3}$ & $-\frac{1}{2}$ & $ 0$ & $\frac{1}{3}$ & $-\frac{1}{2}$ & $-\frac{1}{2}$ & $ 0$ & $\frac{1}{3}$ & $ 0$ & $ 0$ & $ 0$ &  $s^0_{7}$, $s^0_{13}$\\
\MyQSpc & $(\rep{1}, \rep{1}, \rep{1}, \rep{1})$ & $-\frac{1}{6}$ & $\frac{2}{3}$ & $-\frac{1}{2}$ & $ 0$ & $\frac{1}{3}$ & $-\frac{1}{2}$ & $-\frac{1}{2}$ & $ 0$ & $\frac{1}{3}$ & $ 0$ & $ 0$ & $ 0$ &  $s^0_{3}$, $s^0_{9}$\\
$T_{5(0, 0, 0, 0, 1, n_6)}$ \MyQSpc & $(\rep{1}, \rep{2}, \rep{1}, \rep{2})$ & $-\frac{1}{6}$ & $-\frac{1}{3}$ & $-\frac{1}{2}$ & $ 0$ & $\frac{1}{3}$ & $ 0$ & $ 0$ & $ 0$ & $-\frac{1}{6}$ & $ 0$ & $-\frac{1}{2}$ & $ 0$ &  $y_{1}$, $y_{2}$\\
\MyQSpc & $(\rep{1}, \rep{2}, \rep{1}, \rep{1})$ & $-\frac{1}{6}$ & $-\frac{1}{3}$ & $-\frac{1}{2}$ & $ 0$ & $\frac{1}{3}$ & $ 0$ & $ 0$ & $ 0$ & $-\frac{1}{6}$ & $ -1$ & $\frac{1}{2}$ & $ 0$ &  $m_{1}$, $m_{3}$\\
\MyQSpc & $(\rep{1}, \rep{2}, \rep{1}, \rep{1})$ & $-\frac{1}{6}$ & $-\frac{1}{3}$ & $-\frac{1}{2}$ & $ 0$ & $\frac{1}{3}$ & $ 0$ & $ 0$ & $ 0$ & $-\frac{1}{6}$ & $ 1$ & $\frac{1}{2}$ & $ 0$ &  $m_{2}$, $m_{4}$\\
$T_{5(0, 0, 1, 1, 0, n_6)}$ \MyQSpc & $(\rep{1}, \rep{1}, \rep{1}, \rep{2})$ & $-\frac{1}{6}$ & $-\frac{1}{3}$ & $-\frac{1}{2}$ & $ 0$ & $-\frac{1}{6}$ & $ 0$ & $\frac{2}{3}$ & $\frac{5}{6}$ & $\frac{1}{3}$ & $\frac{1}{3}$ & $ 0$ & $-\frac{1}{3}$ &  $\eta_{1}$, $\eta_{2}$\\
\MyQSpc & $(\rep{1}, \rep{1}, \rep{1}, \rep{1})$ & $-\frac{1}{6}$ & $-\frac{1}{3}$ & $-\frac{1}{2}$ & $ 0$ & $-\frac{1}{6}$ & $ 0$ & $\frac{2}{3}$ & $\frac{5}{6}$ & $\frac{1}{3}$ & $-\frac{2}{3}$ & $ 0$ & $\frac{2}{3}$ &  $\bar{n}_{4}$, $\bar{n}_{5}$\\
\MyQSpc & $(\rep{1}, \rep{1}, \rep{1}, \rep{1})$ & $-\frac{1}{6}$ & $-\frac{1}{3}$ & $-\frac{1}{2}$ & $ 0$ & $-\frac{1}{6}$ & $ 0$ & $\frac{2}{3}$ & $\frac{5}{6}$ & $-\frac{2}{3}$ & $-\frac{2}{3}$ & $ 0$ & $-\frac{1}{3}$ &  $n_{1}$, $n_{2}$\\
$T_{5(0, 0, 1, 1, 1, n_6)}$ \MyQSpc & $(\rep{1}, \rep{1}, \rep{1}, \rep{1})$ & $-\frac{1}{6}$ & $-\frac{1}{3}$ & $-\frac{1}{2}$ & $\frac{1}{2}$ & $\frac{1}{3}$ & $ 0$ & $\frac{2}{3}$ & $-\frac{2}{3}$ & $-\frac{1}{6}$ & $\frac{1}{3}$ & $-\frac{1}{2}$ & $-\frac{1}{3}$ &  $s^+_{2}$, $s^+_{4}$\\
\MyQSpc & $(\rep{1}, \rep{1}, \rep{1}, \rep{1})$ & $\frac{5}{6}$ & $-\frac{1}{3}$ & $-\frac{1}{2}$ & $-\frac{1}{2}$ & $-\frac{1}{6}$ & $\frac{1}{2}$ & $\frac{1}{6}$ & $-\frac{1}{6}$ & $-\frac{1}{6}$ & $\frac{1}{3}$ & $-\frac{1}{2}$ & $-\frac{1}{3}$ &  $s^-_{2}$,  $s^-_{4}$\\
\MyQSpc & $(\rep{3}, \rep{1}, \rep{1}, \rep{1})$ & $-\frac{1}{6}$ & $-\frac{1}{3}$ & $-\frac{1}{2}$ & $\frac{1}{6}$ & $\frac{1}{3}$ & $ 0$ & $-\frac{1}{3}$ & $\frac{1}{3}$ & $-\frac{1}{6}$ & $\frac{1}{3}$ & $-\frac{1}{2}$ & $-\frac{1}{3}$ &  $v_{1}$, $v_{2}$\\
\MyQSpc & $(\rep{1}, \rep{1}, \rep{1}, \rep{1})$ & $-\frac{1}{6}$ & $-\frac{1}{3}$ & $-\frac{1}{2}$ & $-\frac{1}{2}$ & $-\frac{1}{6}$ & $\frac{1}{2}$ & $\frac{1}{6}$ & $-\frac{1}{6}$ & $-\frac{1}{6}$ & $\frac{1}{3}$ & $\frac{1}{2}$ & $\frac{2}{3}$ &  $s^-_{1}$, $s^-_{3}$\\
\MyQSpc & $(\rep{1}, \rep{2}, \rep{1}, \rep{1})$ & $-\frac{1}{6}$ & $-\frac{1}{3}$ & $-\frac{1}{2}$ & $ 0$ & $-\frac{1}{6}$ & $-\frac{1}{2}$ & $\frac{1}{6}$ & $\frac{5}{6}$ & $-\frac{1}{6}$ & $\frac{1}{3}$ & $-\frac{1}{2}$ & $-\frac{1}{3}$ &  $m_{5}$, $m_{6}$\\
\MyQSpc & $(\rep{1}, \rep{1}, \rep{1}, \rep{1})$ & $-\frac{1}{6}$ & $-\frac{1}{3}$ & $-\frac{1}{2}$ & $\frac{1}{2}$ & $-\frac{2}{3}$ & $ 0$ & $-\frac{1}{3}$ & $-\frac{2}{3}$ & $-\frac{1}{6}$ & $\frac{1}{3}$ & $-\frac{1}{2}$ & $-\frac{1}{3}$ &  $s^+_{1}$, $s^+_{3}$\\
$T_{5(0, 0, 0, 1, 0, n_6)}$ \MyQSpc & $(\rep{1}, \rep{1}, \rep{4}, \rep{1})$ & $-\frac{1}{6}$ & $-\frac{1}{3}$ & $-\frac{1}{2}$ & $ 0$ & $-\frac{1}{6}$ & $ 0$ & $\frac{1}{3}$ & $-\frac{5}{6}$ & $-\frac{1}{6}$ & $-\frac{1}{3}$ & $\frac{1}{2}$ & $-\frac{1}{6}$ &  $f_{2}$, $f_{3}$\\
\MyQSpc & $(\rep{1}, \rep{1}, \rep{1}, \rep{2})$ & $\frac{5}{6}$ & $-\frac{1}{3}$ & $-\frac{1}{2}$ & $ 0$ & $-\frac{1}{6}$ & $ 0$ & $\frac{1}{3}$ & $-\frac{5}{6}$ & $\frac{1}{3}$ & $-\frac{1}{3}$ & $ 0$ & $\frac{1}{3}$ &  $\bar{\eta}_{1}$,  $\bar{\eta}_{2}$\\
\MyQSpc & $(\rep{1}, \rep{1}, \crep{4}, \rep{1})$ & $-\frac{1}{6}$ & $-\frac{1}{3}$ & $-\frac{1}{2}$ & $ 0$ & $-\frac{1}{6}$ & $ 0$ & $\frac{1}{3}$ & $-\frac{5}{6}$ & $-\frac{1}{6}$ & $-\frac{1}{3}$ & $-\frac{1}{2}$ & $-\frac{1}{6}$ &  $\bar{f}_{2}$, $\bar{f}_{3}$\\
\MyQSpc & $(\rep{1}, \rep{1}, \rep{1}, \rep{1})$ & $\frac{5}{6}$ & $-\frac{1}{3}$ & $-\frac{1}{2}$ & $ 0$ & $-\frac{1}{6}$ & $ 0$ & $\frac{1}{3}$ & $-\frac{5}{6}$ & $\frac{1}{3}$ & $\frac{2}{3}$ & $ 0$ & $-\frac{2}{3}$ &  $n_{3}$, $n_{4}$\\
\MyQSpc & $(\rep{1}, \rep{1}, \rep{1}, \rep{1})$ & $\frac{5}{6}$ & $-\frac{1}{3}$ & $-\frac{1}{2}$ & $ 0$ & $-\frac{1}{6}$ & $ 0$ & $\frac{1}{3}$ & $-\frac{5}{6}$ & $-\frac{2}{3}$ & $\frac{2}{3}$ & $ 0$ & $\frac{1}{3}$ &  $\bar{n}_{6}$, $\bar{n}_{7}$\\
$T_{5(0, 0, 0, 1, 1, n_6)}$ \MyQSpc & $(\crep{3}, \rep{1}, \rep{1}, \rep{1})$ & $-\frac{1}{6}$ & $-\frac{1}{3}$ & $-\frac{1}{2}$ & $-\frac{1}{6}$ & $\frac{1}{3}$ & $ 0$ & $\frac{1}{3}$ & $-\frac{1}{3}$ & $-\frac{1}{6}$ & $-\frac{1}{3}$ & $-\frac{1}{2}$ & $\frac{1}{3}$ &  $\bar{v}_{1}$, $\bar{v}_{2}$\\
\MyQSpc & $(\rep{1}, \rep{1}, \rep{1}, \rep{1})$ & $\frac{5}{6}$ & $-\frac{1}{3}$ & $-\frac{1}{2}$ & $\frac{1}{2}$ & $-\frac{1}{6}$ & $\frac{1}{2}$ & $-\frac{1}{6}$ & $\frac{1}{6}$ & $-\frac{1}{6}$ & $-\frac{1}{3}$ & $-\frac{1}{2}$ & $\frac{1}{3}$ &  $s^+_{6}$, $s^+_{8}$\\
\MyQSpc & $(\rep{1}, \rep{1}, \rep{1}, \rep{1})$ & $-\frac{1}{6}$ & $-\frac{1}{3}$ & $-\frac{1}{2}$ & $-\frac{1}{2}$ & $\frac{1}{3}$ & $ 0$ & $-\frac{2}{3}$ & $\frac{2}{3}$ & $-\frac{1}{6}$ & $-\frac{1}{3}$ & $-\frac{1}{2}$ & $\frac{1}{3}$ &  $s^-_{6}$, $s^-_{8}$\\
\MyQSpc & $(\rep{1}, \rep{1}, \rep{1}, \rep{1})$ & $-\frac{1}{6}$ & $-\frac{1}{3}$ & $-\frac{1}{2}$ & $\frac{1}{2}$ & $-\frac{1}{6}$ & $\frac{1}{2}$ & $-\frac{1}{6}$ & $\frac{1}{6}$ & $-\frac{1}{6}$ & $-\frac{1}{3}$ & $\frac{1}{2}$ & $-\frac{2}{3}$ &  $s^+_{5}$,  $s^+_{7}$\\
\MyQSpc & $(\rep{1}, \rep{2}, \rep{1}, \rep{1})$ & $-\frac{1}{6}$ & $-\frac{1}{3}$ & $-\frac{1}{2}$ & $ 0$ & $-\frac{1}{6}$ & $-\frac{1}{2}$ & $-\frac{1}{6}$ & $-\frac{5}{6}$ & $-\frac{1}{6}$ & $-\frac{1}{3}$ & $-\frac{1}{2}$ & $\frac{1}{3}$ &  $m_{7}$, $m_{8}$\\
\MyQSpc & $(\rep{1}, \rep{1}, \rep{1}, \rep{1})$ & $-\frac{1}{6}$ & $-\frac{1}{3}$ & $-\frac{1}{2}$ & $-\frac{1}{2}$ & $-\frac{2}{3}$ & $ 0$ & $\frac{1}{3}$ & $\frac{2}{3}$ & $-\frac{1}{6}$ & $-\frac{1}{3}$ & $-\frac{1}{2}$ & $\frac{1}{3}$ &  $s^-_{5}$, $s^-_{7}$\\
\end{longtable}
}

\chapter{Group Theory}
\label{app:group}

\section{Weight Lattices}
\label{app:weightlattices}
Given a N-dimensional vector space $V$ with basis vectors $e_i$ a lattice\index{lattice} $\Lambda$ is defined as the set of points
\begin{equation}
\Lambda = \{ \sum_{i=1}^N n_i e_i ,\; n_i \in \mathbbm{Z} \} \subset V\;.
\end{equation}
The dual lattice $\Lambda^*$ is defined by
\begin{equation}
\Lambda^* = \{ p \in V ,\; p \cdot p' \in \mathbbm{Z} \text{ for all } p' \in \Lambda\}\;.
\end{equation}
A lattice is called
\begin{itemize}
\item integral if $p \cdot p' \in \mathbbm{Z}$ for all $p, p' \in \Lambda$ .\index{lattice!integral}
\item even if all $p \in \Lambda$ fulfill $p^2 = \text{even}$ .\index{lattice!even}
\item self-dual if $\Lambda^* = \Lambda$ .\index{lattice!self-dual}
\end{itemize}
More details can be found for example on page 277ff of \cite{Becker:2007zj} and in \cite{Nilles:1987uy}.

\subsubsection{The $\text{E}_8 \times \text{E}_8$ Root Lattice}\index{lattice!$\text{E}_8 \times \text{E}_8$}

The root lattice of $\text{E}_8$ is spanned by the roots
\begin{equation}
\left(\underline{\pm 1, \pm 1, 0^6} \right) \text{ and }
\left(\pm\tfrac{1}{2}, \pm\tfrac{1}{2}, \pm\tfrac{1}{2}, \pm\tfrac{1}{2}, \pm\tfrac{1}{2}, \pm\tfrac{1}{2}, \pm\tfrac{1}{2}, \pm\tfrac{1}{2}\right) \;,
\end{equation}
with an even number of $+$ signs for the latter, the so-called spinorial roots. The root lattice of $\text{E}_8$ is self-dual $\Lambda^* = \Lambda$, i.e. root and weight lattice are identical. Note that the roots fulfill $p^2 = 2$ which is important for the massless spectrum of the heterotic string, see section~\ref{sec:HeteroticString}.

The root lattice of $\text{E}_8 \times \text{E}_8$ is given by the direct sum of two copies of the $\text{E}_8$ root lattice.

\subsubsection{The Weight Lattices $\text{Spin}(32)$ and $\text{Spin}(32)/ \mathbbm{Z}_2$}\index{lattice!$\text{Spin}(32)/ \mathbbm{Z}_2$}

The weight lattice of $\text{Spin}(32)$ can be decomposed into four conjugacy classes ($n_i \in \mathbbm{Z}$):
\begin{itemize}
\item The scalar conjugacy class:
\begin{equation}
\left( n_1, \ldots, n_{16}\right)\qquad \sum n_i = \text{even}
\end{equation}
Note that the roots of $\text{SO}(32)$ form a subset of the this conjugacy class. They are given by weights $p$ of the scalar class fulfilling $p^2 = 2$. Explicitly, these are 240 roots:
\begin{equation} 
\left( \underline{\pm 1, \pm 1, 0^{14}} \right)
\end{equation}

\item The vector conjugacy class:
\begin{equation}
\left( n_1, \ldots, n_{16}\right)\qquad \sum n_i = \text{odd}
\end{equation}
The vector conjugacy class contains the weights of the vector representation $\boldsymbol{32}$.

\item The spinor conjugacy class:
\begin{equation}
\left( n_1 + \tfrac{1}{2}, \ldots, n_{16} + \tfrac{1}{2}\right)\qquad \sum n_i = \text{even}
\end{equation}

\item The antispinor conjugacy class:
\begin{equation}
\left( n_1 + \tfrac{1}{2}, \ldots, n_{16} + \tfrac{1}{2}\right)\qquad \sum n_i = \text{odd}
\end{equation}

\end{itemize}
Here, the conjugacy classes are defined using the equivalence relation that two weights of $\text{Spin}(32)$ are equivalent if they differ by a root of $\text{SO}(32)$.

Note that the only weights of $\text{Spin}(32)$ that fulfill $p^2 = 2$ are the roots. This is important for the massless spectrum of the heterotic string, see section~\ref{sec:HeteroticString}. Finally, the weight lattice of $\text{Spin}(32)/ \mathbbm{Z}_2$ is chosen to be spanned by the scalar and the spinor conjugacy class.

\section[$\SO{8}$ Representations]{$\boldsymbol{\SO{8}}$ Representations}\index{$\SO{8}$}
\label{app:SO8}

We choose a basis such that the simple roots of $\SO{8}$ are
\begin{eqnarray}
\alpha_1 & = & \left(1,-1,0,0\right) \\
\alpha_2 & = & \left(0,1,-1,0\right) \\
\alpha_3 & = & \left(0,0,1,-1\right) \\
\alpha_4 & = & \left(0,0,1,1\right)\;.
\end{eqnarray}
There are three eight-dimensional representations of $\SO{8}$: $\rep{8}_v$, $\rep{8}_s$ and $\rep{8}_c$. They are given by the weights
\begin{eqnarray}
\rep{8}_v & : & \left(\underline{\pm 1, 0, 0, 0}\right) \\
\rep{8}_s & : & \left(\pm\frac{1}{2},\pm\frac{1}{2},\pm\frac{1}{2},\pm\frac{1}{2}\right) \quad \text{even \# of plus signs}\\
\rep{8}_c & : & \left(\pm\frac{1}{2},\pm\frac{1}{2},\pm\frac{1}{2},\pm\frac{1}{2}\right) \quad \text{odd \# of plus signs,}
\end{eqnarray}
such that the highest weights in Dynkin labels\footnote{The Dynkin labels of some weight $p$ are given by $[p\cdot\alpha_1,p\cdot\alpha_2,p\cdot\alpha_3,p\cdot\alpha_4]_\text{DL}$ (using $\alpha_i^2 = 2$).} are $[1, 0, 0, 0]_\text{DL}$, $[0, 0, 0, 1]_\text{DL}$ and $[0, 0, 1, 0]_\text{DL}$, respectively. As a remark, all three eight-dimensional representations of $\text{SO}(8)$ are real. 

\subsubsection{The Tensor Product $\boldsymbol{8_v \times 8_v}$}

The tensor product $\rep{8}_v \times \rep{8}_v$ can be computed by adding all 8 weights of the first $\rep{8}_v$ with the 8 weights of the second one. The $64$ weights are
\begin{eqnarray}
&          & \left(\underline{\pm 2,0,0,0}\right)\\
& 8\; \times & \left(0,0,0,0\right)\\
& 2\; \times & \left(\underline{\pm 1,\pm 1,0,0}\right)\;,
\end{eqnarray}
where the factor gives the multiplicity of the weights, e.g. $8\times\left(0,0,0,0\right)$ means that the zero weight appears eight times. In order to identify the irreducible $\SO{8}$ representations, we have to determine the highest weight of these $64$ weights. It is $[2,0,0,0]_\text{DL}$ and yields the representation $\rep{35}_v$. For the remaining $64 - 35 = 29$ weights, the highest weight can be identified as $[0,1,0,0]_\text{DL}$. This results in the representation $\rep{28}$. The remaining weight $[0,0,0,0]_\text{DL}$ gives clearly a singlet. Thus, the tensor product $\rep{8}_v \times \rep{8}_v$ decomposes as follows
\begin{equation}
\rep{8}_v \times \rep{8}_v =  \rep{1} + \rep{28} + \rep{35}_v\;.
\end{equation}

\subsubsection{The Breaking $\boldsymbol{\SO{8}\rightarrow \SU{4} \times \U{1}}$}

Next, we analyze the breaking
\begin{equation}
\SO{8} \rightarrow \SU{4} \times \U{1}
\end{equation}
where we choose the simple roots $\alpha_1$, $\alpha_2$, $\alpha_3$ of $\SU{4}$ and the $\U{1}$ generator $t$ to be
\begin{eqnarray}
\alpha_1 & = & \left(0,0,1,1\right) \\
\alpha_2 & = & \left(0,1,-1,0\right) \\
\alpha_3 & = & \left(0,0,1,-1\right) \\
t        & = & \left(1,0,0,0\right)\;.
\end{eqnarray}
Consequently, the eight-dimensional $\SO{8}$ representations branch into $\SU{4}\times\U{1}$ representations according to
\begin{equation}
\begin{array}{ccccc}
\left(\underline{\pm 1, 0, 0, 0}\right) & \rightarrow & \left(0,\underline{\pm 1, 0, 0}\right) & \text{and} & \left(\pm 1, 0, 0, 0\right) \\\vspace{0.3cm}
\rep{8}_v & \rightarrow & \rep{6}_0 & + & \rep{1}_1 + \rep{1}_{-1}\\
&&&&\\
\left(\pm\frac{1}{2},\pm\frac{1}{2},\pm\frac{1}{2},\pm\frac{1}{2}\right) & \rightarrow & \left(+\frac{1}{2},\pm\frac{1}{2},\pm\frac{1}{2},\pm\frac{1}{2}\right) & \text{and} & \left(-\frac{1}{2},\pm\frac{1}{2},\pm\frac{1}{2},\pm\frac{1}{2}\right) \\\vspace{0.3cm}
\rep{8}_s & \rightarrow & \rep{4}_{1/2} & + & \crep{4}_{-1/2}\\
&&&&\\
\left(\pm\frac{1}{2},\pm\frac{1}{2},\pm\frac{1}{2},\pm\frac{1}{2}\right) & \rightarrow & \left(+\frac{1}{2},\pm\frac{1}{2},\pm\frac{1}{2},\pm\frac{1}{2}\right) & \text{and} & \left(-\frac{1}{2},\pm\frac{1}{2},\pm\frac{1}{2},\pm\frac{1}{2}\right) \\\vspace{0.3cm}
\rep{8}_c & \rightarrow & \crep{4}_{1/2} & + & \rep{4}_{-1/2}\\
\end{array}
\end{equation}
with an even/odd number of plus signs for $\rep{8}_s$/$\rep{8}_c$. For more details on $\text{SO}(8)$ see e.g. page 282ff of~\cite{Green:1987sp},~\cite{Slansky:1981yr} or appendix B.1 of~\cite{Polchinski:1998rr}.


{
\small
\providecommand{\bysame}{\leavevmode\hbox to3em{\hrulefill}\thinspace}

}

\printindex
\addcontentsline{toc}{chapter}{\numberline{}{Index}}

\end{document}